%% file: main.tex
\documentclass[a4paper,11pt]{article}
\usepackage{jcappub} % for details on the use of the package, please see the JINST-author-manual
\usepackage{subfiles}
\usepackage{lineno}
\usepackage{dcolumn}% Align table columns on decimal point
\usepackage{verbatim}

\usepackage{breakurl}

\usepackage{lmodern}

\pdfoutput=1
\usepackage{graphicx}
\usepackage{bm}
\usepackage{amsmath}
\usepackage{amsfonts}
\usepackage{amssymb}
\usepackage{multirow, makecell}
\usepackage{adjustbox}
\usepackage[capitalise]{cleveref}
\usepackage{acro}
\usepackage[utf8]{inputenc}

\usepackage{longtable}
\usepackage{pdflscape}
\usepackage{subcaption}
\usepackage{caption}
\usepackage{tikz}
\usepackage{pgfplots}
\pgfplotsset{compat=newest,every axis plot/.append style={line width=1.1pt}}

\usepackage{mathrsfs}

\crefname{fig}{Figure}{Figures}
\crefname{sec}{Section}{Sections}
\crefname{subsec}{Subsection}{Subsections}

\newcommand{\Mpc}{{\rm Mpc}}
\newcommand{\be}{{\begin{eqnarray}}}
\newcommand{\ee}{{\end{eqnarray}}}

\newcommand{\overbar}[1]{\mkern 1.5mu\overline{\mkern-1.5mu#1\mkern-1.5mu}\mkern 1.5mu}

\newcommand{\mpl}{m_\mathrm{Pl}}
\newcommand{\Fnl}{F_\mathrm{NL}}
\newcommand{\fnl}{f_\mathrm{NL}}
\newcommand{\Gnl}{G_\mathrm{NL}}
\newcommand{\gnl}{g_\mathrm{NL}}
\newcommand{\Hnl}{H_\mathrm{NL}}
\newcommand{\hnl}{h_\mathrm{NL}}
\newcommand{\tnl}{\tau_\mathrm{NL}}

\newcommand{\abs}[1]{{\left \vert #1 \right \vert}}

\newcommand{\cT}{\mathcal{T}}
\newcommand{\cP}{\mathcal{P}}

\newcommand{\cH}{\mathcal{H}}
\newcommand{\cS}{\mathcal{S}}
\newcommand{\cO}{\mathcal{O}}
\newcommand{\cF}{\mathcal{F}}
\newcommand{\cD}{\mathcal{D}}
\newcommand{\cC}{\mathcal{C}}
\newcommand{\cG}{\mathcal{G}}
\newcommand{\cK}{\mathcal{K}}

\newcommand{\sT}{\mathscr{T}}
\newcommand{\bk}{\mathbf{k}}
\newcommand{\bK}{\mathbf{K}}
\newcommand{\bq}{\mathbf{q}}
\newcommand{\bp}{\mathbf{p}}
\newcommand{\bl}{\mathbf{l}}
\newcommand{\bx}{\mathbf{x}}
\newcommand{\br}{\mathbf{r}}
\newcommand{\bn}{\mathbf{n}}

\newcommand{\ud}{\mathrm{d}}
\newcommand{\ug}{\mathrm{g}}
\newcommand{\ugS}{\mathrm{gS}}
\newcommand{\ugL}{\mathrm{gL}}
\newcommand{\ugX}{\mathrm{gX}}
\newcommand{\uS}{\mathrm{S}}
\newcommand{\uL}{\mathrm{L}}
\newcommand{\uin}{\mathrm{in}}
\newcommand{\uRD}{\mathrm{RD}}

\newcommand{\uGW}{\mathrm{gw}}
\newcommand{\ung}{\mathrm{ng}}
\newcommand{\uc}{\mathrm{c}}
\newcommand{\uth}{\mathrm{th}}
\newcommand{\uR}{\mathrm{R}}
\newcommand{\uP}{\mathrm{P}}
\newcommand{\ca}{\alpha}
\newcommand{\cb}{\beta}
\newcommand{\cc}{\gamma}
\newcommand{\cd}{\delta}
\newcommand{\ce}{\epsilon}
\newcommand{\ci}{\iota}
\newcommand{\Beq}{\begin{align}}
\newcommand{\Eeq}{\end{align}}
\newcommand{\Glm}{\mathcal{G}_{\ell_1 \ell_2 \ell_3}^{m_1 m_2 m_3}}

\DeclareAcronym{GW}{
  short = GW,
  long = gravitational wave ,
  short-plural = s ,
}
\DeclareAcronym{LIGO}{
  short =LIGO ,
  long = Laser Interferometer Gravitational-Wave Observatory ,
  short-plural = ,
}
\DeclareAcronym{LVK}{
  short = LVK ,
  long = {LIGO, Virgo, and KAGRA},
  short-plural = ,
}

\DeclareAcronym{SGWB}{
  short = SGWB ,
  long = stochastic gravitational-wave background ,
  short-plural = s ,
}
\DeclareAcronym{GWB}{
  short = GWB ,
  long = gravitational-wave background ,
  short-plural = s ,
}
\DeclareAcronym{CBC}{
  short = CBC ,
  long = compact binary coalescence ,
  short-plural = s ,
}
\DeclareAcronym{BH}{
  short = BH ,
  long = black hole ,
  short-plural = s ,
}
\DeclareAcronym{BBH}{
  short = BBH ,
  long = binary black hole ,
  short-plural = s ,
}
\DeclareAcronym{PBH}{
  short = PBH ,
  long = primordial black hole ,
  short-plural = s ,
}
\DeclareAcronym{SNR}{
  short = SNR ,
  long = signal-to-noise ratio ,
  short-plural = s ,
}
\DeclareAcronym{IMRPPv2}{
  short = ,
  long = {\normalsize IMRP}{\footnotesize HENOM}{\normalsize P}v2 ,
  short-plural = ,
}
\DeclareAcronym{PTA}{
  short = PTA ,
  long = pulsar timing array ,
  short-plural = s ,
}
\DeclareAcronym{SFR}{
  short = SFR ,
  long = star formation rate ,
  short-plural =  ,
}
\DeclareAcronym{FRW}{
  short = FRW ,
  long = Friedmann-Robertson-Walker ,
  short-plural =  ,
}
\DeclareAcronym{IMR}{
  short = IMR ,
  long = inspiral-merger-ringdown ,
  short-plural =  ,
}
\DeclareAcronym{LISA}{
	short = LISA ,
	long  = Laser Interferometer Space Antenna,
  short-plural =  ,
}
\DeclareAcronym{ET}{
	short = ET ,
	long  = Einstein Telescope,
  short-plural =  ,
}
\DeclareAcronym{CE}{
	short = CE ,
	long  = Cosmic Explorer,
  short-plural =  ,
}
\DeclareAcronym{BBO}{
	short = BBO ,
	long  = Big Bang Observer,
  short-plural =  ,
}
\DeclareAcronym{DECIGO}{
	short = DECIGO ,
	long  = Deci-hertz Interferometer Gravitational wave Observatory,
  short-plural =  ,
}
\DeclareAcronym{ABH}{
	short = ABH ,
	long  = astrophysical black hole,
  short-plural = s ,
}

\DeclareAcronym{PNG}{
	short = PNG ,
	long  = primordial non-Gaussianity ,
  short-plural =  ,
}

\DeclareAcronym{CMB}{
	short = CMB ,
	long  = cosmic microwave background ,
  short-plural =  ,
}
\DeclareAcronym{LSS}{
	short = LSS ,
	long  = large-scale structure ,
  short-plural =  ,
}

\DeclareAcronym{PGW}{
	short = PGW ,
	long  = primordial gravitational wave ,
  short-plural = s ,
}
\DeclareAcronym{SIGW}{
	short = SIGW ,
	long  = scalar-induced gravitational wave ,
  short-plural = s ,
}

\DeclareAcronym{RD}{
	short = RD,
	long  = radiation-dominated ,
  short-plural =  ,
}
\DeclareAcronym{MD}{
	short = MD,
	long  = matter-dominated ,
  short-plural =  ,
}
\DeclareAcronym{eMD}{
	short = eMD,
	long  = early-matter-dominated ,
  short-plural =  ,
}
\DeclareAcronym{SW}{
	short = SW,
	long  = Sachs-Wolfe ,
  short-plural =  ,
}
\DeclareAcronym{ISW}{
	short = ISW,
	long  = integrated Sachs-Wolfe ,
  short-plural =  ,
}

\DeclareAcronym{DM}{
	short = DM,
	long  = dark matter ,
  short-plural =  ,
}

\DeclareAcronym{NANOGrav}{
	short = NANOGrav ,
	long  = North American Nanohertz Observatory for Gravitational Waves ,
  short-plural =  ,
}

\DeclareAcronym{PDF}{
	short = PDF ,
	long  = probability distribution function ,
  short-plural = s ,
}

\DeclareAcronym{SMBH}{
  short = SMBH ,
  long  = supper-massive black hole ,
  short-plural = s ,
}
\DeclareAcronym{SKA}{
	short = SKA ,
	long  = Square Kilometre Array,
  short-plural =  ,
}
\DeclareAcronym{NG15}{
  short = NG15 ,
  long  = NANOGrav 15-year ,
  short-plural =  ,
}

\arxivnumber{2505.16820} % Only if you have one

\title{\boldmath Isotropy, anisotropies and non-Gaussianity in the scalar-induced gravitational-wave background: diagrammatic approach for primordial non-Gaussianity up to arbitrary order}% Force line breaks with \\
\author[a,b,d]{Jun-Peng Li,}
\emailAdd{junpeng.li@nao.ac.jp}
\author[a]{Sai Wang,\footnote{Corresponding author}} 
\emailAdd{wangsai@hznu.edu.cn}
\author[c]{Zhi-Chao Zhao,}
\emailAdd{zhaozc@cau.edu.cn}
\author[d,e,f]{Kazunori Kohri }
\emailAdd{kazunori.kohri@gmail.com}
\affiliation[a]{School of Physics, Hangzhou Normal University, No.2318 Yuhangtang Road, Yuhang District, Hangzhou 311121, China}
\affiliation[b]{Theoretical Physics Division, Institute of High Energy Physics, Chinese Academy of Sciences, 19B Yuquan Road, Shijingshan District, Beijing 100049, China}
\affiliation[c]{Department of Applied Physics, College of Science, China Agricultural University, 17 Qinghua East Road, Haidian District, Beijing 100083, China}
\affiliation[d]{Division of Science, National Astronomical Observatory of Japan (NAOJ), and SOKENDAI, 2-21-1 Osawa, Mitaka, Tokyo 181-8588, Japan}
\affiliation[e]{Theory Center, IPNS, and QUP (WPI), KEK, 1-1 Oho, Tsukuba, Ibaraki 305-0801, Japan}
\affiliation[f]{Kavli IPMU (WPI), UTIAS, The University of Tokyo, Kashiwa, Chiba 277-8583, Japan}

\abstract{
Produced nonlinearly by the enhanced linear cosmological curvature perturbations, the scalar-induced gravitational waves (SIGWs) can serve as a potentially powerful probe of primordial non-Gaussianity (PNG) in the early Universe. 
In this work, we comprehensively investigate the imprints of local-type PNG on the SIGW background beyond the widely used quadratic and cubic approximations. 
We extend the diagrammatic approach to simplify the calculation of the SIGW energy density spectrum with high-order PNG, thereby facilitating systematic analysis for PNG up to arbitrary order. 
Following this approach, we derive semi-analytic formulas for the energy-density fraction spectrum, the angular power spectrum, and the angular bispectrum and trispectrum to describe the isotropic component, anisotropies, and non-Gaussianity of the SIGW background, respectively. 
Particularly, focusing on PNG up to quartic approximation (parameterized by $f_\mathrm{NL}$, $g_\mathrm{NL}$, and $h_\mathrm{NL}$), we numerically compute all contributions to these SIGW spectra. 
We find that PNG can significantly alter the magnitude of the SIGW energy-density spectrum, and can generate substantial anisotropies through the initial inhomogeneities in the SIGW distribution. 
Furthermore, we observe that the SIGW angular bispectrum and trispectrum always vanish when the primordial curvature perturbations are Gaussian; otherwise, they do not, indicating their potential utility as probes of PNG. 
Therefore, we anticipate that the SIGW background will provide essential information about the early Universe.
}

\pgfplotsset{compat=1.18}
\begin{document}
 
\maketitle
\flushbottom

\subfile{TeX/1Introduction}

\subfile{TeX/2ED}

\subfile{TeX/3Isotropy}

\subfile{TeX/4Anisotropy}

\subfile{TeX/5Non-Gaussianity}

\subfile{TeX/6Extended}

\subfile{TeX/7Conclusion}

\subfile{TeX/8Appendix}

\acknowledgments
We appreciate Xian Gao, Yan-Heng Yu, and Jing-Zhi Zhou for helpful discussions. 
S.W. and J.P.L. are supported by the National Natural Science Foundation of China (Grant Nos. 12533001).  Z.C.Z. is supported by the National Key Research and Development Program of China Grant No. 2021YFC2203001. K.K. is supported by KAKENHI Grant Nos. JP22H05270, JP23KF0289, JP24H01825, JP24K07027. 
This study is supported by Advanced Computation Center of Hangzhou Normal University.

\bibliography{biblio}
\bibliographystyle{JHEP}
\end{document}

%% file: TeX/1Introduction.tex
\section{Introduction}\label{sec:intro}

\Acp{GW} arising from the nonlinear interactions of linear cosmological curvature perturbations at horizon reentry, dubbed \acp{SIGW} \cite{Ananda:2006af,Baumann:2007zm,Espinosa:2018eve,Kohri:2018awv,Tomita:1967wkp,Matarrese:1992rp,Matarrese:1993zf,Matarrese:1997ay,Mollerach:2003nq,Carbone:2004iv,Assadullahi:2009jc,Domenech:2021ztg}, emerge as a promising avenue for exploring the early Universe. 
It is well known that prior to the last-scattering surface, the Universe exhibited opacity to electromagnetic waves while remaining transparent to \acp{GW} \cite{Bartolo:2018igk,Flauger:2019cam}. 
Consequently, \acp{SIGW} produced in the early Universe play a pivotal role in conveying valuable information about epochs characterized by a minuscule comoving Hubble horizon, which are inaccessible to observations of the \ac{CMB} and \ac{LSS}. 
The detection of \acp{SIGW} establishes a groundbreaking avenue for understanding the properties and evolution of cosmological scalar metric perturbations at small scales. 
On one hand, it provides insights into the dynamics during the later stages of inflation which predicts enhanced curvature power spectra at small scales (see Refs.~\cite{Ozsoy:2023ryl,LISACosmologyWorkingGroup:2025vdz} and the references therein). 
On the other hand, the detection of \acp{SIGW} sheds light on the nature of cosmological fluids in the early Universe, such as the equation of state \cite{Kohri:2018awv,Inomata:2019zqy,Inomata:2019ivs,Hajkarim:2019nbx,Domenech:2019quo,Domenech:2020kqm,Abe:2020sqb,Domenech:2021ztg,Escriva:2023nzn,Escriva:2024ivo}, adiabaticity \cite{Domenech:2021and,Domenech:2023jve,Chen:2024twp,Yuan:2024qfz,Luo:2025lgr,Yu:2025jgx}, dissipative effects \cite{Yu:2024xmz,Domenech:2025bvr}, and more. 
In particular, \acp{SIGW} can serve as valuable tools in the search for \acp{PBH}, which have been proposed as possible constituents of dark matter \cite{Carr:2020xqk,Green:2020jor,Carr:2021bzv,Carr:2023tpt,Escriva:2022duf}. 
The formation of \acp{PBH} can occur concurrently with the production of \acp{SIGW}, stemming from the gravitational collapse of large overdensities where scalar perturbations achieve sufficient amplitudes \cite{Hawking:1971ei,Carr:1974nx,Carr:1975qj,Chapline:1975ojl}. 
Therefore, the detection of \acp{SIGW} has the potential to impose constraints on the mass distribution of \acp{PBH} and vice versa \cite{Sasaki:2018dmp,LISACosmologyWorkingGroup:2023njw,LISACosmologyWorkingGroup:2025vdz}. 
% Nonetheless, the superposition of \acp{SIGW} and \acp{GW} originating from other sources, both astrophysical and cosmological, creates a stochastic background, posing challenges in distinguishing \ac{SIGW} signals from this composite background \cite{Christensen:2018iqi}. 
Nonetheless, the \ac{SIGW} background is mixed with other \acp{SGWB}, such as the \acp{GW} emitted from binary black holes \cite{LIGOScientific:2016aoc}, posing challenges in distinguishing \ac{SIGW} signals from these superposed \acp{SGWB} \cite{Christensen:2018iqi}. 
A comprehensive analysis of the statistics of the \ac{SIGW} background is imperative, encompassing the energy-density fraction spectrum, angular power spectrum, angular bispectrum, and trispectrum of \acp{SIGW}, which characterize the isotropic component, anisotropies, and non-Gaussianity of \ac{SIGW} energy density, respectively.

\Acp{SIGW} sourced from enhanced and non-Gaussian curvature perturbations at small scales have sparked great interest due to their association with the formation of \acp{PBH}. 
Existing studies have demonstrated that the presence of \ac{PNG} \cite{Maldacena:2002vr, Bartolo:2004if, Allen:1987vq, Bartolo:2001cw, Acquaviva:2002ud, Bernardeau:2002jy, Chen:2006nt, Byrnes:2010em, Wands:2010af, Meerburg:2019qqi, Achucarro:2022qrl} not only has a substantial impact on the abundance of \acp{PBH} significantly \cite{Bullock:1996at,Byrnes:2012yx,Young:2013oia,Bugaev:2013vba,Franciolini:2018vbk,Inomata:2020xad,Kitajima:2021fpq,Ferrante:2022mui,Gow:2022jfb,Franciolini:2023pbf,Iovino:2024tyg,Inui:2024fgk,Passaglia:2018ixg,Atal:2018neu,Atal:2019cdz,Taoso:2021uvl,Meng:2022ixx,Chen:2023lou,Kawaguchi:2023mgk,Choudhury:2023kdb,Yoo:2019pma,Ezquiaga:2019ftu,Carr:2020gox,Riccardi:2021rlf,Escriva:2022pnz,Kehagias:2019eil,Cai:2021zsp,Cai:2022erk,Young:2022phe,Zhang:2021vak,vanLaak:2023ppj,Gow:2023zzp,Franciolini:2023wun,Chen:2024pge,Nakama:2016gzw,Garcia-Bellido:2017aan,Pi:2024jwt,Pritchard:2025yda}, but also leaves remarkable imprints on the \ac{SIGW} background \cite{Adshead:2021hnm,Ragavendra:2021qdu,Abe:2022xur,Yuan:2023ofl,Perna:2024ehx,Li:2023qua,Li:2023xtl,Ruiz:2024weh,Wang:2023ost,Yu:2023jrs,Iovino:2024sgs,Zeng:2024ovg,Cai:2018dig,Unal:2018yaa,Atal:2021jyo,Yuan:2020iwf,Chang:2023aba,Nakama:2016gzw,Garcia-Bellido:2017aan,Ragavendra:2020sop,Zhang:2021rqs,Lin:2021vwc,Chen:2022dqr,Cai:2019elf,Papanikolaou:2024kjb,He:2024luf,Bartolo:2019zvb,Rey:2024giu,Pi:2024jwt,Zhou:2025djn}. 
These studies predominantly concentrate on local-type \ac{PNG} within the quadratic and cubic approximations, characterized by the non-linear parameters $\fnl$ and $\gnl$, respectively. 
Here, we present a concise overview on these studies according to three aspects, i.e., the isotropic component, anisotropies, and non-Gaussianity of \ac{SIGW} energy density, respectively. 
Regarding the isotropic component of \acp{SIGW}, contributions from local-type \ac{PNG} at $\fnl$ order to the energy-density fraction spectrum have been shown in Refs.~\cite{Adshead:2021hnm,Ragavendra:2021qdu,Li:2023qua}, and further investigations for \ac{PNG} up to $\gnl$ order have been carried out in Refs.~\cite{Yuan:2023ofl,Li:2023xtl,Ruiz:2024weh}. 
Refs.~\cite{Abe:2022xur,Perna:2024ehx} have investigated the contributions associated with \ac{PNG} up to quintic approximations, retaining the terms up to quartic order in the amplitude of the primordial curvature power spectrum. 
These studies have illustrated that the consideration of $\fnl$ and $\gnl$ can lead to significant alterations in the energy-density fraction spectrum by orders of magnitude, particularly within the anticipated sensitivity ranges of upcoming or futuristic \ac{GW} detection initiatives \cite{Baker:2019nia,Smith:2019wny,Hu:2017mde,Wang:2021njt,Ren:2023yec,TianQin:2015yph,TianQin:2020hid,Seto:2001qf,Kawamura:2020pcg,Crowder:2005nr,Smith:2016jqs,Capurri:2022lze,Hobbs:2009yy,Demorest:2012bv,Kramer:2013kea,Manchester:2012za,Sesana:2008mz,Thrane:2013oya,Janssen:2014dka,2009IEEEP..97.1482D,Weltman:2018zrl,Moore:2014lga,KAGRA:2013rdx,LIGOScientific:2014pky,Chen:2021nxo}. 
Other related works exploring the effects of \ac{PNG} on the energy-density fraction spectrum of \acp{SIGW} have been shown in Refs.~\cite{Wang:2023ost,Yu:2023jrs,Iovino:2024sgs,Zeng:2024ovg,Cai:2018dig,Unal:2018yaa,Atal:2021jyo,Yuan:2020iwf,Chang:2023aba,Nakama:2016gzw,Garcia-Bellido:2017aan,Ragavendra:2020sop,Zhang:2021rqs,Lin:2021vwc,Chen:2022dqr,Cai:2019elf,Garcia-Saenz:2022tzu,Pi:2024jwt,Papanikolaou:2024kjb,He:2024luf}. 
Moreover, the presence of local-type \ac{PNG} induces large-scale initial inhomogeneities at the \ac{SIGW} emission, resulting in the anisotropies on the skymap that may surpass the propagation effects. 
Refs.~\cite{Bartolo:2019zvb,Li:2023qua} have demonstrated that the \ac{PNG} at $\fnl$ order can induce significant anisotropies in the \ac{SIGW} background, with further investigations extending to \ac{PNG} at $\gnl$ order in Refs.~\cite{Li:2023xtl,Ruiz:2024weh}.  
Regarding non-Gaussianity in the \ac{SIGW} background, Ref.~\cite{Bartolo:2019zvb} provided an approximate expression for the angular bispectrum associated with \ac{PNG} at $\fnl$ order, while a comprehensive analysis of the angular bispectrum and trispectrum for \ac{PNG} up to $\gnl$ order was presented in Ref.~\cite{Li:2024zwx}. 
These studies underscore that the \ac{SIGW} non-Gaussianity results from the squeezed \ac{PNG}, suggesting that the angular bispectrum and trispectrum of \acp{SIGW} could serve as probes of the \ac{PNG}. 
However, \ac{PNG} resulting from distinct physical mechanisms exhibits significant features.
The nonlinear parameters $\fnl$ and $\gnl$ are insufficient for a comprehensive characterization of \ac{PNG}. 
A recent study \cite{Iovino:2024sgs} argues that for logarithmic \ac{PNG}, the amplitude of the \ac{SIGW} energy-density spectrum may significantly deviate from predictions made based on the quadratic approximation.

The imprints of a more general form of \ac{PNG} on the \ac{SIGW} background warrants particular attention, particularly in scenarios where the primordial curvature perturbation $\zeta$ can be represented as a function of its Gaussian component $\zeta_\ug$. 
Our objective is to incorporate \ac{PNG} beyond the commonly employed quadratic ($\fnl$ order) and cubic ($\gnl$ order) approximations, and develop a methodology capable of encompassing a full power-series expansion to all orders based on the diagrammatic approach \cite{Byrnes:2007tm}. 
This endeavor is essential given the applicability of the $\delta N$ formalism \cite{Sasaki:1995aw, Sasaki:1998ug, Lyth:2004gb, Lyth:2005fi} in expressing the primordial curvature perturbation $\zeta$ as a series expansion in terms of its Gaussian component $\zeta_\ug$ in various inflationary models, such as single-field inflation with an ultra-slow-roll phase \cite{Atal:2019cdz} and two-field curvaton models \cite{Sasaki:2006kq, Pi:2021dft}. 
As long as the deviation from the distribution of $\zeta$ from a Gaussian statistics remains within the perturbative regime, the discrepancy put forward in Ref.~\cite{Iovino:2024sgs} is expected to be eliminated in sufficiently high-order series expansion of $\zeta_\ug$. 
Although our present study does not explore concrete scenarios, our methodology is readily applicable to specific inflation models, including cases where the scale dependencies of non-linear parameters are non-negligible. 
In particular, we will present a detailed example of \ac{PNG} up to quartic approximation (i.e., $\hnl$ order), with explicit expressions and numerical results for the energy density spectrum and angular correlation functions provided for analysis.

The paper is structured as follows. 
In Section~\ref{sec:ED}, we review the energy density of \acp{SIGW}, both during production and in the present day, with a basic diagrammatic approach introduced subsequently. 
Section~\ref{sec:Omegabar} delves into the energy-density fraction spectra of \acp{SIGW} at the isotropic background level in the presence of \ac{PNG} by developing the ``renormalized'' diagrammatic approach. 
The angular power spectrum for the anisotropies in \acp{SIGW} is analyzed in Section~\ref{sec:Cl}, considering contributions from initial inhomogeneities and propagation effects at the perturbation level. 
Section~\ref{sec:bl&tl} explores the angular bispectrum and trispectrum of \acp{SIGW} in a manner similar to the angular power spectrum. 
In Section~\ref{sec:discuss}, we broaden the scope of this study to investigate the general infrared behaviors of the \ac{SIGW} energy-density spectra, impacts of the trispectrum of primordial curvature perturbations on \acp{SIGW}, the applicability of our methodology to scenarios involving scale-dependent \ac{PNG} parameters, and the relationship between the \ac{PBH} formation and the \ac{SIGW} background. 
Finally, our conclusions are presented in Section~\ref{sec:conc}. 
In addition, supplementary materials providing a brief summary of Wigner symbols are presented in Appendix~\ref{sec:Wigner}, while further details on the theoretical fundamentals of \acp{SIGW} production can be found in Appendix~\ref{sec:basic}.
Appendix~\ref{sec:FD} provides a detailed list of specific diagrams contributing to the energy-density fraction spectrum and \ac{PNG}-induced inhomogeneities. 
Appendix~\ref{sec:num-int} illustrates the relevant numerical integration techniques, while Appendix~\ref{sec:Boltz} offers specifics on the Boltzmann equation of cosmological \acp{GW}.

%% file: TeX/2ED.tex
\section{Energy density of SIGWs}\label{sec:ED}

In this section, we review the energy density of \acp{SIGW}, both at the time of production and in the present day.  
In particular, we introduce a diagrammatic approach for computing the energy density of \acp{SIGW} in the presence of \ac{PNG}, with detailed calculations and results elaborated in subsequent sections.

In order to characterize the \ac{SIGW} background, we define the energy-density full spectrum $\omega_\uGW$ through the energy density $\rho_\uGW$ as follows
\begin{equation}\label{eq:omega-def}
    \rho_\uGW (\eta,\bx) = \rho_\uc(\eta) \int \ud \ln q \, \ud^{2} \bn \, \omega_\uGW (\eta,\bx,\bq)\ ,
\end{equation}
where $\rho_\uc=3\cH^2\mpl^2/a^2$ represents the critical energy density of the Universe with $\mpl=1/\sqrt{8\pi G}$ denoting the Planck mass and $\cH(\eta)$ representing the conformal Hubble parameter at conformal time $\eta$. 
Moreover, we use $\bq=q\bn$ to represent the 3-momentum of the gravitons, where $q$ is the comoving \ac{SIGW} wavenumber and $\bn$ stands for the \ac{SIGW} propagation direction. 
To relate $\omega_\uGW$ to the observable \ac{SIGW} strain $h_{ij}$, we recall the energy density of \acp{SIGW} on subhorizon scales as presented in Ref.~\cite{Maggiore:1999vm}, i.e.,  
\begin{equation}\label{eq:rho-def}
    \rho_\uGW (\eta,\bx) 
    = \frac{\mpl^2}{16 a^2(\eta)} \overbar{\partial_l h_{ij}(\eta,\bx) \partial_l h_{ij}(\eta,\bx)}\ ,
\end{equation}
where $a(\eta)$ is the scale factor of the Universe and the long overbar labels a spacetime average over a region large compared to the \ac{GW} wavelength but small compared to the horizon scales. 
Furthermore, it is useful to decompose $h_{ij}$ in Fourier modes with different polarization tensors as 
\begin{eqnarray}\label{eq:h-Fourier}
    h_{ij}(\eta,\bx) &=& \int \frac{\ud^3 \bq}{(2\pi)^{3/2}} e^{i\bq\cdot\bx} h_{ij}(\eta, \bq)\nonumber\\
    &=& \sum_{\lambda=+,\times} \int \frac{\ud^3 \bq}{(2\pi)^{3/2}} e^{i\bq\cdot\bx} \epsilon_{ij}^{\lambda}(\bq) h_\lambda(\eta, \bq)\ ,
\end{eqnarray}
where $\epsilon^+_{ij}(\bq) = \left(\epsilon_i(\bq) \epsilon_j(\bq) - \bar{\epsilon}_i(\bq) \bar{\epsilon}_j(\bq)\right) /\sqrt{2}$ and $\epsilon^\times_{ij}(\bq) =  \left(\epsilon_i(\bq) \bar{\epsilon}_j(\bq) + \bar{\epsilon}_i(\bq) \epsilon_j(\bq) \right)/\sqrt{2}$ denote the polarization tensors, with the vectors $\epsilon_{i}(\bq)$, $\bar{\epsilon}_{i}(\bq)$, and $\bn$ forming an orthonormal basis in the three-dimensional momentum space, i.e., $\epsilon_{ij}^\lambda \epsilon_{ij}^{\lambda'} = \delta^{\lambda\lambda'}$ and $q^i\epsilon_{i}(\bq) = q^i\bar{\epsilon}_{i}(\bq) = 0$ by construction. 
The power spectrum of \acp{SIGW} $P_{h_\lambda}$ is then defined as the two-point correlator of $h_\lambda$, i.e., 
\begin{equation}\label{eq:Ph-def} 
    \langle h_\lambda (\eta,\bq) h_{\lambda'} (\eta,\bq')\rangle 
    = \delta_{\lambda\lambda'} \delta^{(3)} (\bq+\bq') P_{h_\lambda} (\eta,q) \ . 
\end{equation}
By combining the above formulae Eqs.~(\ref{eq:omega-def}, \ref{eq:rho-def}, \ref{eq:h-Fourier}), $\omega_\uGW$ can be written in the form of  
\begin{eqnarray}\label{eq:omega-h}
     \omega_\uGW (\eta,\bx,\bq)  
     = - \frac{q^3}{48 \cH^2} \int \frac{\ud^3 \bk}{(2\pi)^{3}} e^{i\bk\cdot\bx} 
        \left[\left(\bk-\bq\right) \cdot \bq \right] \overbar{h_{ij}(\eta,\bk-\bq) h_{ij}(\eta,\bq)}\ . 
\end{eqnarray}
Notably, here and hereinafter, we use $\bk$ as the Fourier counterpart of the position $\bx$.

\subsection{Production of SIGWs}\label{subsec:SIGW&PNG}

First of all, we revisit the calculation of $\omega_\uGW$ at the initial moment $\eta_\uin$, namely the time of \ac{SIGW} production.
Arising from the non-linear interactions of curvature perturbations, \acp{SIGW} are associated with the linear primordial curvature perturbations $\zeta$ through \cite{Ananda:2006af,Baumann:2007zm,Espinosa:2018eve,Kohri:2018awv}
\begin{eqnarray}\label{eq:h}
    h_\lambda(\eta_\uin, \bq) 
    &=& 4 \int \frac{\ud^3 \bq_a}{(2\pi)^{3/2}} 
        \zeta(\bq_a) \zeta(\bq-\bq_a) Q_{\lambda}(\bq,\bq_a) \hat{I} (\abs{\bq - \bq_a},q,\eta_\uin)\ ,
\end{eqnarray}
where $Q_{\lambda}(\bq, \bq_a)$ denotes the projection factor, $\hat{I} (|\mathbf{q} - \mathbf{q}_a|,q,\eta_\uin)$ represents the kernel function, and the subscript $_\uin$ indicates the production time. 
A concise derivation of this formula can be found in Refs.~\cite{Ananda:2006af,Baumann:2007zm,Espinosa:2018eve,Kohri:2018awv}, with the same conventions as in Refs.~\cite{Li:2023qua,Li:2023xtl,Li:2024zwx}. 
The specific expressions for $Q_{\lambda}(\bq, \bq_a)$ and $\hat{I} (|\mathbf{q} - \mathbf{q}_a|,q,\eta_\uin)$ are provided in Eq.~\eqref{eq:Q-def} and Eqs.~(\ref{eq:Ihat-Iuv}, \ref{eq:I-RD}), respectively.
The preceding expressions, Eqs.~(\ref{eq:h-Fourier}, \ref{eq:omega-h}, \ref{eq:h}), suggest that the initial energy-density full spectrum, denoted as $\omega_{\uGW,\uin} (\bq) = \omega_\uGW (\eta_\uin,\bx_\uin,\bq)$, can be schematically represented as $\omega_{\uGW,\uin} (\bq) \sim \langle\zeta^4\rangle_{\bx_\uin}$. 
Here, the angle brackets with the subscript ${}_{\bx_\uin}$ denote an ensemble average over a coarse-grained region centered at the initial position $\bx_\uin$. 
Such coarse graining is necessary because the \acp{SIGW} of interest are emitted at an extremely high redshift ($z > 10^{10}$), meaning the Hubble-horizon is very small at the initial moment. 
Consequently, due to the finite angular resolution of \ac{GW} detectors, any signal observed along a given line-of-sight originates from a superposition of \acp{SIGW} emitted from numerous initial Hubble-horizon patches. 
This implies that the scale of the averaged region is larger than the \ac{GW} wavelength yet significantly smaller than \ac{CMB} scales. 
Furthermore, the initial anisotropies vanish upon averaging, as necessitated by the cosmological principle and the assumption of a statistically isotropic Universe.
Given that $\langle\zeta^4\rangle_{\bx_\uin}$ encompasses the connected trispectrum of $\zeta$, the relation between $\omega_{\uGW,\uin} (\bq)$ and $\langle\zeta^4\rangle_{\bx_\uin}$ implies that $\omega_{\uGW,\uin} (\bq)$ is affected by \ac{PNG}. 

In order to analyze the impacts of \ac{PNG} on $\omega_{\uGW,\uin} (\bq)$, 
we consider the primordial curvature perturbations $\zeta (\bx)$ with small local-type \ac{PNG}, which can generally be expanded around their Gaussian components $\zeta_\ug$ as \cite{Komatsu:2001rj,Okamoto:2002ik,Smidt:2010ra} 
\begin{equation}\label{eq:fnl-gnl-hnl-def}
     \zeta (\bx) = \zeta_\ug (\bx) + \frac{3}{5}\fnl \left( \zeta_\ug^2(\bx) - \langle \zeta_\ug^{2} \rangle \right) + \frac{9}{25}\gnl \zeta_\ug^3(\bx) 
     + \frac{27}{125}\hnl \left( \zeta_\ug^4(\bx) - 3\langle \zeta_\ug^{2} \rangle^2 \right) + \cdots\ .
\end{equation}
For convenience, we recast the non-linear parameters as 
\begin{equation*}
    \Fnl = \frac{3}{5} \fnl \ ,\ 
    \Gnl = \frac{9}{25} \gnl  \ ,\ 
    \Hnl = \frac{27}{125} \hnl \ ,
\end{equation*}
and rewrite the expression of Eq.~\eqref{eq:fnl-gnl-hnl-def} as follows 
\begin{equation}\label{eq:Fnl-Gnl-Hnl-def}
     \zeta (\bx) = \zeta_\ug (\bx) + \Fnl \left( \zeta_\ug^2(\bx) - \langle \zeta_\ug^{2} \rangle \right) + \Gnl \zeta_\ug^3(\bx) + \Hnl \left( \zeta_\ug^4(\bx) - 3\langle \zeta_\ug^{2} \rangle^{2} \right) + \cdots \ .
\end{equation}
Note that this form of expansion has ensured $\langle\zeta\rangle=0$. 
Consequently, the statistics of $\zeta$ can be expressed in terms of the two-point correlators of $\zeta_\ug$. 

We are intrigued by \acp{SIGW} produced during the radiation-dominated era, assuming an enhancement of the primordial curvature power spectrum on the Hubble horizon scales at that epoch. 
To study the statistics of $\omega_{\uGW,\uin}$ on the \ac{CMB} scale, 
we can further decompose $\zeta_\ug$ into short-wavelength mode $\zeta_{\ugS}$ and long-wavelength mode $\zeta_{\ugL}$ \cite{Tada:2015noa}, i.e., 
\begin{equation}\label{eq:S-L-dec}
    \zeta_\ug=\zeta_{\ugS}+\zeta_{\ugL} \ ,
\end{equation}
where $\zeta_{\ugS}$ denotes the enhanced curvature perturbations at the horizon scale, while $\zeta_{\ugL}$ represents the mode extrapolated from the \ac{CMB} scale.
Further, the two-point correlators of $\zeta_{\ugX} (\mathrm{X}=\mathrm{S},\mathrm{L})$ define their respective dimensionless power spectra, namely, 
\begin{equation}\label{eq:PgX-def}
    \langle \zeta_{\ugX} (\bq) \zeta_{\ugX} (\bq') \rangle 
    = \delta^{(3)} (\bq+\bq') \frac{2\pi^2}{q^3} \Delta^2_\mathrm{X} (q)\ .
\end{equation} 
For simplicity, we adopt a scale-invariant spectrum for $\Delta_\uL^{2}$ \cite{Planck:2018vyg} and a normal function with respect to $\ln q$ for $\Delta^2_\uS$, expressed as 
\begin{eqnarray}
    \Delta^2_\uL &=& A_\uL \simeq 2.1\times10^{-9}\ ,\label{eq:Flat}\\
    \Delta^2_\uS (q) &=& \frac{A_\uS}{\sqrt{2\pi\sigma^2}}\exp\left(-\frac{\ln^2 (q/q_\ast)}{2 \sigma^2}\right)\ ,\label{eq:Lognormal}
\end{eqnarray}
where $\sigma$ denotes the spectral width and $A_\uS$ represents the spectral amplitude at the peak wavenumber $q_\ast$ for $\Delta^2_\uS (q)$. 
In this study, we consider values of $A_\uS$ ranging from $10^{-4}$ to $10^{-1}$, corresponding to scenarios of \ac{PBH} formation (see reviews in Refs.~\cite{Green:2020jor,Carr:2020gox,Escriva:2022duf} and references therein). 
Furthermore, the \ac{PNG} parameters in Eq.~\eqref{eq:Fnl-Gnl-Hnl-def}, proposed to be scale-independent for the sake of simplicity, are subject to perturbativity conditions. These conditions are specifically expressed as $|\Fnl| \sqrt{A_\uS} + |\Gnl| A_\uS + |\Hnl| \sqrt{A_\uS^3} < 1$ and $|\Fnl| \sqrt{A_\uS} > |\Gnl| A_\uS > |\Hnl| \sqrt{A_\uS^3}$.

By substituting Eqs.~(\ref{eq:Fnl-Gnl-Hnl-def}, \ref{eq:S-L-dec}) into $\omega_{\uGW,\uin} (\bq) \sim \langle\zeta^4\rangle_{\bx_\uin}$, we can express $\langle\zeta^4\rangle_{\bx_\uin}$ as a polynomial in terms of $\langle\zeta_\ugS^m \zeta_\ugL^n\rangle_{\bx_\uin}$, where $m,n \in \mathbb{N}$ and $m+n \leq 16$ for \ac{PNG} up to $\Hnl$ order. 
The coefficients of this polynomial consist of \ac{PNG} parameters. 
Since the wavelength of $\zeta_{\ugL}$ is significantly larger than the scale of the averaged region, we can extract $\zeta_{\ugL}$ from the average around $\bx_\uin$ when analyzing $\omega_{\uGW,\uin}$ alone. 
In this process, we neglect the wavenumber of $\zeta_{\ugL}$, as it is insignificant compared to that of $\zeta_{\ugS}$. 
Consequently, the sum of all momenta of $\zeta_{\ugS}$ within any averages of this polynomial results in zero due to momentum conservation. 
Notably, we will omit the subscript $_{\bx_\uin}$ from the averages associated with $\zeta_{\ugS}$, because its wavelength is comparable to the comoving Hubble horizon $\cH^{-1}$ and considerably smaller than the averaged region. 
Owing to $A_\uS \gg A_\uL$, we can regard the terms involving $\zeta_{\ugL}$ as corrections to those determined by $\zeta_{\ugS}$, which corresponds to the large-scale inhomogeneities superimposed on the homogeneous component of \acp{SIGW}. 
This will be delineated subsequently. 
After extracting $\zeta_{\ugL}$ and the \ac{PNG} parameters from the correlators, we can approximately expand $\omega_{\uGW,\uin} (\bq) \sim \langle\zeta^4\rangle_{\bx_\uin}$ up to third order in $\zeta_{\ugL}$ in a schematic manner, as shown by 
\begin{eqnarray}\label{eq:omega-expand}
    &&\omega_{\uGW,\uin} (\bq)  
    \sim \langle\zeta^4\rangle_{\bx} \\
    &\sim& \langle\zeta_\uS^4\rangle 
    + \zeta_{\ugL} \bigl(
        \mathcal{O}(\Fnl) \langle\zeta_{\ugS} \zeta_\uS^3\rangle + \mathcal{O}(\Gnl) \langle\zeta_{\ugS}^2 \zeta_\uS^3\rangle + \mathcal{O}(\Hnl) \langle\zeta_{\ugS}^3 \zeta_\uS^3\rangle
    \bigr)\nonumber\\ 
    %\hphantom{\langle\zeta_\uS^4\rangle} 
    &&+ \zeta_{\ugL}^2 \bigl[
        \mathcal{O}(\Fnl^2) \langle\zeta_{\ugS}^2 \zeta_\uS^2\rangle + \mathcal{O}(\Fnl\Gnl) \langle\zeta_{\ugS}^3 \zeta_\uS^2\rangle + \bigl(\mathcal{O}(\Gnl^2) + \mathcal{O}(\Fnl\Hnl)\bigr)\langle\zeta_{\ugS}^4 \zeta_\uS^2\rangle \nonumber\\
        &&\hphantom{+ \zeta_{\ugL}^2 \bigl[} + \mathcal{O}(\Gnl\Hnl) \langle\zeta_{\ugS}^5 \zeta_\uS^2\rangle + \mathcal{O}(\Hnl^2) \langle\zeta_{\ugS}^6 \zeta_\uS^2\rangle + \mathcal{O}(\Gnl) \langle\zeta_{\ugS} \zeta_\uS^3\rangle +  \mathcal{O}(\Hnl) \langle\zeta_{\ugS}^2 \zeta_\uS^3\rangle
    \bigr]\nonumber\\ 
    %\hphantom{\langle\zeta_\uS^4\rangle} 
    &&+ \zeta_{\ugL}^3 \bigl[
        \mathcal{O}(\Fnl^3) \langle\zeta_{\ugS}^3 \zeta_\uS\rangle + \mathcal{O}(\Fnl^2\Gnl) \langle\zeta_{\ugS}^4 \zeta_\uS\rangle + \bigl(\mathcal{O}(\Fnl\Gnl^2) + \mathcal{O}(\Fnl^2\Hnl)\bigr) \langle\zeta_{\ugS}^5 \zeta_\uS\rangle \nonumber\\
        &&\hphantom{+ \zeta_{\ugL}^3 \bigl[} + \bigl(\mathcal{O}(\Gnl^3) + \mathcal{O}(\Fnl\Gnl\Hnl)\bigr) \langle\zeta_{\ugS}^6 \zeta_\uS\rangle
        + \bigl(\mathcal{O}(\Gnl^2\Hnl) + \mathcal{O}(\Fnl\Hnl^2)\bigr) \langle\zeta_{\ugS}^7 \zeta_\uS\rangle \nonumber\\
        &&\hphantom{+ \zeta_{\ugL}^3 \bigl[} + \mathcal{O}(\Gnl\Hnl^2) \langle\zeta_{\ugS}^8 \zeta_\uS\rangle + \mathcal{O}(\Hnl^3) \langle\zeta_{\ugS}^9 \zeta_\uS\rangle + \mathcal{O}(\Fnl\Gnl) \langle\zeta_{\ugS}^2 \zeta_\uS^2\rangle \nonumber\\
        &&\hphantom{+ \zeta_{\ugL}^3 \bigl[} + \bigl(\mathcal{O}(\Gnl^2) + \mathcal{O}(\Fnl\Hnl)\bigr) \langle\zeta_{\ugS}^3 \zeta_\uS^2\rangle + \mathcal{O}(\Gnl\Hnl) \langle\zeta_{\ugS}^4 \zeta_\uS^2\rangle\nonumber\\
        &&\hphantom{+ \zeta_{\ugL}^3 \bigl[} + \mathcal{O}(\Hnl^2) \langle\zeta_{\ugS}^5 \zeta_\uS^2\rangle + \mathcal{O}(\Hnl) \langle\zeta_{\ugS} \zeta_\uS^3\rangle
    \bigr]\ , \nonumber
\end{eqnarray}
where $\zeta_\uS$ denotes the short-wavelength modes of $\zeta$, encompassing both Gaussian and non-Gaussian components. 
Explicitly, the relationship between $\zeta_\uS$ and its Gaussian component $\zeta_\ugS$ is identical to that between $\zeta$ and $\zeta_\ug$ as outlined in Eq.~\eqref{eq:Fnl-Gnl-Hnl-def}. 
By utilizing Wick's theorem in conjunction with Eqs.~(\ref{eq:PgX-def}, \ref{eq:Lognormal}), all multi-point correlators within this polynomial can be contracted and expressed in terms of $\Delta^2_\uS$, which remains independent of the position $\bx_\uin$ and the propagation direction $\bn$. 
Furthermore, by combining Eqs.~(\ref{eq:Ph-def}, \ref{eq:omega-h}, \ref{eq:h}), we can observe that the spatial and directional dependencies of $\omega_{\uGW,\uin} (\bq)$ are dictated solely by $\zeta_{\ugL}$. 
Thus, the leading-order term in Eq.~\eqref{eq:omega-expand} corresponds to the homogeneous component of $\omega_{\uGW,\uin} (\bq)$, while the remaining terms characterize the 1st-, 2nd-, and 3rd-order \ac{PNG}-induced inhomogeneities, respectively. 
We decompose $\omega_{\uGW,\uin} (\bq)$ into its homogeneous component and \ac{PNG}-induced inhomogeneities of varying orders in $\zeta_\ugL$, as given by 
\begin{subequations}\label{eqs:omega-result}
\begin{eqnarray}
    \omega_{\uGW,\uin} (\bq)
    &=& \bar{\omega}_{\uGW,\uin} (q) + \omega_{\ung,\uin}^{(1)}(\bq) + \omega_{\ung,\uin}^{(2)}(\bq) + \omega_{\ung,\uin}^{(3)}(\bq) + \cdots\ ,\label{eq:omega-split}\\
    \omega_{\uGW,\uin}^{(1)}(\bq) &=& \omega_{\ung,\uin}^{(1)}(q) 
    \int \frac{\ud^{3}\bk}{(2\pi)^{3/2}} e^{i\bk\cdot\bx} \zeta_{\ugL}(\bk) \ ,\label{eq:omega-1}\\
    \omega_{\uGW,\uin}^{(2)}(\bq) &=& \omega_{\ung,\uin}^{(2)}(q) \int \frac{\ud^{3}\bk\,\ud^{3}\bp}{(2\pi)^{3}} e^{i\bk\cdot\bx} \zeta_{\ugL}(\bp) \zeta_{\ugL}(\bk - \bp) \ ,\label{eq:omega-2}\\
    \omega_{\uGW,\uin}^{(3)}(\bq) &=& \omega_{\ung,\uin}^{(3)}(q) \int \frac{\ud^{3}\bk\,\ud^{3}\bp\,\ud^{3}\bl}{(2\pi)^{9/2}} e^{i\bk\cdot\bx} \zeta_{\ugL}(\bl)  \zeta_{\ugL}(\bp-\bl) \zeta_{\ugL}(\bk - \bp) \ ,\label{eq:omega-3}
\end{eqnarray}
\end{subequations}
where $\bar{\omega}_{\uGW,\uin} (q) = \langle\omega_{\uGW,\uin}(\bq)\rangle$ denotes the average energy-density spectrum, while $\omega_{\uGW,\uin}^{(1)}(\bq)$, $\omega_{\uGW,\uin}^{(2)}(\bq)$, and $\omega_{\uGW,\uin}^{(3)}(\bq)$ represent the 1st-, 2nd-, and 3rd-order \ac{PNG}-induced inhomogeneities of the \ac{SIGW} energy density, respectively. 
Also, $\omega_{\ung,\uin}^{(1)} (q)$, $\omega_{\ung,\uin}^{(2)} (q)$, and $\omega_{\ung,\uin}^{(3)} (q)$ correspond to their respective large-scale modulation functions. 
Since these large-scale modulation functions are governed by the multi-point correlators of $\zeta_\ugS$, Eqs.~(\ref{eq:omega-1} - \ref{eq:omega-3}) indicate that the \ac{PNG}-induced inhomogeneities of \acp{SIGW} at the time of emission can be attributed to the coupling between short- and long-wavelength modes of curvature perturbations.

The expression for $\bar{\omega}_{\uGW,\uin} (q)$ can be derived by calculating the spatial average of Eq.~\eqref{eq:omega-h}, which is formulated in terms of the power spectrum, as given by \cite{Inomata:2016rbd} 
\begin{equation}\label{eq:omegabar-h} 
    \bar{\omega}_{\uGW,\uin} (q)
    = \frac{q^5}{48\times(2\pi)^3 \cH^2} 
        \overbar{P_{h} (\eta_\uin,q)}\ ,
\end{equation}
where $P_{h} = \sum_{\lambda=+,\times} P_{h_\lambda}$ represents the total power spectrum of \acp{SIGW}.  
Using Eq.~\eqref{eq:h}, we can express $P_{h_\lambda}$ explicitly in terms of the four-point correlator of $\zeta_S$. 
As for the large-scale inhoomogeneities, we introduce the density contrast $\delta_{\uGW,\uin} (\bq)$ to better characterize them, as given by 
\begin{eqnarray}
    \omega_{\uGW,\uin} (\bq)
    &=& \bar{\omega}_{\uGW,\uin} (q) \bigl(1 + \delta_{\uGW,\uin} (\bq) \bigr)\ ,\label{eq:omega-delta}\\
    \delta_{\uGW,\uin} (\bq) &=& \delta_{\uGW,\uin}^{(1)} (\bq) + \delta_{\uGW,\uin}^{(2)} (\bq) + \delta_{\uGW,\uin}^{(3)} (\bq) + \cdots\ ,\label{eq:delta-split}
\end{eqnarray}
where $\delta_{\uGW,\uin}^{(1)} = \omega_{\uGW,\uin}^{(1)} / \bar{\omega}_{\uGW,\uin} \propto \zeta_{\ugL}$, $\delta_{\uGW,\uin}^{(2)} = \omega_{\uGW,\uin}^{(2)} / \bar{\omega}_{\uGW,\uin} \propto \zeta_{\ugL}^2$, and $\delta_{\uGW,\uin}^{(3)} = \omega_{\uGW,\uin}^{(3)} / \bar{\omega}_{\uGW,\uin} \propto \zeta_{\ugL}^3$ represent the  density contrast of different orders in $\zeta_{\ugL}$. 
The equations presented, namely Eqs.~(\ref{eq:omega-expand} - \ref{eq:delta-split}), indicate that the short-wavelength mode is responsible for the production of \acp{SIGW}, while the long-wavelength mode modulates the distribution of their energy density on large scales. 
Consequently, we can utilize $\bar{\omega}_{\uGW,\uin}$ and the correlations of $\delta_{\uGW,\uin}$ to describe the average and other statistical properties of \acp{SIGW} at their production moment. 
Their explicit expressions will be explored in the following sections.

\subsection{Present-day energy density of SIGWs}\label{subsec:SIGW&0}

In this subsection, we aim to investigate the transfer of energy density of \acp{SIGW} from the time of emission to the present observation. 
Similar to the average energy-density spectrum $\bar{\omega}_{\uGW,\uin} (q)$ and the density contrast $\delta_{\uGW,\uin} (\bq)$ defined at the initial moment $\eta_\uin$ in Eqs.~(\ref{eqs:omega-result}, \ref{eq:omega-delta}), we extend these definitions to any given moment $\eta$ as follows  
\begin{equation}\label{eq:omega-delta-0}
    \omega_\uGW (\eta,\bx,\bq)
    = \bar{\omega}_\uGW (\eta,q) \bigl(1 + \delta_\uGW (\eta,\bx,\bq) \bigr)\ .
\end{equation}
Here, $\bar{\omega}_{\uGW,\uin} (q)$ characterizes the homogeneous and isotropic background, while $\delta_\uGW (\eta,\bx,\bq)$ measures the fluctuations on this background. 
This relationship implies that $\bar{\omega}_\uGW (\eta,q) = \langle\omega_\uGW (\eta,\bx,\bq)\rangle$ and that the average of the density contrast satisfies $\langle\delta_\uGW\rangle = 0$.
Moreover, the homogeneous and isotropic background is typically described by the energy-density fraction spectrum $\bar{\Omega}_\uGW (\eta,q)$. 
This spectrum is defined in terms of the spatially-averaged energy density $\bar{\rho}_\uGW (\eta)$. 
It is expressed as $\bar{\rho}_\uGW (\eta) = \rho_\uc(\eta) \int \ud \ln q \, \bar{\Omega}_{\uGW} (\eta,q)$, where the overbar signifies spatially-averaged physical quantities. 
By comparing this with definition with Eq.~\eqref{eq:omega-def}, we have  
\begin{equation}\label{eq:O-o}
    \bar{\Omega}_\uGW (\eta,q) = 4 \pi \bar{\omega}_\uGW (\eta,q)\ .
\end{equation}

It is essential to investigate the relationship between the present-day energy-density fraction spectrum, denoted as $\bar{\Omega}_{\uGW,0} (\nu) = \bar{\Omega}_\uGW (\eta_0,2\pi\nu)$, and the average energy-density spectrum at the time of emission, $\bar{\omega}_{\uGW,\uin} (q)$. 
Here, the subscript $_0$ denotes the current epoch of observation. 
The present-day \ac{SIGW} frequency $\nu$ is determined by the comoving momentum $q$ (specifically, $\nu = q / (2\pi)$), as $q$ is conserved during propagation at the background level—a result derived in Appendix~\ref{sec:Boltz}. 
Regarding the amplitude, we focus on \acp{SIGW} produced during the radiation-dominated era, as discussed previously. 
Accounting for the dilution of radiation components and the evolution of the effective number of relativistic degrees of freedom in the Universe, we combine Eq.~\eqref{eq:O-o} to yield the expression for $\bar{\Omega}_{\uGW,0} (\nu)$ \cite{Wang:2019kaf} 
\begin{eqnarray}\label{eq:Omega0}
    h^2\bar{\Omega}_{\uGW,0} (\nu) 
    &=& 4 \pi h^2\Omega_{\mathrm{rad}, 0} \bar{\omega}_{\uGW,\uin} (q) \left(\frac{g_{*,\rho}(T_\uin)}{g_{*,\rho}(T_0)} \right)
        \left(\frac{g_{*,s}(T_0)}{g_{*,s}(T_\uin)} \right)^{4/3} \ ,
\end{eqnarray}
where the relation between the initial temperature $T_\uin$ and the \ac{SIGW} frequency $\nu$ is approximately given by \cite{Wang:2019kaf,Zhao:2022kvz} 
\begin{eqnarray}\label{eq:nu-in}
    \frac{\nu}{\mathrm{nHz}} 
    = 26.5 \left(\frac{T_\uin}{\mathrm{GeV}}\right) \left(\frac{g_{*,\rho}(T_\uin)}{106.75}\right)^{1/2} \left(\frac{g_{*,s}(T_\uin)}{106.75}\right)^{-1/3} \ .
\end{eqnarray} 
Here, $h^2\Omega_{\mathrm{rad},0} = 4.2 \times 10^{-5}$ represents the energy-density fraction of radiation in the present Universe, and $h$ denotes the dimensionless Hubble constant \cite{Planck:2018vyg}. 
$g_{*,\rho}$ and $g_{*,s}$ characterize the effective numbers of relativistic degrees of freedom, tabulated as functions of the cosmic temperature $T$ (see Ref.~\cite{Saikawa:2018rcs} for concrete tables). 
% Moreover, the subscript $_\mathrm{eq}$ labels the epoch of matter-radiation equality, while the subscript $_0$ indicates the present observation time. 
% Eqs.~\eqref{eq:Omega0} and \eqref{eq:nu-in} illustrate that the effective numbers of relativistic degrees of freedom not only impact the amplitude of $h^2\bar{\Omega}_{\uGW,0}$ but also influence the frequency of the present-day \acp{SIGW}. 
% This influence effectively acts as a multiplicative prefactor for $h^2\bar{\Omega}_{\uGW,0} (\nu)$ when we consider $\nu=q/2\pi$ in Eq.~\eqref{eq:Omega0}. 
% In this study, we concentrate on the milli-Hertz band, where this prefactor remains constant at 0.39 for $\nu \gtrsim 10^{-5}$~Hz. 
% For the nano-Hertz band, this prefactor increases as the frequency decreases, reaching 0.77 at $\nu = 10^{-9}$~Hz and becoming 1 for $\nu \lesssim 10^{-12}$~Hz. 

The present-day density contrast, denoted as $\delta_{\uGW,0}(\bq) = \delta_{\uGW}(\eta_{0},\bx_{0},\bq)$, can be derived by solving the Boltzmann equation for gravitons \cite{Contaldi:2016koz,Bartolo:2019oiq,Bartolo:2019yeu,Schulze:2023ich}. 
The solution includes both the initial inhomogeneities and propagation effects, as given by 
\begin{equation}\label{eq:delta-0}
    \delta_{\uGW,0}(\bq) = \delta_{\uGW,\uin}(\bq) + \left(6 - n_{\uGW} (\nu)\right) \Phi (\eta_\uin, \bx_\uin)\ ,
\end{equation}
where $\Phi(\eta_\uin,\bx_\uin)$ denotes the Bardeen potential at the time of \acp{SIGW} production, while $n_{\uGW}$ represents the spectral index of $\bar{\Omega}_{\uGW,0}$. 
A brief review of the derivation process of Eq.~\eqref{eq:delta-0} is provided in Appendix~\ref{sec:Boltz}. 
Notably, in our analysis of propagation effects, we focus exclusively on the \ac{SW} effect, since it is the predominant phenomenon among all propagation effects. 
The formula presented here differs slightly from the results reported in Refs.~\cite{Bartolo:2019zvb,Li:2023qua,Li:2023xtl,Wang:2023ost}. 
This discrepancy arises from our inclusion of initial inhomogeneity induced by large-scale scalar perturbations, which is referred to as inflationary initial conditions in Refs.~\cite{Mierna:2024pkh, ValbusaDallArmi:2023nqn, ValbusaDallArmi:2024hwm}. 
This is regarded as a general feature of cosmological \acp{GW} produced via non-thermal and non-adiabatic mechanisms. 
% This discrepancy arises from our inclusion of the initial inhomogeneity induced by non-linear effects \cite{Mierna:2024pkh}, traditionally referred to as the adiabatic contribution to the initial inhomogeneity \cite{Schulze:2023ich, ValbusaDallArmi:2023nqn,ValbusaDallArmi:2024hwm}. 
\footnote{
When accounting for \ac{GW} energy density modified by large-scale scalar perturbations~\cite{Mierna:2024pkh, ValbusaDallArmi:2023nqn,ValbusaDallArmi:2024hwm}, the initial energy density of \acp{SIGW} is modified as given in Appendix~\ref{sec:Boltz}, i.e.,
\begin{equation}\label{eq:rho-modify}
    \tilde{\rho}_\uGW (\eta_\uin,\bx_\uin)
    = \frac{\mpl^2 (1 + 2 \Phi_\uL (\eta_\uin, \bx_\uin))}{16 a^2(\eta_\uin)} \overbar{\partial_l h_{ij}(\eta_\uin,\bx_\uin) \partial_l h_{ij}(\eta_\uin,\bx_\uin)}\ ,
\end{equation}
where $\Phi_\uL (\eta_\uin, \bx_\uin)$ denotes the large-scale scalar perturbations.
Despite this modification, we continue to use Eq.~\eqref{eq:rho-def} for the calculation of $\omega_{\uGW,\uin}(\bq)$ and $\delta_{\uGW,\uin}(\bq)$ for convenience. 
To indicate the inclusion of this modification, we place a tilde above these quantities.
}
For the first terms in Eq.~\eqref{eq:delta-0}, we derive the initial density contrast $\delta_{\uGW,\uin}(\bq)$ based on the framework outlined in Subsection~\ref{subsec:SIGW&PNG}, deferring the detailed derivation to later sections. 
The second term includes the \ac{SW} effect, expressed as $\left(4 - n_{\uGW} (\nu)\right) \Phi (\eta_\uin, \bx_\uin)$, along with the inhomogeneity arising from initial inflationary condition \cite{Mierna:2024pkh, ValbusaDallArmi:2023nqn,ValbusaDallArmi:2024hwm}, given by $2 \Phi (\eta_\uin, \bx_\uin)$. 
When calculating the correlators of $\delta_{\uGW,0}$, we only take into account long-wavelength scalar perturbations that re-entered the Hubble horizon during matter domination, owing to the angular resolution limitations of current and upcoming \ac{GW} detectors. 
Therefore, the long-wavelength Bardeen potential at the production time, $\Phi(\eta_\uin,\bx_\uin)$, up to third order in $\zeta_{\ugL}$, is given by 
\begin{eqnarray}\label{eq:Phi}
    \Phi (\eta_\uin,\bx_\uin) 
    = \frac{3}{5} \int \frac{\ud^{3}\bk}{(2\pi)^{3/2}} e^{i\bk\cdot\bx} 
    &\biggl(&
        \zeta_{\ugL}(\bk) + \Fnl \int \frac{\ud^{3}\bp}{(2\pi)^{3/2}} \zeta_{\ugL}(\bp) \zeta_{\ugL}(\bk - \bp)\nonumber\\
        &&+ \Gnl \int \frac{\ud^{3}\bp\,\ud^{3}\bl}{(2\pi)^{3}} \zeta_{\ugL}(\bl) \zeta_{\ugL}(\bp - \bl) \zeta_{\ugL}(\bk - \bp) + \cdots
    \biggr)\ .
\end{eqnarray}
Moreover, the spectral index of $\bar{\Omega}_{\uGW,0}$ is defined as 
\begin{equation}\label{eq:ngw-def}
    n_{\uGW} (\nu) = \frac{\partial\ln \bar{\Omega}_{\uGW,0} (\nu)}{\partial\ln \nu} \simeq \frac{\partial\ln \bar{\omega}_{\uGW,\uin} (q)}{\partial\ln q}\Big|_{q=2\pi\nu} \ .
\end{equation}
In this analysis, we neglect the time evolution of $n_{\uGW}$ during propagation, as the influence of the numbers of relativistic species on the frequency remains constant in Eq.~\eqref{eq:Omega0} for the milli-Hertz band under consideration. 
These two equations, Eqs.~(\ref{eq:Phi}, \ref{eq:ngw-def}), indicate that the \ac{SW} effect can be expressed as a series expansion in terms of $\zeta_{\ugL}$, with the leading term being of $\cO (\zeta_{\ugL})$ order. 
Given that the initial density contrast $\delta_{\uGW,\uin}(\bq)$ is similarly expressed in Eq.~\eqref{eq:delta-split}, we can represent the present-day density contrast $\delta_{\uGW,0} (\bq)$ as a power-series expansion in $\zeta_{\ugL}$, namely 
\begin{equation}\label{eq:delta0-123}
    \delta_{\uGW,0}(\bq) = \delta_{\uGW,0}^{(1)}(\bq) + \delta_{\uGW,0}^{(2)}(\bq) + \delta_{\uGW,0}^{(3)}(\bq) + \cdots\ ,
\end{equation}
where we introduce three quantities of the form 
\begin{eqnarray}
    \delta_{\uGW,0}^{(1)}(\bq) 
        &=& \biggl[
            \frac{\omega_{\ung,\uin}^{(1)} (q)}{\bar{\omega}_{\uGW,\uin} (q)}
            + \frac{3}{5} \bigl(6 - n_{\uGW} (\nu)\bigr)
        \biggr]
        \int \frac{\ud^{3}\bk}{(2\pi)^{3/2}} e^{i\bk\cdot\bx} \zeta_{\ugL}(\bk)\ ,\label{eq:delta0-1}\\
    \delta_{\uGW,0}^{(2)}(\bq) 
        &=& \biggl[
            \frac{\omega_{\ung,\uin}^{(2)} (q)}{\bar{\omega}_{\uGW,\uin} (q)}
            + \frac{3}{5} \Fnl \bigl(6 - n_{\uGW} (\nu)\bigr)
        \biggr]
        \int \frac{\ud^{3}\bk\,\ud^{3}\bp}{(2\pi)^{3}} e^{i\bk\cdot\bx} \zeta_{\ugL}(\bp) \zeta_{\ugL}(\bk - \bp)\ ,\label{eq:delta0-2}\\
    \delta_{\uGW,0}^{(3)}(\bq) 
        &=& \biggl[
            \frac{\omega_{\ung,\uin}^{(3)} (q)}{\bar{\omega}_{\uGW,\uin} (q)}
            + \frac{9}{25} \Gnl \bigl(6 - n_{\uGW} (\nu)\bigr)
        \biggr]
        \int \frac{\ud^{3}\bk\,\ud^{3}\bp\,\ud^{3}\bl}{(2\pi)^{9/2}} e^{i\bk\cdot\bx} \zeta_{\ugL}(\bl) \zeta_{\ugL}(\bp - \bl) \zeta_{\ugL}(\bk - \bp)\ .\nonumber\\ \label{eq:delta0-3}
\end{eqnarray}
Consequently, the angular correlation functions of $\delta_{\uGW,0} (\bq)$, which encapsulate the statistical information regarding anisotropies in \acp{SIGW}, can be expressed in terms of the correlators of $\zeta_{\ugL}$. 
Furthermore, we offer a supplementary explanation regarding other propagation effects. 
The \ac{ISW} effect is relatively less significant than the \ac{SW} effect for $\ell \lesssim 10^2$, as demonstrated in \cite{Bartolo:2019zvb,ValbusaDallArmi:2020ifo}. 
The gravitational lensing effect is of higher order compared to the \ac{SW} effect \cite{Bartolo:2019oiq,Bartolo:2019yeu} and is anticipated to primarily influence $\delta_{\uGW,0}$ at higher multipoles \cite{Dodelson:2003ft}. 
If necessary, our analysis can be readily extended to incorporate all these effects.

As a result, we are able to characterize the isotropy of \ac{SIGW} background using $\bar{\Omega}_{\uGW,0}$, while the statistics of the anisotropies in the \ac{SIGW} background are described through the angular correlations of $\delta_{\uGW,0}$. 
All of these quantities can be expressed in terms of the multi-point correlators of $\zeta$. 
Although the physics underlying the production and propagation of \acp{SIGW} has been clarified, a significant challenge remains in calculating the multi-point correlators of $\zeta$ for non-Gaussian cases. 
Addressing this issue is essential for establishing a association between the \ac{SIGW} background and inflationary physics. 
Our proposed solution to this problem will be presented in the following subsection.

\subsection{Feynman-like diagrammatic approach}\label{subsec:FD-approach}

\begin{figure*}[htbp]
    \centering
    \includegraphics[width =\textwidth]{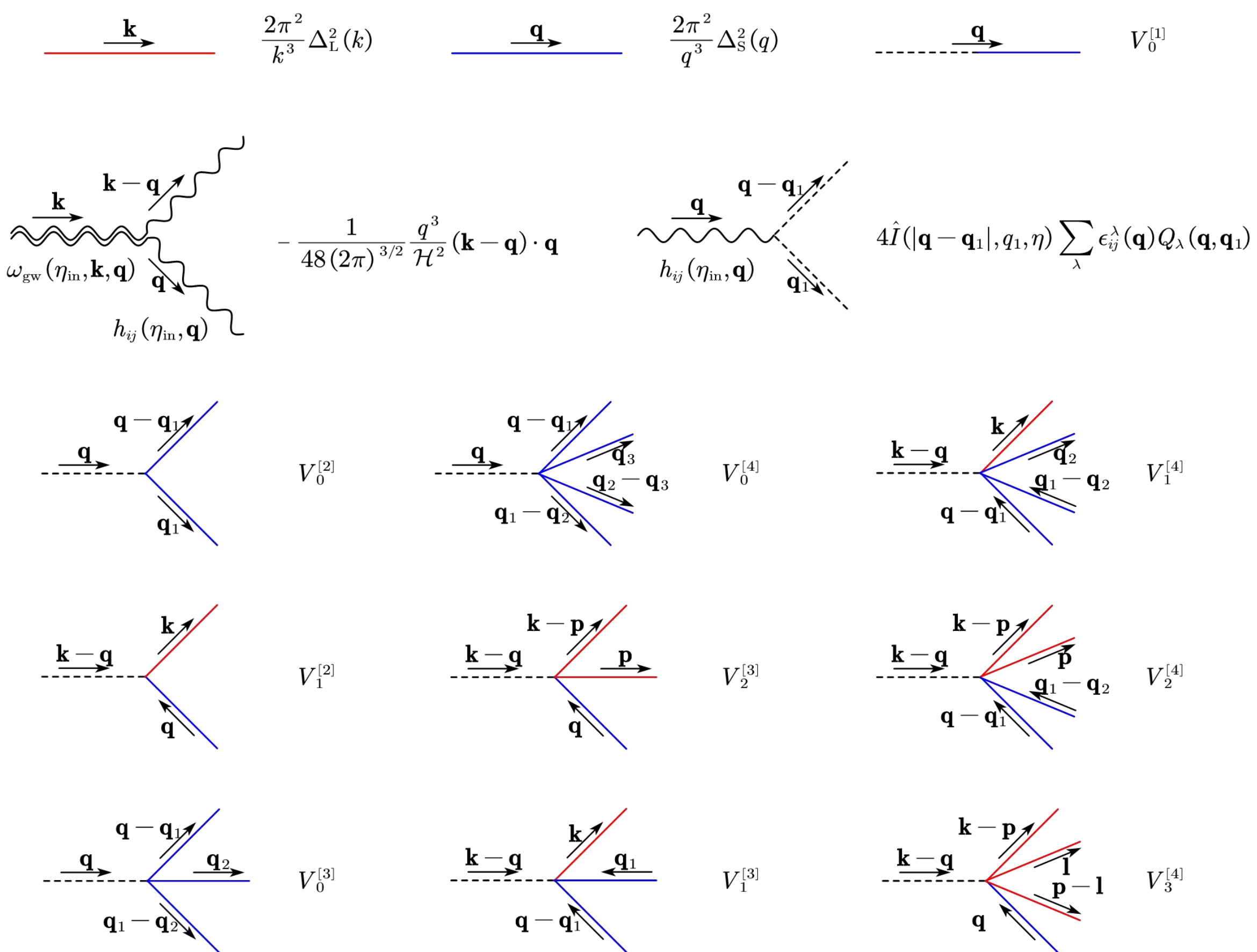}
    \caption{ Feynman-like rules for evaluation of Eq.~(\ref{eq:omegabar-zeta}) by incorporating Eq.~(\ref{eq:Fnl-Gnl-Hnl-def}). Wavy lines represent GWs, double wavy lines denote $\omega_{\uGW}$, dashed lines denote the transfer functions, and (red/blue) solid lines stand for the (long-/short-wavelength) primordial curvature power spectra. Arrows indicate the flow of comoving 3-momenta, which are conserved at each vertex. The total 3-momentum vanishes for each individual diagram, and all the loop 3-momenta should be integrated over. 
}\label{fig:FD-Rules}
\end{figure*}

As elaborated in the previous two subsections, the present-day energy-density fraction spectrum $\bar{\Omega}_{\uGW,0}$ of \acp{SIGW} is expressed in terms of $\bar{\omega}_{\uGW,\uin}$ in Eq.~\eqref{eq:Omega0}, while the present-day density contrast $\delta_{\uGW,0} (q)$ for \acp{SIGW} is formulated in terms of $\omega_{\ung,\uin}^{(1)} (q)$, $\omega_{\ung,\uin}^{(2)} (q)$, and $\omega_{\ung,\uin}^{(3)} (q)$ in Eqs.~(\ref{eq:delta0-123} - \ref{eq:delta0-3}). 
To derive their explicit expressions, the principal challenge lies in evaluating the numerous multi-point correlators of $\zeta_{\ug}$, as presented in Eqs.~(\ref{eq:omega-expand}, \ref{eqs:omega-result}). 
In the context of local-type \ac{PNG} introduced in Eq.~\eqref{eq:Fnl-Gnl-Hnl-def} up to quartic approximation (i.e., $\Hnl$ order), the calculation of $\bar{\omega}_{\uGW,\uin}$ necessitates addressing up to 16-point correlators of $\zeta_{\ugS}$. 
Furthermore, the calculation of the angular two-, three-, and four-point correlators of $\bar{\omega}_{\uGW,\uin}$ requires up to 14-point correlators of $\zeta_{\ugS}$ as well as 9-point correlators of $\zeta_{\ugL}$. 
Although Wick's theorem can be applied to facilitate the contractions of these correlators, the large number of two-point correlators involved presents a significant challenge.

Fortunately, the diagrammatic approach has proven to be a helpful tool for addressing this challenge in previous works \cite{Unal:2018yaa, Atal:2021jyo, Adshead:2021hnm,Ragavendra:2021qdu,Abe:2022xur,Li:2023qua,Li:2023xtl,Perna:2024ehx, Ruiz:2024weh}. 
This approach is a classical method for handling multi-point correlation functions of non-Gaussian fields. 
Its application to primordial curvature perturbations was initiated by Ref.~\cite{Byrnes:2007tm}.
Subsequent works, Refs.~\cite{Unal:2018yaa, Atal:2021jyo}, began applying it to compute the \ac{SIGW} energy density with \ac{PNG}. 
A milestone was Ref.~\cite{Adshead:2021hnm}, which first established a complete set of Feynman-like rules for calculating the \ac{SIGW} fractional energy spectrum at $\fnl$ order. 
This framework was subsequently applied by Ref.~\cite{Ragavendra:2021qdu} to specific inflationary models, while Ref.~\cite{Li:2023xtl} extended the diagrammatic approach to compute the \ac{SIGW} energy density fraction in the presence of $\gnl$-order \ac{PNG}. 
Further developments were made in Refs.~\cite{Abe:2022xur, Perna:2024ehx}, where the diagrammatic approaches were employed to systematically investigate contributions to the \ac{SIGW} energy density fraction up to $\mathcal{O}(A_S^4)$ for polynomially expanded \ac{PNG}. 
Recently, Ref.~\cite{Ruiz:2024weh} adapted these Feynman-like rules to address the renormalization of the one-loop scalar power spectrum in ultra-slow-roll inflation models \cite{Ballesteros:2024zdp}. 
A comparison of the number of diagrams involved in these various works is presented in Appendix~\ref{sec:compare}. 
Further, our earlier works \cite{Li:2023qua, Li:2023xtl, Li:2024zwx} advanced the diagrammatic approach by extending it to study inhomogeneities induced by \ac{PNG}, enabling the treatment of angular correlations in $\omega_{\mathrm{GW},\mathrm{in}}$. 
This extension was conceptually inspired by Ref.~\cite{Bartolo:2019zvb}, which used Feynman-like diagrams schematically to illustrate how \ac{PNG} may induce angular anisotropies of the \ac{SIGW} background. 

Similar to these works, we introduce appropriate Feynman-like rules (see \cref{fig:FD-Rules}), which then enable the diagrammatic representation of $\omega_{\uGW,\uin}$ across various contraction terms. 
In this set of Feynman-like rules, wavy lines denote the \ac{GW} strain, while double wavy lines represent $\omega_{\uGW}$. 
Dashed lines indicate the transfer functions, and (red/blue) solid lines stand for the propagators of (long-/short-wavelength) primordial curvature perturbations, corresponding to their power spectra. 
Arrows adjacent to these lines signify the flow of comoving 3-momenta, which are conserved at each vertex.
It is important to note that, as discussed in Subsection~\ref{subsec:SIGW&PNG}, we will disregard the 3-momenta contributed by the red lines when calculating the sum of all momenta at any vertex, since they are significantly smaller than those contributed by the blue lines. 
The sum of all momenta of $\zeta_{\ugS}$ at any vertex results in zero due to momentum conservation. 
Furthermore, the total 3-momentum for each individual diagram vanishes, and all loop 3-momenta should be integrated over. 
In particular, while vertices with identical solid-line counts but distinct red-line counts are treated as distinct entities, we assign them the same \ac{PNG} parameters. 
This treatment is consistent, as our present analysis is confined to the scale-independent \ac{PNG} scenario.
Specifically, according to \cref{fig:FD-Rules}, while the Gaussian-vertex means $V_0^{[1]}=1$ by definition, we introduce the $\Fnl$-vertices with $V_0^{[2]}=V_1^{[2]}=\Fnl$, the $\Gnl$-vertices with $V_0^{[3]}=V_1^{[3]}=V_2^{[3]}=\Gnl$, and the $\Hnl$-vertices with $V_0^{[4]}=V_1^{[4]}=V_2^{[4]}=V_3^{[4]}=\Hnl$. 
Utilizing these Feynman-like rules, we can draw diagrams that represent $\omega_{\uGW, \uin}$ and its multi-point correlators. 
The various configurations of solid lines correspond to different contraction terms of the correlators of $\zeta$. 
This visualization highlights the symmetry among certain contraction terms, which significantly simplifies the computation of numerous contraction terms performed directly using Wick's theorem.

\begin{figure}[htbp]
    \centering
    \includegraphics[width = 0.3 \columnwidth]{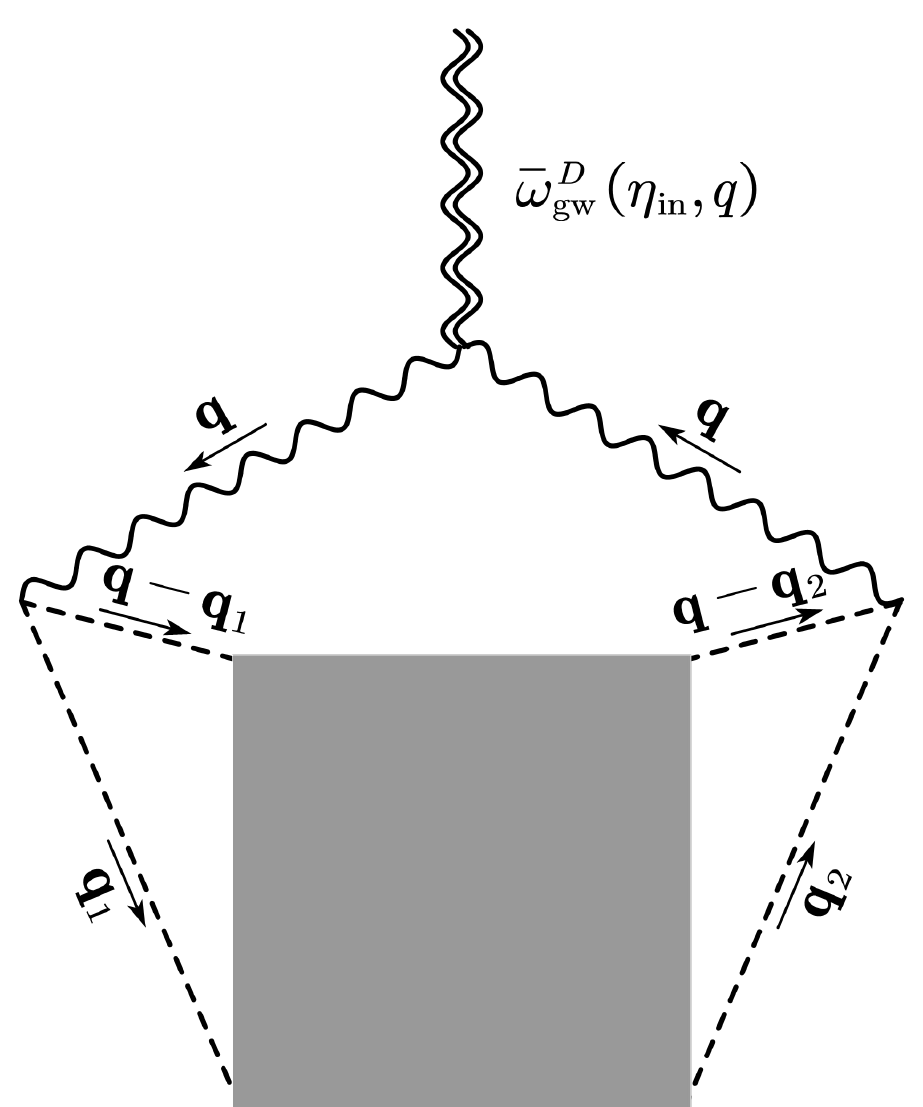}
    \caption{Common structure of the diagrams for the SIGW average energy-density spectrum $\bar{\omega}_{\uGW,\uin}$, where the shaded square should be replaced with one of the panels from \cref{fig:G-like-FD} - \cref{fig:PN-like-FD} for a specific diagram. }\label{fig:omegabar-FD_Frame}
\end{figure}

\begin{figure*}[htbp]
    \centering
    \includegraphics[width =0.32 \textwidth]{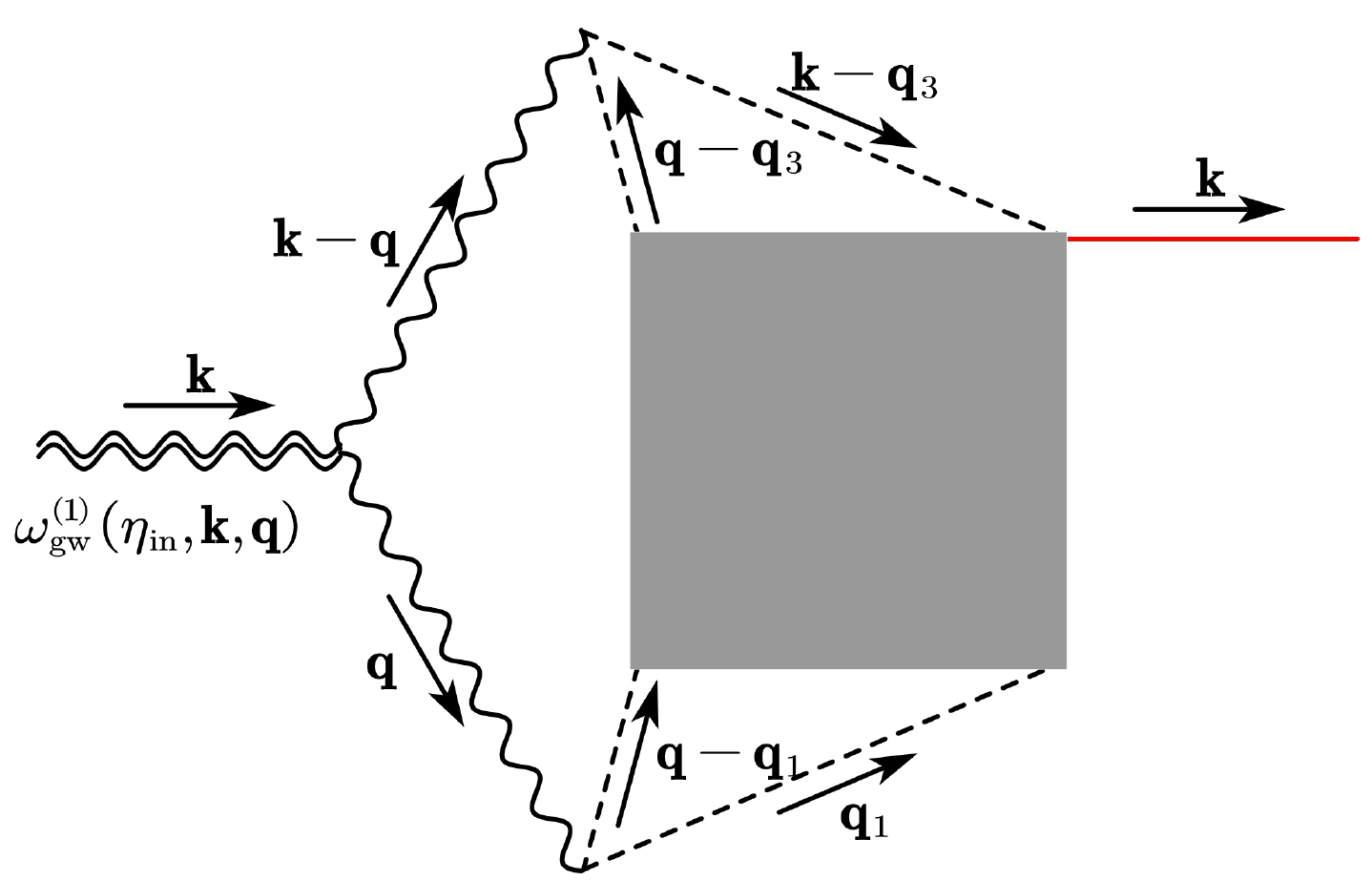}
    \hfil
    \includegraphics[width =0.32 \textwidth]{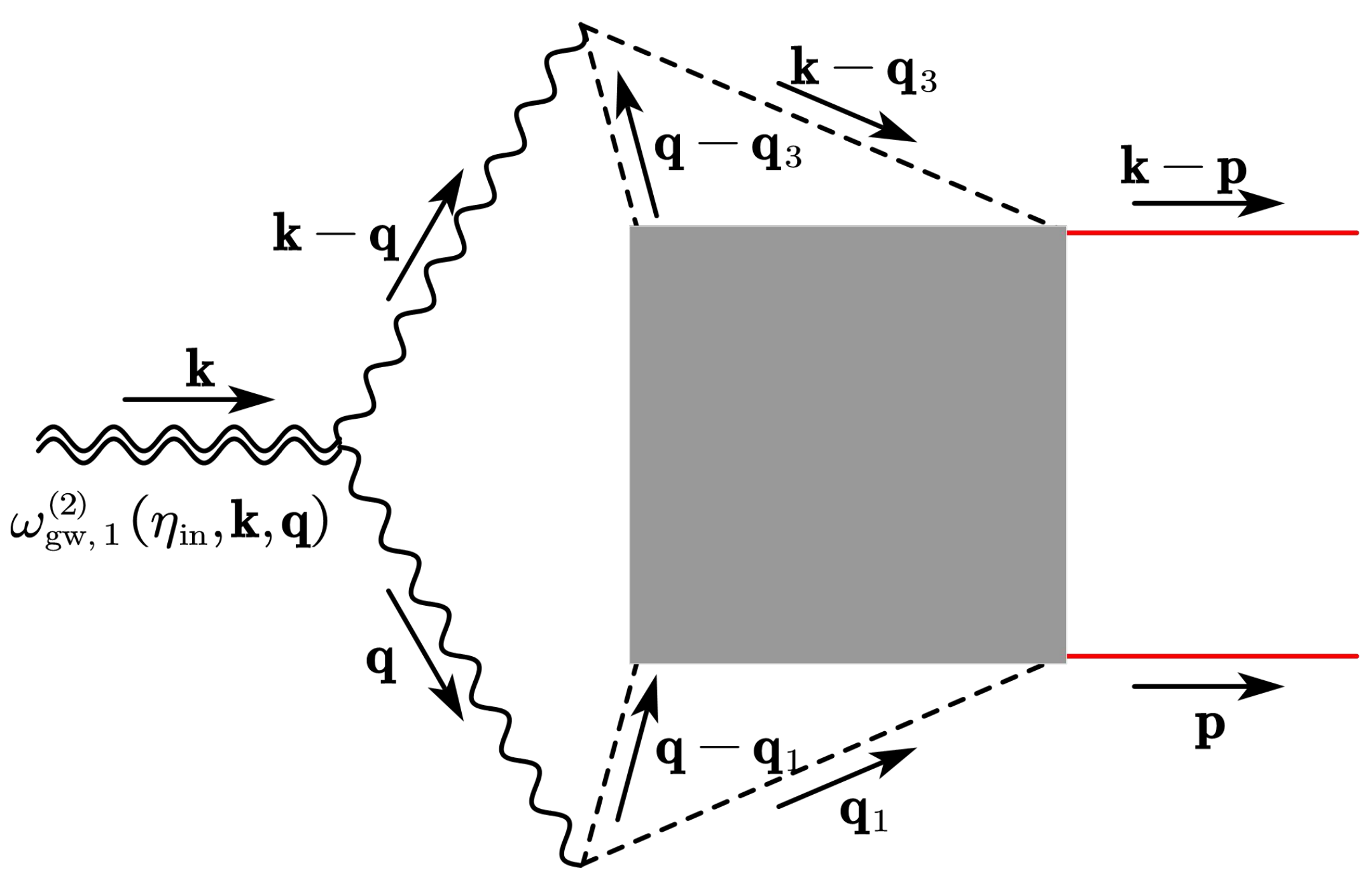}
    \hfil
    \includegraphics[width =0.32 \textwidth]{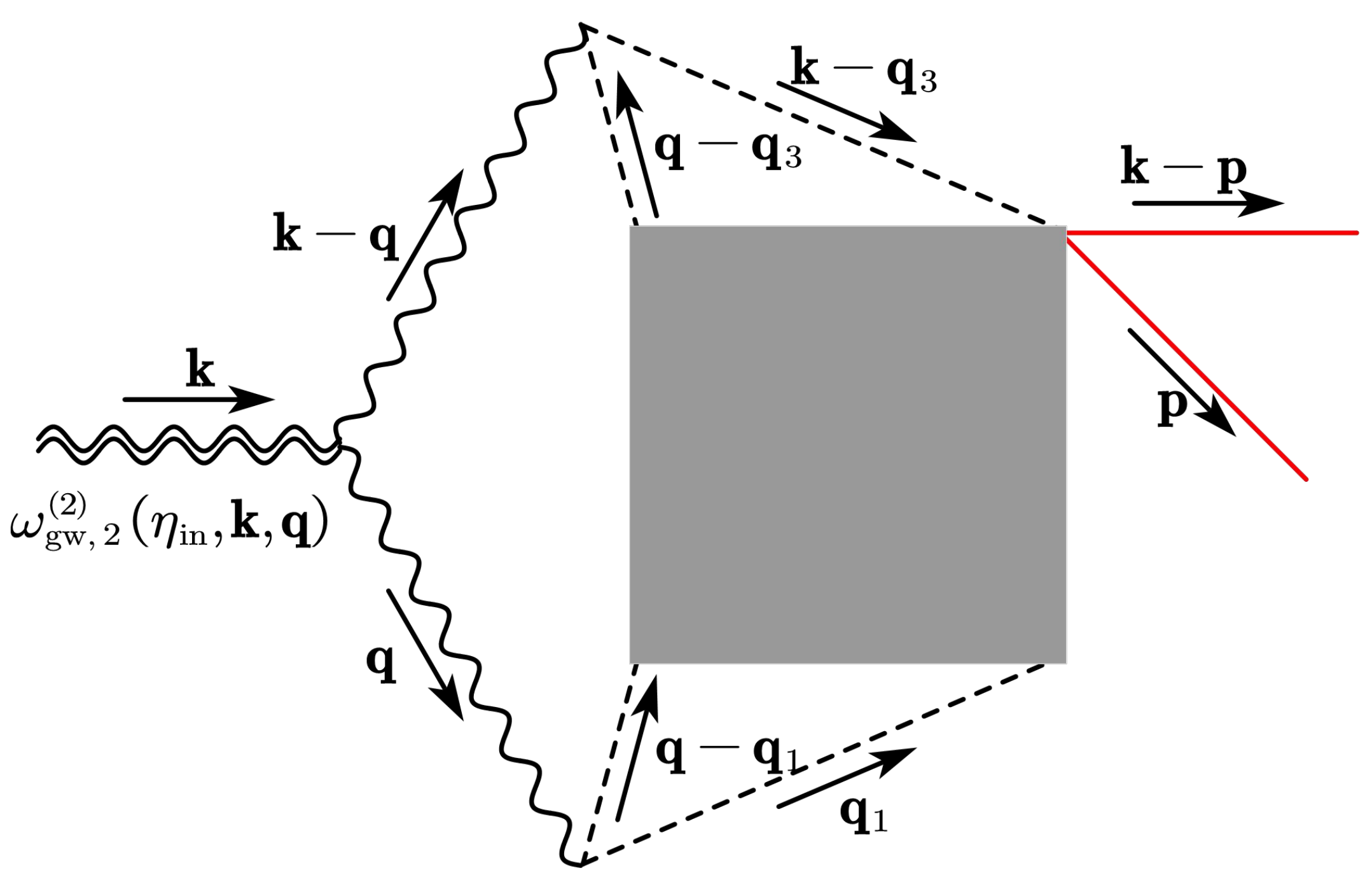}
    \hfil \\ \vspace{0.3cm}
    \includegraphics[width =0.32 \textwidth]{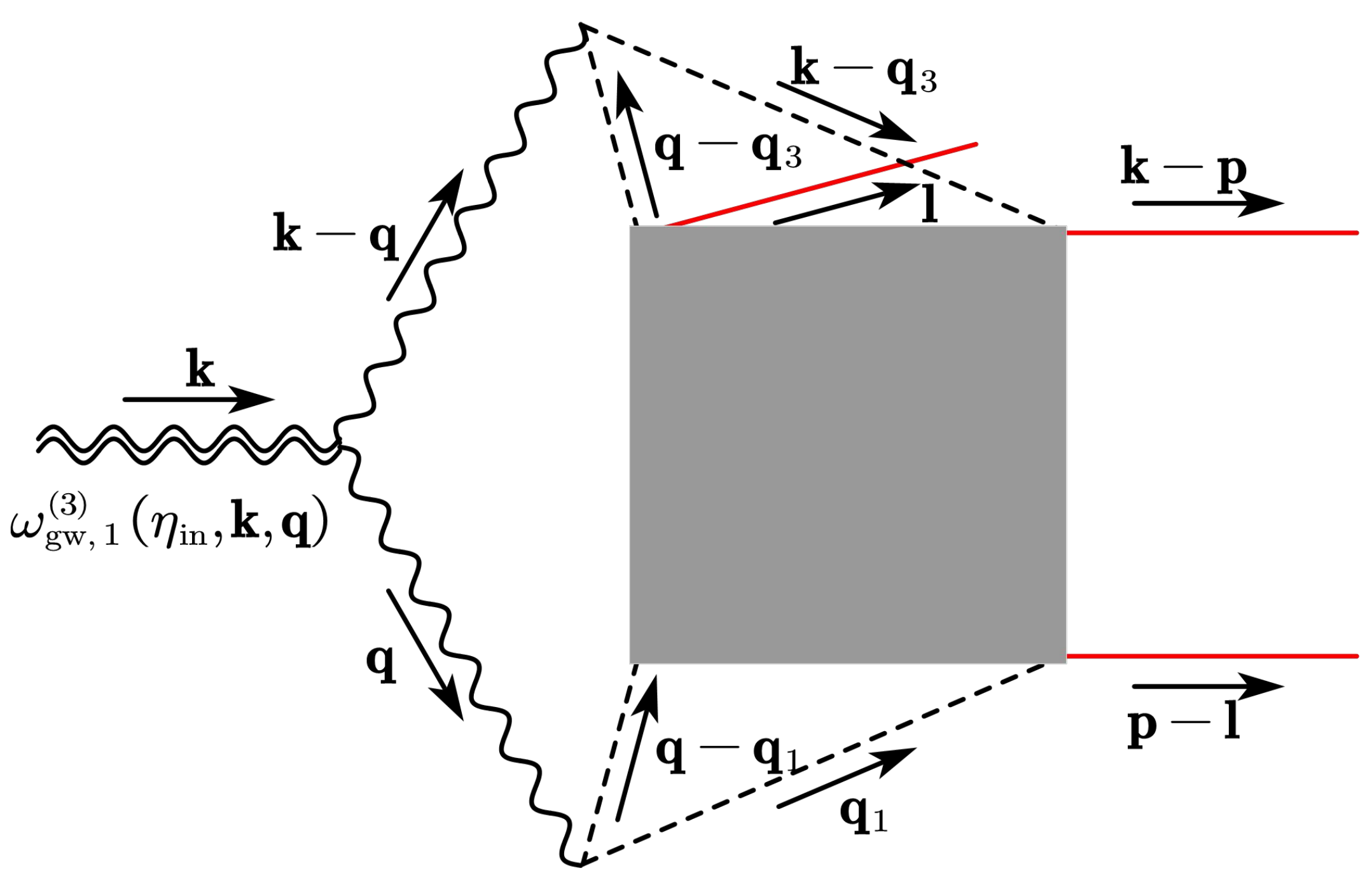}
    \hfil
    \includegraphics[width =0.32 \textwidth]{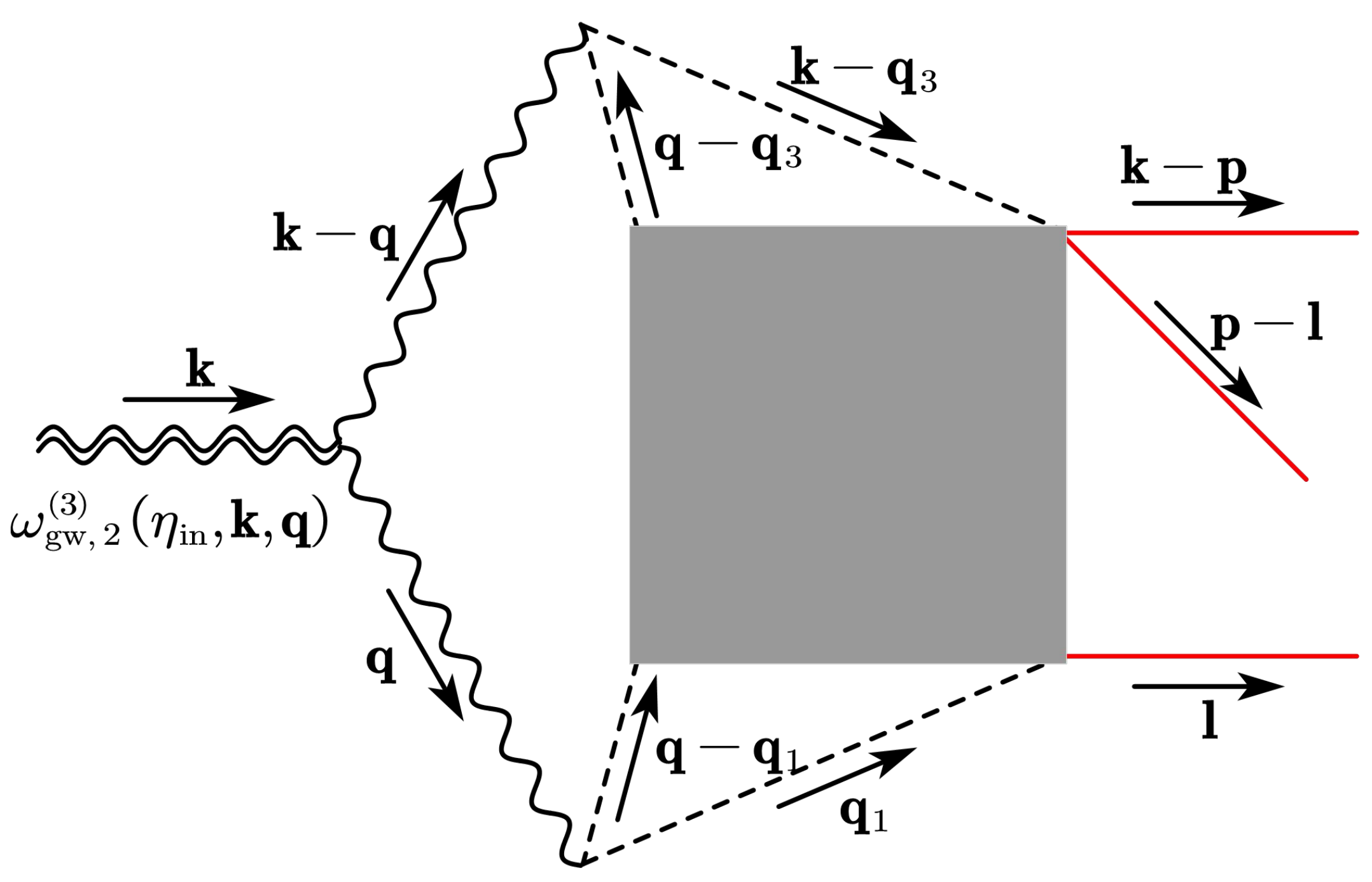}
    \hfil
    \includegraphics[width =0.32 \textwidth]{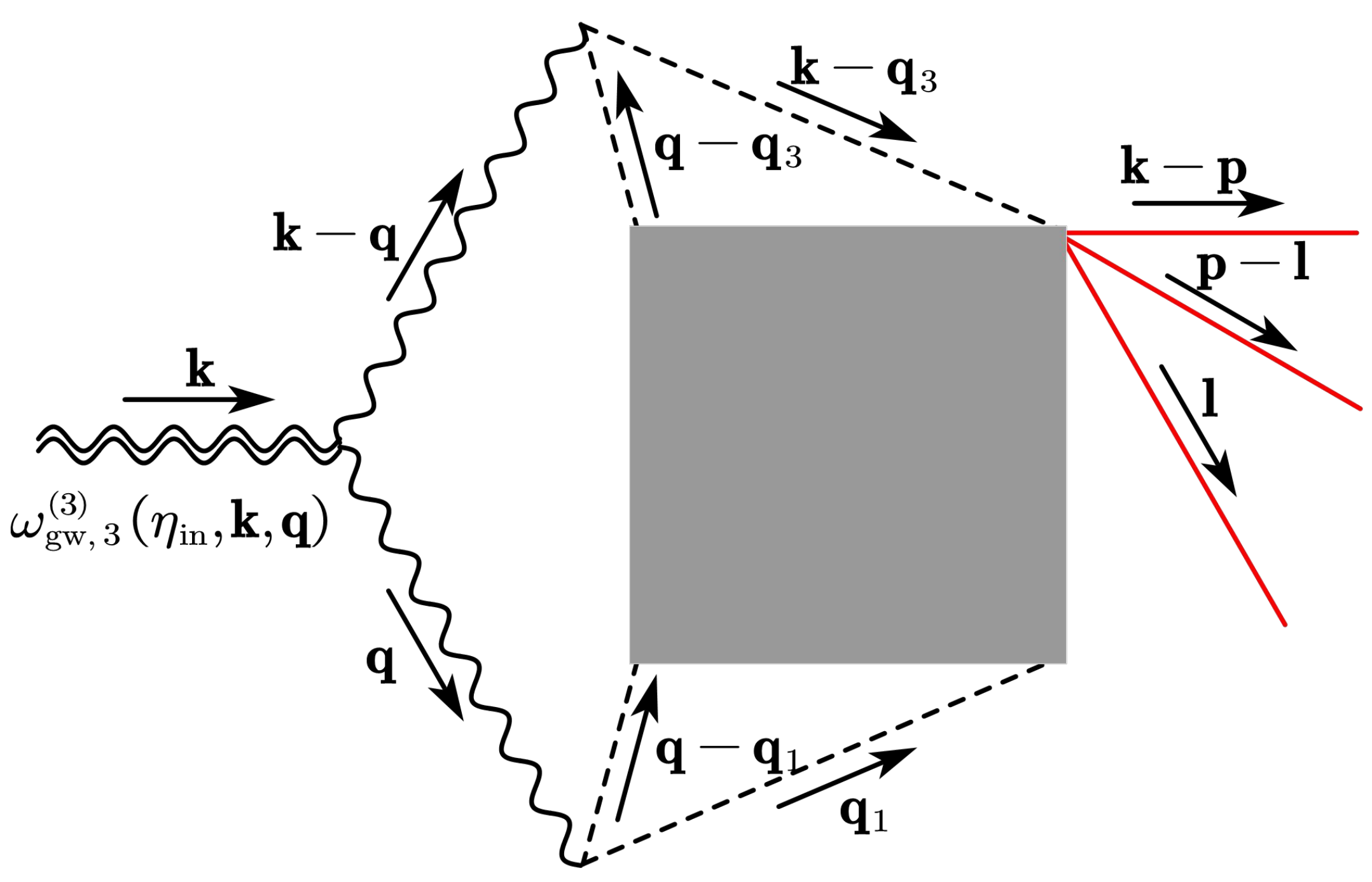}
    \caption{
    \Ac{PNG}-induced inhomogeneities of $\omega_{\uGW,\uin} (\bq)$ up to third order in $\zeta_{\ugL}$, where the shaded squares should be replaced with one of the panels from \cref{fig:G-like-FD} to \cref{fig:PN-like-FD} for a specific diagram according to \cref{tab:order-FD}. From left to right and top to bottom, the six panels represent $\omega_{\uGW,\uin}^{(1)} (\bq)$, $\omega_{\uGW,\uin,1}^{(2)} (\bq)$, $\omega_{\uGW,\uin,2}^{(2)} (\bq)$, $\omega_{\uGW,\uin,1}^{(3)} (\bq)$, $\omega_{\uGW,\uin,2}^{(3)} (\bq)$, and $\omega_{\uGW,\uin,3}^{(3)} (\bq)$, respectively. }\label{fig:FD_Frame}
\end{figure*}

Utilizing the diagrammatic rules in \cref{fig:FD-Rules}, we construct the common structure of the diagrams for $\bar{\omega}_{\uGW,\uin}$ in \cref{fig:omegabar-FD_Frame} for conciseness, where the shaded square covers the various configurations of the blue solid lines introduced in \cref{fig:FD-Rules}. 
The shaded square should be replaced with panels from \cref{fig:G-like-FD} to \cref{fig:PN-like-FD} in Appendix~\ref{sec:FD}, whereas the detailed calculation process for all diagrams pertaining to $\bar{\omega}_{\uGW,\uin}$ is presented in Section~\ref{sec:Omegabar}. 
We further present the diagrams for the \ac{PNG}-induced inhomogeneities of $\omega_{\uGW,\uin} (\bq)$ on large-scales, as shown in \cref{fig:FD_Frame}. 
These six diagrams correspond to attaching extensional red solid lines to the diagrams of $\bar{\omega}_{\uGW,\uin}$ in \cref{fig:omegabar-FD_Frame}. 
The upper left panel depicts the diagram of the 1st-order \ac{PNG}-induced inhomogeneity $\omega_{\uGW,\uin}^{(1)}$ in $\zeta_{\ugL}$, as defined in Eq.~\eqref{eq:omega-1}. 
The other two upper panels, labeled $\omega_{\uGW,\uin,1}^{(2)} (\bq)$ and $\omega_{\uGW,\uin,2}^{(2)} (\bq)$ from left to right, represent the diagrams of the two components of the 2nd-order \ac{PNG}-induced inhomogeneity $\omega_{\uGW,\uin}^{(2)}$  in $\zeta_{\ugL}$, as defined in Eq.~\eqref{eq:omega-2}. 
The three lower panels, labeled $\omega_{\uGW,\uin,1}^{(3)} (\bq)$, $\omega_{\uGW,\uin,2}^{(3)} (\bq)$, and $\omega_{\uGW,\uin,3}^{(3)} (\bq)$ from left to right, illustrate the diagrams of the three components of the 3rd-order \ac{PNG}-induced inhomogeneity $\omega_{\uGW,\uin}^{(3)}$ in $\zeta_{\ugL}$, as defined in Eq.~\eqref{eq:omega-3}. 
In other words, the 2nd- and 3rd-order \ac{PNG}-induced inhomogeneities are the sums of all related components, expressed as 
\begin{eqnarray}
    \omega_{\uGW,\uin}^{(2)} (\bq) &=& \omega_{\uGW,\uin,1}^{(2)} (\bq) + \omega_{\uGW,\uin,2}^{(2)} (\bq) \ ,\\
    \omega_{\uGW,\uin}^{(3)} (\bq) &=& \omega_{\uGW,\uin,1}^{(3)} (\bq) + \omega_{\uGW,\uin,2}^{(3)} (\bq) + \omega_{\uGW,\uin,3}^{(3)} (\bq) \ .
\end{eqnarray}
It is important to note that, although the extensional red solid lines are attached to vertices on the same side in the diagrams for $\omega_{\uGW,\uin,2}^{(2)}$ and $\omega_{\uGW,\uin,2}^{(3)}$ in \cref{fig:FD_Frame}, this is not a necessary requirement. 
Any two vertices can be utilized as long as, after attaching the red solid lines, the total number of solid lines attached to these vertices does not exceed four. 
Similar to \cref{fig:omegabar-FD_Frame}, the shaded squares represent the configurations of blue solid lines, which correspond to the correlators of $\zeta_{\ugS}$ in Eq.~\eqref{eq:omega-expand}. 
These shaded squares should be replaced with specific squares (or equivalent ones) after a $90^\circ$ rotation according to the Gaussian-vertex count, $\Fnl$-vertex count, $\Gnl$-vertex count, and $\Hnl$-vertex count in each panel from \cref{fig:G-like-FD} to \cref{fig:PN-like-FD} in Appendix~\ref{sec:FD}. 
Specifically, for any panel from \cref{fig:G-like-FD} to \cref{fig:PN-like-FD}, a Gaussian-vertex can attach up to three red solid lines, a $\Fnl$-vertex can have two, a $\Gnl$-vertex can have one, and a $\Hnl$-vertex cannot attach any red solid lines. 
Furthermore, we provide a table that indicates whether a certain square from \cref{fig:G-like-FD} to \cref{fig:PN-like-FD} can be used to replace these shaded squares.

Calculating the \ac{PNG}-induced inhomogeneities is quite tedious if we replace the squares individually. 
However, based on the results for $\bar{\omega}_{\uGW,\uin}$, we can derive explicit expressions for $\omega_{\ung,\uin}^{(1)}$, $\omega_{\ung,\uin}^{(2)}$, and $\omega_{\ung,\uin}^{(3)}$ in a simpler manner. 
We leverage the fact that the respective numbers of vertices in a specific diagram $D$ correspond to the powers of the \ac{PNG} parameters to which $\bar{\omega}_{\uGW,\uin}^D$ is proportional. 
For convenience, we classify $\bar{\omega}_{\uGW,\uin}$ into 19 distinct categories based on the powers of the three \ac{PNG} parameters. 
Each category, denoted as $\bar{\omega}_{\uGW,\uin}^{(a,b,c)}$, represents the component proportional to $\Fnl^a \Gnl^b \Hnl^c A^{(a+2b+3c+4)/2}$, where $a$, $b$, and $c$ are natural numbers. 
Specifically, the classification can be expressed as
\begin{equation}\label{eq:omegabar-abc-total}
    \bar{\omega}_{\uGW,\uin} (q) = \sum_{(a,b,c)} \bar{\omega}_{\uGW,\uin}^{(a,b,c)} (q)\ ,
\end{equation}
where the superscripts $a$, $b$, and $c$ satisfy the conditions $a+b+c \leq 4$ and $a,b,c,(a+c)/2\in\mathbb{N}$. 
The requirement that $(a+c)$ is even reflects a sign degeneracy between $\Fnl$ and $\Hnl$. 
We can then ascertain the numbers of each vertex type: the number of $\Fnl$-vertices (denoted as $V_0^{[2]}$) is $N_{2} = a$, the number of $\Gnl$-vertices (denoted as $V_0^{[3]}$) is $N_{3} = b$, and the number of $\Hnl$-vertices (denoted as $V_0^{[4]}$) is $N_{4} = c$.
The relation $\bar{\omega}_{\mathrm{GW},\mathrm{in}} \sim \langle\zeta_{\mathrm{gS}}^4\rangle$ indicates that a total of four vertices are involved in the propagators of $\zeta_{\mathrm{gS}}$, yielding the Gaussian-vertex count $N_{\mathrm{1}} = 4 - a - b - c$.
With these counts determined, we can assign the corresponding categories to the six replacement squares (shaded) in \cref{fig:FD_Frame}. 
For the shaded squares in the diagram of $\omega_{\uGW,\uin}^{(1)}$, the replacement squares must belong to categories where $c \leq 3$. 
Similarly, for $\omega_{\uGW,\uin,1}^{(2)}$, it requires $c \leq 2$, and for $\omega_{\uGW,\uin,1}^{(3)}$, $c \leq 1$. 
In the case of $\omega_{\uGW,\uin,2}^{(2)}$, the replacement squares should belong to categories where $b+c \leq 3$. 
The conditions for the categories that can replace the shaded squares of $\omega_{\uGW,\uin,2}^{(3)}$ are $b+c \leq 3$ and $c \leq 2$. 
It is necessary for the replacement squares related to $\omega_{\uGW,\uin,3}^{(3)}$ to contain at least one Gaussian-vertex, associated with categories where $a+b+c \leq 3$. 
The specific results for the replacements across all categories are also provided in \cref{tab:order-FD}. 
The categories contributing to the large-scale modulations $\omega_{\ung,\uin}^{(1)}$, $\omega_{\ung,\uin}^{(2)}$, and $\omega_{\ung,\uin}^{(3)}$, along with the coefficients for these categories, are discussed in detail in Section~\ref{sec:Cl} and Section~\ref{sec:bl&tl}.

%% file: TeX/3Isotropy.tex
\section{Isotropic component of the SIGW background}\label{sec:Omegabar}

In this section, we will provide a detailed derivation of the \ac{SIGW} energy-density fraction spectrum for \ac{PNG} up to all orders. 
In particular, we will present the theoretical results for \ac{PNG} up to $\Hnl$ order and elaborate on the associated numerical findings. 

\subsection{``Renormalized'' Feynman-like rules}\label{subsec:Renorm-FD}

In the previous section, we have discussed the calculation of the average energy-density spectrum $\bar{\omega}_{\uGW,\uin} (q)$ using a diagrammatic approach based on the Feynman-like rules illustrated in \cref{fig:FD-Rules}. 
Previous studies \cite{Adshead:2021hnm,Li:2023qua} indicate that for \ac{PNG} up to quadratic approximation in $\zeta_{\ugS}$ ($\Fnl$ order), a total of 7 diagrams are required to capture all contractions in the calculation of $\bar{\omega}_{\uGW,\uin} (q)$. 
For \ac{PNG} up to cubic approximation (or $\Gnl$ order), 49 diagrams are necessary \cite{Li:2023xtl}. 
However, deriving $\bar{\omega}_{\uGW,\uin} (q)$ for \ac{PNG} up to quartic approximation (or $\Hnl$ order) necessitates considering 236 diagrams, as shown in \cref{fig:G-like-FD} - \cref{fig:PN-like-FD} in Appendix~\ref{sec:FD}. 
Should this study be extended to include higher orders of \ac{PNG}, a substantially larger number of diagrams will be required, corresponding to a significantly greater workload. 
Therefore, it is crucial to investigate the patterns among the diagrams in \cref{fig:G-like-FD} - \cref{fig:PN-like-FD} in Appendix~\ref{sec:FD} to alleviate this workload. 

\begin{figure*}
    \centering
    \includegraphics[width = \textwidth]{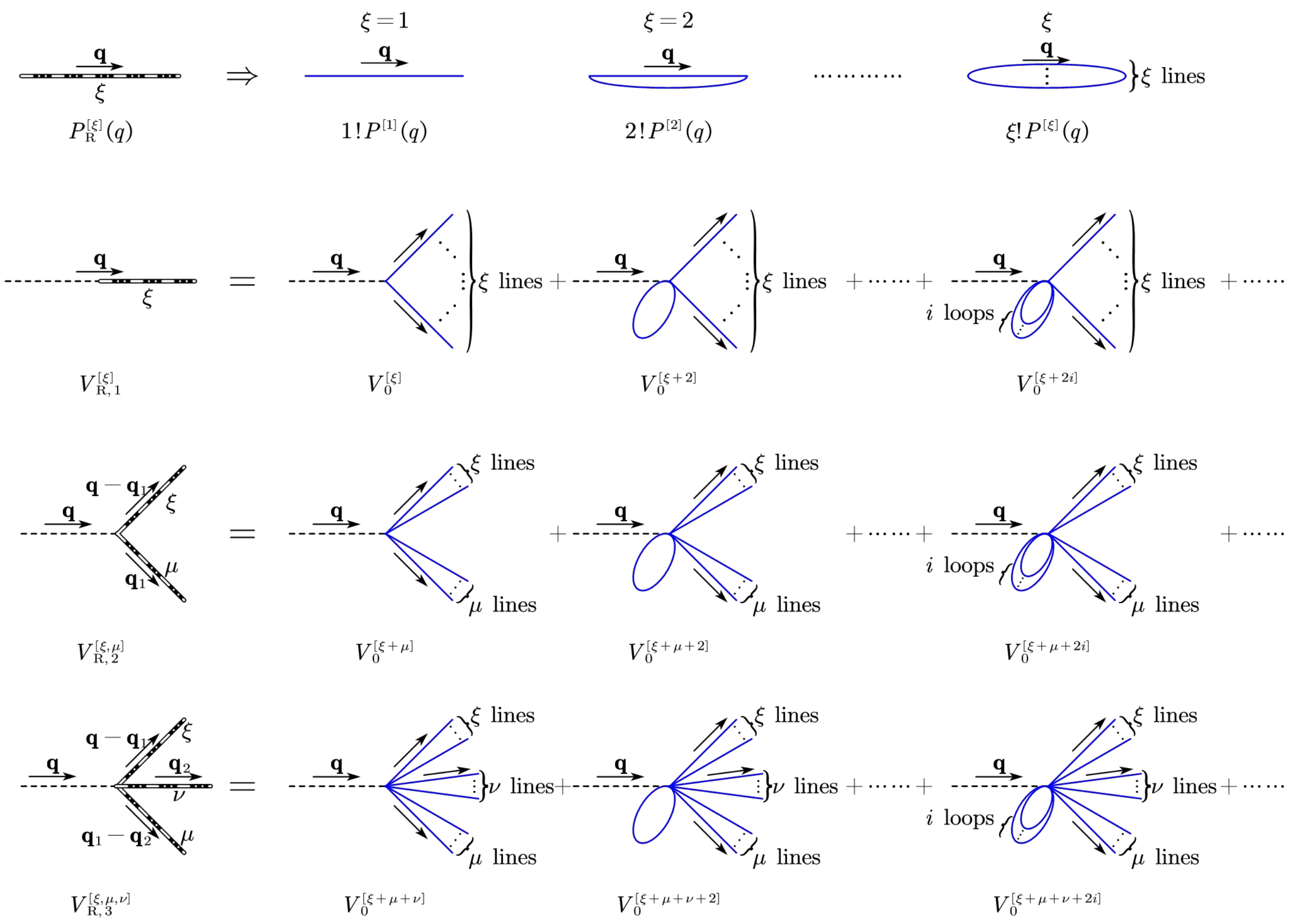}
    \caption{  ``Renormalized'' Feynman-like rules. }\label{fig:FD-Rules-new}
\end{figure*}

By comparing these diagrams, we find that among the hundreds of diagrams, they can be classified into nine distinct topologically independent families, which we denote as ``$G$-like'', ``$C$-like'', ``$Z$-like'', ``$P$-like'', ``$N$-like'', ``$CZ$-like'', ``$PZ$-like'', ``$NC$-like'', and ``$PN$-like'', i.e., 
\begin{eqnarray}\label{eq:omegabar-X-total}
    \bar{\omega}_{\uGW,\uin} (q)
    = \sum_X \bar{\omega}_{\uGW,\uin}^{X-\mathrm{like}} (q)\ ,
\end{eqnarray}
where $\bar{\omega}_{\uGW,\uin}^{X-\mathrm{like}} (q)$ denotes the contributions from each family to the average energy-density spectrum, with $X = G, C, Z, P, N, CZ, PZ, NC$, and $PN$. 
In the Feynman-like diagrams of the same family, the vertices are connected in similar ways, with differences lying in the loop structures at the vertices or the propagators of $\zeta_{\ugS}$. 
Leveraging this similarity, we propose the ``renormalized'' Feynman-like rules, which incorporate loop structures into the convolved propagators and the ``renormalized''  vertices, as illustrated in \cref{fig:FD-Rules-new}.
The convolved propagator, represented as a thick-dashed line with a Greek letter $\xi$ (or $\mu$, $\nu$) adjacent, consists of $\xi$ (or $\mu$, $\nu$) solid lines, where the Greek letter denotes a positive integer. 
The ``renormalized'' vertices attached by these thick-dashed lines represent the sum of the vertices that encompass all possible loop structures at both the propagators and the vertices themselves.
Additionally, to represent the four-point correlator $\langle\zeta_\uS(\bk-\bq-\bq_1)\zeta_\uS(\bq_1)\zeta_\uS(\bq-\bq_2)\zeta_\uS(\bq_2)\rangle$, any ``renormalized'' vertex can be attached by up to three thick-dashed lines. 
Hence, there are three distinct types of ``renormalized'' vertices classified by the number of attached thick-dashed lines, as illustrated in \cref{fig:FD-Rules-new}.
Introducing these ``renormalized'' Feynman-like rules allows us to perform the integration over loop momenta in advance, thereby reducing the dimensions of the multiple integrals and simplifying the calculation of symmetric factors. 
We will provide detailed explanations regarding these convolved propagators and ``renormalized'' vertices in the subsequent paragraphs.

The convolved propagator composed of $\xi$ propagators of $\zeta_\ugS$ is referred to as a $(\xi-1)$-fold convolved propagator, where $\xi$ is a positive integer. 
The $(\xi-1)$ loops in this convolved propagator correspond to the $(\xi-1)$-fold convolutions of the power spectrum of $\zeta_\ugS$. 
To maintain consistency, we introduce $P^{[\xi]} (q)$ for $\xi \in \mathbb{N}_+$ as follows 
\begin{eqnarray}
    P^{[1]} (q) &=& \frac{2\pi^2}{q^3} \Delta^2_\uS(q)\ ,\label{eq:P1-def}\\
    P^{[\xi]} (q) &=& \int \frac{\ud^3 \tilde{\bq}}{(2 \pi)^3} P^{[\xi-1]} (\tilde{q}) P^{[1]} (\abs{\bq - \tilde{\bq}})\ ,\ \text{where } \xi \geq 2, \xi \in \mathbb{N}_+\ .\label{eq:Pxi-def}
\end{eqnarray}
As illustrated in \cref{fig:FD-Rules-new}, the symmetric factor contributed by a $(\xi-1)$-fold convolved propagator is obtained by counting all possible connections of the blue solid lines, which is precisely the permutation of $\xi$. 
By multiplying this symmetric factor by $P^{[\xi]}$, we can express the $(\xi-1)$-fold convolved propagator as follows 
\begin{equation}\label{eq:P_redef}
    P_{\uR,a}^{[\xi]} = \xi! P_a^{[\xi]}\ ,
\end{equation}
where the subscript $_a$ denotes the momentum. 
This convention will facilitate the subsequent calculations involving the symmetric factor, in which we will employ identical subscripts to represent the momenta of the symmetric convolved propagators. 
Notably, in this context, the symmetry between two convolved propagators within a diagram indicates that an equivalent diagram can be obtained by interchanging these two convolved propagators.

We direct our focus to the ``renormalized'' vertices, each of which can be decomposed into vertices characterized by different numbers of ``self-closed loops''. 
For scale-independent \ac{PNG}, the loop integral over the momentum of any involved ``self-closed loop'' of $\zeta_\ugS$ yields 
\begin{equation}\label{eq:loop-int}
    \int\frac{\ud^3 \tilde{\bq}}{(2\pi)^3} P^{[1]} (q) = A_\uS \ .
\end{equation}
Moreover, $i$ ``self-closed loops'' at the vertex also transform the \ac{PNG} parameter into a $2i$-order higher one and correspondingly alter the symmetric factors. 
For example, when evaluating the symmetric factor contributed by the vertex $V_0^{[2i+\xi]}$ that includes $i$ ``self-closed loops'', the permutation $(2i)!$ arising from the loops must be divided by the overcount $2^i i!$. 
The overcount of $2^i$ accounts for coincident endpoints of the $i$ ``self-closed loops'', while the overcount of $i!$ arises from the equivalence among all ``self-closed loops''. 
Furthermore, if a $(\xi-1)$-fold convolved propagator connects this vertex to another vertex, such as those contributing to the ``renormalized'' vertex $V_{\uR,1}^{[\xi]}$ in \cref{fig:FD-Rules-new}, then the selection of $\xi$ blue solid lines attached to this vertex from a total of $(2i+\xi)$ also contributes to the overall symmetric factor. 
Similarly, for the vertex $V_0^{[2i+\xi+\mu]}$ that contributes to the ``renormalized'' vertex $V_{\uR,2}^{[\xi,\mu]}$, a $(\xi-1)$-fold convolved propagator connects this vertex to another vertex, a $(\mu-1)$-fold convolved propagator connects it to yet another vertex, and $2i$ solid lines form $i$ ``self-closed loops''. 
The symmetric factor can be obtained by multiplying the symmetric factor of vertex $V_0^{[2i+\xi]}$ by the number of ways to select $\mu$ from a total of $(2i+\xi+\mu)$.
The symmetric factor of vertex $V_0^{[2i+\xi+\mu+\nu]}$ that contributes to the ``renormalized'' vertex $V_{\uR,3}^{[\xi,\mu,\nu]}$, follows a similar pattern and can be obtained by multiplying the symmetric factor of the aforementioned vertex $V_0^{[2i+\xi+\mu]}$ by the number of ways to select $\nu$ from a total of $(2i+\xi+\mu+\nu)$. 
Thus, for the four-point correlator $\langle\zeta_\uS(\bk-\bq-\bq_1)\zeta_\uS(\bq_1)\zeta_\uS(\bq-\bq_2)\zeta_\uS(\bq_2)\rangle$, the relevant ``renormalized'' vertices for \ac{PNG} up to $o$-th order approximation is defined as follows 
\begin{subequations}\label{eqs:V_redef}
\begin{eqnarray}
    V_{\uR,1}^{[\xi]} &=& \sum_{i=0}^{\lfloor (o-\xi)/2 \rfloor} \frac{(2i)!}{2^i i!} \binom{2i+\xi}{\xi} A_\uS^i V_0^{[2i+\xi]}\ ,\\
    V_{\uR,2}^{[\xi,\mu]} &=& \sum_{i=0}^{\lfloor (o-\xi-\mu)/2 \rfloor} \frac{(2i)!}{2^i i!} \binom{2i+\xi+\mu}{\mu} \binom{2i+\xi}{\xi} A_\uS^i V_0^{[2i+\xi+\mu]}\ ,\\
    V_{\uR,3}^{[\xi,\mu,\nu]} &=& \sum_{i=0}^{\lfloor (o-\xi-\mu-\nu)/2 \rfloor} \frac{(2i)!}{2^i i!} \binom{2i+\xi+\mu+\nu}{\nu} \binom{2i+\xi+\mu}{\mu} \binom{2i+\xi}{\xi} A_\uS^i V_0^{[2i+\xi+\mu+\nu]}\ ,
\end{eqnarray}
\end{subequations}
where the symbols $\lfloor\cdots\rfloor$ represent the floor function, and the parentheses containing two numbers indicate combinations.

\begin{figure*}[htbp]
\centering
    \subcaptionbox*{$G$-like.}{\includegraphics[width = 0.25\columnwidth]{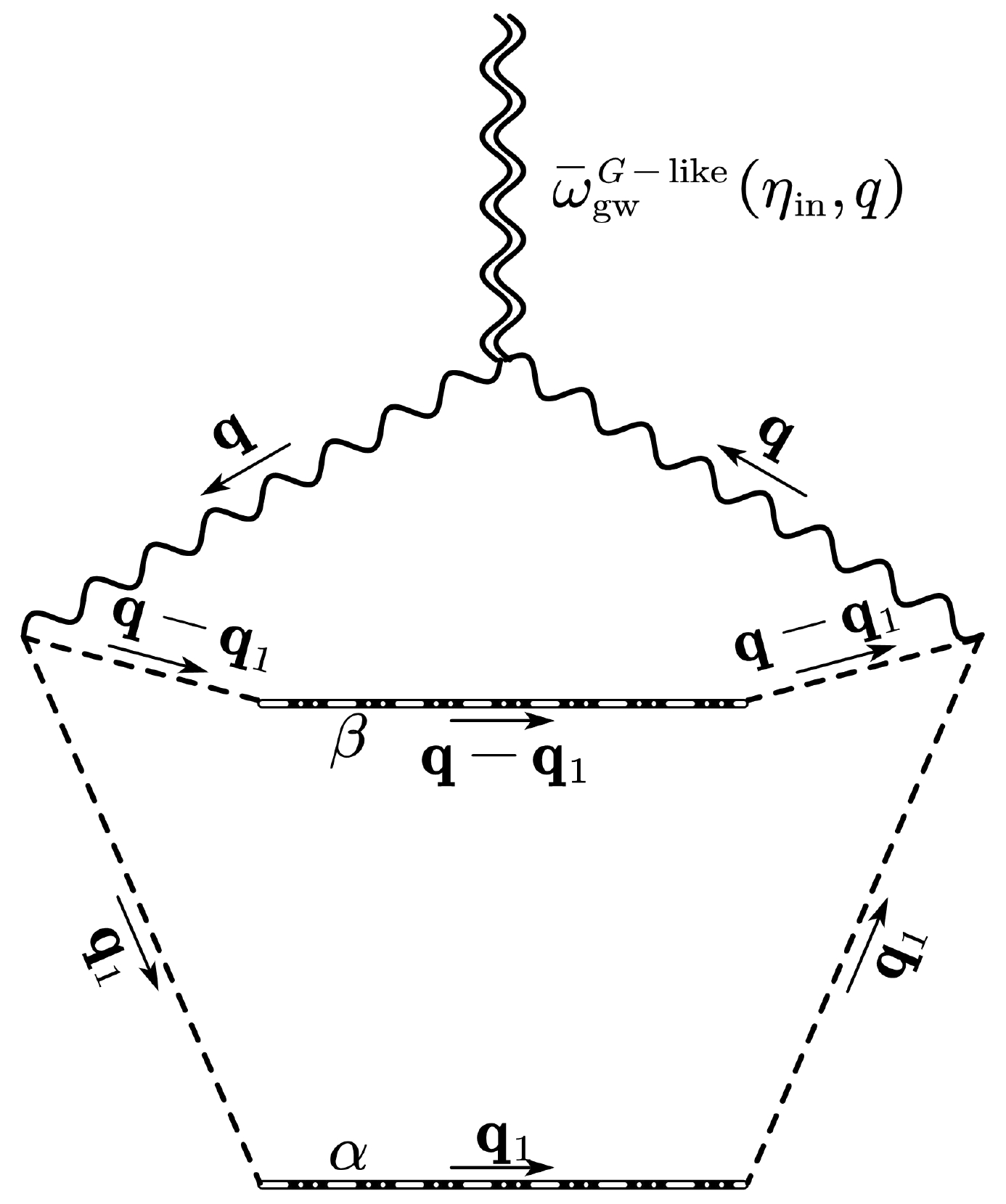}}\hfil
    \subcaptionbox*{$C$-like.}{\includegraphics[width = 0.25\columnwidth]{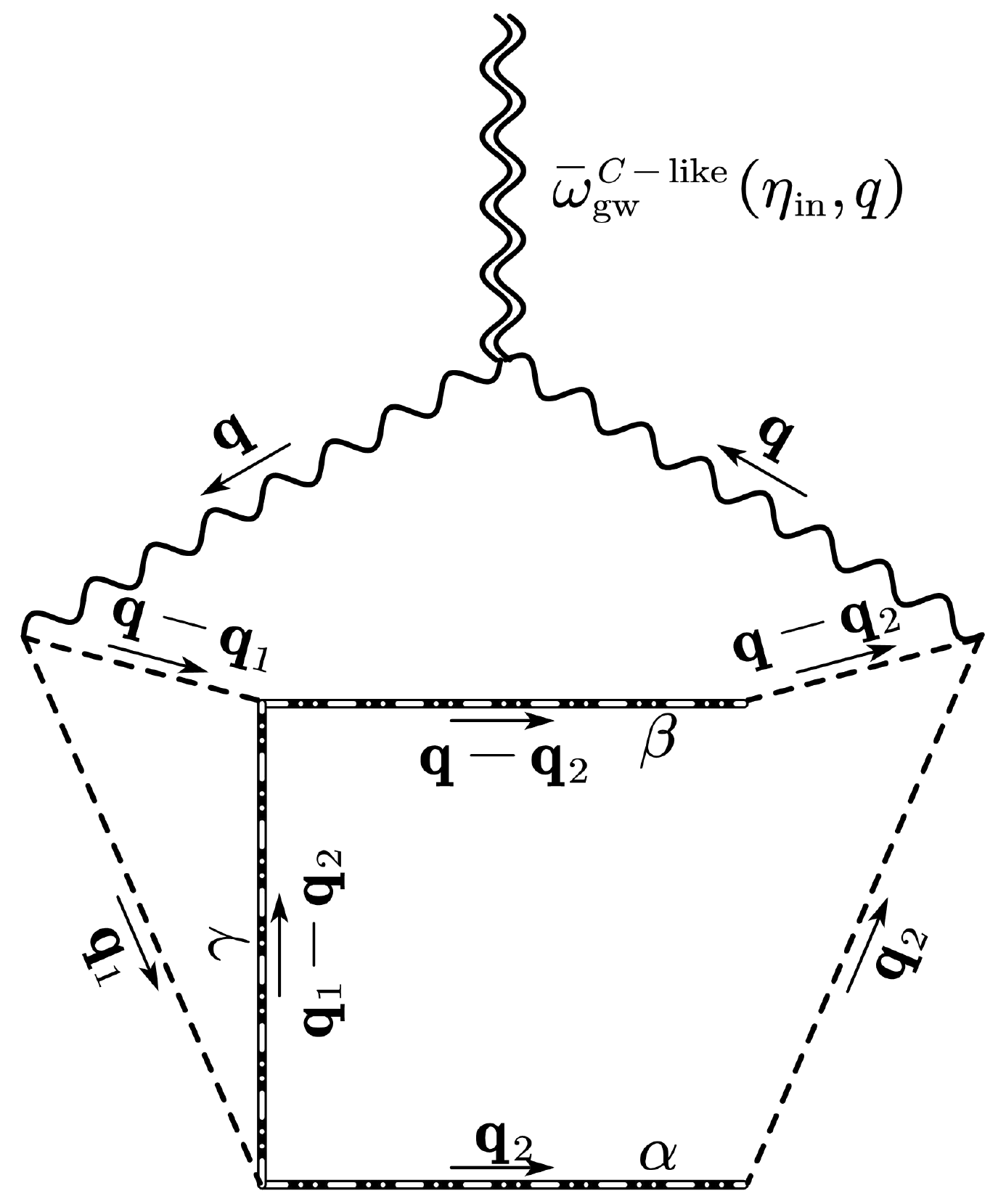}}\hfil
    \subcaptionbox*{$Z$-like.}{\includegraphics[width = 0.25\columnwidth]{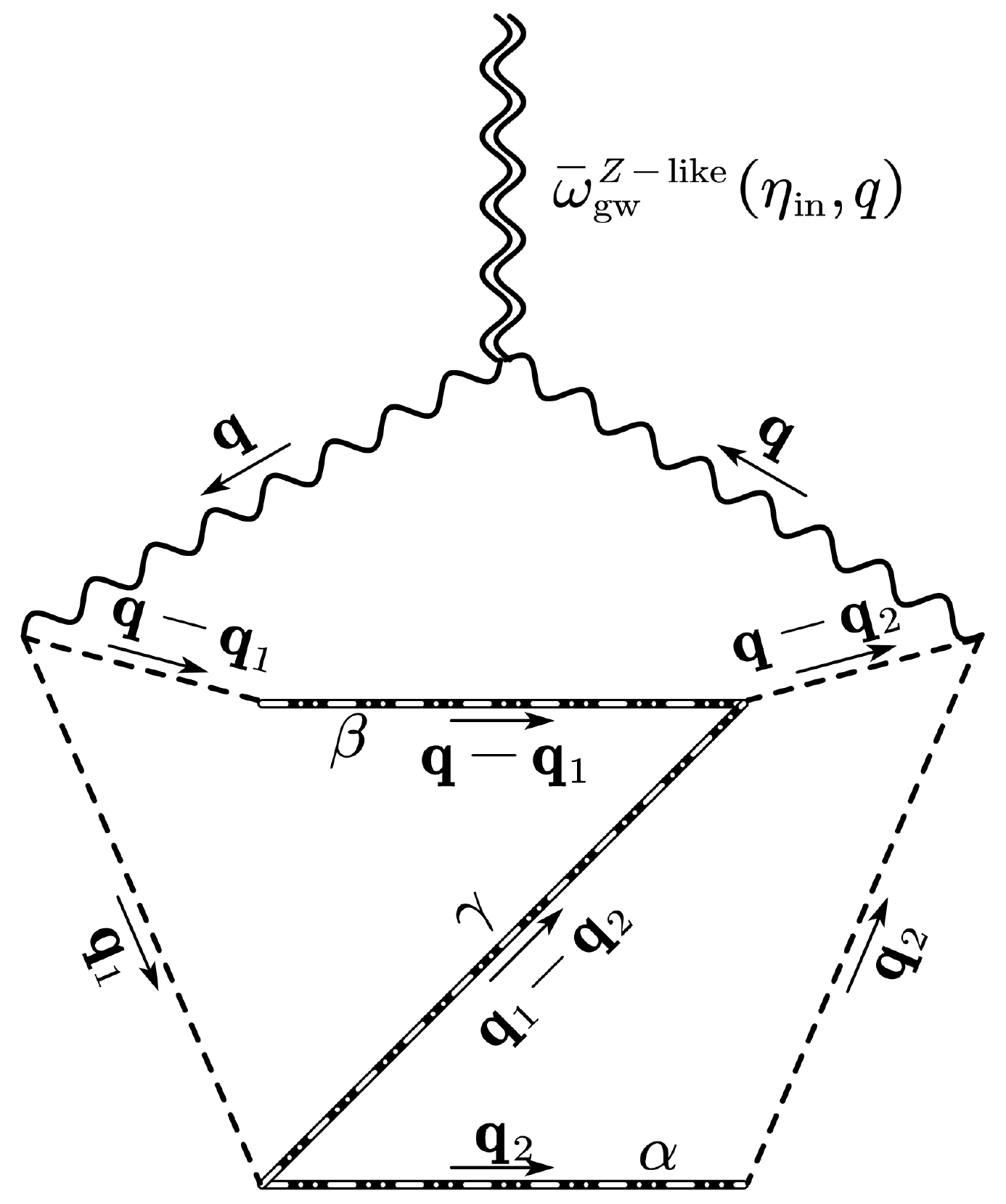}}\\
    \vspace{0.6cm}
    \subcaptionbox*{$P$-like.}{\includegraphics[width = 0.25\columnwidth]{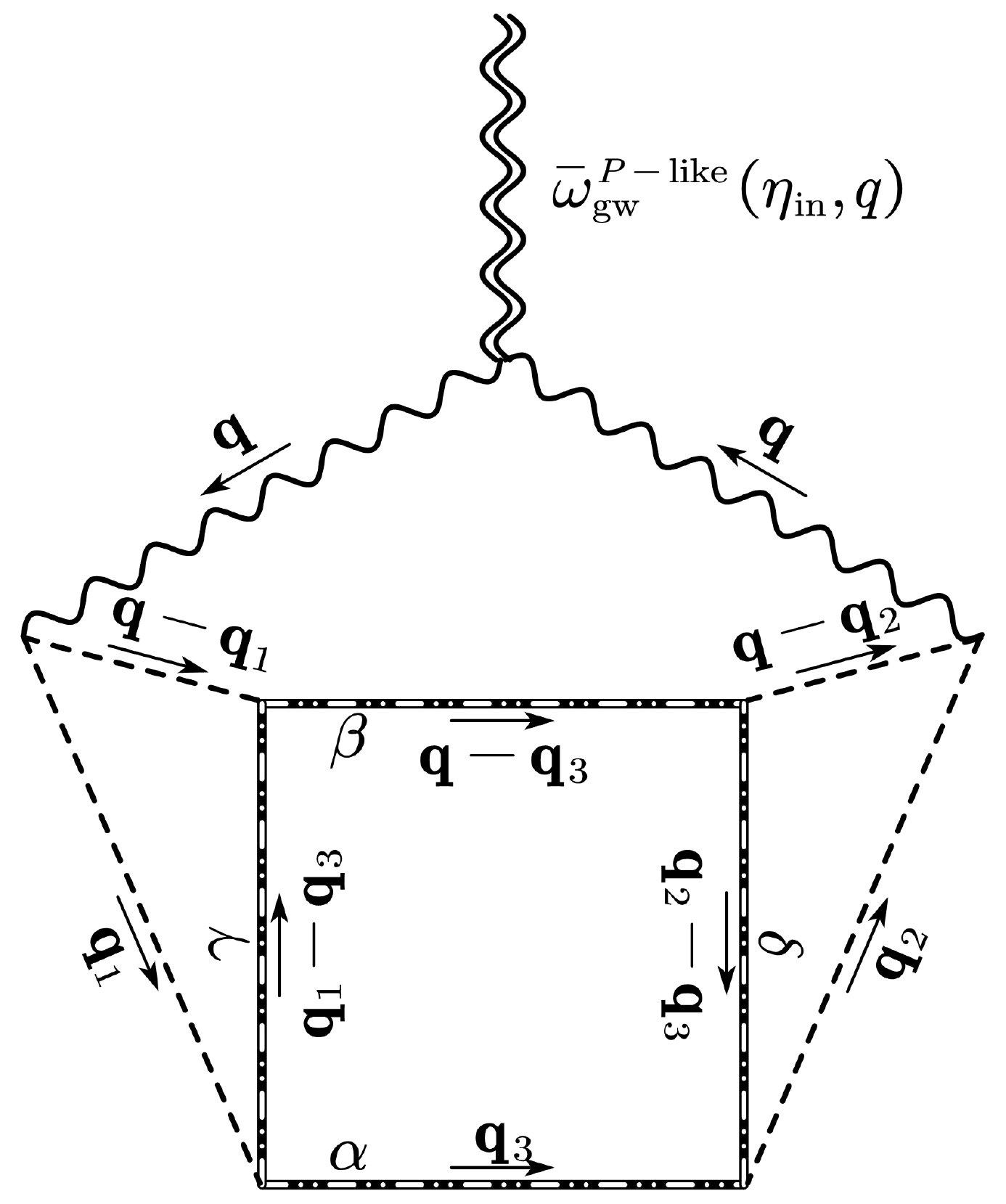}}\hfil
    \subcaptionbox*{$N$-like.}{\includegraphics[width = 0.25\columnwidth]{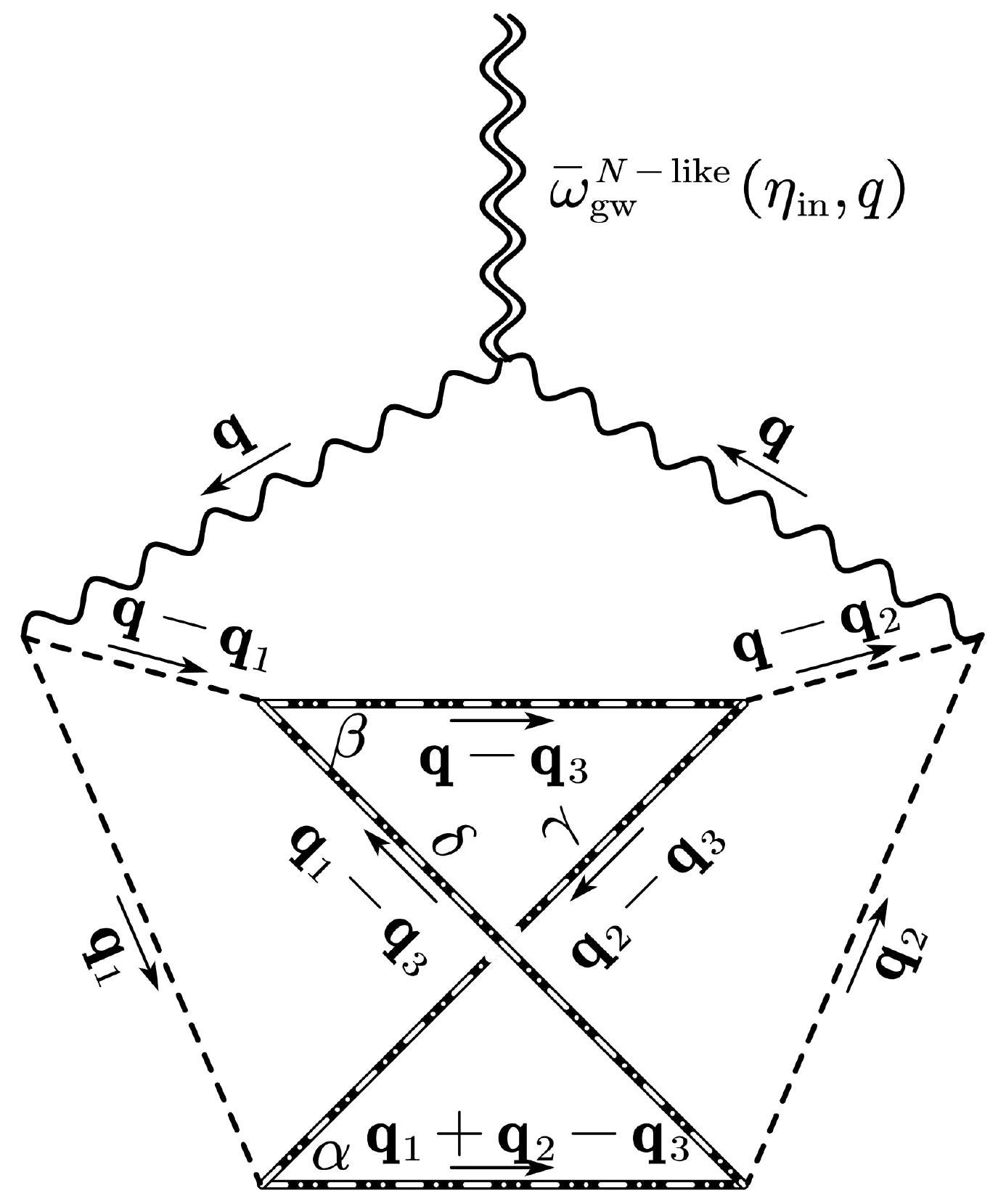}}\hfil
    \subcaptionbox*{$CZ$-like.}{\includegraphics[width = 0.25\columnwidth]{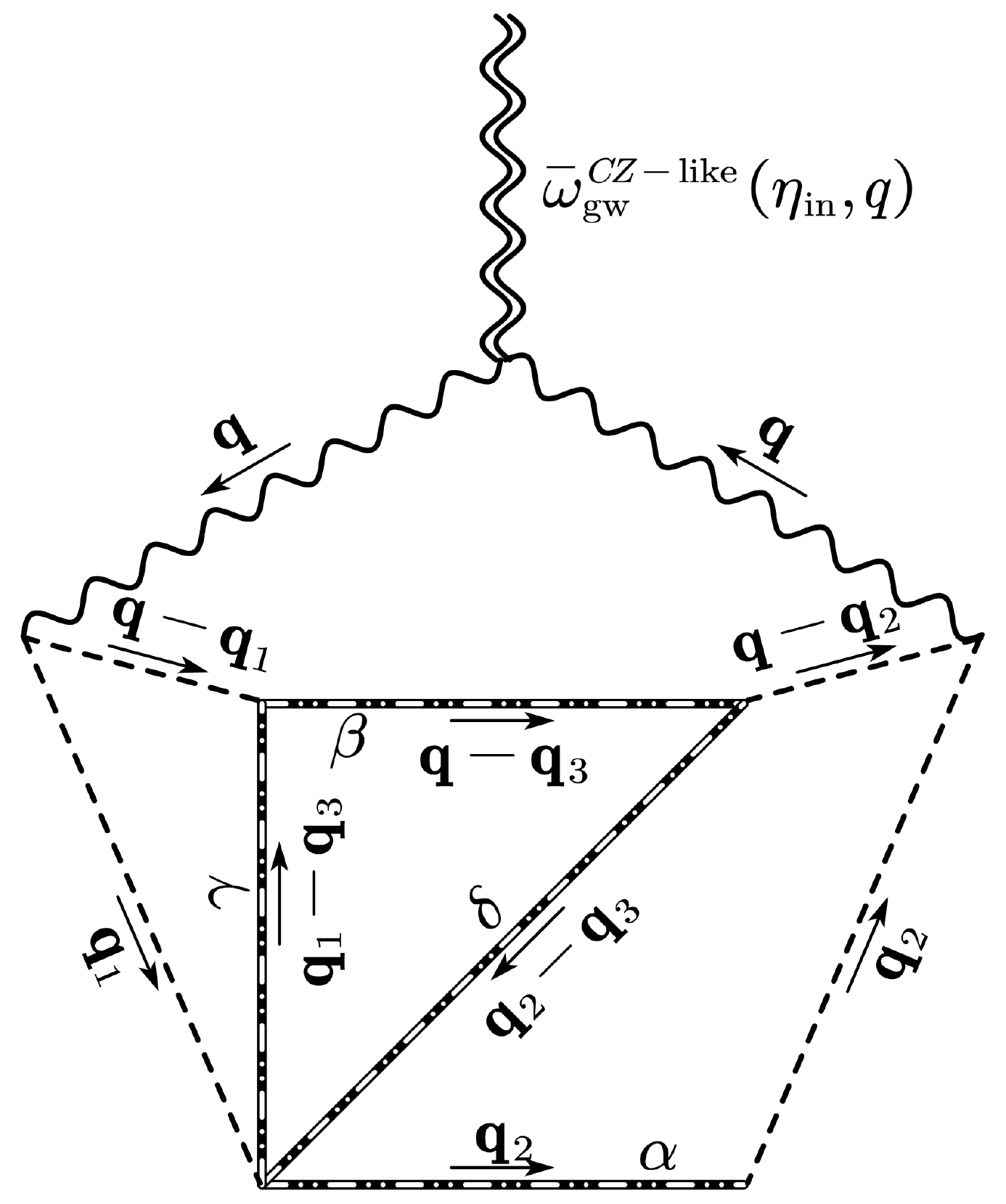}}\\
    \vspace{0.6cm}
    \subcaptionbox*{$PZ$-like.}{\includegraphics[width = 0.25\columnwidth]{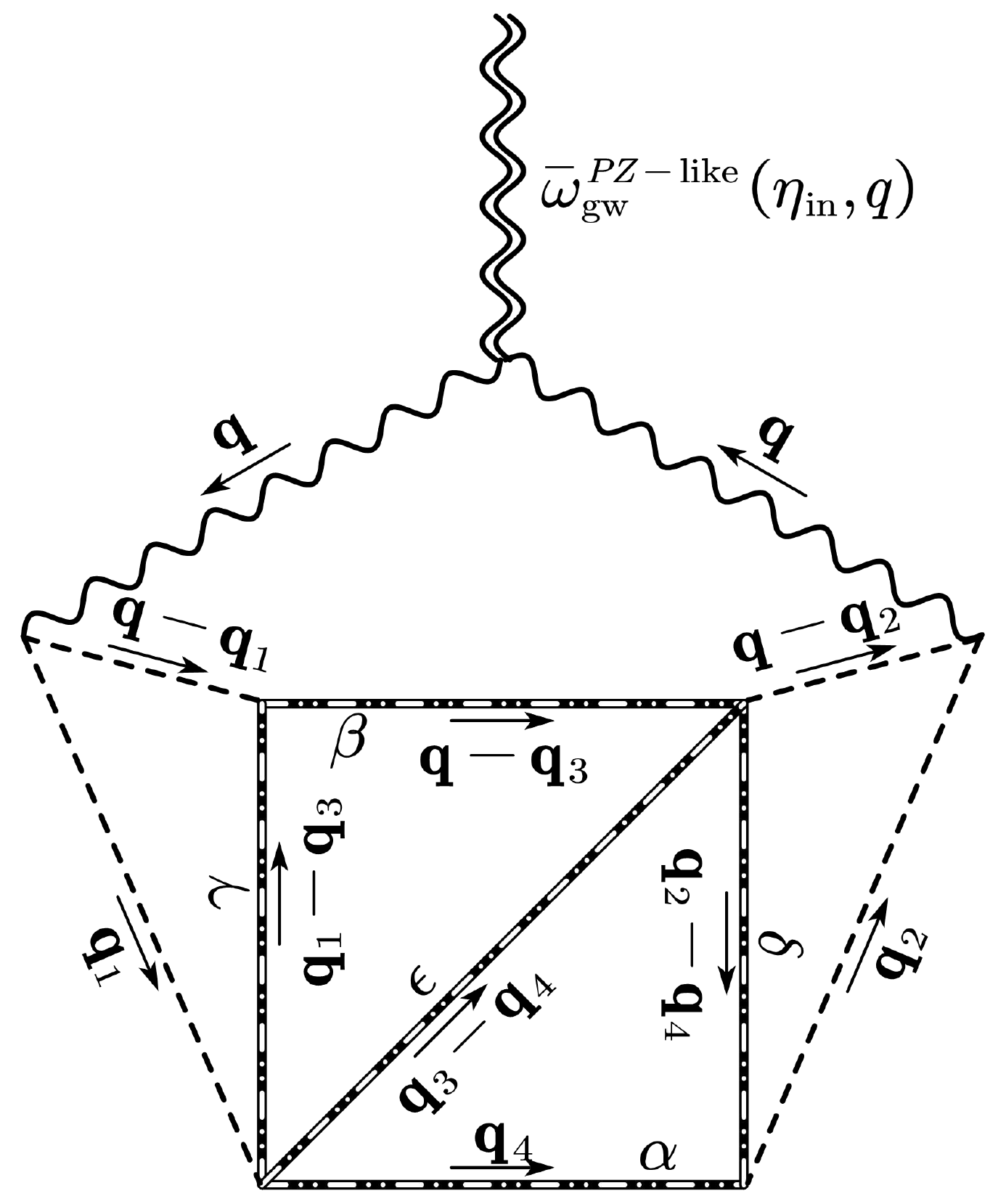}}\hfil
    \subcaptionbox*{$NC$-like.}{\includegraphics[width = 0.25\columnwidth]{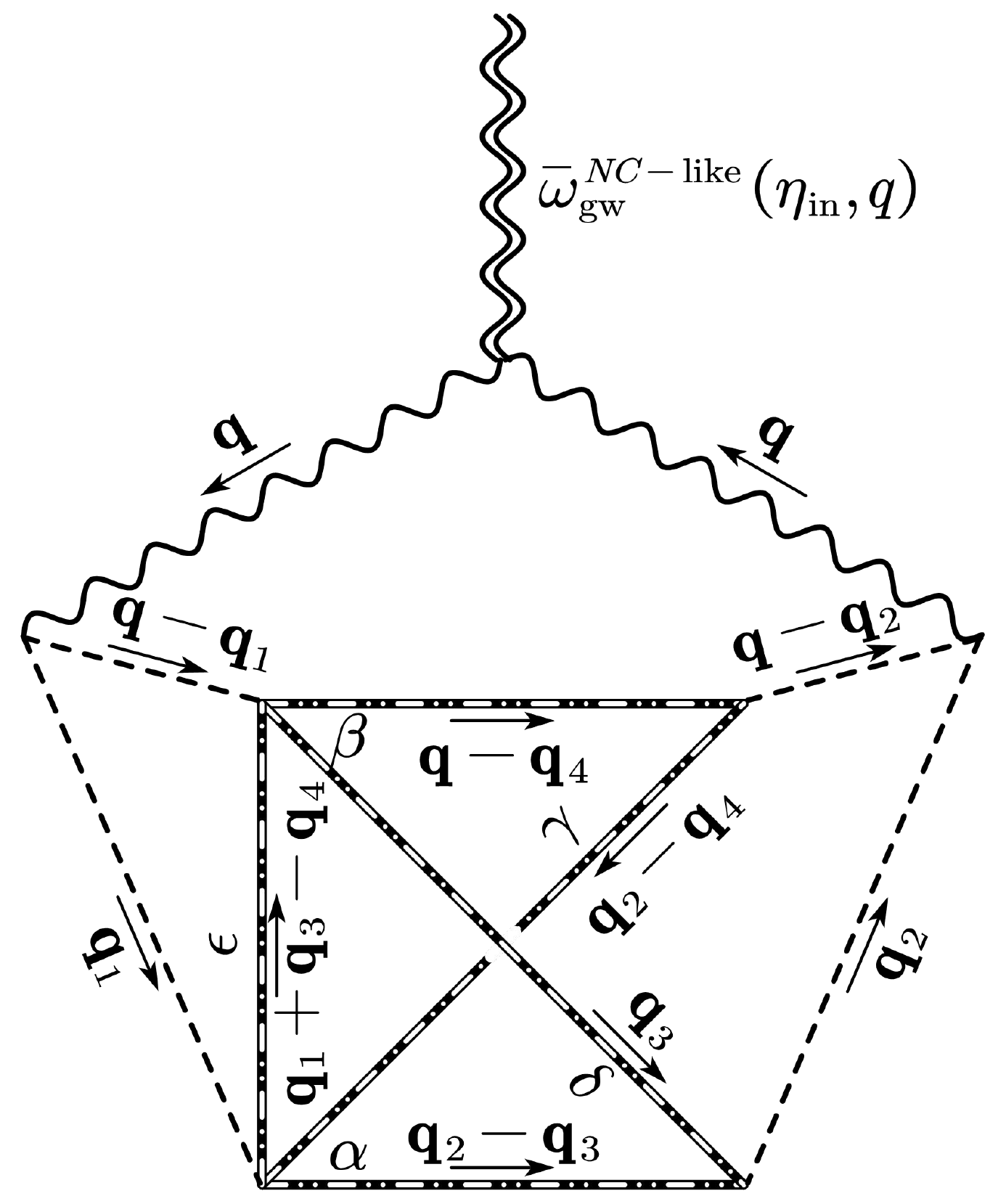}}\hfil
    \subcaptionbox*{$PN$-like.}{\includegraphics[width = 0.25\columnwidth]{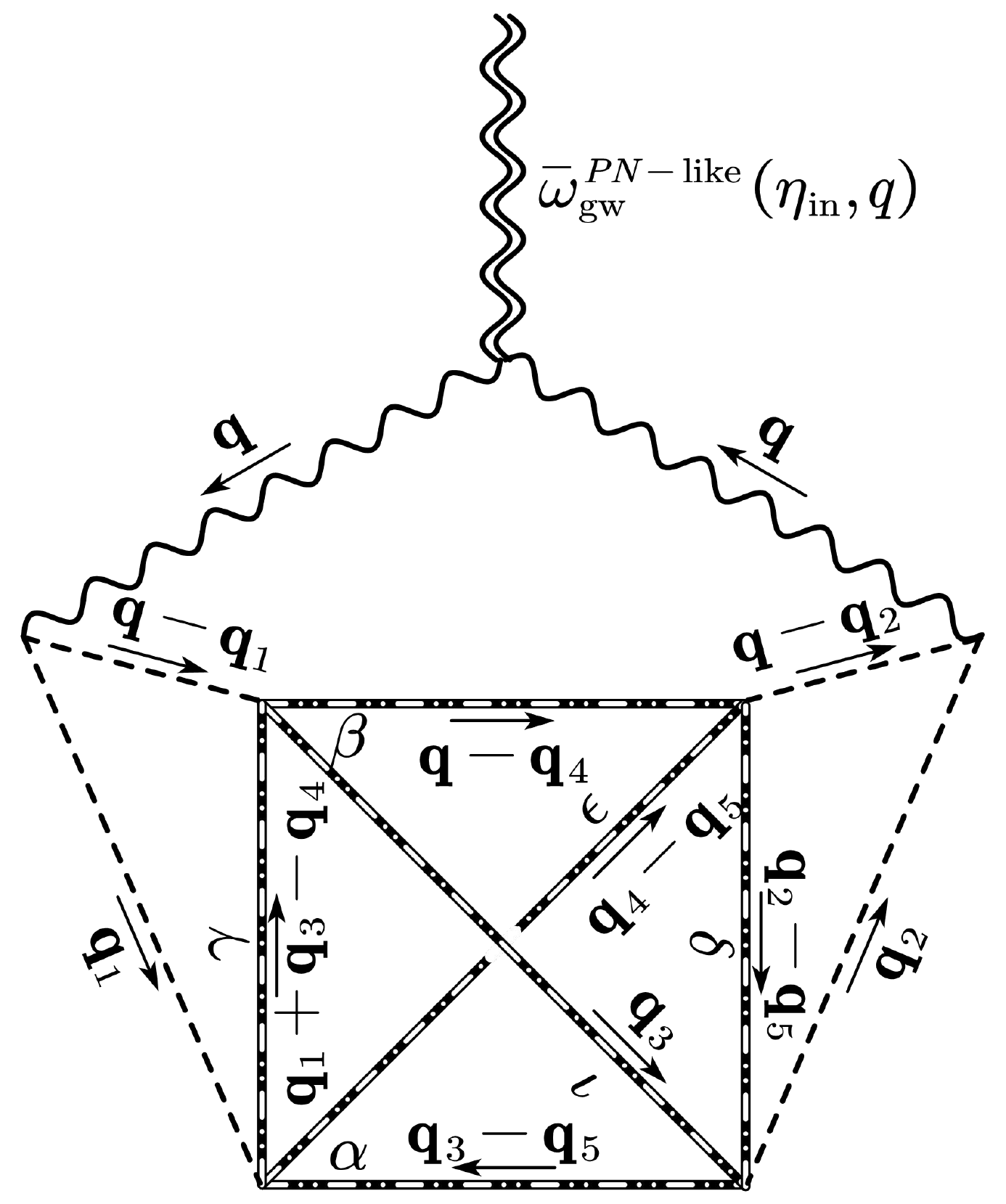}}
\caption{``Renormalized'' Feynman-like diagrams for the nine families of $\bar{\omega}_{\uGW,\uin} (q)$.}\label{fig:Feynman_Diagrams}
\end{figure*}

We utilize the ``renormalized'' Feynman-like rules depicted in \cref{fig:FD-Rules-new} to reconstruct the diagrams corresponding to $\bar{\omega}_{\uGW,\uin}$. 
Specifically, we replace the solid lines with thick-dashed lines in the diagrams labeled as ``$G$'', ``$C$'', ``$Z$'', ``$P$'', ``$N$'', ``$CZ$'', ``$PZ$'', ``$NC$'', and ``$PN$'' in \cref{fig:G-like-FD} - \cref{fig:PN-like-FD} in Appendix~\ref{sec:FD}, and subsequently substitute these into the shaded squares in \cref{fig:omegabar-FD_Frame}. 
Through this method, we obtain nine ``renormalized'' diagrams in \cref{fig:Feynman_Diagrams}, each encompassing all diagrams within the corresponding family. 
By employing the Feynman-like rules presented in \cref{fig:FD-Rules} alongside the ``renormalized'' rules in \cref{fig:FD-Rules-new}, in conjunction with Eqs.~(\ref{eq:P_redef}, \ref{eqs:V_redef}), we can compute the contributions from each family to the average energy-density spectrum based on the respective ``renormalized'' diagram. 
By aggregating their individual contributions as expressed in Eq.~\eqref{eq:omegabar-X-total}, we can determine the total average energy-density spectrum $\bar{\omega}_{\uGW,\uin} (q)$. 
The detailed calculation process is outlined in the following subsection.

\subsection{Energy-density fraction spectrum}\label{subsec:ED-int}

Based on the common structure of the diagrams representing the \ac{SIGW} average energy-density spectrum $\bar{\omega}_{\uGW,\uin}$, as illustrated in \cref{fig:omegabar-FD_Frame}, and the Feynman-like rules detailed in \cref{fig:FD-Rules}, we combine Eqs.~(\ref{eq:Ph-def}, \ref{eq:h}, \ref{eq:omegabar-h}) to express the contribution to $\bar{\omega}_{\uGW,\uin}$ from a certain family designated as $X$-like in the form of  
\begin{eqnarray}\label{eq:omegabar-zeta}
    \bar{\omega}_{\uGW,\uin}^{X-\mathrm{like}} %(\eta_\uin,q) 
    &=& \frac{q^5}{3(2\pi)^3 \cH^2}
        \Bigg\{\prod_{i=1}^n \int \frac{\ud^3 \bq_i}{(2\pi)^{3}}\Bigg\} \cP^{X-\mathrm{like}} \\
        &&\hphantom{\frac{q^5}{3(2\pi)^3 \cH^2}
        \Bigg\{\prod_{i=1}^n } \times \Bigg\{\vspace{-5pt}\sum_{\lambda = +,\times} Q_{\lambda}(\bq, \bq_1) Q_{\lambda}(\bq, \bq_j) 
        \overbar{\hat{I} (\abs{\bq - \bq_1}, q_1, \eta_\uin) 
        \hat{I} (\abs{\bq - \bq_j}, q_j, \eta_\uin)}\Bigg\} \ ,\nonumber
\end{eqnarray}
where we introduce $\cP^{X-\mathrm{like}}$ to differentiate between the various families, each composed of the convolved propagators and ``renormalized'' vertices defined in Eqs.~(\ref{eq:P_redef}, \ref{eqs:V_redef}). 
Furthermore, $n$ represents the number of undetermined momenta to be integrated over, specifically: $n=1$ for the $G$-like family, $n=2$ for the $C$-like and $Z$-like families, $n=3$ for the $P$-like, $N$-like, and $CZ$-like families, $n=4$ for the $PZ$-like and $NC$-like families, and $n=5$ for the $PN$-like family. 
Notably, for the variables of the projection factor $Q_\lambda$ and the kernel functions $\hat{I}$ associated with the $G$-like family, the subscript $_j$ is set to $1$, while for the other families, $j$ is set to $2$.

Upon observing the ``renormalized'' diagrams in \cref{fig:Feynman_Diagrams}, we find that swapping the Greek letter labels adjacent to specific pairs of convolved propagators within a diagram results in an equivalent diagram as the original. 
Propagators meeting this criterion are termed symmetric convolved propagators. 
As stated below Eqs.~\eqref{eq:P_redef}, we will use the same subscript to represent the momenta of these symmetric convolved propagators, which allows for the ready aggregation of all symmetric factors associated with equivalent diagrams.

For the ``renormalized'' $G$-like diagram (the upper left panel in \cref{fig:Feynman_Diagrams}), we can write down the expression of $\cP^{G-\mathrm{like}}$ based on the ``renormalized'' Feynman-like rules in \cref{fig:FD-Rules-new}, namely,  
\begin{eqnarray}\label{eq:G-like}
    \cP^{G-\mathrm{like}} = 2 \Biggl(\sum_{\alpha=1}^o V_{\uR,1}^{[\alpha]} P_{\uR,a}^{[\alpha]} V_{\uR,1}^{[\alpha]}\Biggr) \Biggl(\sum_{\beta=1}^o V_{\uR,1}^{[\beta]} P_{\uR,a}^{[\beta]} V_{\uR,1}^{[\beta]}\Biggr)\ . 
\end{eqnarray}
After expanding this expression through Eqs.~(\ref{eq:P_redef}, \ref{eqs:V_redef}), each term is proportional to $P_a^{[\ca]} P_a^{[\cb]}$, where $\ca,\cb$ must satisfy the conditions $(\ca,\cb \leq o, \text{ and } \ca \geq \cb)$. 
Subsequently, we substitute the two subscripts $a$ with the momenta $q_1$ and $|\bq-\bq_1|$, transforming $P_a^{[\ca]} P_a^{[\cb]}$ into $P^{[\ca]}(q_1) P^{[\cb]}(|\bq-\bq_1|)$. 
In particular, for scale-independent \ac{PNG} up to $\Hnl$ order (i.e., $o=4$), we introduce the shorthand notation $P_G^{[\ca,\cb]} = P^{[\ca]}(q_1) P^{[\cb]}(|\bq-\bq_1|)$ for convenience, and $\cP^{G-\mathrm{like}}$ reads  
\begin{eqnarray}\label{eq:G-like-HNL}
    \cP^{G-\mathrm{like}} &=& 2 \left(1 + 3 A_\uS \Gnl\right)^4 P_G^{[1,1]} + 8 \left(1 + 3 A_\uS \Gnl\right)^2 \left(\Fnl + 6 A_\uS \Hnl\right)^2 P_G^{[2,1]} \nonumber\\ 
    && + 24 \Gnl^2 \left(1 + 3 A_\uS \Gnl\right)^2 P_G^{[3,1]} + 96 \Hnl^2 \left(1 + 3 A_\uS \Gnl\right)^2 P_G^{[4,1]} \nonumber\\
    && + 8 \left(\Fnl + 6 A_\uS \Hnl\right)^4 P_G^{[2,2]} + 48 \Gnl^2 \left(\Fnl + 6 A_\uS \Hnl\right)^2 P_G^{[3,2]} \nonumber\\
    && + 192 \Hnl^2 \left(\Fnl + 6 A_\uS \Hnl\right)^2 P_G^{[4,2]} + 72 \Gnl^4 P_G^{[3,3]} + 576 \Gnl^2 \Hnl^2 P_G^{[4,3]} \nonumber\\
    && + 1152 \Hnl^4 P_G^{[4,4]} \ .
\end{eqnarray}
This expression comprises 10 terms that are proportional to $P_G^{[\ca,\cb]}$, corresponding to 10 sets of diagrams featuring the same convolved propagators. 
Within any set of diagrams, one diagram without ``self-closed loops'' accounts for the portion of the term that is independent of $A_\uS$, while the other diagrams with $i$ ``self-closed loops'' represent the parts proportional to $A_\uS^i$ in that term. 
To explore the contribution of each individual diagram in detail, we can further expand this expression and substitute each term into Eq.~\eqref{eq:omegabar-zeta} individually. 
The symmetric factors for the 39 diagrams shown in \cref{fig:G-like-FD} can be obtained from the prefactor of the corresponding expressions. 
Following this, the correspondences between $\cP^{X-\mathrm{like}}$ for other families and their corresponding diagrams in Appendix~\ref{sec:FD} are analogous, as showcased in the subsequent paragraphs.

The ``renormalized'' $C$-like diagram (the upper middle panel in \cref{fig:Feynman_Diagrams}) yields 
\begin{eqnarray}\label{eq:C-like}
    \cP^{C-\mathrm{like}} = 4 \sum_{\gamma=1}^{o-1} \sum_{\beta=1}^{o-\gamma} \sum_{\alpha=1}^{o-\gamma} V_{\uR,1}^{[\alpha]} P_{\uR,a}^{[\alpha]} V_{\uR,2}^{[\alpha,\gamma]} P_{\uR,b}^{[\gamma]} V_{\uR,2}^{[\gamma,\beta]} P_{\uR,a}^{[\beta]} V_{\uR,1}^{[\beta]}\ .
\end{eqnarray}
After expanding this expression through Eqs.~(\ref{eq:P_redef}, \ref{eqs:V_redef}), each term is proportional to $P_a^{[\ca]} P_a^{[\cb]} P_{b}^{[\cc]}$, where the positive integers $\ca, \cb, \cc$ must satisfy the conditions $(\ca+\cc, \cb+\cc \leq o, \text{ and } \ca \geq \cb)$. 
Subsequently, we replace the two subscripts $a$ with the momenta $q_2$ and $|\bq-\bq_2|$, and the two subscripts $b$ with the momentum $|\bq_1-\bq_2|$, transforming $P_a^{[\ca]} P_a^{[\cb]} P_{b}^{[\cc]}$ into $P^{[\ca]}(q_2) P^{[\cb]}(|\bq-\bq_2|) P^{[\cc]}(|\bq_1-\bq_2|)$. 
In particular, for scale-independent \ac{PNG} up to $\Hnl$ order (i.e., $o=4$), we introduce the shorthand notation $P_C^{[\ca,\cb;\cc]}=P^{[\ca]}(q_2) P^{[\cb]}(|\bq-\bq_2|) P^{[\cc]}(|\bq_1-\bq_2|)$ for convenience, leading to the expression for $\cP^{C-\mathrm{like}}$ as 
\begin{eqnarray}\label{eq:C-like-HNL}
    \cP^{C-\mathrm{like}} &=& 16 \left(1 + 3 A_\uS \Gnl\right)^2 \left(\Fnl + 6 A_\uS \Hnl\right)^2 P_C^{[1,1;1]} \nonumber\\
    && + 96 \Gnl \left(1 + 3 A_\uS \Gnl\right) \left(\Fnl + 6 A_\uS \Hnl\right)^2 P_C^{[2,1;1]} \nonumber\\
    && + 384 \Gnl \Hnl \left(1 + 3 A_\uS \Gnl\right) \left(\Fnl + 6 A_\uS \Hnl\right) P_C^{[3,1;1]} \nonumber\\
    && + 144 \Gnl^2 \left(\Fnl + 6 A_\uS \Hnl\right)^2 P_C^{[2,2;1]} + 1152 \Gnl^2 \Hnl \left(\Fnl + 6 A_\uS \Hnl\right) P_C^{[3,2;1]} \nonumber\\
    && + 2304 \Gnl^2 \Hnl^2  P_C^{[3,3;1]} + 72 \Gnl^2 \left(1 + 3 A_\uS \Gnl\right)^2 P_C^{[1,1;2]} \nonumber\\
    && + 576 \Gnl \Hnl \left(1 + 3 A_\uS \Gnl\right) \left(\Fnl + 6 A_\uS \Hnl\right) P_C^{[2,1;2]} \nonumber\\
    && + 1152 \Hnl^2 \left(\Fnl + 6 A_\uS \Hnl\right)^2 P_C^{[2,2;2]} + 384 \Hnl^2 \left(1 + 3 A_\uS \Gnl\right)^2 P_C^{[1,1;3]} \ .
\end{eqnarray}
This expression consists of 10 terms that are proportional to $P_C^{[\ca,\cb;\cc]}$, corresponding to 10 sets of diagrams featuring the same convolved propagators. 
The contribution from each diagram can be evaluated by fully expanding this expression and substituting each term into Eq.~\eqref{eq:omegabar-zeta} individually, allowing us to deduce the symmetric factors for the 41 diagrams depicted in \cref{fig:C-like-FD} from the prefactors.

The ``renormalized'' $Z$-like diagram (the upper right panel in \cref{fig:Feynman_Diagrams}) yields  
\begin{eqnarray}\label{eq:Z-like}
    \cP^{Z-\mathrm{like}} = 4 \sum_{\gamma=1}^{o-1} \sum_{\beta=1}^{o-\gamma} \sum_{\alpha=1}^{o-\gamma} V_{\uR,1}^{[\alpha]} P_{\uR,a}^{[\alpha]} V_{\uR,2}^{[\alpha,\gamma]} P_{\uR,c}^{[\gamma]} V_{\uR,2}^{[\gamma,\beta]} P_{\uR,a}^{[\beta]} V_{\uR,1}^{[\beta]}\ ,
\end{eqnarray}
After expanding this expression through Eqs.~(\ref{eq:P_redef}, \ref{eqs:V_redef}), each term is proportional to $P_a^{[\ca]} P_a^{[\cb]} P_{c}^{[\cc]}$, where the positive integers $\ca, \cb, \cc$ must satisfy the conditions $(\ca+\cc, \cb+\cc \leq o, \text{ and } \ca \geq \cb)$. 
We then replace the two subscripts $a$ with the momenta $q_2$ and $|\bq-\bq_1|$, and the two subscripts $c$ with the momentum $|\bq_1-\bq_2|$, transforming $P_a^{[\ca]} P_a^{[\cb]} P_{c}^{[\cc]}$ into $P^{[\ca]}(q_2) P^{[\cb]}(|\bq-\bq_1|) P^{[\cc]}(|\bq_1-\bq_2|)$. 
In particular, for scale-independent \ac{PNG} up to $\Hnl$ order (i.e., $o=4$), we introduce the shorthand notation $P_Z^{[\ca,\cb;\cc]}=P^{[\ca]}(q_2) P^{[\cb]}(|\bq-\bq_1|) P^{[\cc]}(|\bq_1-\bq_2|)$ for convenience, leading to the expression for $\cP^{Z-\mathrm{like}}$ as 
\begin{eqnarray}\label{eq:Z-like-HNL}
    \cP^{Z-\mathrm{like}} &=& 16 \left(1 + 3 A_\uS \Gnl\right)^2 \left(\Fnl + 6 A_\uS \Hnl\right)^2 P_Z^{[1,1;1]} \nonumber\\
    && + 96 \Gnl \left(1 + 3 A_\uS \Gnl\right) \left(\Fnl + 6 A_\uS \Hnl\right)^2 P_Z^{[2,1;1]} \nonumber\\
    && + 384 \Gnl \Hnl \left(1 + 3 A_\uS \Gnl\right) \left(\Fnl + 6 A_\uS \Hnl\right) P_Z^{[3,1;1]} \nonumber\\
    && + 144 \Gnl^2 \left(\Fnl + 6 A_\uS \Hnl\right)^2 P_Z^{[2,2;1]} + 1152 \Gnl^2 \Hnl \left(\Fnl + 6 A_\uS \Hnl\right) P_Z^{[3,2;1]} \nonumber\\
    && + 2304 \Gnl^2 \Hnl^2  P_Z^{[3,3;1]} + 72 \Gnl^2 \left(1 + 3 A_\uS \Gnl\right)^2 P_Z^{[1,1;2]} \nonumber\\
    && + 576 \Gnl \Hnl \left(1 + 3 A_\uS \Gnl\right) \left(\Fnl + 6 A_\uS \Hnl\right) P_Z^{[2,1;2]} \nonumber\\
    && + 1152 \Hnl^2 \left(\Fnl + 6 A_\uS \Hnl\right)^2 P_Z^{[2,2;2]} + 384 \Hnl^2 \left(1 + 3 A_\uS \Gnl\right)^2 P_Z^{[1,1;3]} \ .
\end{eqnarray}
This expression consists of 10 terms proportional to $P_Z^{[\ca,\cb;\cc]}$, corresponding to 10 sets of diagrams with the same convolved propagators. 
The contribution from each diagram can be evaluated by fully expanding this expression and substituting each term into Eq.~\eqref{eq:omegabar-zeta} individually, allowing us to deduce the symmetric factors for the 41 diagrams depicted in \cref{fig:Z-like-FD} from the prefactors.

The ``renormalized'' $P$-like diagram (the middle left panel in \cref{fig:Feynman_Diagrams}) yields  
\begin{eqnarray}\label{eq:P-like}
    \cP^{P-\mathrm{like}} = 2 \sum_{\delta=1}^{o-1} \sum_{\gamma=1}^{o-1} \sum_{\beta=1}^{\min(o-\gamma,o-\delta)} \sum_{\alpha=1}^{\min(o-\gamma,o-\delta)} P_{\uR,a}^{[\alpha]} V_{\uR,2}^{[\alpha,\gamma]} P_{\uR,b}^{[\gamma]} V_{\uR,2}^{[\gamma,\beta]} P_{\uR,a}^{[\beta]} V_{\uR,2}^{[\beta,\delta]} P_{\uR,b}^{[\delta]} V_{\uR,2}^{[\delta,\alpha]}\ ,
\end{eqnarray}
After expanding this expression through Eqs.~(\ref{eq:P_redef}, \ref{eqs:V_redef}), each term is proportional to $P_a^{[\ca]} P_a^{[\cb]} P_b^{[\cc]} P_b^{[\cd]}$, where the positive integers $\ca,\cb,\cc,\cd$ must satisfy the conditions $(\ca+\cc,\ca+\cd,\cb+\cc,\cb+\cd \leq o, \ca \geq \cb, \text{ and } \cc \geq \cd)$. 
We then replace the two subscripts $a$ with the momenta $q_3$ and $|\bq-\bq_3|$, and the two subscripts $b$ with the momenta $|\bq_1-\bq_3|$ and  $|\bq_2-\bq_3|$, transforming $P_a^{[\ca]} P_a^{[\cb]} P_b^{[\cc]} P_b^{[\cd]}$ to $P^{[\ca]}(q_3) P^{[\cb]}(|\bq-\bq_3|)P^{[\cc]}(|\bq_1-\bq_3|) P^{[\cd]}(|\bq_2-\bq_3|)$. 
In particular, for scale-independent \ac{PNG} up to $\Hnl$ order (i.e., $o=4$), we introduce the shorthand notation $P_P^{[\ca,\cb;\cc,\cd]}=P^{[\ca]}(q_3) P^{[\cb]}(|\bq-\bq_3|)P^{[\cc]}(|\bq_1-\bq_3|) P^{[\cd]}(|\bq_2-\bq_3|)$ for convenience, leading to the expression for $\cP^{P-\mathrm{like}}$ as 
\begin{eqnarray}\label{eq:P-like-HNL}
    \cP^{P-\mathrm{like}} &=& 32 \left(\Fnl + 6 A_\uS \Hnl\right)^4 P_P^{[1,1;1,1]} + 288 \Gnl^2 \left(\Fnl + 6 A_\uS \Hnl\right)^2 \left(P_P^{[2,1;1,1]} + P_P^{[1,1;2,1]}\right) \nonumber\\
    && + 1536 \Hnl^2 \left(\Fnl + 6 A_\uS \Hnl\right)^2 \left(P_P^{[3,1;1,1]} + P_P^{[1,1;3,1]}\right) \nonumber\\
    && + 648 \Gnl^4 \left(P_P^{[2,2;1,1]} + P_P^{[1,1;2,2]}\right) + 6912 \Gnl^2 \Hnl^2 \left(P_P^{[3,2;1,1]} + P_P^{[1,1;3,2]}\right) \nonumber\\
    && + 18432 \Hnl^4 \left(P_P^{[3,3;1,1]} + P_P^{[1,1;3,3]}\right) + 3456 \Gnl^2 \Hnl \left(\Fnl + 6 A_\uS \Hnl\right) P_P^{[2,1;2,1]} \nonumber\\
    && + 10368 \Gnl^2 \Hnl^2 \left(P_P^{[2,2;2,1]} + P_P^{[2,1;2,2]}\right) + 41472 \Hnl^4 P_P^{[2,2;2,2]} \ .
\end{eqnarray}
This expression consists of 15 terms proportional to $P_P^{[\ca,\cb;\cc,\cd]}$, corresponding to 15 sets of diagrams with the same convolved propagators. 
The contribution from each diagram can be evaluated by fully expanding this expression and substituting each term into Eq.~\eqref{eq:omegabar-zeta} individually, allowing us to deduce the symmetric factors for the 30 diagrams depicted in \cref{fig:P-like-FD} from the prefactors.

The ``renormalized'' $N$-like diagram (the center panel in \cref{fig:Feynman_Diagrams}) yields  
\begin{eqnarray}\label{eq:N-like}
    \cP^{N-\mathrm{like}} = \sum_{\delta=1}^{o-1} \sum_{\gamma=1}^{o-1} \sum_{\beta=1}^{\min(o-\gamma,o-\delta)} \sum_{\alpha=1}^{\min(o-\gamma,o-\delta)} P_{\uR,a}^{[\alpha]} V_{\uR,2}^{[\alpha,\gamma]} P_{\uR,c}^{[\gamma]} V_{\uR,2}^{[\gamma,\beta]} P_{\uR,a}^{[\beta]} V_{\uR,2}^{[\beta,\delta]} P_{\uR,c}^{[\delta]} V_{\uR,2}^{[\delta,\alpha]}\ ,
\end{eqnarray}
After expanding this expression through Eqs.~(\ref{eq:P_redef}, \ref{eqs:V_redef}), each term is proportional to $P_a^{[\ca]} P_a^{[\cb]} P_c^{[\cc]} P_c^{[\cd]}$, where the positive integers $\ca, \cb, \cc, \cd$ must satisfy the conditions $(\ca+\cc, \ca+\cd, \cb+\cc, \cb+\cd \leq o, \ca \leq \cb, \text{ and } \cc \geq \cd)$. 
Moreover, upon examining the ``renormalized'' $N$-like diagram, we notice that swapping the Greek letter labels $\ca$ and $\cb$ adjacent to the two symmetric convolved propagators with the Greek letter labels $\cc$ and $\cd$ adjacent to the other two symmetric convolved propagators. 
Thereby, the term proportional to $P_a^{[\ca]} P_a^{[\cb]} P_c^{[\cc]} P_c^{[\cd]}$ can be merged with the term proportional to $P_a^{[\cc]} P_a^{[\cd]} P_c^{[\ca]} P_c^{[\cb]}$, which can be simplified under the constraint $\ca+\cb \geq \cc+\cd$. 
We then substitute the two subscripts $a$ with the momenta $|\bq_1+\bq_2-\bq_3|$ and $|\bq-\bq_3|$, and the two subscripts $c$ with $|\bq_1-\bq_3|$ and $|\bq_2-\bq_3|$, transforming $P_a^{[\ca]} P_a^{[\cb]} P_c^{[\cc]} P_c^{[\cd]}$ into $P^{[\ca]}(|\bq_1+\bq_2-\bq_3|) P^{[\cb]}(|\bq-\bq_3|) P^{[\cc]}(|\bq_2-\bq_3|) P^{[\cd]}(|\bq_1-\bq_3|)$. 
In particular, for scale-independent \ac{PNG} up to $\Hnl$ order (i.e., $o=4$), we introduce the shorthand notation $P_N^{[\ca,\cb;\cc,\cd]} = P^{[\ca]}(|\bq_1+\bq_2-\bq_3|) P^{[\cb]}(|\bq-\bq_3|) P^{[\cc]}(|\bq_2-\bq_3|) P^{[\cd]}(|\bq_1-\bq_3|)$ for convenience, leading to the expression for $\cP^{N-\mathrm{like}}$ as 
\begin{eqnarray}\label{eq:N-like-HNL}
    \cP^{N-\mathrm{like}} &=& 16 \left(\Fnl + 6 A_\uS \Hnl\right)^4 P_N^{[1,1;1,1]} + 288 \Gnl^2 \left(\Fnl + 6 A_\uS \Hnl\right)^2 P_N^{[1,2;1,1]} \nonumber\\
    && + 1536 \Hnl^2 \left(\Fnl + 6 A_\uS \Hnl\right)^2 P_N^{[1,3;1,1]} + 648 \Gnl^4 P_N^{[2,2;1,1]} + 6912 \Gnl^2 \Hnl^2 P_N^{[2,3;1,1]} \nonumber\\
    && + 18432 \Hnl^4 P_N^{[3,3;1,1]} + 1728 \Gnl^2 \Hnl \left(\Fnl + 6 A_\uS \Hnl\right) P_N^{[1,2;2,1]} \nonumber\\
    && + 10368 \Gnl^2 \Hnl^2 P_N^{[2,2;2,1]} + 20736 \Hnl^4 P_N^{[2,2;2,2]} \ .
\end{eqnarray}
This expression comprises 9 terms proportional to $P_N^{[\ca,\cb;\cc,\cd]}$, corresponding to 9 sets of diagrams featuring the same convolved propagators. 
The contribution from each diagram can be evaluated by fully expanding this expression and substituting each term into Eq.~\eqref{eq:omegabar-zeta} individually, allowing us to deduce the symmetric factors for the 19 diagrams depicted in \cref{fig:N-like-FD} from the prefactors.

The ``renormalized'' $CZ$-like diagram (the middle right panel in \cref{fig:Feynman_Diagrams}) yields  
\begin{eqnarray}\label{eq:CZ-like}
    \cP^{CZ-\mathrm{like}} = 8 \sum_{\delta=1}^{o-2} \sum_{\gamma=1}^{o-2} \sum_{\alpha=1}^{o-\gamma-\delta} \sum_{\beta=1}^{\min(o-\gamma,o-\delta)} P_{\uR,a}^{[\alpha]} V_{\uR,1}^{[\alpha]} P_{\uR,b}^{[\beta]} V_{\uR,2}^{[\beta,\gamma]} P_{\uR,c}^{[\gamma]} V_{\uR,3}^{[\alpha,\gamma,\delta]} P_{\uR,d}^{[\delta]} V_{\uR,2}^{[\delta,\beta]} \ ,
\end{eqnarray}
After expanding this expression through Eqs.~(\ref{eq:P_redef}, \ref{eqs:V_redef}), each term is proportional to $P_a^{[\ca]} P_b^{[\cb]} P_c^{[\cc]} P_d^{[\cd]}$, where the positive integers $\ca, \cb, \cc, \cd$ must satisfy the conditions $(\ca+\cc+\cd, \cb+\cc, \text{ and } \cb+\cd \leq o)$. 
We then replace the subscripts: $a$ with the momentum $q_2$, $b$ with $|\bq-\bq_3|$, $c$ with $|\bq_1-\bq_3|$, and $d$ with $|\bq_2-\bq_3|$, transforming $P_a^{[\ca]} P_b^{[\cb]} P_c^{[\cc]} P_d^{[\cd]}$ into $P^{[\ca]}(q_2) P^{[\cb]}(|\bq-\bq_3|) P^{[\cc]}(|\bq_1-\bq_3|) P^{[\cd]}(|\bq_2-\bq_3|)$. 
In particular, for scale-independent \ac{PNG} up to $\Hnl$ order (i.e., $o=4$), we introduce the shorthand notation $P_{CZ}^{[\ca;\cb;\cc;\cd]}=P^{[\ca]}(q_2) P^{[\cb]}(|\bq-\bq_3|) P^{[\cc]}(|\bq_1-\bq_3|) P^{[\cd]}(|\bq_2-\bq_3|)$ for convenience, leading to the expression for $\cP^{CZ-\mathrm{like}}$ as 
\begin{eqnarray}\label{eq:CZ-like-HNL}
    \cP^{CZ-\mathrm{like}} &=& 192 \Gnl \left(1 + 3 A_\uS \Gnl\right) \left(\Fnl + 6 A_\uS \Hnl\right)^2 P_{CZ}^{[1;1;1;1]} + 864 \Gnl^3 \left(1 + 3 A_\uS \Gnl\right) P_{CZ}^{[1;2;1;1]} \nonumber\\
    && + 4608 \Gnl \Hnl^2 \left(1 + 3 A_\uS \Gnl\right) P_{CZ}^{[1;3;1;1]} + 768 \Hnl \left(\Fnl + 6 A_\uS \Hnl\right)^3 P_{CZ}^{[2;1;1;1]} \nonumber\\
    && + 3456 \Gnl^2 \Hnl \left(\Fnl + 6 A_\uS \Hnl\right) P_{CZ}^{[2;2;1;1]} + 18432 \Hnl^3 \left(\Fnl + 6 A_\uS \Hnl\right) P_{CZ}^{[2;3;1;1]} \nonumber\\
    && + 1152 \Gnl \Hnl \left(1 + 3 A_\uS \Gnl\right) \left(\Fnl + 6 A_\uS \Hnl\right) \left(P_{CZ}^{[1;1;2;1]} + P_{CZ}^{[1;1;1;2]}\right)  \nonumber\\
    && + 6912 \Gnl \Hnl^2 \left(1 + 3 A_\uS \Gnl\right) \left(P_{CZ}^{[1;2;2;1]} + P_{CZ}^{[1;2;1;2]}\right) \ .
\end{eqnarray}
This expression consists of 10 terms proportional to $P_{CZ}^{[\ca;\cb;\cc;\cd]}$, corresponding to 10 sets of diagrams with the same convolved propagators. 
The contribution from each diagram can be evaluated by fully expanding this expression and substituting each term into Eq.~\eqref{eq:omegabar-zeta} individually, allowing us to deduce the symmetric factors for the 36 diagrams depicted in \cref{fig:CZ-like-FD} from the prefactors.

The ``renormalized'' $PZ$-like diagram (the lower left panel in \cref{fig:Feynman_Diagrams}) yields  
\begin{align}\label{eq:PZ-like}
    \cP^{PZ-\mathrm{like}} = 4 \sum_{\epsilon=1}^{o-2} \sum_{\delta=1}^{o-\epsilon-1} \sum_{\gamma=1}^{o-\epsilon-1} \sum_{\beta=1}^{\min(o-\gamma,o-\delta-\epsilon)} &\sum_{\alpha=1}^{\min(o-\gamma-\epsilon,o-\delta)} P_{\uR,a}^{[\alpha]} V_{\uR,3}^{[\alpha,\gamma,\epsilon]} P_{\uR,b}^{[\gamma]} V_{\uR,2}^{[\beta,\gamma]} P_{\uR,a}^{[\beta]} V_{\uR,3}^{[\beta,\delta,\epsilon]} \nonumber\\
    &\qquad\qquad\quad \times P_{\uR,b}^{[\delta]} V_{\uR,2}^{[\alpha,\delta]} P_{\uR,c}^{[\epsilon]}\ ,
\end{align}
After expanding this expression through Eqs.~(\ref{eq:P_redef}, \ref{eqs:V_redef}), each term is proportional to $P_a^{[\ca]} P_a^{[\cb]} P_b^{[\cc]} P_b^{[\cd]} P_c^{[\ce]}$, where the positive integers $\ca, \cb, \cc, \cd, \ce$ must satisfy the conditions $(\ca+\cc+\ce, \ca+\cd, \cb+\cc, \cb+\cd+\ce \leq o, \ca \geq \cb, \text{ and } \cc \geq \cd)$. 
We then replace the subscripts: $a$ with the momenta $q_4$ and $|\bq-\bq_3|$, $b$ with the momenta $|\bq_1-\bq_3|$ and $|\bq_2-\bq_4|$, and $c$ with the momentum $|\bq_3-\bq_4|$. 
The replacements transform $P_a^{[\ca]} P_a^{[\cb]} P_b^{[\cc]} P_b^{[\cd]} P_c^{[\ce]}$ into $P^{[\ca]}(q_4) P^{[\cb]}(|\bq-\bq_3|) P^{[\cc]}(|\bq_1-\bq_3|) P^{[\cd]}(|\bq_2-\bq_4|) P^{[\ce]}(|\bq_3-\bq_4|)$. 
In particular, for scale-independent \ac{PNG} up to $\Hnl$ order (i.e., $o=4$), we introduce the shorthand notation $P_{PZ}^{[\ca,\cb;\cc,\cd;\ce]}=P^{[\ca]}(q_4) P^{[\cb]}(|\bq-\bq_3|) P^{[\cc]}(|\bq_1-\bq_3|) P^{[\cd]}(|\bq_2-\bq_4|) P^{[\ce]}(|\bq_3-\bq_4|)$ for convenience, leading to the expression for $\cP^{PZ-\mathrm{like}}$ as 
\begin{eqnarray}\label{eq:PZ-like-HNL}
    \cP^{PZ-\mathrm{like}} &=& 576 \Gnl^2 \left(\Fnl + 6 A_\uS \Hnl\right)^2 P_{PZ}^{[1,1;1,1;1]} \nonumber\\
    && + 6912 \Gnl^2 \Hnl \left(\Fnl + 6 A_\uS \Hnl\right) \left(P_{PZ}^{[2,1;1,1;1]} + P_{PZ}^{[1,1;2,1;1]}\right) \nonumber\\
    && + 20736 \Gnl^2 \Hnl^2 \left(P_{PZ}^{[2,2;1,1;1]} + P_{PZ}^{[1,1;2,2;1]}\right) + 55296 \Hnl^3 \left(\Fnl + 6 A_\uS \Hnl\right) P_{PZ}^{[2,1;2,1;1]} \nonumber\\
    && + 4608 \Hnl^2 \left(\Fnl + 6 A_\uS \Hnl\right)^2 P_{PZ}^{[1,1;1,1;2]} \ .
\end{eqnarray}
This expression comprises 7 terms proportional to $P_{PZ}^{[\ca,\cb;\cc,\cd;\ce]}$, corresponding to 7 sets of diagrams that feature the same convolved propagators. 
The contribution from each diagram can be evaluated by fully expanding this expression and substituting each term into Eq.~\eqref{eq:omegabar-zeta} individually, allowing us to deduce the symmetric factors for the 14 diagrams depicted in \cref{fig:PZ-like-FD} from the prefactors.

The ``renormalized'' $NC$-like diagram (the lower middle panel in \cref{fig:Feynman_Diagrams}) yields  
\begin{align}\label{eq:NC-like}
    \cP^{NC-\mathrm{like}} = 2 \sum_{\epsilon=1}^{o-2} \sum_{\delta=1}^{o-\epsilon-1} \sum_{\gamma=1}^{o-\epsilon-1} \sum_{\beta=1}^{\min(o-\gamma,o-\delta-\epsilon)} &\sum_{\alpha=1}^{\min(o-\gamma-\epsilon,o-\delta)} P_{\uR,a}^{[\alpha]} V_{\uR,3}^{[\alpha,\gamma,\epsilon]} P_{\uR,c}^{[\gamma]} V_{\uR,2}^{[\beta,\gamma]} P_{\uR,a}^{[\beta]} V_{\uR,3}^{[\beta,\delta,\epsilon]} \nonumber\\
    &\qquad\qquad\quad \times P_{\uR,c}^{[\delta]} V_{\uR,2}^{[\alpha,\delta]} P_{\uR,b}^{[\epsilon]}\ ,
\end{align}
After expanding this expression through Eqs.~(\ref{eq:P_redef}, \ref{eqs:V_redef}), each term is proportional to $P_a^{[\ca]} P_a^{[\cb]} P_c^{[\cc]} P_c^{[\cd]} P_b^{[\ce]}$, where the positive integers $\ca, \cb, \cc, \cd, \ce$ must satisfy the conditions $(\ca+\cc+\ce, \ca+\cd, \cb+\cc, \cb+\cd+\ce \leq o, \ca \geq \cb, \text{ and } \cc \geq \cd)$. 
Similar to the ``renormalized'' $N$-like diagram, we observe that swapping the Greek letter labels $\ca$ and $\cb$ adjacent to the two symmetric convolved propagators with the Greek letter labels $\cc$ and $\cd$ adjacent to the other two symmetric convolved propagators yields an equivalent diagram. 
Thereby, the term proportional to $P_a^{[\ca]} P_a^{[\cb]} P_c^{[\cc]} P_c^{[\cd]} P_b^{[\ce]}$ can be merged with the term proportional to $P_a^{[\cc]} P_a^{[\cd]} P_c^{[\ca]} P_c^{[\cb]} P_b^{[\ce]}$, which can be simplified under the constraint $\ca+\cb \geq \cc+\cd$. 
Next, we replace the subscripts: $a$ with the momenta $|\bq_2-\bq_3|$ and $|\bq-\bq_4|$, $b$ with the momentum $|\bq_1+\bq_3-\bq_4|$, and $c$ with the momenta $q_3$ and $|\bq_2-\bq_4|$. 
The replacements transform $P_a^{[\ca]} P_a^{[\cb]} P_c^{[\cc]} P_c^{[\cd]} P_b^{[\ce]}$ into $P^{[\ca]}(|\bq_2-\bq_3|) P^{[\cb]}(|\bq-\bq_4|) P^{[\cc]}(|\bq_2-\bq_4|) P^{[\cd]}(q_3) P^{[\ce]}(|\bq_1+\bq_3-\bq_4|)$. 
In particular, for scale-independent \ac{PNG} up to $\Hnl$ order (i.e., $o=4$), we introduce the shorthand notation $P_{NC}^{[\ca,\cb;\cc,\cd;\ce]}=P^{[\ca]}(|\bq_2-\bq_3|) P^{[\cb]}(|\bq-\bq_4|) P^{[\cc]}(|\bq_2-\bq_4|) P^{[\cd]}(q_3) P^{[\ce]}(|\bq_1+\bq_3-\bq_4|)$ for convenience, leading to the expression for $\cP^{NC-\mathrm{like}}$ as 
\begin{eqnarray}\label{eq:NC-like-HNL}
    \cP^{NC-\mathrm{like}} &=& 288 \Gnl^2 \left(\Fnl + 6 A_\uS \Hnl\right)^2 P_{NC}^{[1,1;1,1;1]} + 6912 \Gnl^2 \Hnl \left(\Fnl + 6 A_\uS \Hnl\right) P_{NC}^{[2,1;1,1;1]} \nonumber\\
    && + 20736 \Gnl^2 \Hnl^2 P_{NC}^{[2,2;1,1;1]} + 27648 \Hnl^3 \left(\Fnl + 6 A_\uS \Hnl\right) P_{NC}^{[2,1;2,1;1]} \nonumber\\
    && + 2304 \Hnl^2 \left(\Fnl + 6 A_\uS \Hnl\right)^2 P_{NC}^{[1,1;1,1;2]} \ .
\end{eqnarray}
This expression comprises 5 terms proportional to $P_{NC}^{[\ca,\cb;\cc,\cd;\ce]}$, corresponding to 5 sets of diagrams that feature the same convolved propagators. 
The contribution from each diagram can be evaluated by fully expanding this expression and substituting each term into Eq.~\eqref{eq:omegabar-zeta} individually, allowing us to deduce the symmetric factors for the 11 diagrams depicted in \cref{fig:NC-like-FD} from the prefactors.

The ``renormalized'' $PN$-like diagram (the lower right panel in \cref{fig:Feynman_Diagrams}) yields  
\begin{align}\label{eq:PN-like}
    \cP^{PN-\mathrm{like}} = \sum_{\iota=1}^{o-2} \sum_{\epsilon=1}^{o-2} \sum_{\delta=1}^{\min(o-\epsilon,o-\iota)-1} &\sum_{\gamma=1}^{\min(o-\epsilon,o-\iota)-1} \sum_{\beta=1}^{\min(o-\gamma-\epsilon,o-\delta-\iota)} \sum_{\alpha=1}^{\min(o-\gamma-\iota,o-\delta-\epsilon)} 
    P_{\uR,a}^{[\alpha]} V_{\uR,3}^{[\alpha,\gamma,\epsilon]} \nonumber\\
    &\qquad\quad \times P_{\uR,b}^{[\gamma]} V_{\uR,3}^{[\beta,\gamma,\iota]} P_{\uR,a}^{[\beta]} V_{\uR,3}^{[\beta,\delta,\epsilon]} P_{\uR,b}^{[\delta]} V_{\uR,3}^{[\alpha,\delta,\iota]} P_{\uR,c}^{[\epsilon]} P_{\uR,c}^{[\iota]}\ ,
\end{align}
After expanding this expression through Eqs.~(\ref{eq:P_redef}, \ref{eqs:V_redef}), each term is proportional to $P_a^{[\ca]} P_a^{[\cb]} P_b^{[\cc]} P_b^{[\cd]} P_c^{[\ce]} P_c^{[\ci]}$, where the positive integers $\ca, \cb, \cc, \cd, \ce, \ci$ must satisfy the conditions $(\ca+\cc+\ce, \ca+\cd+\ci, \cb+\cc+\ci, \cb+\cd+\ce \leq o, \ca \geq \cb, \text{ and } \cc \geq \cd, \ce \geq \ci)$. 
Similar to the ``renormalized'' $N$-like diagram, we observe that swapping the indices $\ca$ and $\cb$ of the two symmetric convolved propagators with the indices $\ce$ and $\ci$ of the other two symmetric convolved propagators yields an equivalent diagram. 
Thereby, the term proportional to $P_a^{[\ca]} P_a^{[\cb]} P_b^{[\cc]} P_b^{[\cd]} P_c^{[\ce]} P_c^{[\ci]}$ can be merged with the term proportional to $P_a^{[\ce]} P_a^{[\ci]} P_b^{[\cc]} P_b^{[\cd]} P_c^{[\ca]} P_c^{[\cb]}$, which can be simplified under the constraint $\ca+\cb \geq \ce+\ci$. 
Next, we replace the subscripts: $a$ with the momenta $|\bq_3-\bq_5|$ and $|\bq-\bq_4|$, $b$ with the momenta $|\bq_1+\bq_3-\bq_4|$ and $|\bq_2-\bq_5|$, and $c$ with the momenta $q_3$ and $|\bq_4-\bq_5|$. 
This replacements transform $P_a^{[\ca]} P_a^{[\cb]} P_b^{[\cc]} P_b^{[\cd]} P_c^{[\ce]} P_c^{[\ci]}$ into $P^{[\ca]}(|\bq_3-\bq_5|) P^{[\cb]}(|\bq-\bq_4|) P^{[\cc]}(|\bq_1+\bq_3-\bq_4|) P^{[\cd]}(|\bq_2-\bq_5|) P^{[\ce]}(|\bq_4-\bq_5|) P^{[\ci]}(q_3)$. 
In particular, for scale-independent \ac{PNG} up to $\Hnl$ order (i.e., $o=4$), we introduce the shorthand notation $P_{PN}^{[\ca,\cb;\cc,\cd;\ce,\ci]}=P^{[\ca]}(|\bq_3-\bq_5|) P^{[\cb]}(|\bq-\bq_4|) P^{[\cc]}(|\bq_1+\bq_3-\bq_4|) P^{[\cd]}(|\bq_2-\bq_5|) P^{[\ce]}(|\bq_4-\bq_5|) P^{[\ci]}(q_3)$ for convenience, leading to the expression for $\cP^{PN-\mathrm{like}}$ as 
\begin{eqnarray}\label{eq:PN-like-HNL}
    \cP^{PN-\mathrm{like}} &=& 1296 \Gnl^4 P_{PN}^{[1,1;1,1;1,1]} + 20736 \Gnl^2 \Hnl^2 \left(2 P_{PN}^{[2,1;1,1;1,1]} + P_{PN}^{[1,1;2,1;1,1]}\right) \nonumber\\
    && + 82944 \Hnl^4 \left(2 P_{PN}^{[2,2;1,1;1,1]} + P_{PN}^{[1,1;2,2;1,1]}\right) \ .
\end{eqnarray} 
This expression comprises 5 terms proportional to $P_{PN}^{[\ca,\cb;\cc,\cd;\ce,\ci]}$. 
The contribution from each diagram can be evaluated by substituting each term into Eq.~\eqref{eq:omegabar-zeta} individually, allowing us to deduce the symmetric factors for the 5 diagrams depicted in \cref{fig:PN-like-FD} from the prefactors.

To summarize briefly, for the \ac{PNG} up to $o$-th order approximation, we can derive the integrals for $\bar{\omega}_{\uGW,\uin}^{X-\mathrm{like}}$ across all nine families by substituting Eqs.~(\ref{eq:G-like}, \ref{eq:C-like}, \ref{eq:Z-like}, \ref{eq:P-like}, \ref{eq:N-like}, \ref{eq:CZ-like}, \ref{eq:PZ-like}, \ref{eq:NC-like}, \ref{eq:PN-like}) into Eq.~\eqref{eq:omegabar-zeta} respectively. 
A specific case where $o=4$ is detailed in Eqs.~(\ref{eq:G-like-HNL}, \ref{eq:C-like-HNL}, \ref{eq:Z-like-HNL}, \ref{eq:P-like-HNL}, \ref{eq:N-like-HNL}, \ref{eq:CZ-like-HNL}, \ref{eq:PZ-like-HNL}, \ref{eq:NC-like-HNL}, \ref{eq:PN-like-HNL}). 
Furthermore, in this example, in order to analyze the specific contributions from the 236 independent diagrams depicted in \cref{fig:G-like-FD} - \cref{fig:PN-like-FD}, we can substitute individual terms from these equations, rather than the complete equations, into Eq.~\eqref{eq:omegabar-zeta}, with the symmetric factor for each specific diagram being readily inferred from the corresponding coefficients. 
Since we have carried out the integrals of the convolutions of the power spectrum for the cases of $\xi=2,3,4$ in Eq.~\eqref{eq:Pxi-def} under the condition of scale-independent \ac{PNG}, this process encompasses a total of 81 integrals. 
In this manner, $\bar{\omega}_{\uGW,\uin}^{(a,b,c)}$, which signifies the component proportional to $\Fnl^a \Gnl^b \Hnl^c A_\uS^{(a+2b+3c)/2}$ as introduced in Eq.~\eqref{eq:omegabar-abc-total}, can be directly obtained using these individual calculations. 
Based on the results from these 19 categories, we are able to calculate the homogeneous component and \ac{PNG}-induced inhomogeneities of the initial \ac{SIGW} energy-density spectrum, which will be presented in the following.

\subsection{Numerical results}\label{subsec:Omegarbar-num}

To analyze the influence of the various model parameters on the isotropic component of the \ac{SIGW} background, we perform numerical calculations of the 81 integrals for all contributing diagrams individually. 
Specifically, with the spectral width set at $\sigma = 1$ and constant coefficients, such as symmetric factors, scale-independent \ac{PNG} parameters, and the spectral amplitude $A_\uS$, excluded from the integrands, we utilize \texttt{vegas} to carry out numerical integration for the unscaled spectra for diagrams that do not contain ``self-closed loops''. 
While the spectral width $\sigma$ is fixed for simplicity, other values can be readily accommodated if required. 
Although variations in $\sigma$ may affect the numerical results, they do not alter our conclusions. 
A detailed explanation of the process for transforming the integrals into a numerically advantageous form is provided in Appendix~\ref{sec:num-int}, with a focus on the radiation-dominated era of the early Universe. 
By subsequently multiplying the calculated results by the corresponding constant coefficients, we obtain $\bar{\omega}_{\uGW,\uin}^D$, where $D$ labels the specific diagram as shown in \cref{fig:G-like-FD} - \cref{fig:PN-like-FD}. 
The sum of $\bar{\omega}_{\uGW,\uin}^D$ within a specific family leads to $\bar{\omega}_{\uGW,\uin}^{X-\mathrm{like}}$, while the sum of $\bar{\omega}_{\uGW,\uin}^Y$ within a specific category results in $\bar{\omega}_{\uGW,\uin}^{(a,b,c)}$.
The total energy-density fraction spectrum $\bar{\omega}_{\uGW,\uin}(q)$ is obtained by summing over all categories of $\bar{\omega}_{\uGW,\uin}^{(a,b,c)}$ or all families of $\bar{\omega}_{\uGW,\uin}^{X-\mathrm{like}}$, akin to Eq.~\eqref{eq:omegabar-abc-total} and Eq.~\eqref{eq:omegabar-X-total}. 
Following this, we apply Eq.~\eqref{eq:Omega0} to derive the present-day energy density fraction spectrum $\bar{\Omega}_{\uGW,0}(\nu)$. 
In this subsection, we will present these numerical results and demonstrate the effects of \ac{PNG} on the isotropic component of the \ac{SIGW} background.

\begin{figure*}[htbp]
    \centering
    \includegraphics[width = 1 \textwidth]{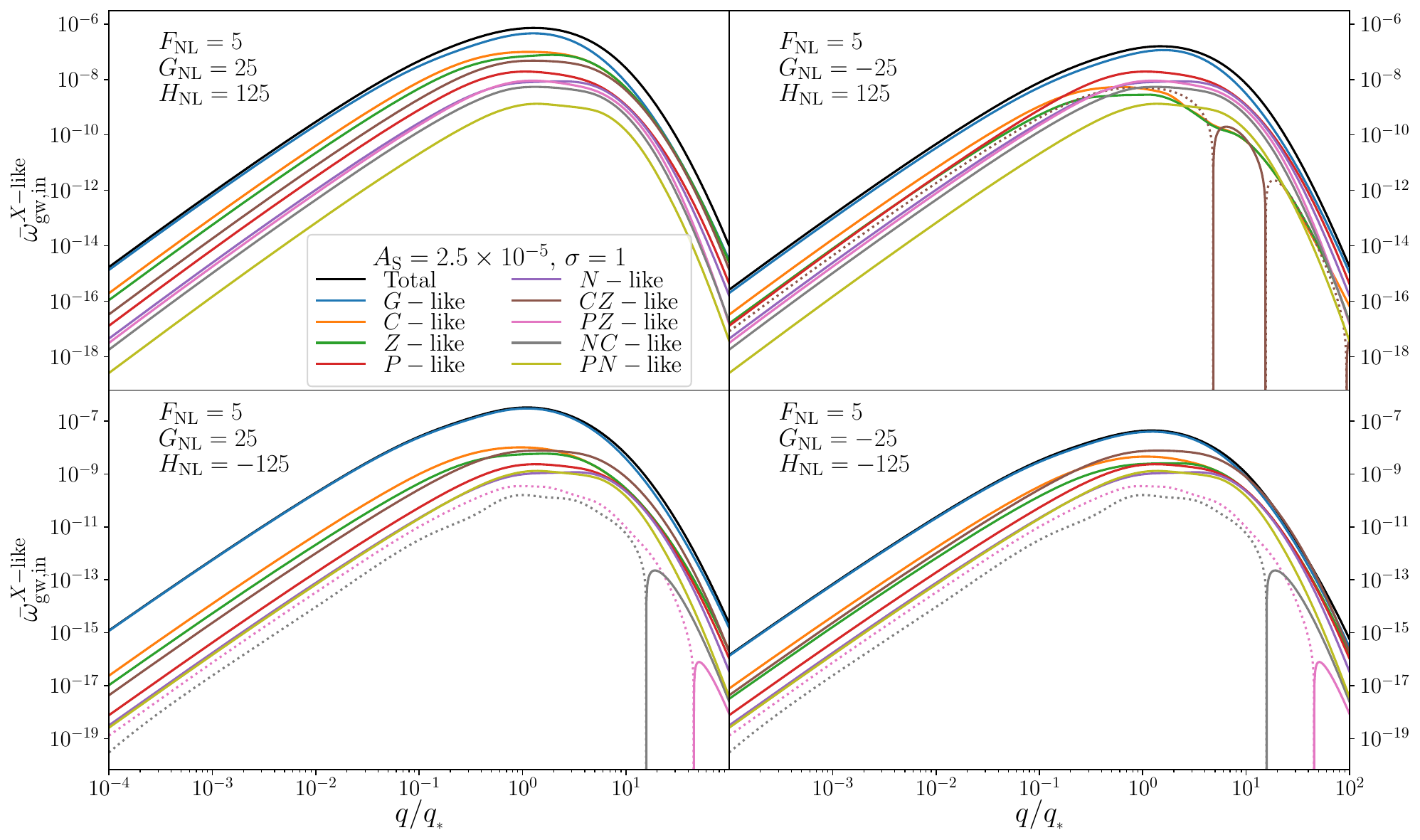}
    \caption{Contributions from the nine families of \acp{SIGW} energy-density fraction spectrum. Dotted lines indicate where the values are negative. }\label{fig:Omega-like}
\end{figure*}

To compare the contributions of the nine families to the isotropic component of the \ac{SIGW} background produced during the radiation-dominated era, we present their spectra $\bar{\omega}_{\uGW,\uin}^{X-\mathrm{like}}$ alongside the total average energy-density spectrum $\bar{\omega}_{\uGW,\uin}$ (indicated in black) in \cref{fig:Omega-like}. 
We consider the scenarios where $A_\uS=3.6\times10^{-3}$ and $|\Hnl|A_\uS^{3/2} = 0.3|\Gnl|A_\uS = 0.3^2|\Fnl|A_\uS^{1/2} = 0.3^3$.
Here, we specifically focus on the effects of varying the signs of $\Gnl$ and $\Hnl$, while keep the sign of $\Fnl$ positive due to the sign degeneracy between $\Fnl$ and $\Hnl$ in $\bar{\omega}_{\uGW,\uin}$. 
In more detail, the upper panels correspond to positive values of $\Fnl \Hnl$, while the left panels depict positive $\Gnl$. 
Conversely, the lower panels illustrate negative $\Fnl \Hnl$, and the right panels show negative $\Gnl$. 
Across all scenarios, the ``$G$-like'' contributions are generally dominant among the nine families, while contributions from other families related to the connected components of the four-point correlator of $\zeta_\uS$ are typically subdominant. 
In fact, the ``$G$-like'' spectra are significantly larger than their Gaussian counterparts, indicating that loop corrections arising from non-Gaussian effects are substantial. 
Excluding the ``$G$-like'' contributions, the contributions from the ``$C$-like'', ``$Z$-like'', and ``$CZ$-like'' families are usually more significant, unless $\Gnl$ is negative and $\Fnl \Hnl$ is positive; in such cases, the ``$P$-like'' contribution becomes the most prominent. 
Notably, these contributions are more pronounced in the ultraviolet regime compared to the ``$G$-like'' contributions. 
These features underscore the necessity of studying all families, as this may aid in determining the values of \ac{PNG} parameters. 
Furthermore, for negative \ac{PNG} parameters, it is important to note that contributions from certain families, such as ``$CZ$-like'', ``$PZ$-like'', and ``$NC$-like'', could also be negative within specific frequency bands, although they are often subdominant relative to contributions from other families in those ranges. 
Additionally, the exploration of other scenarios, including various values of $\sigma$ and additional epochs in the early Universe, can be readily extended for further investigation.

\begin{figure}[htbp]
    \centering
    \includegraphics[width = 0.8 \columnwidth]{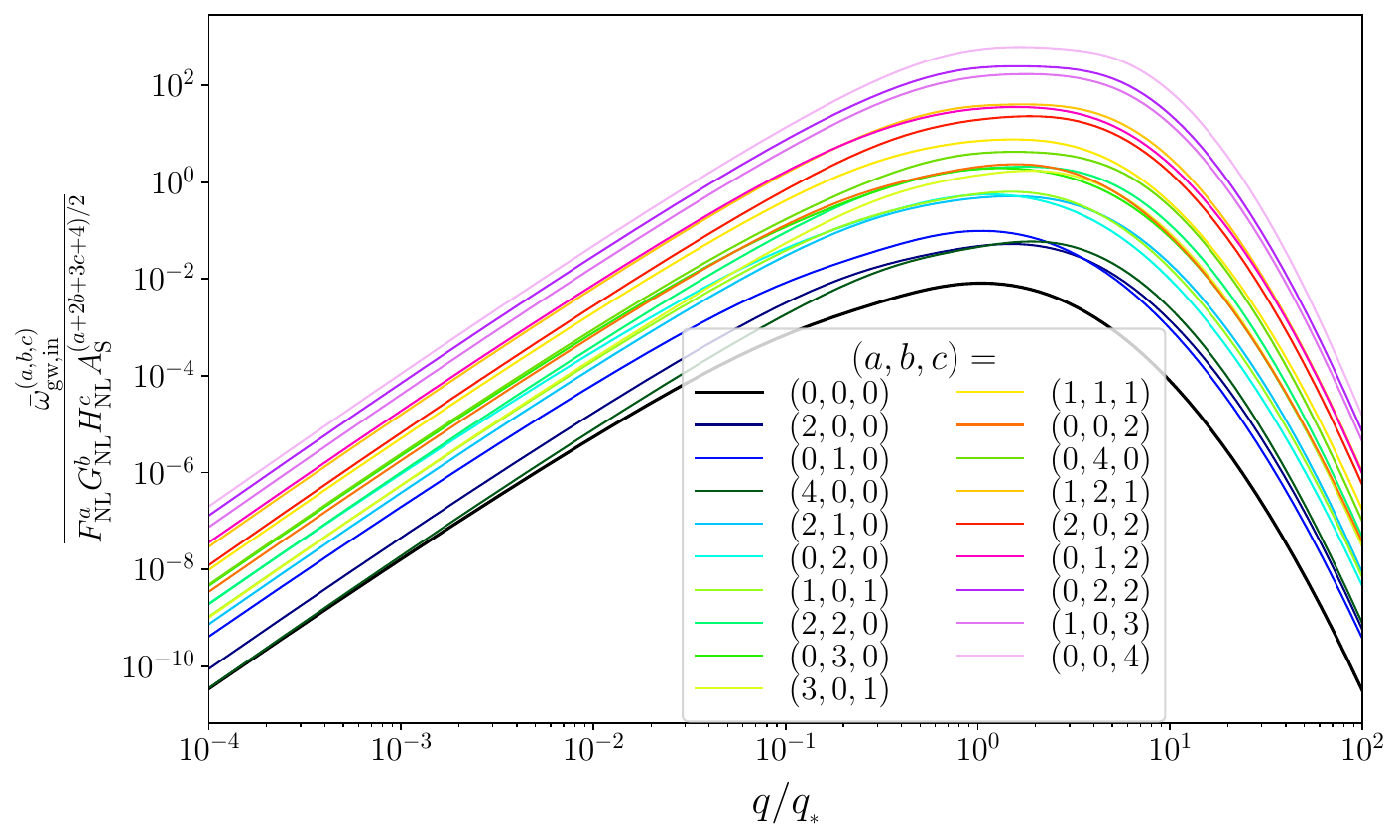}
    \caption{Unscaled (or equivalently, $A_\uS=1$, $\Fnl=1$, $\Gnl=1$ and $\Hnl=1$) components of the energy-density fraction spectra of \acp{SIGW} in powers of the $\Fnl$, $\Gnl$ and $\Hnl$. }\label{fig:Unscaled_Omegabar}
\end{figure}

In addition to the decomposition into nine families based on the topology of the diagrams shown in \cref{fig:G-like-FD} - \cref{fig:PN-like-FD}, $\bar{\omega}_{\uGW,\uin}$ is also classified into 19 distinct categories according to the powers of the three \ac{PNG} parameters, as introduced in Eq.~\eqref{eq:omegabar-abc-total}. 
The numerical results presented in \cref{fig:Unscaled_Omegabar} illustrate the unscaled components of the energy-density fraction spectra, denoted as $(\Fnl^a \Gnl^b \Hnl^c A_{S}^{(a+2b+3c+4)/2})^{-1}\bar{\omega}_{\uGW,\uin}^{(a,b,c)}$. 
Specifically, $\bar{\omega}_{\uGW,\uin}^{(0,0,0)}$ corresponds to the energy-density fraction spectrum in the Gaussian case, while $\bar{\omega}_{\uGW,\uin}^{(0,1,0)} = 24 A_\uS \Gnl \bar{\omega}_{\uGW,\uin}^{(0,0,0)}$ and $\bar{\omega}_{\uGW,\uin}^{(2,0,0)}$ are the leading-order categories in $A_\uS$ arising from \ac{PNG}. 
These results facilitate an efficient scan of the parameter space for $A_\uS$ and \ac{PNG} parameters by allowing for the subsequent rescaling of the unscaled spectra, which is particularly advantageous for isolating the impacts of each parameter on the total spectrum. 
The 19 profiles illustrated in the figure illustrate that as the power of $A_\uS$ (i.e., $(a+2b+3c+4)/2$) increases, the unscaled spectrum $\bar{\omega}_{\uGW,\uin}^{(a,b,c)}$ generally displays larger amplitudes.
For instance, the unscaled spectral amplitudes of $\bar{\omega}_{\uGW,\uin}^{(2,0,0)}$ and $\bar{\omega}_{\uGW,\uin}^{(0,1,0)}$ are nearly an order of magnitude greater than that of $\bar{\omega}_{\uGW,\uin}^{(0,0,0)}$, suggesting that the individual contributions from $\bar{\omega}_{\uGW,\uin}^{(2,0,0)}$ and $\bar{\omega}_{\uGW,\uin}^{(0,1,0)}$ become comparable to the Gaussian component for $\Fnl^2 A_\uS \sim \Gnl A_\uS \sim 1$ when $A_\uS \gtrsim 0.1$. 
Likewise, other categories with higher powers of $A_\uS$ cannot be neglected if $A_\uS$ and the \ac{PNG} parameters are sufficiently large. 
For categories of the same order in $A_\uS$, most unscaled spectral amplitudes are comparable, with the exception of $\bar{\omega}_{\uGW,\uin}^{(4,0,0)}$, $\bar{\omega}_{\uGW,\uin}^{(0,4,0)}$, and $\bar{\omega}_{\uGW,\uin}^{(1,1,1)}$. 
The unscaled spectral amplitudes of $\bar{\omega}_{\uGW,\uin}^{(4,0,0)}$ and $\bar{\omega}_{\uGW,\uin}^{(0,4,0)}$ are significantly smaller than those of other categories at $\mathcal{O}(A_\uS^6)$, whereas the unscaled spectral amplitude of $\bar{\omega}_{\uGW,\uin}^{(1,1,1)}$ is notably larger than the amplitudes of other categories at $\mathcal{O}(A_\uS^5)$. 
Therefore, when the \ac{PNG} parameters are of the same order of magnitude, the figure clearly illustrates that the largest contribution at the $\mathcal{O}(A_\uS^5)$ level arises from $\bar{\omega}_{\uGW,\uin}^{(1,1,1)}$.

\begin{figure}[htbp]
    \centering
    \includegraphics[width = 0.8 \columnwidth]{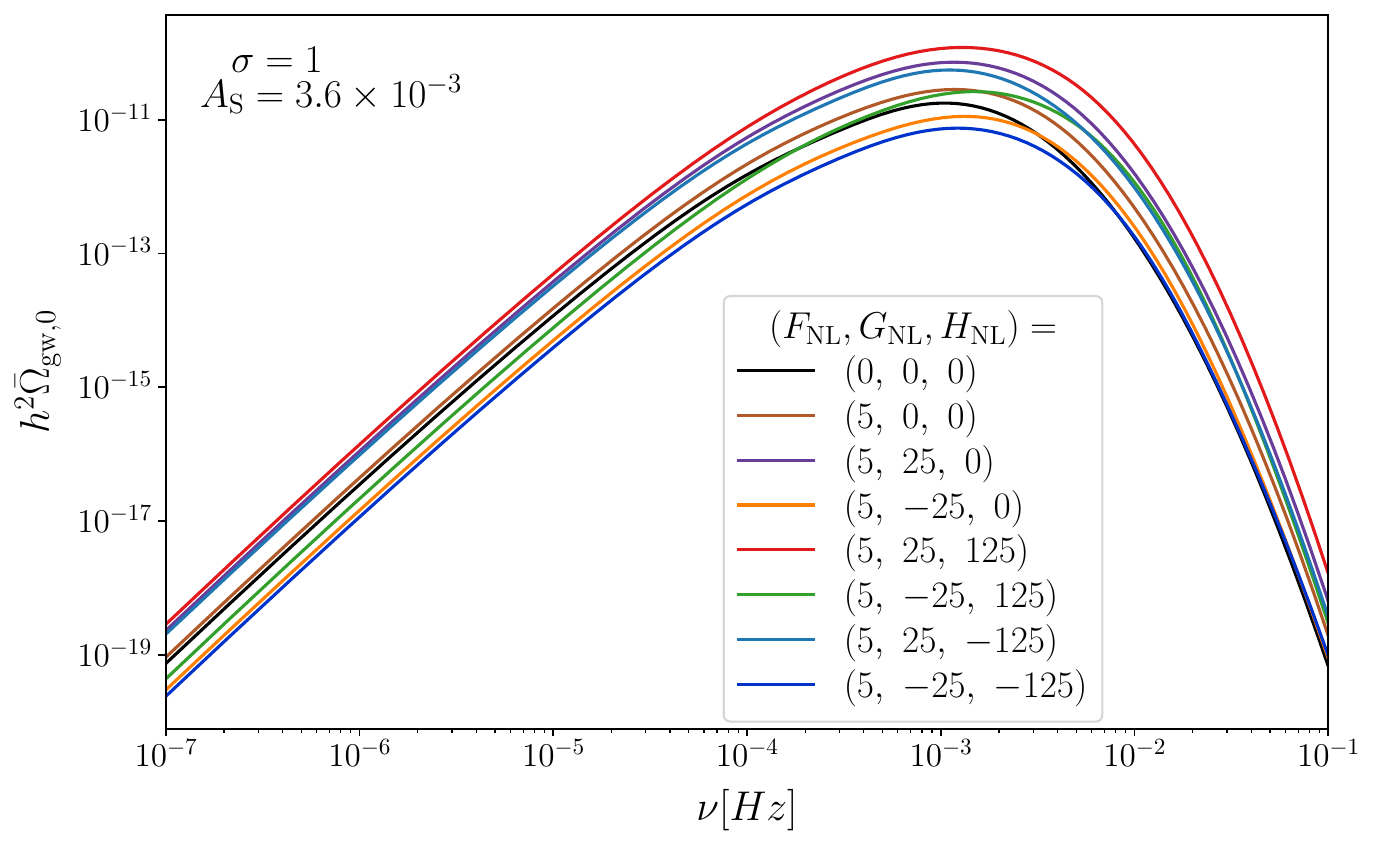}
    \caption{Energy-density fraction spectra of \acp{SIGW} in the present Universe, where $\nu_\ast=q_\ast/2\pi$. }\label{fig:Total_Omegabar}
\end{figure}

To investigate the impact of various orders of \ac{PNG} and their respective signs on the energy density of the \ac{SIGW}, we present the present-day energy-density fraction spectrum of \acp{SIGW}, denoted as $h^2 \bar{\Omega}_{\uGW,0} (\nu)$, in \cref{fig:Total_Omegabar}. 
The model parameters, including the spectral amplitude $A_\uS$, the spectral width $\sigma$ of the primordial curvature spectrum, and the modes of the \ac{PNG} parameters (when non-zero), are consistent with those from \cref{fig:Omega-like}. 
In comparison to the Gaussian case (depicted in black), it is observed that positive (negative) \ac{PNG} enhances (suppresses) the spectral magnitude of $h^2 \bar{\Omega}_{\uGW,0} (\nu)$, which span nearly two orders of magnitude. 
The black, brown, violet, and red curves illustrate that the presence of higher orders of positive \ac{PNG} results in an increase in spectral magnitude. 
In contrast, the spectral magnitudes featuring negative $\Gnl$ with positive or zero $\Hnl$ are smaller than those of the Gaussian case, particularly in the low-frequency range. 
This figure reveals that the sign of $\Gnl$ has a more substantial impact on the energy-density spectrum than the sign of $\Fnl \Hnl$. 
This difference can be understood by referring to \cref{fig:Unscaled_Omegabar}. 
In fact, altering the signs of $\Gnl$ or $\Hnl$ only affects the signs of categories with odd $b$ or odd $c$, respectively. 
As indicated in \cref{fig:Unscaled_Omegabar}, the dominant contribution from categories with odd $b$ is $\omega_\uGW^{(0,1,0)}$ at $\mathcal{O}(A_\uS^3)$ order, while the dominant contribution from categories with odd $c$ is $\omega_\uGW^{(1,0,1)}$ at $\mathcal{O}(A_\uS^4)$ order. 
Given that we have fixed $A_\uS = 3.6 \times 10^{-3} \ll 10^{-1}$, the contribution from $\omega_\uGW^{(0,1,0)}$ to the total spectrum is significantly larger than that from $\omega_\uGW^{(1,0,1)}$.
Additionally, it is noteworthy that the spectral shape of $h^2 \bar{\Omega}_{\uGW,0} (\nu)$ remains nearly invariant across various levels of \ac{PNG}. 
Specifically, as the frequency increases from $\nu/\nu_\ast = 10^{-4}$ to $\nu/\nu_\ast = 10^2$, the spectral index decreases from approximately $3$ to $-9$ (as will be demonstrated in Subsection~\ref{subsec:IR}). 
This characteristic will be crucial for investigating the anisotropies in the \ac{SIGW} background in subsequent sections.

\begin{figure*}[htbp]
    \centering
    \includegraphics[width = 1 \textwidth]{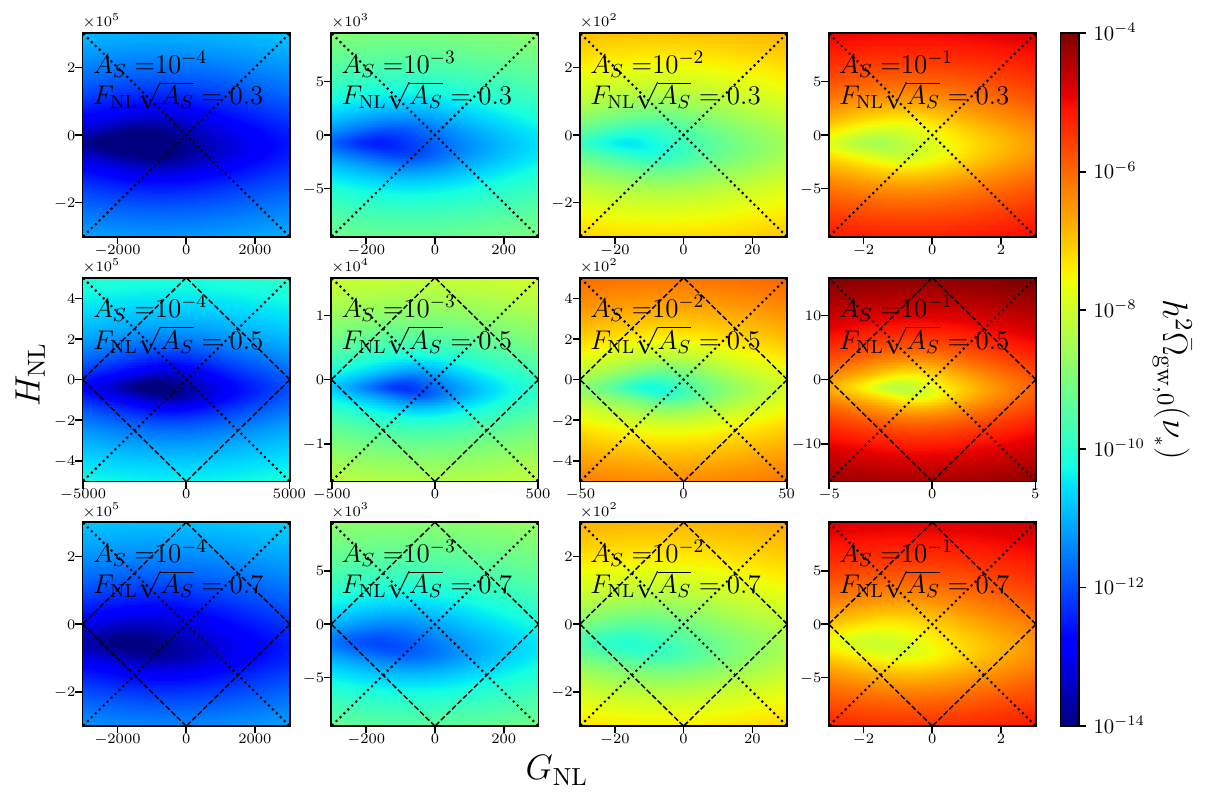}
    \caption{Present-day energy-density fraction spectrum of SIGWs with respect to the $\Fnl$, $\Gnl$ and $\Hnl$ with the frequency set to $\nu=\nu_\ast$. Dotted lines represent $\Hnl = \pm \Gnl / \sqrt{A_\uS}$, and dashed lines represent $|\Fnl|\sqrt{A_\uS} + |\Gnl|A_\uS + |\Hnl|\sqrt{A_\uS^3} = 1$. }\label{fig:Omega_G-H}
\end{figure*}

\cref{fig:Omega_G-H} displays the dependence of the present-day energy-density fraction spectrum of \acp{SIGW} on \ac{PNG} parameters and $A_\uS$. 
A $3 \times 4$ array of contour plots is presented for $h^2 \bar{\Omega}_{\uGW,0}$ at the frequency $\nu = \nu_\ast$ in relation to $\Gnl$ and $\Hnl$. 
From left to right, the values of $A_\uS$ increase from $10^{-4}$ to $10^{-1}$ in each column. 
The three rows, arranged from top to bottom, correspond to values of $\Fnl\sqrt{A_\uS}$ of 0.3, 0.5, and 0.7, respectively. 
Within each row of contour plots, specific values of $\Gnl A_\uS$ and $\Hnl A_\uS^{3/2}$ correspond to the same position. 
Each contour plot is divided into four triangular regions by dotted lines: the left and right regions indicate where $\Hnl \sqrt{A_\uS^3}$ is less than $\Gnl A_\uS$, whereas the top and bottom regions represent the converse. 
Furthermore, the area enclosed by dashed lines signifies the condition $|\Fnl|\sqrt{A_\uS} + |\Gnl|A_\uS + |\Hnl|\sqrt{A_\uS^3} < 1$. 
It is noteworthy that the plots in the first row do not contain dashed lines as they satisfy this condition at all positions. 
The intersection regions of the left and right triangular regions with the area bounded by dashed lines in each panel represent the ranges of \ac{PNG} parameters that meet perturbativity conditions. 
As anticipated, the contour plots in each row display similar patterns. 
The primary difference lies in the spectral magnitudes of $h^2 \bar{\Omega}_{\uGW,0}(\nu)$, which consistently increase with $A_\uS$.
Additionally, for each contour plot, the spectral magnitude attains its maximum value when $\Gnl A_\uS \simeq \Hnl \sqrt{A_\uS^3}$ reaches its peak values within the perturbativity ranges, while it reaches the minimum value when $\Gnl$ and $\Hnl$ assume small negative values due to the positive value of $\Fnl$. 
It is evident that $h^2 \bar{\Omega}_{\uGW,0} (\nu)$ exhibits greater sensitivity to $\Hnl \sqrt{A_\uS^3}$ than to $\Gnl A_\uS$, suggesting that the contributions from the $\Hnl$ order may be more significant than those from the $\Gnl$ order within the perturbativity ranges.

\begin{figure}[htbp]
    \centering
    \includegraphics[width = 1 \columnwidth]{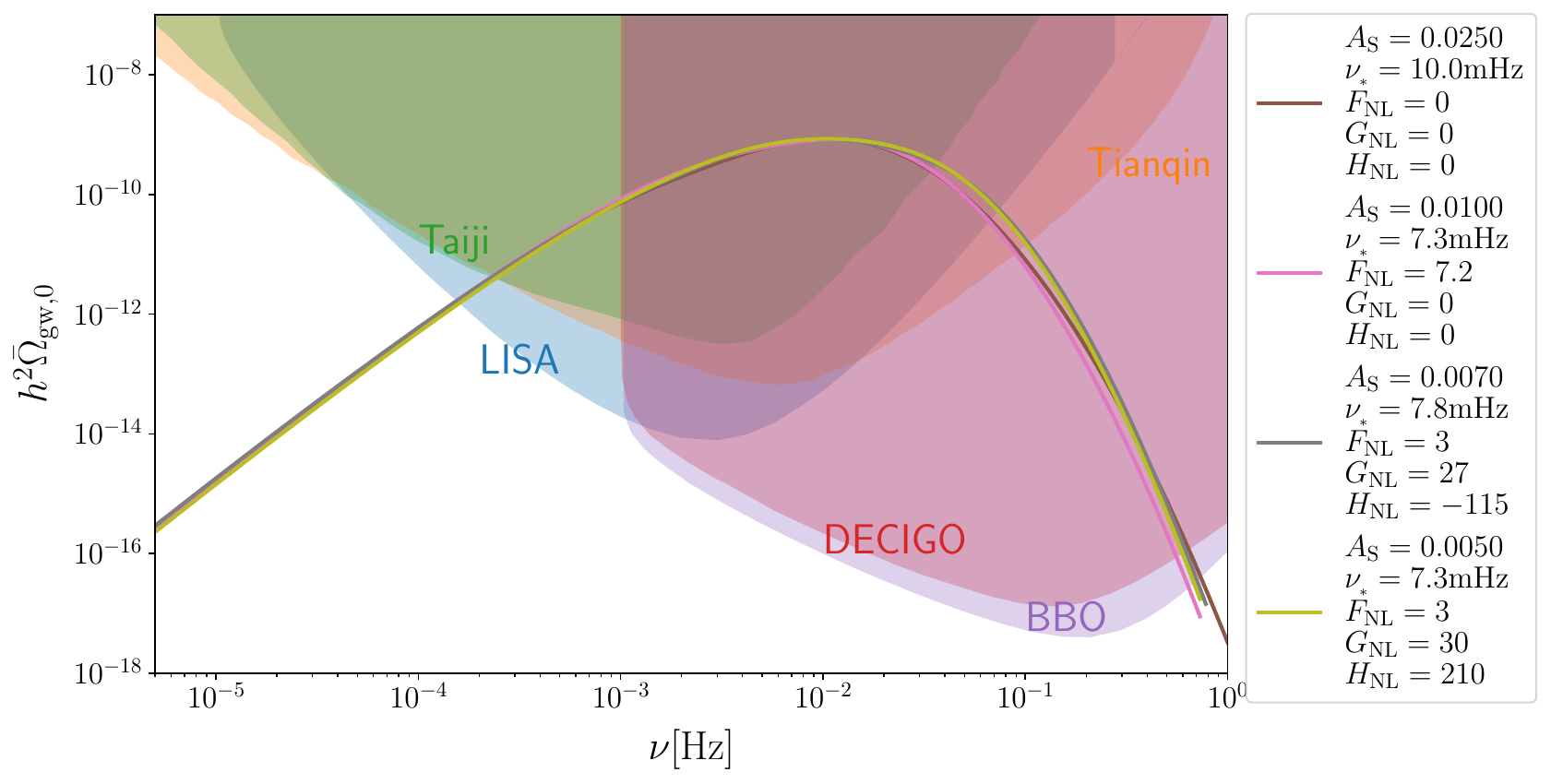}
    \caption{Energy-density fraction spectra of \acp{SIGW} in the present Universe compared with the sensitivity curves of \ac{LISA} (blue shaded region) \cite{LISA:2024hlh, Baker:2019nia, Smith:2019wny}, Tianqin (orange shaded region) \cite{TianQin:2015yph,TianQin:2020hid,Zhou:2023rop}, Taiji (green shaded region) \cite{Hu:2017mde,Ren:2023yec}, \ac{DECIGO} (red shaded region) \cite{Seto:2001qf, Kawamura:2020pcg, Schmitz:2020syl}, and \ac{BBO} (purple shaded region) \cite{Crowder:2005nr, Smith:2016jqs, Schmitz:2020syl}. 
    For all the cases, $\sigma=1$ is assumed for the spectra.
    }\label{fig:Omega_degeneracy}
\end{figure}

For model parameters within appropriate ranges, the energy-density fraction spectrum of \acp{SIGW} predicted by this work is theoretically detectable by upcoming space-based \ac{GW} detectors and their networks, such as Taiji \cite{Hu:2017mde,Ren:2023yec}, Tianqin \cite{TianQin:2015yph,TianQin:2020hid,Zhou:2023rop}, \ac{LISA}  \cite{LISA:2024hlh, Baker:2019nia, Smith:2019wny}, \ac{BBO} \cite{Crowder:2005nr,Harry:2006fi}, \ac{DECIGO} \cite{Sato:2017dkf,Kawamura:2020pcg}, and others. 
We select four sets of model parameters and plot a range of predicted spectra against the sensitivity curves of these detectors in \cref{fig:Omega_degeneracy}. 
The comparisons indicate that \ac{LISA}, Taiji, and Tianqin possess the capability to detect \acp{SIGW} when $A_\uS$ reaches $10^{-3}$, particularly in cases of positive \ac{PNG}. 
Moreover, \ac{BBO} and \ac{DECIGO} exhibit higher sensitivity in detecting \acp{GW}, enabling them to detect \acp{SIGW} at smaller values of $A_\uS$. 
Notably, detectable \acp{SIGW} by these space-borne detectors may be accompanied by the production of \acp{PBH}, which could potentially account for all dark matter (see reviews in Refs.~\cite{Green:2020jor,Carr:2020gox,Escriva:2022duf} and references therein). 
Specifically, apart from the amplitude of the primordial curvature power spectrum $A_\uS$, the abundance of \acp{PBH} is also significantly influenced by \ac{PNG} \cite{Bullock:1996at,Byrnes:2012yx,Young:2013oia,Bugaev:2013vba,Nakama:2016gzw,Franciolini:2018vbk,Inomata:2020xad,Kitajima:2021fpq,Ferrante:2022mui,Gow:2022jfb,Franciolini:2023pbf,Iovino:2024tyg,Inui:2024fgk,Passaglia:2018ixg,Atal:2018neu,Atal:2019cdz,Taoso:2021uvl,Meng:2022ixx,Chen:2023lou,Kawaguchi:2023mgk,Choudhury:2023kdb,Garcia-Bellido:2017aan,Yoo:2019pma,Ezquiaga:2019ftu,Carr:2020gox,Riccardi:2021rlf,Escriva:2022pnz,Kehagias:2019eil,Cai:2021zsp,Cai:2022erk,Young:2022phe,Zhang:2021vak,vanLaak:2023ppj,Gow:2023zzp,Franciolini:2023wun,Chen:2024pge}. 
Thus, evaluating the model parameters in relation to \ac{SIGW} signals is crucial. 
However, it is important to acknowledge the significant degeneracies among the model parameters that affect the energy-density fraction spectra. 
The four spectra in \cref{fig:Omega_degeneracy} are closely aligned, yet they correspond to entirely distinct combinations of parameters. 
Therefore, the \ac{SIGW} energy-density fraction spectrum characterizing the isotropic component may not sufficiently capture all statistical features of primordial curvature perturbations. 
A further analysis on the anisotropies in the \ac{SIGW} background is essential to extract more information about the early Universe.

%% file: TeX/4Anisotropy.tex
\section{Anisotropies in the SIGW background}\label{sec:Cl}

In Section~\ref{sec:ED}, we have conducted an analysis of the energy-density full spectrum of \acp{SIGW}, which includes both the underlying isotropic component and the superimposed fluctuations. 
The examination of the isotropic component is comprehensively presented in Section~\ref{sec:Omegabar}. 
Building upon the methodologies and findings of these investigations, we aim to scrutinize the statistics of fluctuations on the \ac{SIGW} background in the subsequent sections. 
In this section, we focus on investigating the two-point correlation function of density contrast in \acp{SIGW}, a fundamental quantity for characterizing the anisotropies in the \ac{SIGW} background.

\subsection{Angular power spectrum: two-point angular correlation}

To characterize the anisotropies in the \ac{SIGW} background, we adopt the standard treatment for \ac{CMB} anisotropies and introduce the reduced angular power spectrum of \acp{SIGW} using the two-point angular correlator of the present-day density contrast $\delta_{\uGW,0}$. 
This spectrum describes the statistical properties of the \ac{SIGW} energy density from any two directions. 
For convenience, we denote $\bq = q\bn$ and utilize spherical harmonics to decompose $\delta_{\uGW,0} (\bq)$ into $\delta_{\uGW,0,\ell m} (2\pi\nu)$, as given by 
\begin{eqnarray}\label{eqs:spher-harm}
    \delta_{\uGW,0}(\bq) = \sum_{\ell m} \delta_{\uGW,0,\ell m}(2\pi\nu) Y_{\ell m} (\bn)\, , \ 
    \text{where } \delta_{\uGW,0,\ell m}(2\pi\nu) = \int \ud^2 \bn\, \delta_{\uGW,0}(\bq) Y_{\ell m}^\ast (\bn)\ .
\end{eqnarray}  
By applying the cosmological principle, which assumes statistical homogeneity and isotropy on large scales, we achieve rotation-invariance in the \ac{GW} background. 
Consequently, the two-point correlator of $\delta_{\uGW,0,\ell m}$ defines the reduced angular power spectrum in the form of 
\begin{equation}\label{eq:Ct-def}
    \left\langle\prod_{i=1}^2\delta_{\uGW,0,\ell_i m_i}(2\pi\nu)\right\rangle
    = \delta_{\ell_1 \ell_2} \delta_{m_1 m_2} (-1)^{m_1} \tilde{C}_{\ell_1} (\nu)\ .
\end{equation}
Previous works in Refs.~\cite{Bartolo:2019zvb,Li:2023qua,Li:2023xtl,ValbusaDallArmi:2020ifo,Dimastrogiovanni:2021mfs,LISACosmologyWorkingGroup:2022kbp,LISACosmologyWorkingGroup:2022jok,Unal:2020mts,Malhotra:2020ket,Carr:2020gox,Cui:2023dlo,Malhotra:2022ply,ValbusaDallArmi:2023nqn,LISACosmologyWorkingGroup:2023njw} have extensively explored the reduced angular power spectrum of \acp{GW} originating from cosmological sources \cite{Caprini:2018mtu}. 
Due to the limitations in the angular resolutions of ongoing and planned \ac{GW} detectors \cite{Baker:2019ync,Capurri:2022lze,Gair:2015hra,Romano:2016dpx,LISACosmologyWorkingGroup:2022kbp,LIGOScientific:2016nwa,LIGOScientific:2019gaw,KAGRA:2021mth,NANOGrav:2023tcn}, these investigations primarily focus on the low multipoles $\ell$, which aligns with the focus of our own work. 
Nevertheless, the extension of our analysis to higher multipoles $\ell$ can be theoretically accomplished with relative ease if necessary.

\begin{figure}[htbp]
    \centering
    \includegraphics[width = 0.8 \columnwidth]{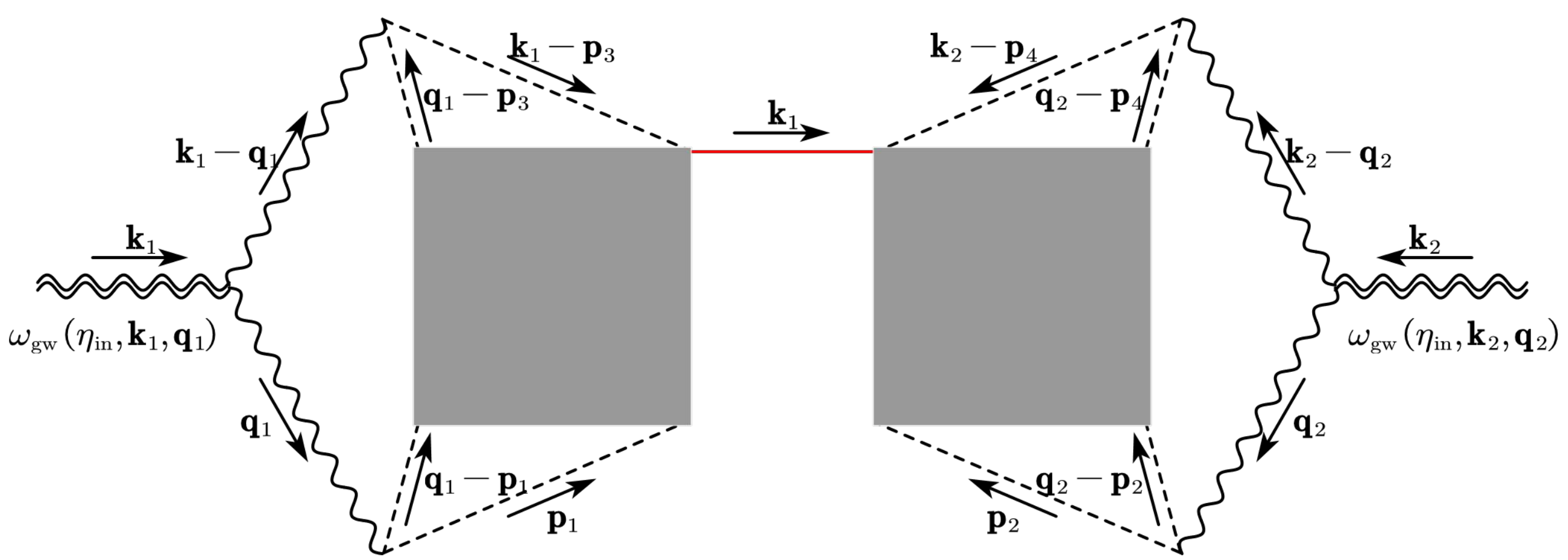}
    \caption{Feynman-like diagram for the two-point correlator of $\omega_\uGW^{(1)} (\eta_\uin,\bk,\bq)$. }\label{fig:2-correlator}
\end{figure}

The derivation process of $\tilde{C}_\ell (\nu)$ for \ac{SIGW} has been previously discussed in Refs.~\cite{Bartolo:2019yeu,Bartolo:2019oiq,Bartolo:2019zvb,Li:2023qua,Li:2023xtl}, paralleling the derivation process for the \ac{CMB} at low multipoles. 
We will provide a concise overview of this process. 
According to Eqs.~(\ref{eq:omega-split}, \ref{eq:omega-delta}, \ref{eq:delta-0}), the calculation of $\tilde{C}_{\ell} (\nu) \sim \langle\delta_{\uGW,0} \delta_{\uGW,0}\rangle$ involves the two-point correlator of $\omega_{\uGW,\uin} (\bq)$. 
As emphasized in Section~\ref{sec:ED}, Eqs.~\eqref{eqs:omega-result} are applicable for evaluating this correlator, suggesting that $\delta_{\uGW,0}$ can be expressed as a polynomial in powers of $\zeta_{\ugL}$, as indicated in Eq.~\eqref{eq:delta0-123}. 
Thus, the reduced angular power spectrum $\tilde{C}_\ell$ is a polynomial in powers of $A_\uL$. 
Given that $A_\uL$ is a small quantity of order $\cO(10^{-9})$, we focus on the leading order in $A_\uL$. 
We approximate the density contrast as $\delta_{\uGW,0}^{(1)}$, which encompasses the initial inhomogeneities and the \ac{SW} effect at order $\cO(\zeta_{\ugL})$ in Eq.~\eqref{eq:delta0-1}. 
The Feynman-like diagram for the associated two-point correlator $\langle\omega_{\uGW,\uin}^{(1)} \omega_{\uGW,\uin}^{(1)}\rangle$ is illustrated in \cref{fig:2-correlator}, based on \cref{fig:FD-Rules} and \cref{fig:FD_Frame}. 
Below, we briefly review the derivation process for the explicit expression of $\tilde{C}_\ell (\nu)$. 
Following the decomposition of $\delta_{\uGW,0}^{(1)}$ using spherical harmonics, as outlined in Eqs.~\eqref{eqs:spher-harm}, we express the two-point correlator $\langle\delta_{\uGW,0,\ell_1 m_1}^{(1)} \delta_{\uGW,0,\ell_2 m_2}^{(1)}\rangle$ in Eq.~(\ref{eq:Ct-def}) as an integral, with integration variables $\bk_1,\bk_2,\bn_1,\bn_2$. 
Subsequently, after moving integration-independent factors (variables only dependent on $\eta$ and $\nu$) outside the integral, the integrand consists of the spherical harmonics $Y_{\ell m}^\ast (\bn)$ and $e^{i\bk_1\cdot\bx_1 + \bk_2\cdot\bx_2} \langle \zeta_{\ugL}(\bk_1) \zeta_{\ugL}(\bk_2) \rangle$. 
The latter can be reduced to $\delta^{(3)} (\bk_1+\bk_2) e^{i\bk_1\cdot(\bx_1-\bx_2)} 2\pi^2 \Delta^2_\uL / k_1^3$ according to Eqs.~\eqref{eq:PgX-def} and \eqref{eq:Flat}. 
Moreover, in the leading-order approximation of the Boltzmann equation, as discussed in Appendix~\ref{sec:Boltz} and Refs.~\cite{Contaldi:2016koz,Bartolo:2019oiq,Bartolo:2019yeu}, the line-of-sight relation $\bx_1-\bx_0 = (\eta_\uin-\eta_0) \bn_1$ is effective, resulting in $\bx_1-\bx_2 = (\eta_\uin-\eta_0)(\bn_1-\bn_2)$. 
Therefore, the exponential factor $e^{i\bk_1\cdot(\bx_1-\bx_2)} \sim e^{i\bk_1\cdot(\bn_2-\bn_1)(\eta_0-\eta_\uin)}$ can be rewritten in terms of spherical harmonics and spherical Bessel functions using the identity 
\begin{equation}\label{eq:exp-jYY}
    e^{i r (\bn_1 \cdot \bn_2)} = 4 \pi \sum_{\ell m}  i^\ell j_\ell (r) Y^\ast_{\ell m} (\bn_1) Y_{\ell m} (\bn_2)\ ,
\end{equation}  
where $\bn_1$ and $\bn_2$ denote unit vectors and $r \in \mathbb{R}$. 
It is crucial to note that $j_\ell (-\mu) = (-1)^\ell j_\ell (\mu)$.
By utilizing the properties of spherical harmonics and the integral relation 
\footnote{The properties of spherical harmonics relevant to the derivation processes in this work include the conjugate transformation $(-1)^\ell Y_{\ell m} (\bn) = Y_{\ell m} (-\bn) = Y_{\ell m}^\ast (\bn) = (-1)^m Y_{\ell -m} (\bn)$, orthogonality $\int \ud^2 \bn\, Y_{\ell_1 m_1}(\bn) Y_{\ell_2 m_2}^\ast (\bn) = \delta_{\ell_1 \ell_2} \delta_{m_1 m_2}$, and completeness $\sum_{\ell m} Y_{\ell m}^\ast (\bn_1) Y_{\ell m}(\bn_2) = \delta^{(2)} (\bn_1-\bn_2)$. 
}
\begin{equation}\label{eq:jl-2-int}
    \int_0^\infty \ud \ln k\, j_{\ell}^2 (k \left(\eta_0 - \eta_\uin\right)) = \frac{1}{2 \ell (\ell + 1)} \ , 
\end{equation} 
we successfully derive the fundamental formula for the reduced angular power spectrum, leading to 
\begin{equation}\label{eq:Ct}
    \tilde{C}_\ell (\nu) 
    = \frac{2 \pi A_\uL}{\ell (\ell+1)} 
        \left[
            \frac{\omega_{\ung,\uin}^{(1)} (q)}{\bar{\omega}_{\uGW,\uin} (q)}
            + \frac{3}{5} \left(6 - n_{\uGW} (\nu)\right)
        \right]^2\ ,
\end{equation}
where the relationship between $\nu$ and $q$ is illustrated in Eq.~\eqref{eq:nu-in} and the subsequent discussion therein. 
In this formula, the first term in the square brackets corresponds to the 1st-order \ac{PNG}-induced inhomogeneity of \acp{SIGW} at the production time, while the second term corresponds to the \ac{SW} effect and the initial inhomogeneity arising from non-adiabaticity due to large-scale scalar perturbations. 
Thus, this formula indicates that the anisotropies in \acp{SIGW} are attributed to both the presence of \ac{PNG} and the large-scale inhomogeneity of cosmological fluids during propagation. 

While the average energy-density spectrum $\bar{\omega}_{\uGW,\uin} (q)$ has been derived in Section~\ref{sec:Omegabar} and the spectral index $n_\uGW$ can be obtained directly from Eq.~\eqref{eq:ngw-def}, the remaining task is to derive the specific expression for the 1st-order large-scale modulation $\omega_{\ung,\uin}^{(1)} (q)$ introduced in Eqs.~\eqref{eqs:omega-result}. 
As illustrated in \cref{fig:2-correlator}, the connected diagram of $\langle \omega_{\uGW,\uin}^{(1)} \omega_{\uGW,\uin}^{(1)} \rangle$ consists of two diagrams of $\omega_{\uGW,\uin}^{(1)}$ (the first panel of \cref{fig:FD_Frame}), with their extensional red solid lines connected to form a single diagram, referred to as a ``non-Gaussian bridge''. 
Compared to the diagram of $\bar{\omega}_{\uGW,\uin}$ in \cref{fig:omegabar-FD_Frame}, the inclusion of the ``non-Gaussian bridge'' in \cref{fig:2-correlator} results in a transformation of the vertices and symmetry factors. 
Due to the symmetry between the two diagrams of $\omega_{\uGW,\uin}^{(1)}$ within \cref{fig:2-correlator}, we can derive the specific expression for $\omega_{\uGW,\uin}^{(1)}$, or equivalently for $\omega_{\ung,\uin}^{(1)}$ according to Eq~\eqref{eq:omega-1}, via the first (upper right) panel of \cref{fig:FD_Frame}. 
Based on the Feynman-like rules in \cref{fig:FD-Rules}, the attachment of a red solid line transforms a $V_0^{[i]}$-vertex into a $V_{1}^{[i+1]}$-vertex, where $i \leq o-1$. 
Furthermore, the attachment also alters the symmetric factor of the diagram, involving a change in the permutation factor from $i!$ to $(i+1)!$ along with selecting $1$ out of $N_i$ $V_0^{[i]}$-vertices. 
Consequently, for \ac{PNG} up to $\Hnl$ order (i.e., $o=4$), the large-scale modulation at $\cO (\zeta_{\ugL})$ order, $\omega_{\ung,\uin}^{(1)}$, is given by 
\begin{eqnarray}\label{eq:Ong1-dp-def}
    \omega_{\ung,\uin}^{(1)}(q) &=& \sum_{c=0}^4 \sum_{b=0}^{\lfloor 4-c \rfloor} \sum_{a=0}^{\lfloor 4-b-c \rfloor} \bar{\omega}_{\uGW,\uin}^{(a,b,c)} (q) \sum_{i=1}^{o-1=3} \left(\frac{(i+1)!}{i!}\frac{V_1^{[i+1]}}{V_0^{[i]}}\right)^1 \binom{N_i}{1}\ .
\end{eqnarray}
Recall that $N_i$ denotes the number of vertices $V_0^{[i]}$, where $N_1 = 4-a-b-c$, $N_2 = a$, $N_3 = b$, and $N_4 = c$, as established in Subsection~\ref{subsec:FD-approach}. 
This formulation is readily extensible to include higher-order \ac{PNG}; one need only expand the superscript $(a,b,c)$ to incorporate the powers of higher-order \ac{PNG} parameters and update the counts $N_i$ accordingly. 
In particular, for scale-independent \ac{PNG} parameters, namely, $V_j^{[1]}=\Fnl$, $V_j^{[2]}=\Gnl$, and $V_j^{[3]}=\Hnl$, we have 
\begin{eqnarray}\label{eq:Ong1}
    \omega_{\ung,\uin}^{(1)} &=& 2 \Fnl \Bigl(4 \bar{\omega}_{\uGW,\uin}^{(0,0,0)} + 2 \bar{\omega}_{\uGW,\uin}^{(2,0,0)} + 3 \bar{\omega}_{\uGW,\uin}^{(0,1,0)} + \bar{\omega}_{\uGW,\uin}^{(2,1,0)} + 2 \bar{\omega}_{\uGW,\uin}^{(0,2,0)} + 2 \bar{\omega}_{\uGW,\uin}^{(1,0,1)} + \bar{\omega}_{\uGW,\uin}^{(0,3,0)} + \bar{\omega}_{\uGW,\uin}^{(1,1,1)} \nonumber\\
    &&\quad + 2 \bar{\omega}_{\uGW,\uin}^{(0,0,2)} + \bar{\omega}_{\uGW,\uin}^{(0,1,2)} \Bigr)
    + \frac{3\Gnl}{\Fnl} \Bigl(2 \bar{\omega}_{\uGW,\uin}^{(2,0,0)} + 4 \bar{\omega}_{\uGW,\uin}^{(4,0,0)} + 2 \bar{\omega}_{\uGW,\uin}^{(2,1,0)} + \bar{\omega}_{\uGW,\uin}^{(1,0,1)} + 2 \bar{\omega}_{\uGW,\uin}^{(2,2,0)} \nonumber\\
    &&\quad + 3 \bar{\omega}_{\uGW,\uin}^{(3,0,1)} + \bar{\omega}_{\uGW,\uin}^{(1,1,1)} + \bar{\omega}_{\uGW,\uin}^{(1,2,1)} + 2 \bar{\omega}_{\uGW,\uin}^{(2,0,2)} + \bar{\omega}_{\uGW,\uin}^{(1,0,3)}\Bigr)
    + \frac{4\Hnl}{\Gnl} \Bigl(2 \bar{\omega}_{\uGW,\uin}^{(2,0,0)} + \bar{\omega}_{\uGW,\uin}^{(0,1,0)} \nonumber\\
    &&\quad + \bar{\omega}_{\uGW,\uin}^{(2,1,0)} + 2 \bar{\omega}_{\uGW,\uin}^{(0,2,0)} + 3 \bar{\omega}_{\uGW,\uin}^{(0,3,0)} + \bar{\omega}_{\uGW,\uin}^{(1,1,1)} + 4 \bar{\omega}_{\uGW,\uin}^{(0,4,0)} + 2 \bar{\omega}_{\uGW,\uin}^{(1,2,1)} + \bar{\omega}_{\uGW,\uin}^{(0,1,2)} + 2 \bar{\omega}_{\uGW,\uin}^{(0,2,2)} \Bigr)\ .\nonumber\\
\end{eqnarray}
By substituting Eq.~\eqref{eq:Ong1} into Eq.~\eqref{eq:Ct}, we obtain the complete expression for the \ac{SIGW} reduced angular power spectrum $\tilde{C}_\ell (\nu)$. 
% Interestingly, since both $\omega_{\ung,\uin}^{(1)}$ and $\bar{\omega}_{\uGW,\uin}$ can be represented as linear combinations of the 19 categories $\bar{\omega}_{\uGW,\uin}^{(a,b,c)}$, the common feature for all categories described by Eqs.~(\ref{eq:Omega-infrared}, \ref{eq:ngw-infrared}) indicates that the reduced angular power spectrum expressed by Eq.~\eqref{eq:Ct} is nearly scale-invariant in the infrared regime, regardless of the magnitude of \ac{PNG}. 
% the common feature for all categories described by Eq.~(\ref{eq:Omega-infrared}) indicates that $\omega_{\ung,\uin}^{(1)} / \bar{\omega}_{\uGW,\uin}$ is scale-invariant in the infrared regime. Furthermore, as confirmed by Eq.~\eqref{eq:ngw-infrared}, there exists a nearly scale-invariant spectral index in this regime. Thus, we anticipate that the reduced angular power spectrum articulated in Eq.~\eqref{eq:Ct} will also be nearly scale-invariant in the infrared regime, regardless of the presence of \ac{PNG}. 

Eq.~\eqref{eq:Ct} demonstrates that the multipole dependence of the reduced angular power spectrum yields $\tilde{C}_\ell (\nu) \propto [\ell(\ell+1)]^{-1}$, similar to the behavior observed for the \ac{CMB} at low multipoles, as reported in previous studies \cite{Bartolo:2019zvb, Li:2023qua, Li:2023xtl, Wang:2023ost, Dimastrogiovanni:2022eir}. 
In contrast, the frequency dependence of $\tilde{C}_\ell (\nu)$, described by $\ell(\ell+1) \tilde{C}_{\ell}(\nu)$, is peculiar and primarily due to the initial inhomogeneity arising from \ac{PNG}. 
While the frequency dependence of the second term in the square brackets of Eq.~\eqref{eq:Ct} is entirely determined by the energy-density fraction spectrum, the frequency dependence of the first term, $\omega_{\ung,\uin}^{(1)} (q) / \bar{\omega}_{\uGW,\uin} (q)$, is non-trivial. 
Thus, we anticipate that the angular power spectrum will serve as a novel tool for extracting valuable information from the \ac{SIGW} background. 
Both the multipole and frequency dependencies are vital for distinguishing \ac{SIGW} signals from various sources \cite{LISACosmologyWorkingGroup:2022kbp}, as \acp{GW} produced by different sources may exhibit diverse features in their multipole and frequency dependencies. 
For \acp{GW} generated by binary black holes, the multipole dependence of angular power spectra is roughly proportional to $(\ell + 1/2)^{-1}$ \cite{Cusin:2018rsq, Cusin:2017fwz, Cusin:2019jhg, Cusin:2019jpv, Jenkins:2018kxc, Jenkins:2018uac, Jenkins:2019nks, Contaldi:2016koz, Wang:2021djr, Mukherjee:2019oma, Bavera:2021wmw, Bellomo:2021mer}, while for \acp{GW} originating from cosmic string loops, it behaves as $\ell^{0}$ \cite{Jenkins:2018nty, Kuroyanagi:2016ugi, Olmez:2011cg}. 
However, the multipole dependence of the angular power spectra for certain cosmological \acp{GW} is expected to behave similarly to that of \acp{SIGW}, including sources such as inflation \cite{Adshead:2020bji, Dimastrogiovanni:2021mfs, Dimastrogiovanni:2019bfl, Jeong:2012df, ValbusaDallArmi:2023nqn}, domain walls \cite{Liu:2020mru}, first-order phase transitions \cite{Li:2022svl, Li:2021iva, Domcke:2020xmn, Jinno:2021ury, Geller:2018mwu, Kumar:2021ffi, Racco:2022bwj}, and preheating \cite{Bethke:2013aba, Bethke:2013vca, Caprini:2018mtu}. 
The common feature among these cosmological \acp{GW} is that their density contrasts can be decomposed into direction-independent initial inhomogeneity terms and identical propagation terms. 
To distinguish \acp{SIGW} from these sources, in addition to analyzing frequency and multipole dependencies, cross-correlations between \ac{GW} background and various observables, such as the \ac{CMB} \cite{Dimastrogiovanni:2021mfs, Cusin:2018rsq, Ricciardone:2021kel, Malhotra:2020ket, Braglia:2021fxn, Capurri:2021prz, Dimastrogiovanni:2022eir, Galloni:2022rgg, Ding:2023xeg, Cyr:2023pgw, Zhao:2024gan}, \ac{LSS} \cite{Cusin:2018rsq, Canas-Herrera:2019npr, Alonso:2020mva, Yang:2020usq, Yang:2023eqi, Libanore:2023ovr, Bosi:2023amu, Balaudo:2022znx, Bravo:2025csu}, and 21 cm lines \cite{Scelfo:2021fqe, Seto:2005tq}, are also beneficial.

We direct our attention to the behavior of $\ell(\ell+1) \tilde{C}_{\ell}(\nu)$ for various values of the three \ac{PNG} parameters. 
First, we consider a scenario where the \ac{PNG}, characterized by $(|\Fnl|A_\uS^{1/2} + |\Gnl|A_\uS + |\Hnl|A_\uS^{3/2})$, is sufficiently small such that the Gaussian component of the energy-density fraction spectrum, $\bar{\omega}_{\uGW,\uin}^{(0,0,0)}$, dominates among all categories of the \ac{SIGW} energy-density fraction spectrum. 
In Eq.~\eqref{eq:Ong1}, the leading-order term of $\omega_{\ung,\uin}^{(1)}$ in $A_\uS$ is $8\Fnl\bar{\omega}_{\uGW,\uin}^{(0,0,0)}$, leading to an approximate expression for the frequency dependence of the reduced angular power spectrum as 
\begin{equation}\label{eq:Ct-NG-min}
    \ell (\ell+1) \tilde{C}_\ell (\nu) 
    = 2 \pi A_\uL 
        \left[
            8 \Fnl + \frac{3}{5} \left(6 - n_{\uGW} (\nu)\right)
        \right]^2\ .
\end{equation}
As discussed with reference to \cref{fig:Total_Omegabar}, since $ n_{\uGW} $ approximately varies from $ 3 $ to $ -9 $ in the frequency range $ (10^{-4} \nu_\ast, 10^2 \nu_\ast) $, the second term in the square brackets of this formula roughly changes from $3$ to $15$. 
Therefore, if $\Fnl$ is of order one, the first term could become comparable to the second term, resulting in notable alterations in the profile of $\ell (\ell+1) \tilde{C}_\ell (\nu)$ when the sign of $\Fnl$ is changed. 
Moreover, the sum of the powers of $\Fnl$ and $\Hnl$ is no longer an even integer in $\ell (\ell + 1) \tilde{C}_\ell (\nu)$. 
This implies that the sign degeneracy between $\Fnl$ and $\Hnl$ in the energy-density fraction spectrum, as discussed in Subsection~\ref{subsec:FD-approach} and Subsection~\ref{subsec:Omegarbar-num}, is broken by the reduced angular power spectrum. 
Although the energy-density spectrum closely resembles the Gaussian case, the reduced angular power spectrum for the non-Gaussian case exhibits significant deviations from its Gaussian counterpart. 
Consequently, analyzing the \ac{SIGW} angular power spectrum is crucial in the pursuit of \ac{PNG}. 
For large \ac{PNG}, the first term, $\omega_{\ung,\uin}^{(1)} (q) / \bar{\omega}_{\uGW,\uin} (q)$, varies with different values of the \ac{PNG} parameters, potentially resulting in distinct features in the reduced angular power spectrum. 
This will be elaborated upon in the numerical results presented subsequently. 
It is possible to infer the model parameters by combining insights from the energy-density fraction spectrum and the reduced angular power spectrum.

The reduced angular power spectrum $\tilde{C}_\ell (\nu)$ quantifies the covariance of the relative anisotropies in the \ac{SIGW} background. 
To evaluate the detectability of these anisotropies in \acp{SIGW}, we can de-normalize $\tilde{C}_\ell (\nu)$ by multiplying it by the square of the average energy-density spectrum, thereby yielding the angular power spectrum as follows \begin{equation}\label{eq:C-def}
    C_{\ell}(\nu) =  \bar{\omega}_{\uGW,0}^2 (\nu) \tilde{C}_{\ell}(\nu)  \ .
\end{equation}
This angular power spectrum is defined in such a manner that it characterizes the covariance of the absolute anisotropies in the \ac{SIGW} background. 
It can be directly compared to the noise spectrum of upcoming \ac{GW} detectors, facilitating the assessment of the detectability of the corresponding \acp{SIGW} signals.

\begin{figure*}[htbp]
    \centering
    \includegraphics[width =\textwidth]{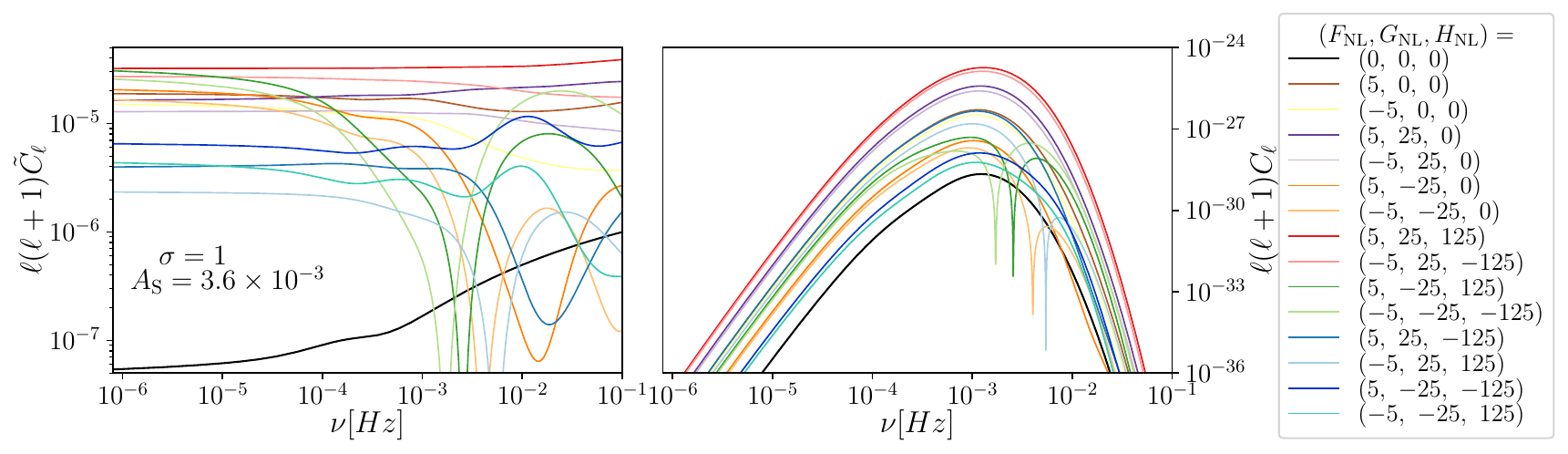}
    \caption{ Angular power spectrum of SIGWs. }\label{fig:C-Ct}
\end{figure*}

Based on the numerical results for the 19 categories of the \ac{SIGW} energy-density fraction spectrum illustrated in \cref{fig:Unscaled_Omegabar}, we compute the 1st-order large-scale modulation through Eq.~\eqref{eq:Ong1} and the corresponding (reduced) angular power spectrum of \acp{SIGW} via Eq.~\eqref{eq:Ct}. 
The results for $\ell(\ell+1) \tilde{C}_{\ell}(\nu)$ (left panel) and $\ell(\ell+1) C_{\ell} (\nu)$ (right panel) concerning frequency $\nu$ are presented in \cref{fig:C-Ct}. 
For consistency, we utilize the same sets of model parameters as those in \cref{fig:Total_Omegabar}, while also considering variations in the signs of $\Fnl$. 
This represents scenarios where \ac{PNG} is pronounced. 
By combining this figure with Eq.~\eqref{eq:Ct-NG-min}, which pertains to cases of small \ac{PNG}, we investigate the effects of varying magnitudes of \ac{PNG} on the anisotropies in the \ac{SIGW} background. 
Specifically, we depict the (reduced) angular power spectrum for the Gaussian scenario using solid black curves for comparison. 
Given that the \ac{SW} effect and the initial inhomogeneity arising from non-adiabaticity remain relatively unchanged in non-Gaussian scenarios due to the similar profiles of energy-density fraction spectra shown in \cref{fig:Total_Omegabar}, the black solid curve can also be regarded as these effects term for other scenarios. 
Comparing it with the colored curves representing non-Gaussian scenarios, we observe that the presence of \ac{PNG} typically leads to a significant enhancement in the spectral amplitude of $C_\ell (\nu)$ ($\tilde{C}_\ell (\nu)$). 
Specifically, unless there is a balance between the contributions of the initial inhomogeneities terms and the propagation term across specific frequency bands, the anisotropies in \acp{SIGW} are usually more prominent in non-Gaussian scenarios compared to Gaussian scenarios, regardless of the signs of the \ac{PNG} parameters. 
In the scenarios depicted in the figure, when either $\Gnl < 0$ or $\Fnl\Hnl < 0$, the magnitudes of the (reduced) angular power spectra in non-Gaussian scenarios are lower than in the Gaussian scenario across specific frequency bands. 
The angular power spectra $\ell (\ell+1) C_\ell (\nu)$ exhibit characteristic dips in the profiles of these scenarios. 
Moreover, around the pivot frequency, the black solid curve in the left panel of \cref{fig:C-Ct} reveals that $\ell (\ell+1) \tilde{C}_\ell (\nu)$ in the Gaussian case displays a blue-tilted shape. 
In contrast, the colored curves and Eq.~\eqref{eq:Ct-NG-min} demonstrate a different slope for $\ell (\ell+1) \tilde{C}_\ell (\nu)$ in non-Gaussian scenarios, as long as the \ac{PNG} is not negligible. 
This finding suggests that the presence of \ac{PNG} can be determined by measuring the anisotropies in the \ac{SIGW} background. 
% With regard to the scale-invariant behavior in the infrared regime for $\ell (\ell+1) \tilde{C}_\ell (\nu)$ across all scenarios, this characteristic may be beneficial for reducing noise in future \ac{GW} detection. 
Additionally, comparing the right panel of \cref{fig:C-Ct} with \cref{fig:Total_Omegabar}, it is noteworthy that large magnitudes of $h^2 \bar{\Omega}_{\uGW,0} (\nu)$ do not necessarily result in large magnitudes of $\ell (\ell+1) C_\ell (\nu)$. 
For instance, in the case of $(|\Fnl|=5, \Gnl=\Hnl=0)$, the angular power spectrum is significantly enhanced, surpassing the magnitudes observed in the cases of $(\Fnl=5, \Gnl=25, \Hnl=125)$ and $(\Fnl=-5, \Gnl=25, \Hnl=-125)$. 
This property may prove useful for determining model parameters and breaking parameter degeneracies, as shown in \cref{fig:Omega_degeneracy}. 
The sign degeneracy between $\Fnl$ and $\Hnl$ is disrupted by the second term in the square brackets of Eq.~\eqref{eq:Ct}, because it is independent of the presence of \ac{PNG}. 
In other words, if the \ac{PNG}-induced inhomogeneity dominates and the other effects are negligible, the reduced angular power spectra for different signs of $\Fnl$ and $\Hnl$ remain similar and nearly flat.

\begin{figure*}[htbp]
    \centering
    \includegraphics[width = 1 \textwidth]{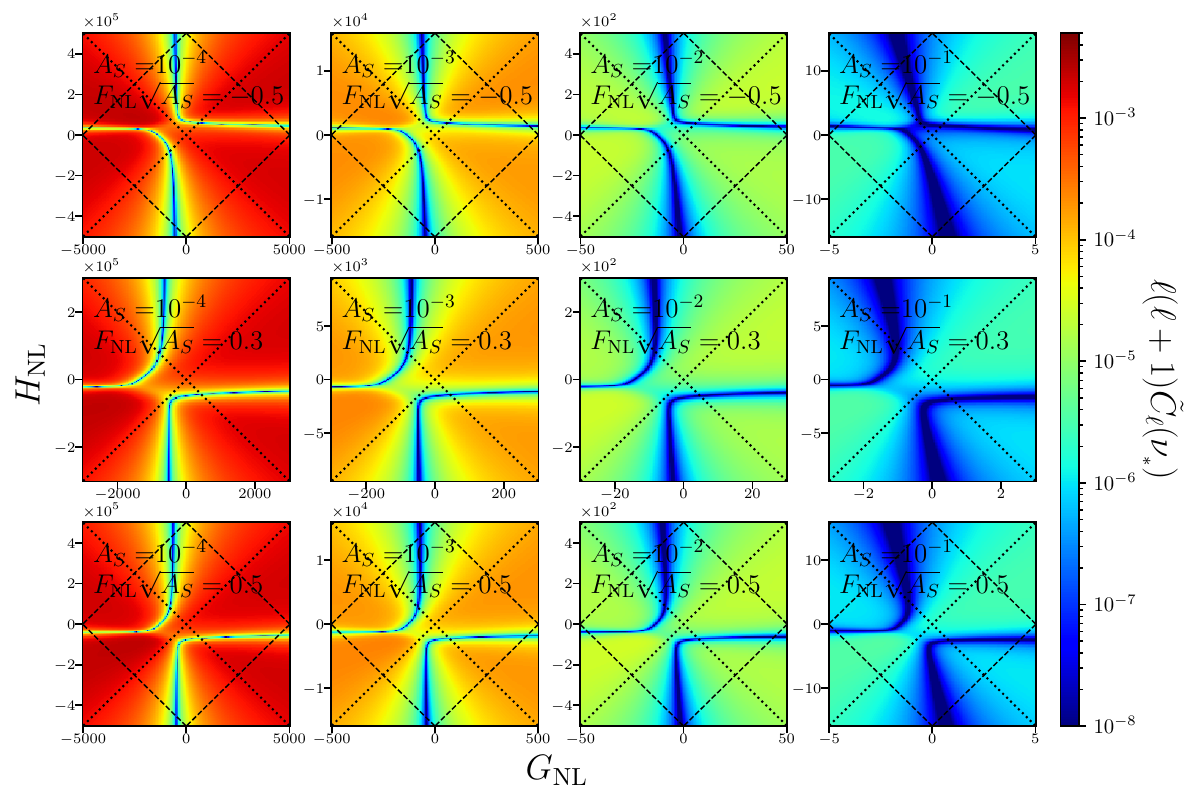}
    \caption{ Reduced angular power spectrum $\ell(\ell+1) \tilde{C}_\ell (\nu)$ of SIGWs with respect to the $\Fnl$, $\Gnl$ and $\Hnl$, where we fix $\nu/\nu_\ast=1$ and $\sigma=1$. The conventions for the dotted and dashed lines are the same as those in Figure~\ref{fig:Omega_G-H}. }\label{fig:Ct_G-H}
\end{figure*}

In \cref{fig:Ct_G-H}, we illustrate the dependence of the \ac{SIGW} angular power spectrum on $\Gnl$ and $\Hnl$ by presenting a $3 \times 4$ array of contour plots for $\ell (\ell+1) \tilde{C}_\ell (\nu)$ at the frequency $\nu = \nu_\ast$, reflecting the variations with respect to $\Gnl$ and $\Hnl$. 
In this array, we vary $\Fnl\sqrt{A_\uS}$ across the three rows of contour plots as $-0.5, 0.3, 0.5$ from top to bottom, and we vary $A_\uS$ across the four columns as $10^{-4}, 10^{-3}, 10^{-2}, 10^{-1}$ from left to right. 
One can identify areas that satisfy the perturbativity conditions from the regions demarcated by dotted and dashed lines in each contour plot, with the conventions for these lines aligning with those in \cref{fig:Omega_G-H}. 
Notably, the magnitudes of the reduced angular power spectra decrease as $A_\uS$ increases. 
By mirroring each contour plot of the first row along the axis of $\Hnl=0$ and comparing it with the contour plot in the third row, we observe that the two corresponding plots are not entirely identical. 
The differences become more pronounced as $A_\uS$ increases, indicating that the sign degeneracy between $\Fnl$ and $\Hnl$ in $h^2 \bar{\Omega}_{\uGW,0} (\nu)$ is broken in $\ell (\ell+1) \tilde{C}_\ell (\nu)$, particularly for larger values of $A_\uS$. 
Furthermore, it is notable that the position of the largest magnitude in $\ell (\ell+1) \tilde{C}_\ell (\nu_\ast)$ is close to that of the smallest magnitude in each contour plot. 
These positions occur within the ranges where both $\Fnl\Hnl$ and $\Gnl$ are negative, and $\Hnl A_\uS^{3/2}$ remains relatively small.

\begin{figure}[htbp]
    \centering
    \includegraphics[width = 0.8 \columnwidth]{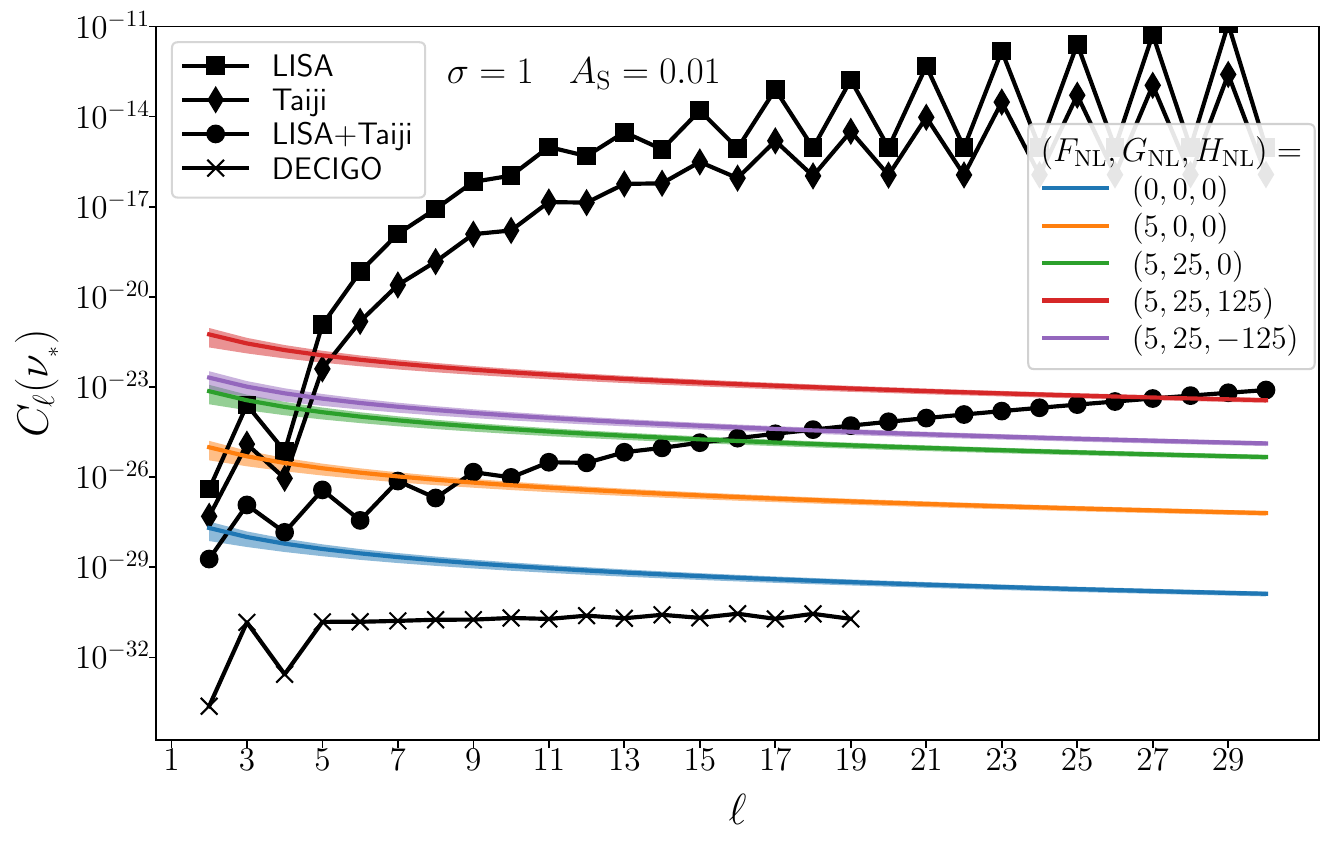}
    \caption{Comparison between the theoretical angular power spectra and the noise angular power spectra of \ac{LISA}, Taiji and their networks at the milli-Hz band \cite{Zhao:2024yau}, as well as \ac{DECIGO} at the deci-Hz band \cite{Capurri:2022lze}. Shaded regions represent the uncertainties (68\% confidence level) due to the cosmic variance, i.e., $\Delta C_{\ell}/C_{\ell}=\sqrt{2/(2\ell+1)}$.   }
    \label{fig:Cl_sensitivity}
\end{figure}

Within the relevant parameter regimes, the anisotropies in \acp{SIGW} can potentially be detected at low multipoles by upcoming space-borne \ac{GW} detectors, such as \ac{LISA}, Taiji, and \ac{DECIGO}. 
\cref{fig:Cl_sensitivity} presents a comparison of the angular sensitivity curves of these \ac{GW} detectors with the theoretical angular power spectra of \acp{SIGW} at the pivot frequency $\nu_\ast$ as a function of multipoles. 
Notably, the sensitivity regime of \ac{LISA} and Taiji falls within the milli-Hz band, while that of \ac{DECIGO} operates in the deci-Hz band. 
For the angular power spectra of \acp{SIGW} at the frequency $\nu = \nu_\ast$, we set $A_\uS = 10^{-2}$, which is relevant to scenarios involving \ac{PBH} formation \cite{Carr:2020gox}, and vary the \ac{PNG} parameters. 
The shaded regions along each angular power spectrum curve represent uncertainties due to cosmic variance at the $68\%$ confidence level. 
It is observed that the angular power spectra of \acp{SIGW} can be detected by \ac{LISA} and Taiji individually at $\ell \leq 5$, and their combined efforts can more effectively assess the properties of \acp{SIGW}, potentially detecting their anisotropies up to $\ell \leq 19$. 
Although the anisotropies in \ac{SIGW} under the Gaussian scenario exceed the detection capabilities of \ac{LISA} and Taiji, they could be probed by the more advanced \ac{DECIGO}.

%% file: TeX/5Non-Gaussianity.tex
\section{Non-Gaussianity in the SIGW background}\label{sec:bl&tl}

In this section, we extend our investigation to the non-Gaussianity in the \ac{SIGW} background, characterized by the higher-order statistics of the density contrast. 
We focus on the three- and four-point correlation functions that define the angular bispectrum and trispectrum, respectively. 
While the angular power spectrum studied in the previous section refers to the covariance of energy density fluctuations superimposed on the \ac{SIGW} background, the angular bispectrum and trispectrum serve as natural extensions of skewness and kurtosis for random fluctuations. 
Consequently, these spectra are fundamental quantities representing the non-Gaussianity in the anisotropic \ac{SIGW} background, providing valuable insights into the nature of the early Universe.

\subsection{Angular bispectrum: three-point angular correlation}\label{subsec:b}

As is customary in \ac{CMB} studies, we define the \ac{SIGW} angular bispectrum \cite{Bartolo:2019zvb} for the rotation-invariant \ac{GW} background through the three-point correlator of $\delta_{\uGW,0,\ell m}(2\pi\nu)$, denoted as 
\begin{eqnarray}\label{eq:Bl-def}
    \left\langle\prod_{i=1}^3\delta_{\uGW,0,\ell_i m_i}(2\pi\nu)\right\rangle 
    = \begin{pmatrix}
         \ell_1 & \ell_2 & \ell_3 \\
         m_1    & m_2    & m_3   
    \end{pmatrix}
    B_{\ell_1 \ell_2 \ell_3} (\nu)\ ,
\end{eqnarray} 
where the matrix enclosed in parentheses labels the Wigner 3-$j$ symbol.
\footnote{
It is also common to define the reduced angular bispectrum in an alternative manner \cite{Komatsu:2001rj,Planck:2013wtn,Bartolo:2019zvb}, denoted as 
\begin{eqnarray*}%\label{eq:btilde-def}
    \left\langle\prod_{i=1}^3\delta_{\uGW,0,\ell_i m_i}(2\pi\nu)\right\rangle 
    = \Glm 
    \tilde{b}_{\ell_1 \ell_2 \ell_3} (\nu)\ ,
\end{eqnarray*} 
where the Gaunt integral $\Glm$, defined later in Eq.~\eqref{eq:Glm-def}, ensures the assumptions of statistical isotropy and parity invariance. 
Compared to the definition provided in Eq.~\eqref{eq:Bl-def}, this formulation factorizes $h_{\ell_1 \ell_2 \ell_3}$, thereby ensuring parity invariance with $h_{\ell_1 \ell_2 \ell_3}=0$ for odd parity.
}
Furthermore, it is important to note that the tetrahedral domain of multipole triplets must satisfy both the triangular inequalities ($\ell_1 \leq \ell_2 + \ell_3$, $\ell_2 \leq \ell_1 + \ell_3$, $\ell_3 \leq \ell_1 + \ell_2$, and $m_1 + m_2 + m_3 = 0$), which are upheld by the Wigner 3-$j$ symbols. 
A brief overview of the properties of Wigner 3-$j$ symbols is presented in Appendix~\ref{sec:Wigner}.  
Additionally, parity invariance imposes the condition ($\ell_1 + \ell_2 + \ell_3 = 2n$, where $n \in \mathbb{N}$). 
These conditions are also consistent with those of the \ac{CMB} angular bispectrum \cite{Planck:2013wtn,Komatsu:2001ysk}.

\begin{figure*}[htbp]
    \centering
    \includegraphics[width =0.45 \textwidth]{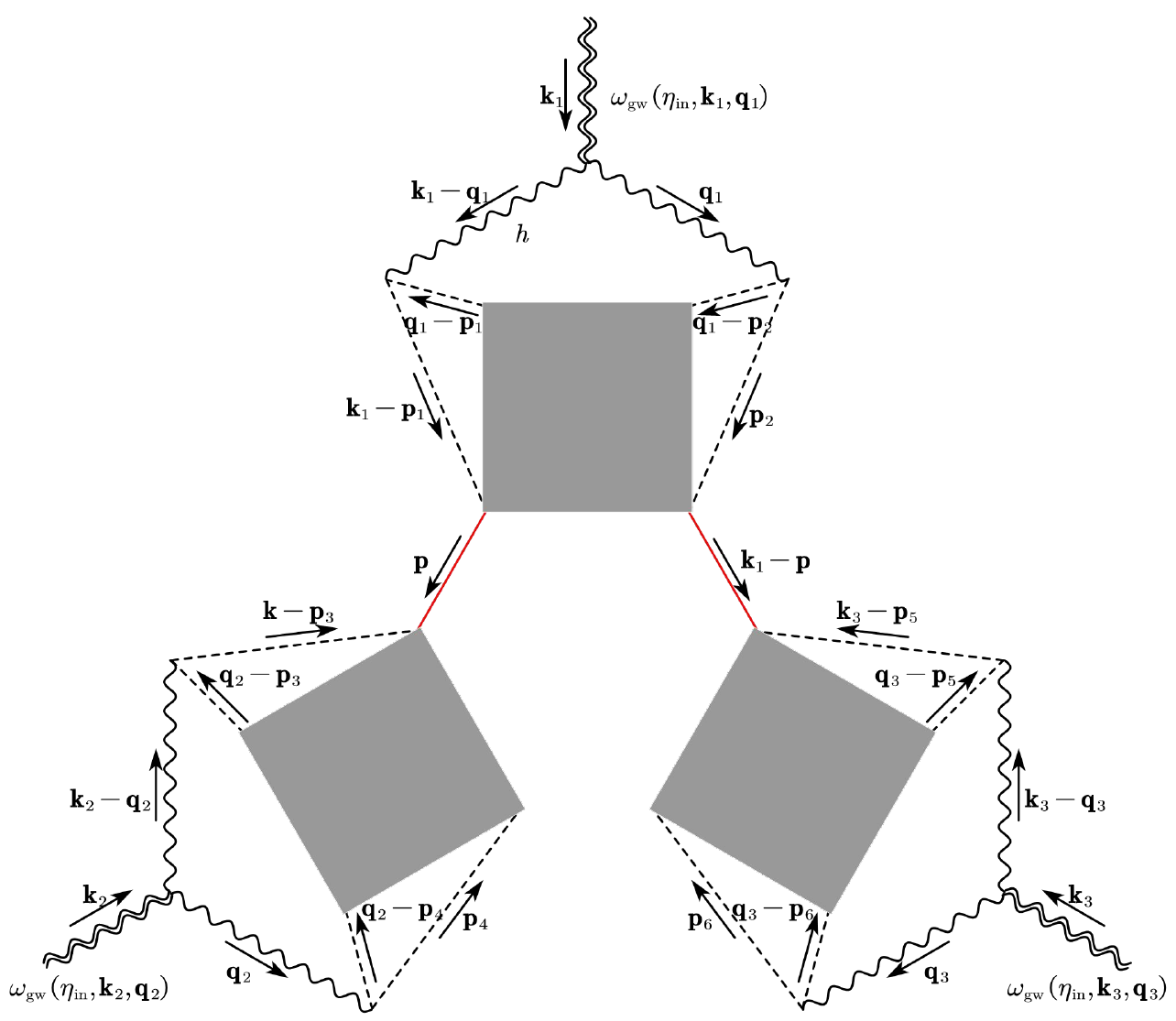}
    \qquad
    \includegraphics[width =0.45 \textwidth]{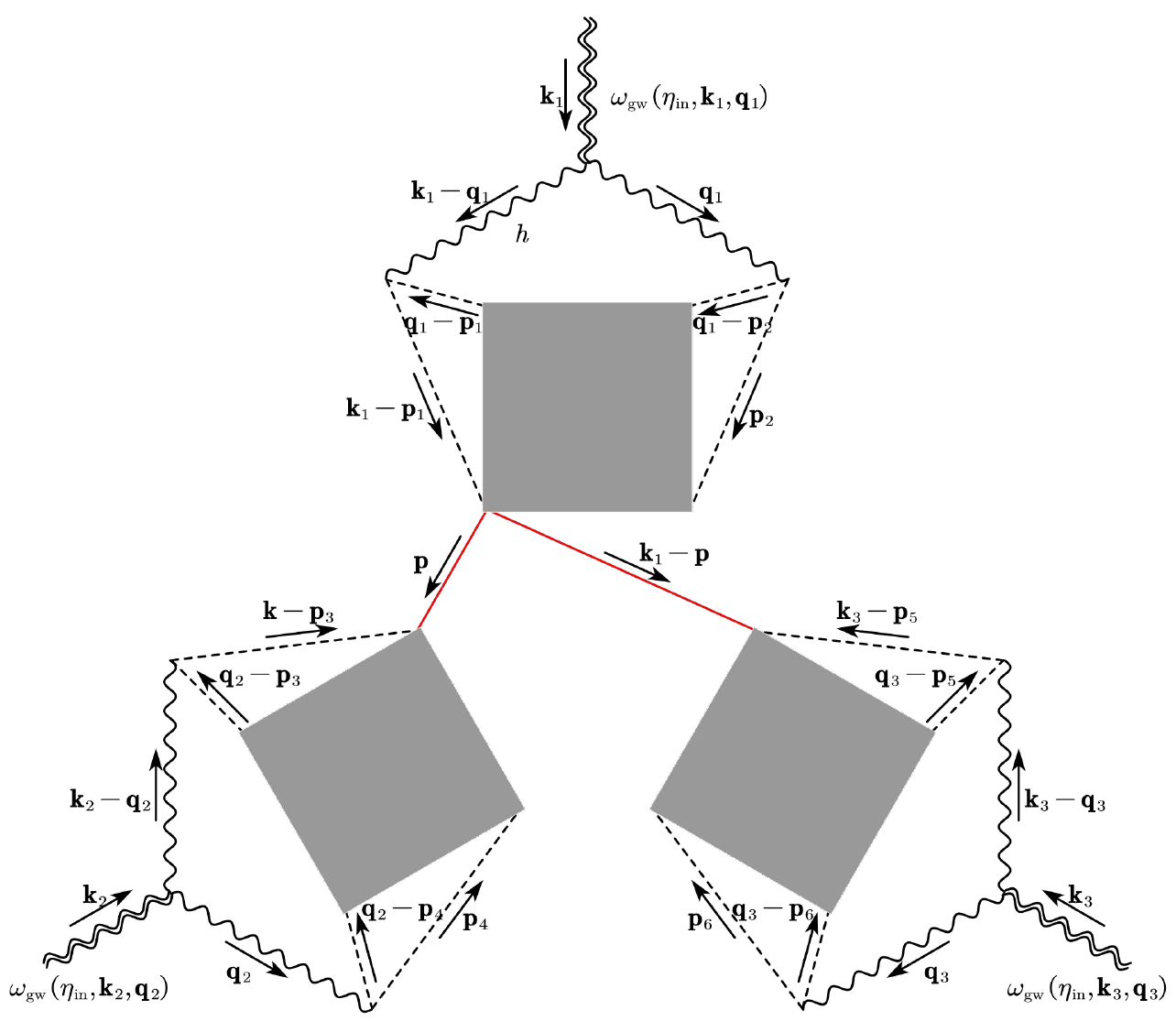}
    \caption{Feynman-like diagrams for the three-point correlator of $\omega_\uGW (\eta_\uin,\bk,\bq)$ at leading order in $A_\uL$. }\label{fig:3-correlator}
\end{figure*}

The angular bispectrum quantifies the mixed central third-order moment, corresponding to skewness, of the fluctuations on the \ac{SIGW} background. 
Previous studies have provided relevant insights into \ac{PNG} at $\Fnl$ order \cite{Bartolo:2019zvb,Bartolo:2019yeu}, extending to $\Gnl$ order \cite{Li:2024zwx}. 
In this work, we extrapolate this analysis to include \ac{PNG} up to all orders, offering a specific expression for \ac{PNG} at $\Hnl$ order. 
The methodology for studying the angular bispectrum parallels that employed for investigating the angular power spectrum in Section~\ref{sec:Cl}. 
The calculation of the angular bispectrum, denoted as $B_{\ell_1 \ell_2 \ell_3}$, involves the three-point correlator of the initial energy-density full spectrum, expressed as $\langle\omega_{\uGW,\uin} \omega_{\uGW,\uin} \omega_{\uGW,\uin} \rangle$. 
Based on \cref{fig:FD-Rules} and \cref{fig:FD_Frame}, we construct the Feynman-like diagrams for this three-point correlator by connecting three diagrams of $\omega_{\uGW,\uin}$ using ``non-Gaussian bridges''. 
To determine the leading-order connected contribution in $A_\uL$, it is crucial to minimize the presence of ``non-Gaussian bridges'' in the diagrams. 
As shown in \cref{fig:3-correlator}, we consider only the diagrams with two ``non-Gaussian bridges'', corresponding to the connected contribution at $\cO(A_\uL^2)$ order, while excluding diagrams with a higher number of ``non-Gaussian bridges''. 
Therefore, we aim to evaluate the three-point correlator $\langle\delta_{\uGW,0}^{(2)} \delta_{\uGW,0}^{(1)} \delta_{\uGW,0}^{(1)}\rangle$, with the expressions for $\delta_{\uGW,0}^{(1)}$ and $\delta_{\uGW,0}^{(2)}$ provided in Eq.~\eqref{eq:delta0-1} and Eq.~\eqref{eq:delta0-2}, respectively.

We provide a concise overview of the derivation process for the explicit expression of $B_{\ell_1 \ell_2 \ell_3} (\nu)$ here. 
Following the decomposition utilizing spherical harmonics, as demonstrated in Eqs.~\eqref{eqs:spher-harm}, we substitute $\delta_{\uGW,0,\ell m}^{(1)}$ and $\delta_{\uGW,0,\ell m}^{(2)}$ into the three-point correlator, resulting in an integral for $B_{\ell_1 \ell_2 \ell_3} (\nu)$ as outlined in Eq.~\eqref{eq:Bl-def}. 
Within the integrand, the correlator simplifies to $\langle \zeta_{\ugL}(\bp) \zeta_{\ugL}(\bk_1 - \bp) \zeta_{\ugL}(\bk_2) \zeta_{\ugL}(\bk_3) \rangle \sim \delta^{(3)}(\bk_1 + \bk_2 + \bk_3) (2 \pi^2 \Delta^2_\uL)^2 / (k_2 k_3)^3 + 2 \,\text{permutations}$, and the exponential factor $e^{i\bk_1 \cdot \bx_1 + \bk_2 \cdot \bx_2 + \bk_3 \cdot \bx_3}$ transforms to $\prod_{i=1}^3 e^{-i(\bk_i \cdot \bn_i)(\eta_0 - \eta_\uin)}$ due to the line-of-sight relation and momentum conservation. 
Furthermore, the Fourier counterpart of the momentum-conservation factor reads 
\begin{equation}\label{eq:Diracdelta-Fourier}
    \delta^{(3)} (\bk_1+\bk_2+\bk_3) = \int \frac{\ud^3 \br}{(2\pi)^3}\, \prod_{i=3}^3 e^{i \bk_i \cdot \br} \ ,
\end{equation}
which allows us to express it in terms of an exponential function. 
By combining this with the previously mentioned exponential factor, we utilize the identity in Eq.~\eqref{eq:exp-jYY} to express both in terms of spherical harmonics and spherical Bessel functions. 
To address the multiple spherical harmonics present in the integral, we leverage their properties and introduce the Gaunt integral as follows  
\begin{eqnarray}\label{eq:Glm-def}
    \Glm = \int \ud^2 \bn\, Y_{\ell_1 m_1} (\bn) Y_{\ell_2 m_2} (\bn) Y_{\ell_3 m_3} (\bn) 
    = h_{\ell_1 \ell_2 \ell_3} 
    \begin{pmatrix}
         \ell_1 & \ell_2 & \ell_3 \\
         m_1    & m_2    & m_3   
    \end{pmatrix}\ ,
\end{eqnarray}
where $h_{\ell_1 \ell_2 \ell_3}$ is defined as  
\begin{equation}\label{eq:hlll}
    h_{\ell_1 \ell_2 \ell_3} 
    = \sqrt{\frac{(2\ell_1 + 1)(2\ell_2 + 1)(2\ell_3 + 1)}{4\pi}}
    \begin{pmatrix}
         \ell_1 & \ell_2 & \ell_3 \\
         0      & 0      & 0     
    \end{pmatrix}\ .
\end{equation} 
It is important to note that $h_{\ell_1 \ell_2 \ell_3}$ is non-zero only if $(\ell_1 + \ell_2 + \ell_3)$ is even, ensuring the parity invariance. 
Consequently, due to orthogonality, all spherical harmonics vanish after integration over all directional variables. 
Regarding the spherical Bessel functions, we employ the integral relation in Eq.~(\ref{eq:jl-2-int}) and the closure relation for spherical Bessel functions 
\begin{eqnarray}\label{eq:jl-closure}
    \frac{2}{\pi} \int_0^\infty \ud k\, k^2 j_{\ell} (k (\eta_0 - \eta_\uin)) j_{\ell} (k r) = \frac{\delta (\eta_0 - \eta_\uin - r)}{r^2}\ ,
\end{eqnarray}
which ultimately enables us to integrate over $r$ as seen in Eq.~\eqref{eq:Diracdelta-Fourier}. 
After completing all integrations, we successfully derive the formula for $B_{\ell_1 \ell_2 \ell_3}$ defined in Eq.~\eqref{eq:Bl-def} as follows 
\begin{eqnarray}\label{eq:Bl-res}
    B_{\ell_1 \ell_2 \ell_3} (\nu)
    &=& b (\nu) h_{\ell_1 \ell_2 \ell_3}
    \biggl[
        \frac{1}{\ell_1 \ell_2 \left(\ell_1 + 1\right)\left(\ell_2 + 1\right)} 
        + (\ell_1\leftrightarrow\ell_3) + (\ell_2\leftrightarrow\ell_3)
        % \frac{1}{\ell_2 \ell_3 \left(\ell_2 + 1\right)\left(\ell_3 + 1\right)} + \frac{1}{\ell_3 \ell_1 \left(\ell_3 + 1\right)\left(\ell_1 + 1\right)}
    \biggr]\ ,
\end{eqnarray}
introducing its frequency-dependent function $b (\nu)$ as 
\begin{eqnarray}\label{eq:b-def}
    b (\nu) &=& 2 \bigl(2 \pi A_\uL\bigr)^2 
    \biggl[
        \frac{\omega_{\mathrm{ng},\uin}^{(1)} (q)}{\bar{\omega}_{\uGW,\uin} (q)}
        + \frac{3}{5} \bigl(6 - n_{\uGW} (\nu)\bigr)
    \biggr]^2 \biggl[
        \frac{\omega_{\mathrm{ng},\uin}^{(2)} (q)}{\bar{\omega}_{\uGW,\uin} (q)}
        + \frac{3}{5} \Fnl \bigl(6 - n_{\uGW} (\nu)\bigr)
    \biggr]\ .
\end{eqnarray}
Aside from the 2nd-order large-scale modulation $\omega_{\ung,\uin}^{(2)}$ introduced in Eqs.~\eqref{eqs:omega-result}, all other physical quantities have been obtained in the preceding sections. 
Therefore, the next objective is to determine the specific expression for $\omega_{\ung,\uin}^{(2)}$.

As illustrated in the second (upper middle) and third (upper left) panels of \cref{fig:FD_Frame}, the diagrams representing $\omega_{\uGW,\uin}^{(2)}$ can be viewed as attaching two extended red solid lines to certain diagrams of $\bar{\omega}_{\uGW,\uin}$ shown in \cref{fig:omegabar-FD_Frame}. 
These two extensional solid lines may be connected to either the same vertex or different vertices. 
In another perspective, scenarios in which the attachment involves only one type of vertex are classified into one group, while those involving different types of vertices are categorized into another group.
\begin{itemize}
    \item In the first group, the attachment of the red solid lines converts $n$ $V_0^{[i]}$ vertices into $n$ $V_{l}^{[i+l]}$ vertices, where the positive integers $n$, $l$, and $i$ satisfy $nl = 2$ and $i \leq o-l$. 
    This attachment also alters the symmetric factor by changing $n$ permutations of $i!$ to $(i+1)!$ and selecting $n$ vertices from $N_i$ $V_0^{[i]}$-vertices.
    \item In the second group, the attachment transforms a $V_0^{[i]}$-vertex into a $V_{1}^{[i+1]}$-vertex and a $V_0^{[j]}$-vertex into a $V_{1}^{[j+1]}$-vertex, where the positive integers $i$ and $j$ satisfy $i \neq j$ and $i,j \leq o-1$. 
    The attachment alters the symmetric factor by changing the permutations of $i!$ and $j!$ to $(i+1)!$ and $(j+1)!$, respectively, while selecting one vertex from $N_i$ $V_0^{[i]}$-vertices and one vertex from $N_j$ $V_0^{[j]}$-vertices.
\end{itemize}
Consequently, for \ac{PNG} up to $\Hnl$ order (i.e., $o=4$), the large-scale modulation at the $\cO (\zeta_{\ugL}^2)$ order, $\omega_{\ung,\uin}^{(2)}$, is expressed as 
\begin{subequations}\label{eqs:Ong2-dp-def}
\begin{eqnarray}
    \omega_{\ung,\uin}^{(2)} (q) 
    &=& \sum_{c=0}^4 \sum_{b=0}^{\lfloor 4-c \rfloor} \sum_{a=0}^{\lfloor 4-b-c \rfloor} \bar{\omega}_{\uGW,\uin}^{(a,b,c)} (q) \Bigl(\sT_1^{(2)} + \sT_2^{(2)}\Bigr)\ ,\\
    \sT_1^{(2)} &=& \sum_{l=1}^2 \sum_{n=1}^2 \sum_{i=1}^{o-l} \left(\frac{(i+l)!}{i!}\frac{V_l^{[i+l]}}{V_0^{[i]}}\right)^n \binom{N_i}{n} \delta_{ln,2}\ ,\\ 
    \sT_2^{(2)} &=& \sum_{i=1}^{4-1} \sum_{j=1}^{\min(i-1,o-1)} \left(\frac{(i+1)!}{i!}\frac{V_1^{[i+1]}}{V_0^{[i]}}\right)^1 \binom{N_i}{1} \left(\frac{(j+1)!}{j!}\frac{V_1^{[j+1]}}{V_0^{[j]}}\right)^1 \binom{N_j}{1}\ .
\end{eqnarray}
\end{subequations} 
Similar to Eq.~(\ref{eq:Ong1-dp-def}), this formulation is readily extensible to include higher-order \ac{PNG}. 
Specifically, for scale-independent \ac{PNG} parameters, we have 
\begin{eqnarray}\label{eq:Ong2}
    \omega_{\ung,\uin}^{(2)}
    &=& 4\Fnl^2 \Bigl(6 \bar{\omega}_{\uGW,\uin}^{(0,0,0)} + \bar{\omega}_{\uGW,\uin}^{(2,0,0)} + 3 \bar{\omega}_{\uGW,\uin}^{(0,1,0)} + \bar{\omega}_{\uGW,\uin}^{(0,2,0)} + \bar{\omega}_{\uGW,\uin}^{(1,0,1)} + \bar{\omega}_{\uGW,\uin}^{(0,0,2)}\Bigr) \nonumber\\
    && + 6 \Gnl \Bigl(4 \bar{\omega}_{\uGW,\uin}^{(0,0,0)} + 6 \bar{\omega}_{\uGW,\uin}^{(2,0,0)} + 3 \bar{\omega}_{\uGW,\uin}^{(0,1,0)} + 3 \bar{\omega}_{\uGW,\uin}^{(2,1,0)} + 2 \bar{\omega}_{\uGW,\uin}^{(0,2,0)} + 4 \bar{\omega}_{\uGW,\uin}^{(1,0,1)} + \bar{\omega}_{\uGW,\uin}^{(0,3,0)}\nonumber\\
    &&\quad + 2 \bar{\omega}_{\uGW,\uin}^{(1,1,1)} + 2 \bar{\omega}_{\uGW,\uin}^{(0,0,2)} + \bar{\omega}_{\uGW,\uin}^{(0,1,2)} \Bigr) + \frac{12 \Hnl}{\Fnl} \Bigl(2 \bar{\omega}_{\uGW,\uin}^{(2,0,0)} + 4 \bar{\omega}_{\uGW,\uin}^{(4,0,0)} + 4 \bar{\omega}_{\uGW,\uin}^{(2,1,0)}\nonumber\\ 
    &&\quad + \bar{\omega}_{\uGW,\uin}^{(1,0,1)} + 6 \bar{\omega}_{\uGW,\uin}^{(2,2,0)} + 3 \bar{\omega}_{\uGW,\uin}^{(3,0,1)} + 2 \bar{\omega}_{\uGW,\uin}^{(1,1,1)} + 3 \bar{\omega}_{\uGW,\uin}^{(1,2,1)} + 2 \bar{\omega}_{\uGW,\uin}^{(2,0,2)} + \bar{\omega}_{\uGW,\uin}^{(1,0,3)} \Bigr) \nonumber\\
    && + \frac{9\Gnl^2}{\Fnl^2} \Bigl(\bar{\omega}_{\uGW,\uin}^{(2,0,0)} + 6 \bar{\omega}_{\uGW,\uin}^{(4,0,0)} + \bar{\omega}_{\uGW,\uin}^{(2,1,0)} + \bar{\omega}_{\uGW,\uin}^{(2,2,0)} + 3 \bar{\omega}_{\uGW,\uin}^{(3,0,1)} + \bar{\omega}_{\uGW,\uin}^{(2,0,2)}\Bigr) \nonumber\\
    && + \frac{8 \Fnl \Hnl}{\Gnl} \Bigl(3 \bar{\omega}_{\uGW,\uin}^{(0,1,0)} + \bar{\omega}_{\uGW,\uin}^{(2,1,0)} + 4 \bar{\omega}_{\uGW,\uin}^{(0,2,0)} + 3 \bar{\omega}_{\uGW,\uin}^{(0,3,0)} + \bar{\omega}_{\uGW,\uin}^{(1,1,1)} + \bar{\omega}_{\uGW,\uin}^{(0,1,2)}\Bigr)\nonumber\\
    && + \frac{16 \Hnl^2}{\Gnl^2} \Bigl(\bar{\omega}_{\uGW,\uin}^{(0,2,0)} + \bar{\omega}_{\uGW,\uin}^{(2,2,0)} + 3 \bar{\omega}_{\uGW,\uin}^{(0,3,0)} + \bar{\omega}_{\uGW,\uin}^{(1,2,1)} + 6 \bar{\omega}_{\uGW,\uin}^{(0,4,0)} + \bar{\omega}_{\uGW,\uin}^{(0,2,2)}\Bigr)\ .
\end{eqnarray} 
We substitute this expression for $\omega_{\ung,\uin}^{(2)}$ and $\omega_{\ung,\uin}^{(1)}$, described in Eq.~\eqref{eq:Ong1}, into Eq.~\eqref{eq:b-def} to derive the specific formula for the angular bispectrum $B_{\ell_1 \ell_2 \ell_3} (\nu)$. 
The multipole dependence exhibits similarities to that of the \ac{CMB} in the low-multipole region, as reported in prior studies \cite{Bartolo:2019zvb,Li:2024zwx}. 
Therefore, we will focus on the frequency dependence of $B_{\ell_1 \ell_2 \ell_3} (\nu)$, denoted as $b (\nu)$, in the following analysis.

\begin{figure*}[htbp]
    \centering
    \includegraphics[width = \textwidth]{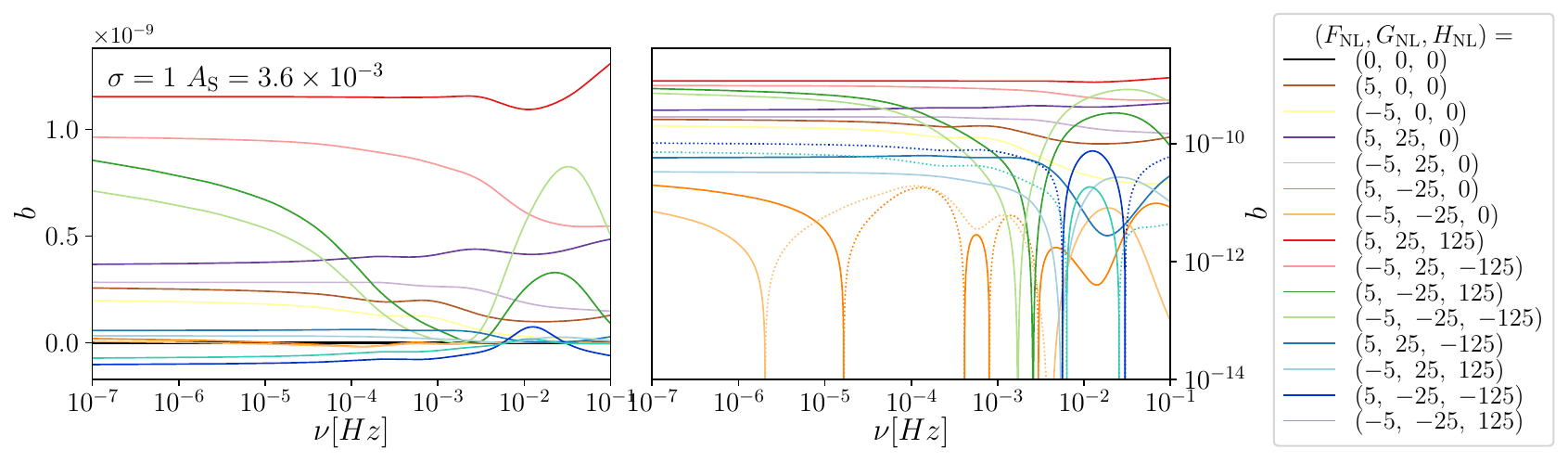}
    \caption{Frequency dependence of the angular bispectrum of SIGWs, with the left and right panels displaying numerical results on linear and logarithmic vertical scales, respectively. 
    The dotted lines in the right panel indicate negative sign regions.  }\label{fig:b_tilde}
\end{figure*}
    
% Similar to the reduced angular power spectrum in Eq.~\eqref{eq:Ct}, $b (\nu)$ is also expected to exhibit nearly scale-invariant behavior in the infrared regime, irrespective of the magnitude of \ac{PNG}. 
Here, we analyze $b (\nu)$ across various scenarios parameterized by different values of $\Fnl$, $\Gnl$, and $\Hnl$. 
Our focus is on the frequency dependence around the pivot frequency. 
First, in the scenario of Gaussian $\zeta$, the large-scale modulations $\omega_{\ung,\uin}^{(1)} = \omega_{\ung,\uin}^{(2)} = 0$ indicate that $b(\nu) = 0$, and the terms in the two square brackets of Eq.~\eqref{eq:b-def} also become zero in this scenario. 
This aligns with the expectation that the \acp{GW} produced over large distances are Gaussian since they are independently induced by incoherent primordial curvature perturbations. 
Therefore, the observation of a non-zero angular bispectrum of \acp{SIGW} can serve as compelling evidence for the presence of \ac{PNG}.
Second, in scenarios with minimal \ac{PNG}, the leading-order term in $A_\uS$ of $\omega_{\ung,\uin}^{(2)}$ is approximately $24(\Fnl^2 + \Gnl) \bar{\omega}_{\uGW,\uin}^{(0,0,0)}$.
Assuming the Gaussian component $\bar{\omega}_{\uGW,\uin}^{(0,0,0)}$ dominates in $\bar{\omega}_{\uGW,\uin}$, we combine this term with the leading-order term of $\omega_{\ung,\uin}^{(1)}$ to provide an approximate expression for $b (\nu)$ as 
\begin{eqnarray}\label{eq:b-NG-min}
    b (\nu) &\simeq& 8\pi^2 A_\uL^2 
    \biggl[
        8 \Fnl + \frac{3}{5} \bigl(6 - n_{\uGW} (\nu)\bigr)
    \biggr]^2 \biggl[
        24 \bigl(\Fnl^2 + \Gnl\bigr) + \frac{3}{5} \Fnl \bigl(6 - n_{\uGW} (\nu)\bigr)
    \biggr]\nonumber\\
    &=& 4 \pi A_\uL \ell (\ell+1) \tilde{C}_\ell (\nu) 
    \biggl[
        24 \bigl(\Fnl^2 + \Gnl\bigr) + \frac{3}{5} \Fnl \bigl(6 - n_{\uGW} (\nu)\bigr)
    \biggr]\ .
\end{eqnarray} 
Under this approximation, we examine the behavior of $ b (\nu) $ for various magnitudes of $ \Fnl $, $ \Gnl $, and $ \bigl(6 - n_{\uGW} (\nu)\bigr) $. 
Although the sign degeneracy between $ \Fnl $ and $ \Hnl $ cannot be resolved by the angular bispectrum when $ \Fnl $ dominates over $ \bigl(6 - n_{\uGW} (\nu)\bigr)$ (i.e., $ |\Fnl| \gg 1 $), this situation can only occur when $ A_\uS $ is small enough to validate this approximation, resulting in an unobservable angular bispectrum in the near future.
Conversely, in an extreme case where the \ac{SIGW} signals are prominent and the \ac{PNG} parameters are sufficiently small ($ |\Fnl| \ll 1 $ and $ \Gnl / \Fnl \ll 1 $), the factor $ \bigl(6 - n_{\uGW} (\nu)\bigr) $ dominates over $ \Fnl $ and $ \bigl(\Fnl + \Gnl/\Fnl\bigr) $ in certain frequency bands, leading to $ b (\nu) $ being approximately proportional to $ \Fnl \bigl[\ell (\ell+1) \tilde{C}_\ell (\nu)\bigr]^{3/2} $ in those bands. 
Consequently, it is anticipated that the angular bispectrum of \acp{SIGW} will clarify the signs of $ \Fnl $ and $ \Hnl $ for future detected \ac{SIGW} signals. 
Compared to the angular power spectrum, $ b (\nu) $ exhibits similarities with $ \ell (\ell+1) \tilde{C}_\ell (\nu) $ when $ \Fnl $ or $ \Gnl $ dominates among $ \Fnl $, $ \Gnl $, and $ \bigl(6 - n_{\uGW} (\nu)\bigr) $. 
Otherwise, the differences between $ b (\nu) $ and $ \ell (\ell+1) \tilde{C}_\ell (\nu) $ can be used to improve the accuracy of the inferred values of the \ac{PNG} parameters from the energy-density fraction spectrum and the angular power spectrum, suggesting that the angular bispectrum can serve as a complementary tool for determining model parameters.
Moreover, for large \ac{PNG}, the unique features arising from $\omega_{\ung,\uin}^{(2)} (q) / \bar{\omega}_{\uGW,\uin} (q)$ are anticipated to provide additional insights beyond the angular power spectrum. 
We elaborate on cases of substantial \ac{PNG}, with numerical results depicted in \cref{fig:b_tilde}, where the employed model parameters are identical to those in \cref{fig:C-Ct}. 
Both panels of this figure illustrate the same angular bispectra, with the left and right panels displaying numerical results on linear and logarithmic vertical scales, respectively. 
Notably, the dotted lines in the right panel indicate where the sign is negative, which appears in cases of negative $\Gnl$ and non-positive $\Fnl\Hnl$. 
It is important to highlight that while Gaussian $\zeta$ always results in a Gaussian \ac{SIGW} background, the presence of \ac{PNG} does not consistently lead to a non-zero angular bispectrum across all frequency ranges, as shown in the left panel of \cref{fig:b_tilde}. 
Furthermore, this figure demonstrates that the presence of a positive (negative) $\Fnl\Hnl$ does not necessarily correlate with a (smaller) larger magnitude of $b(\nu)$ compared to cases lacking $\Hnl$. 
The complex and variable behaviors of $b(\nu)$ suggest that an analysis of the \ac{SIGW} angular bispectrum is valuable in the search for \ac{PNG}.

\begin{figure*}[htbp]
    \centering
    \includegraphics[width = 1 \textwidth]{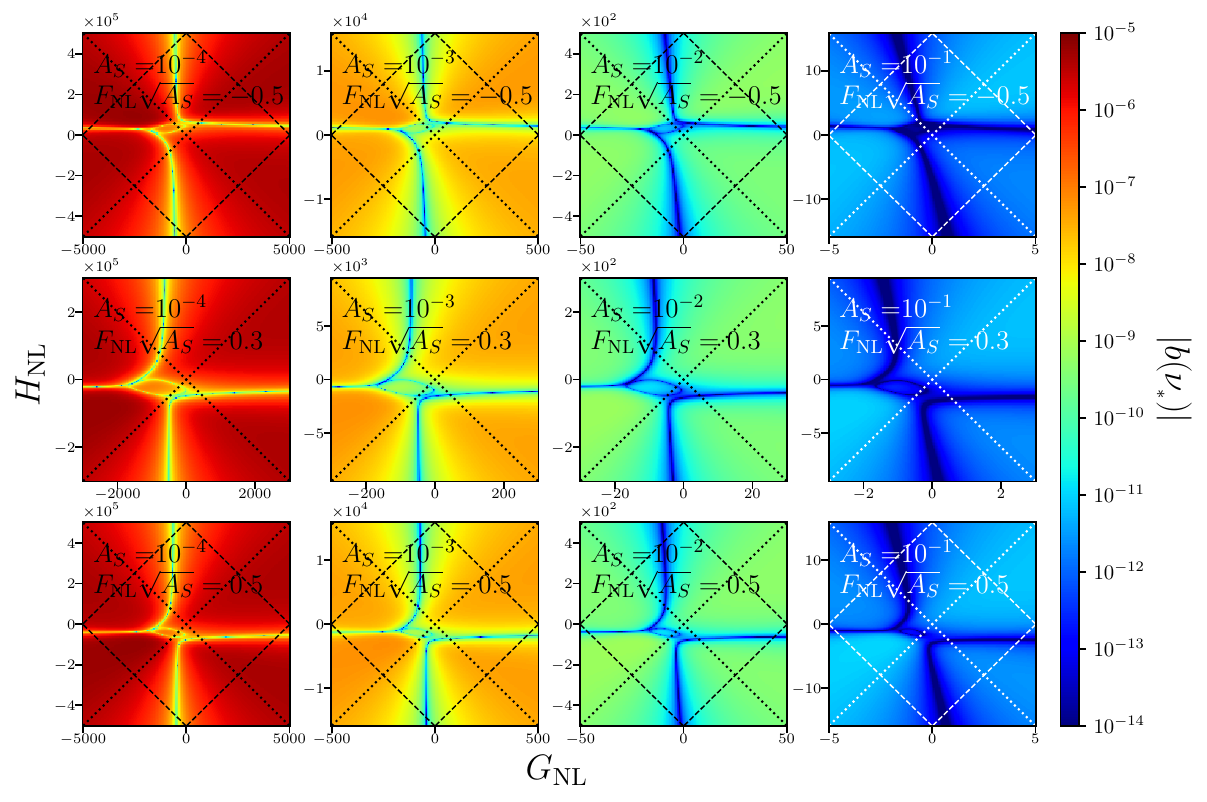}
    \caption{ Angular bispectrum of SIGWs with respect to the $\Fnl$, $\Gnl$ and $\Hnl$. The conventions for the dotted and dashed lines are the same as those in \cref{fig:Ct_G-H}. }\label{fig:bt_G-H}
\end{figure*}

To comprehensively illustrate the dependence of $b (\nu)$ on \ac{PNG} parameters and $A_\uS$, we present an array of contour plots in \cref{fig:bt_G-H}, with model parameters identical to those in \cref{fig:Ct_G-H}. 
The conventions for the dotted and dashed lines are also consistent with those in \cref{fig:Ct_G-H}. 
Notably, each contour plot features a ``loop'' near the center. 
Compared to the results for \ac{PNG} up to $\Gnl$ order in Ref.~\cite{Li:2024zwx}, the ``loops'' in this figure are deflected and deviate from the center. 
It is important to emphasize that the values of the angular bispectrum within the ``loop'' in each panel are indeed negative. 
In other words, when both $\Fnl\Hnl$ and $\Gnl$ are negative and their absolute values are relatively mild, the angular bispectrum turns negative. 
Furthermore, a comparison across different panels reveals that larger values of $A_\uS$ tend to reduce the magnitude of the reduced bispectrum, while a small value of $A_\uS$ ($10^{-4}$) can result in a reduced bispectrum value as high as $10^{-5}$. 
We can also observe that the sign degeneracy is broken by comparing the contour plots in the third row with those in the first row after mirroring them along the axis of $\Hnl = 0$, similar to \cref{fig:Ct_G-H}.

\subsection{Angular trispectrum: four-point angular correlation}\label{subsec:t}

Analogous to the four-point angular correlation function of the \ac{CMB} \cite{Hu:2001fa,Komatsu:2001ysk,Komatsu:2010hc,Kogo:2006kh,Okamoto:2002ik}, the four-point correlator of $\delta_{\uGW,0,\ell m}$ for the rotation-invariant \ac{GW} background yields 
\begin{eqnarray}\label{eq:Tl-def}
    \left\langle\prod_{i=1}^4 \delta_{\uGW,0,\ell_i m_i}(2\pi \nu)\right\rangle
    &=& \sum_{LM} (-1)^M 
    \begin{pmatrix}
         \ell_1 & \ell_2 & L \\
         m_1    & m_2    & -M   
    \end{pmatrix}
    \begin{pmatrix}
         \ell_3 & \ell_4 & L \\
         m_3    & m_4    & M   
    \end{pmatrix}
    T^{\ell_1 \ell_2}_{\ell_3 \ell_4} (L,\nu) \ ,
\end{eqnarray} 
where $T^{\ell_1 \ell_2}_{\ell_3 \ell_4}(L,\nu)$ represents the angular averaged trispectrum. 
This angular averaged trispectrum describes the mixed central fourth-order moment of $\delta_{\uGW,0,\ell,m}$, containing information about the kurtosis of fluctuations on the \ac{SIGW} background. 
Geometrically, the Wigner 3-$j$ symbols indicate that the quadrilateral $\{\ell_1, \ell_2, \ell_3, \ell_4\}$ partitions into two triangles ($\{\ell_1, \ell_2, L\}$ and $\{\ell_3, \ell_4, L\}$) divided by a diagonal of length $L$, thus ensuring the triangular inequalities  ($|\ell_1 - \ell_2| \leq L \leq \ell_1 + \ell_2$, $|\ell_3 - \ell_4| \leq L \leq \ell_3 + \ell_4$, and $m_1 + m_2 - M = m_3 + m_4 + M = 0$). 
Furthermore, parity invariance necessitates the condition $(\ell_1 + \ell_2 + \ell_3 + \ell_4 = 2n$, where $n \in \mathbb{N}$). 

Unlike the two- and three-point correlators of $\delta_{\uGW,0}$, where the disconnected components are zero due to $\langle\delta_{\uGW,0}\rangle = 0$, the disconnected components of the four-point correlators are non-zero. 
Consequently, we decompose $T^{\ell_1 \ell_2}_{\ell_3 \ell_4}(L,\nu)$ into the disconnected piece, denoted as ${T_\mathrm{G}}^{\ell_1 \ell_2}_{\ell_3 \ell_4}(L,\nu)$, and the connected piece, denoted as ${T_\mathrm{c}}^{\ell_1 \ell_2}_{\ell_3 \ell_4}(L,\nu)$, namely 
\begin{equation}
    T^{\ell_1 \ell_2}_{\ell_3 \ell_4} (L,\nu) 
    = {T_\mathrm{G}}^{\ell_1 \ell_2}_{\ell_3 \ell_4} (L,\nu) + {T_\mathrm{c}}^{\ell_1 \ell_2}_{\ell_3 \ell_4} (L,\nu)\ .
\end{equation} 
The former is driven by the Gaussian statistics of $\delta_{\uGW,0}$ and can thus be expressed in terms of $\tilde{C}_\ell(\nu)$. 
Its explicit form is identical to that of \ac{CMB}, as provided by \cite{Hu:2001fa,Komatsu:2010hc,Kogo:2006kh} 
\begin{eqnarray}\label{eq:TGl-def}
    {T_\mathrm{G}}^{\ell_1 \ell_2}_{\ell_3 \ell_4} (L,\nu) 
    &=& (-1)^{\ell_1 + \ell_3} \sqrt{(2\ell_1 + 1)(2\ell_3 + 1)} \tilde{C}_{\ell_1} (\nu) \tilde{C}_{\ell_3} (\nu) \delta_{\ell_1 \ell_2} \delta_{\ell_3 \ell_4} \delta_{L 0} \nonumber\\ 
    && + (2 L + 1) \tilde{C}_{\ell_1} (\nu) \tilde{C}_{\ell_2} (\nu) \Bigl((-1)^{\ell_1 + \ell_2 + L} \delta_{\ell_1 \ell_3} \delta_{\ell_2 \ell_4} + \delta_{\ell_1 \ell_4} \delta_{\ell_2 \ell_3}\Bigr)\ .
\end{eqnarray}
This piece is non-zero only when $L = 0$ (with $\ell_1 = \ell_2$ and $\ell_3 = \ell_4$ simultaneously) or $\ell_1 = \ell_2 = \ell_3 = \ell_4$. 
Conversely, when $\ell_1$, $\ell_2$, $\ell_3$, and $\ell_4$ are not equal, $T^{\ell_1 \ell_2}_{\ell_3 \ell_4}(L,\nu)$ is non-zero only if the connected piece is present. 
We consider the disconnected piece, ${T_\mathrm{G}}^{\ell_1 \ell_2}_{\ell_3 \ell_4}(L,\nu)$, to be redundant, as the information it conveys is already encapsulated within the reduced angular power spectrum $\tilde{C}_{\ell}(\nu)$. 
Consequently, this study focuses on the connected piece, ${T_\mathrm{c}}^{\ell_1 \ell_2}_{\ell_3 \ell_4}(L,\nu)$, which characterizes the non-Gaussianity inherent in the \ac{SIGW} background.

The derivation process of ${T_\mathrm{c}}^{\ell_1 \ell_2}_{\ell_3 \ell_4}(L,\nu)$ closely mirrors that of \ac{CMB} \cite{Hu:2001fa,Komatsu:2010hc,Kogo:2006kh}. 
By construction, ${T_\mathrm{c}}^{\ell_1 \ell_2}_{\ell_3 \ell_4}(L,\nu)$ is associated with two triangles formed by the pairs ${(\ell_1,\ell_2), (\ell_3,\ell_4)}$ and the shared side $L$. 
However, given the symmetry of $(\ell_1,\ell_2,\ell_3,\ell_4)$ in the four-point correlation, one can opt for alternative pairs ${(\ell_1,\ell_3), (\ell_2,\ell_4)}$ or ${(\ell_1,\ell_4), (\ell_2,\ell_3)}$ to construct other triangles, which should yield the same angular trispectrum. 
Therefore, we enforce permutation symmetry by introducing a new quantity $P^{\ell_1 \ell_2}_{\ell_3 \ell_4}(L,\nu)$, defined as follows 
\begin{eqnarray}\label{eq:Pl-def}
    \left\langle\prod_{i=1}^4 \delta_{\uGW,0,\ell_i m_i}(2\pi \nu)\right\rangle_\mathrm{c}
    &=& \sum_{LM} (-1)^M 
    \begin{pmatrix}
         \ell_1 & \ell_2 & L \\
         m_1    & m_2    & -M   
    \end{pmatrix}
    \begin{pmatrix}
         \ell_3 & \ell_4 & L \\
         m_3    & m_4    & M   
    \end{pmatrix}
    P^{\ell_1 \ell_2}_{\ell_3 \ell_4} (L,\nu)\nonumber\\ 
    && + (\ell_2 \leftrightarrow \ell_3) + (\ell_2 \leftrightarrow \ell_4) \ ,
\end{eqnarray}
where the subscript $_\mathrm{c}$ emphasizes the connected piece, and the disconnected component has been subtracted from this correlator. 
Comparing this with Eq.~\eqref{eq:Tl-def}, we can project the latter two pairs onto $(\ell_1,\ell_2)$ by using the Wigner 6-$j$ symbols, such that ${T_\mathrm{c}}^{\ell_1 \ell_2}_{\ell_3 \ell_4}(L,\nu)$ can be expressed as 
\begin{align}\label{eq:Tl-Pl}
    {T_\mathrm{c}}^{\ell_1 \ell_2}_{\ell_3 \ell_4} (L,\nu) 
    = P^{\ell_1 \ell_2}_{\ell_3 \ell_4} (L,\nu) 
    + (2L + 1) \sum_{L'} 
    \Biggl(
        &(-1)^{\ell_2 + \ell_3}
        \begin{Bmatrix}
            \ell_1 & \ell_2 & L \\
            \ell_4 & \ell_3 & L'
        \end{Bmatrix} 
        P^{\ell_1 \ell_3}_{\ell_2 \ell_4} (L',\nu) \nonumber\\ 
        & + (-1)^{L + L'}
        \begin{Bmatrix}
            \ell_1 & \ell_2 & L \\
            \ell_3 & \ell_4 & L'
        \end{Bmatrix} 
        P^{\ell_1 \ell_4}_{\ell_3 \ell_2} (L',\nu)
    \Biggr)\ .
\end{align}
The basic properties of the Wigner 6-$j$ symbols are also presented in Appendix~\ref{sec:Wigner}. 
Furthermore, due to the symmetry of the Wigner 3-$j$ symbols in Eq.~\eqref{eq:Pl-def}, we can further decompose $P^{\ell_1 \ell_2}_{\ell_3 \ell_4}(L,\nu)$ into four permutations as follows 
\begin{eqnarray}\label{eq:tl-def}
    P^{\ell_1 \ell_2}_{\ell_3 \ell_4} (L,\nu) 
    &=& t^{\ell_1 \ell_2}_{\ell_3 \ell_4} (L,\nu) 
    + (-1)^{\ell_1 + \ell_2 + L} t^{\ell_2 \ell_1}_{\ell_3 \ell_4} (L,\nu) \nonumber\\
    && + (-1)^{\ell_3 + \ell_4 + L} t^{\ell_1 \ell_2}_{\ell_4 \ell_3} (L,\nu)
    + (-1)^{\ell_1 + \ell_2 + \ell_3 + \ell_4 + 2L} t^{\ell_2 \ell_1}_{\ell_4 \ell_3} (L,\nu)\ ,
\end{eqnarray}
where $t^{\ell_1 \ell_2}_{\ell_3 \ell_4}(L,\nu)$ is referred to as the reduced angular trispectrum of \acp{SIGW}. 
Notably, although there are four terms on the right-hand side of Eq.~\eqref{eq:tl-def}, only two are independent due to the even parity condition, namely $t^{\ell_1 \ell_2}_{\ell_3 \ell_4}(L,\nu) = t^{\ell_2 \ell_1}_{\ell_4 \ell_3}(L,\nu)$ and $t^{\ell_2 \ell_1}_{\ell_3 \ell_4}(L,\nu) = t^{\ell_1 \ell_2}_{\ell_4 \ell_3}(L,\nu)$. 
In particular, as discussed below Eq.~\eqref{eq:TGl-def}, when all four multipoles satisfy the aforementioned triangular inequalities and parity invariance and are unequal, the disconnected piece of the four-point angular correlators contributes nothing. 
This permits a direct reflection of the contribution from the angular trispectrum.

\begin{figure*}[htbp]
    \centering
    \includegraphics[width =0.45\textwidth]{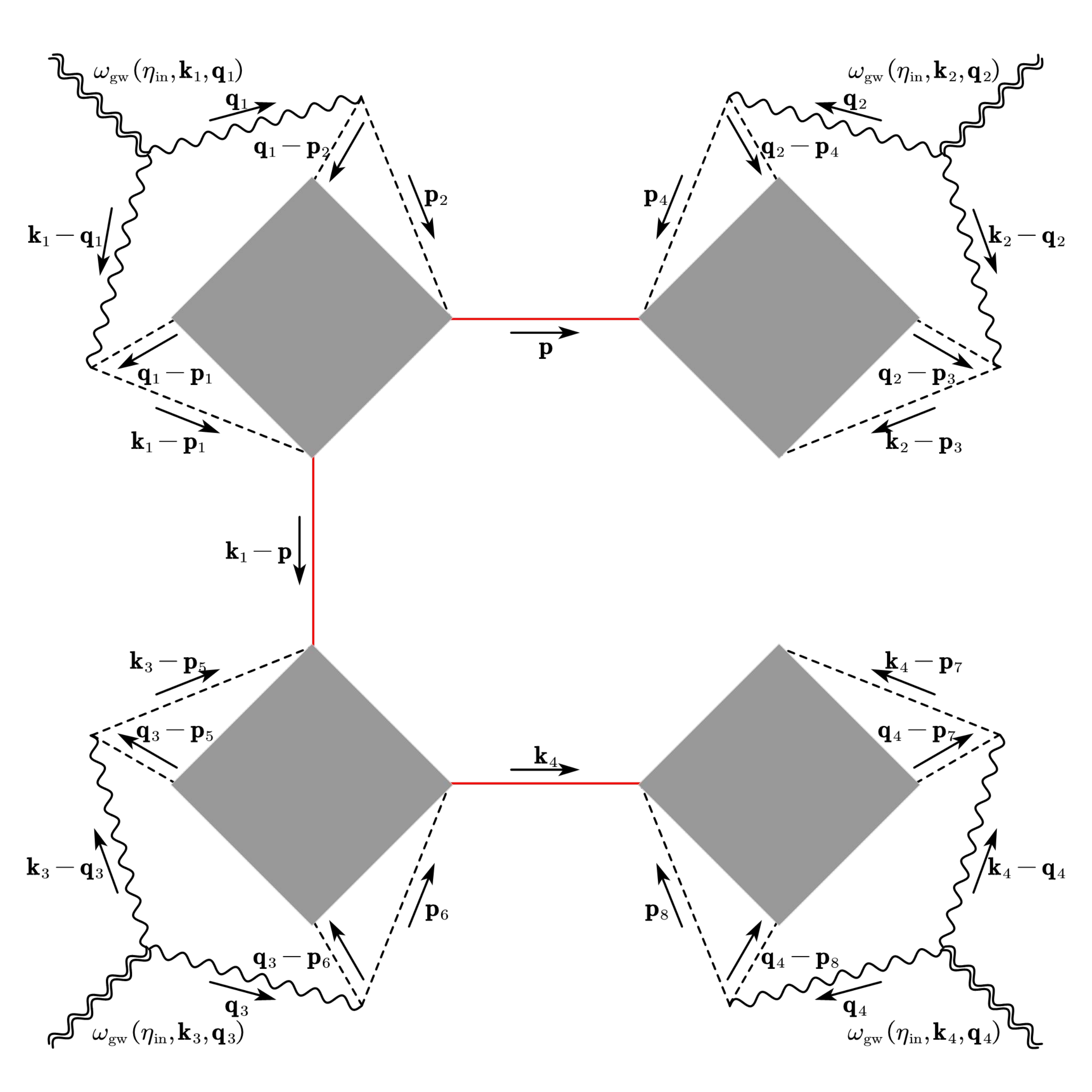}
    \hfill 
    \includegraphics[width =0.45\textwidth]{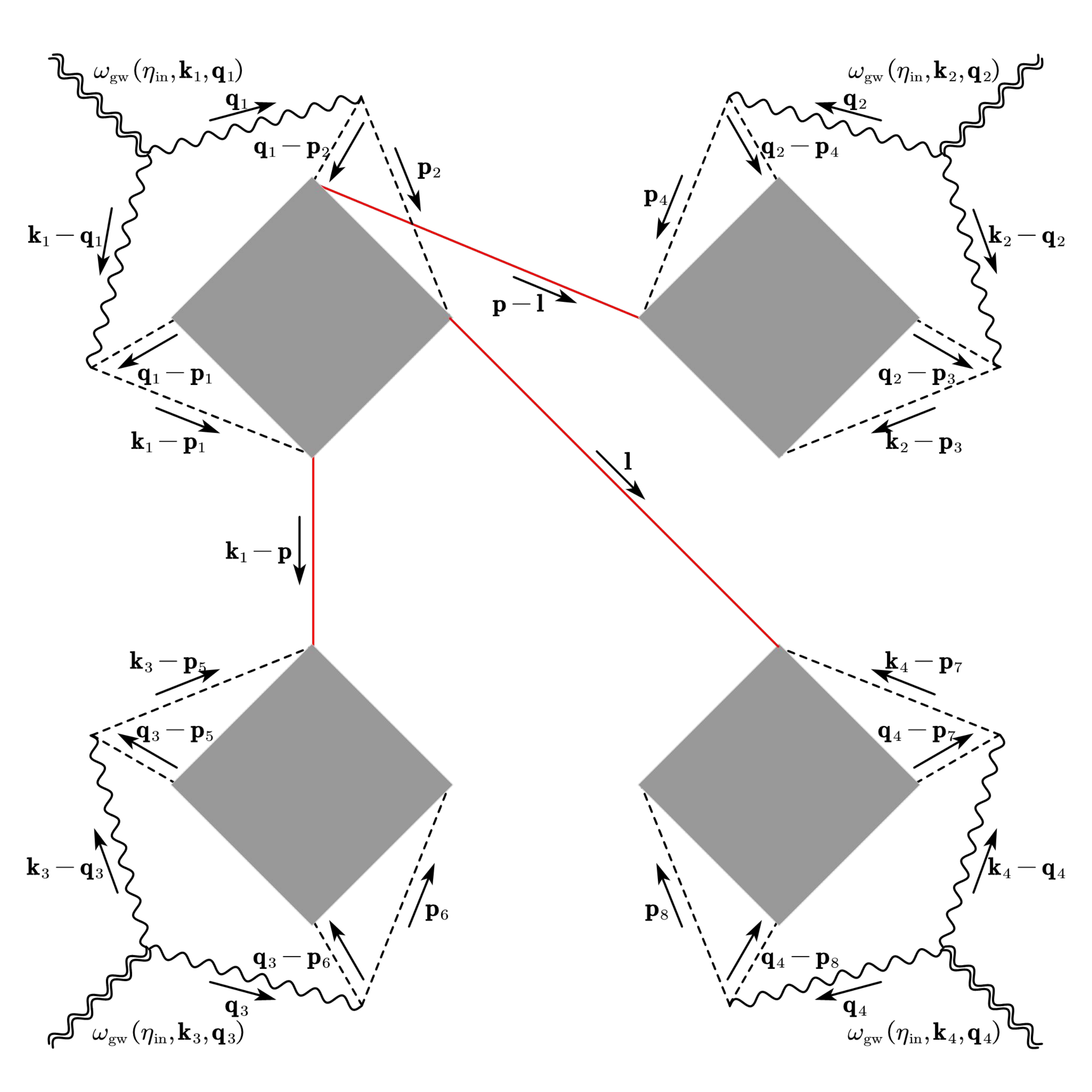}
    \\ \vspace{0.5cm}
    \includegraphics[width =0.45\textwidth]{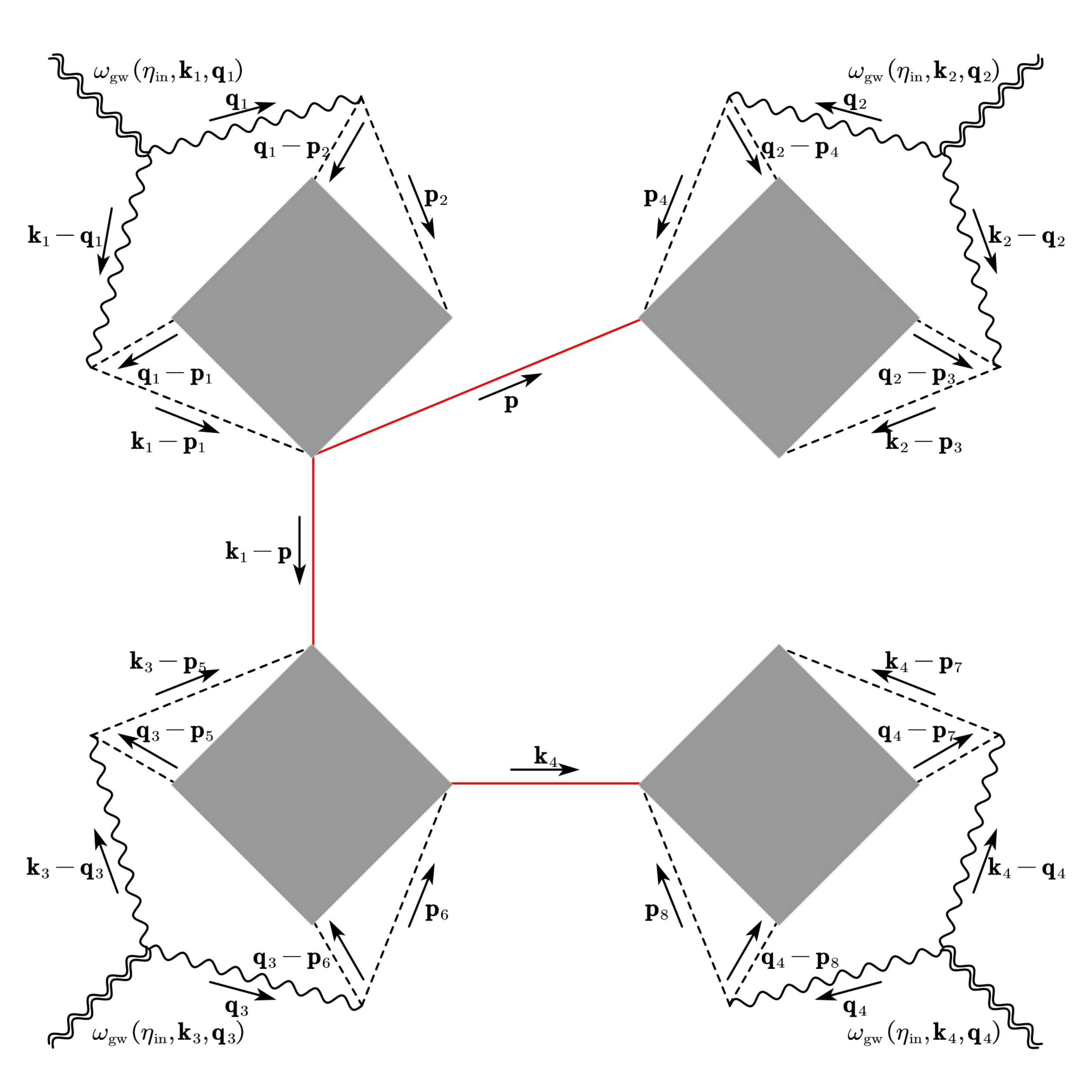}
    \hfill 
    \includegraphics[width =0.45\textwidth]{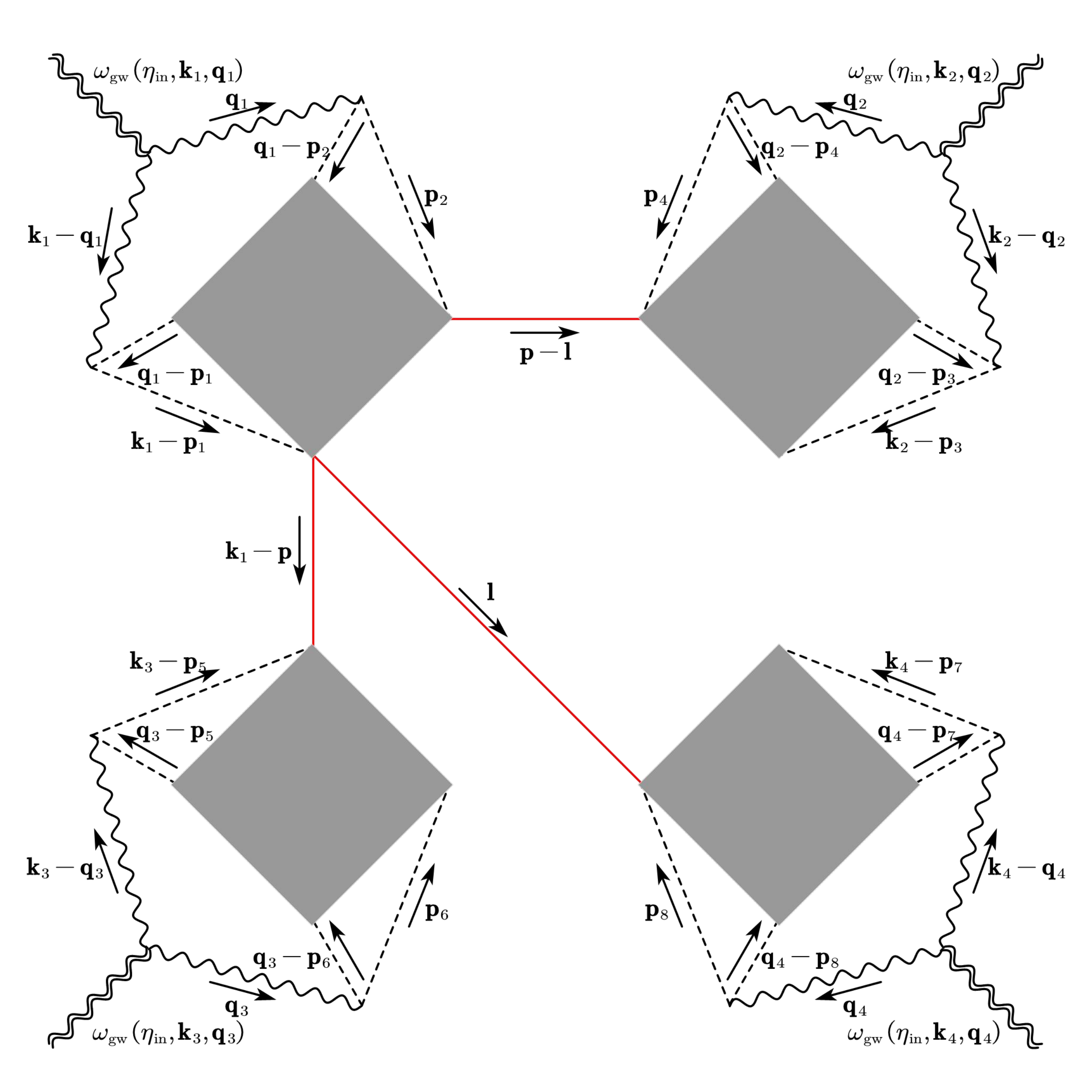}
    \\ \vspace{0.5cm} 
    \includegraphics[width =0.45\textwidth]{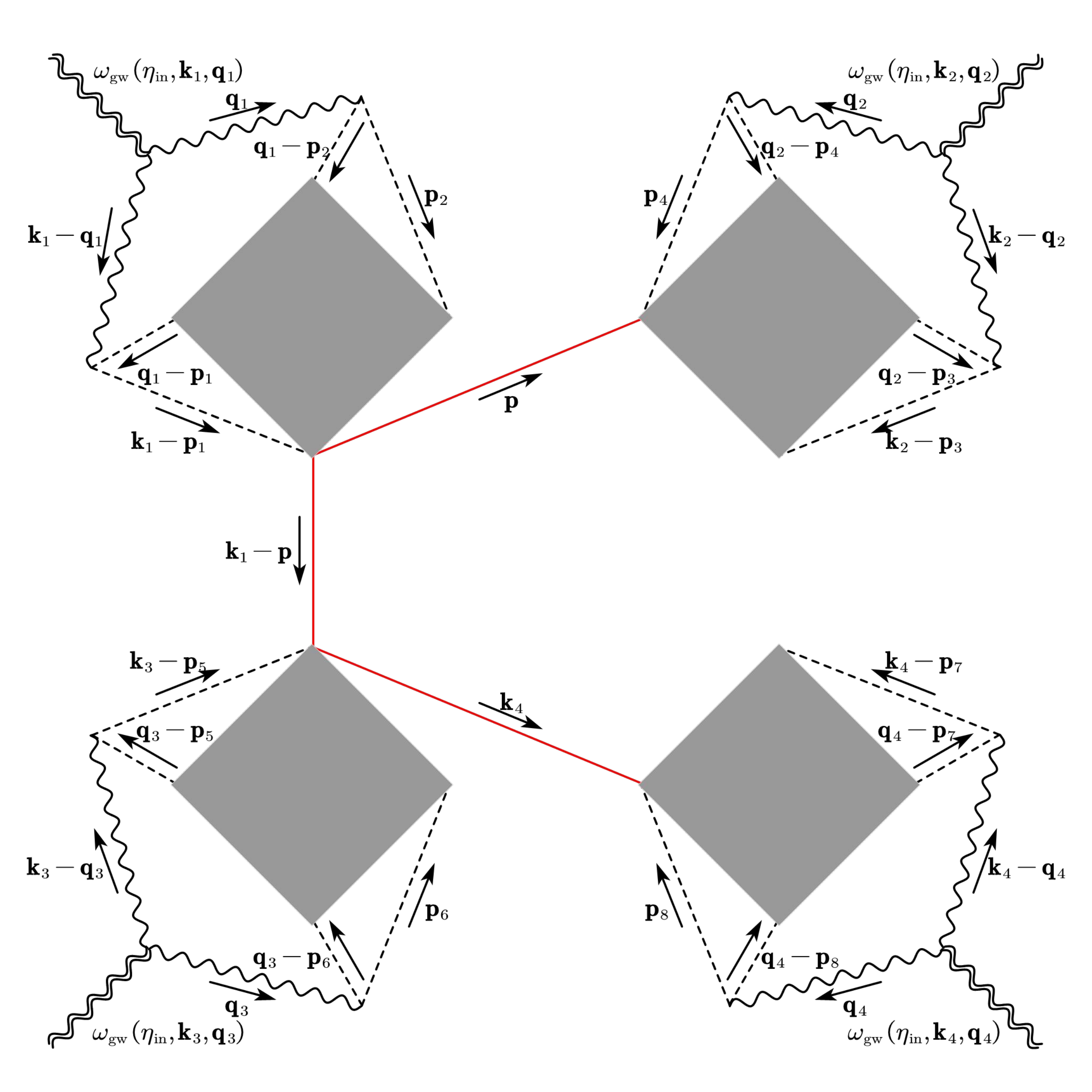}
    \hfill
    \includegraphics[width =0.45\textwidth]{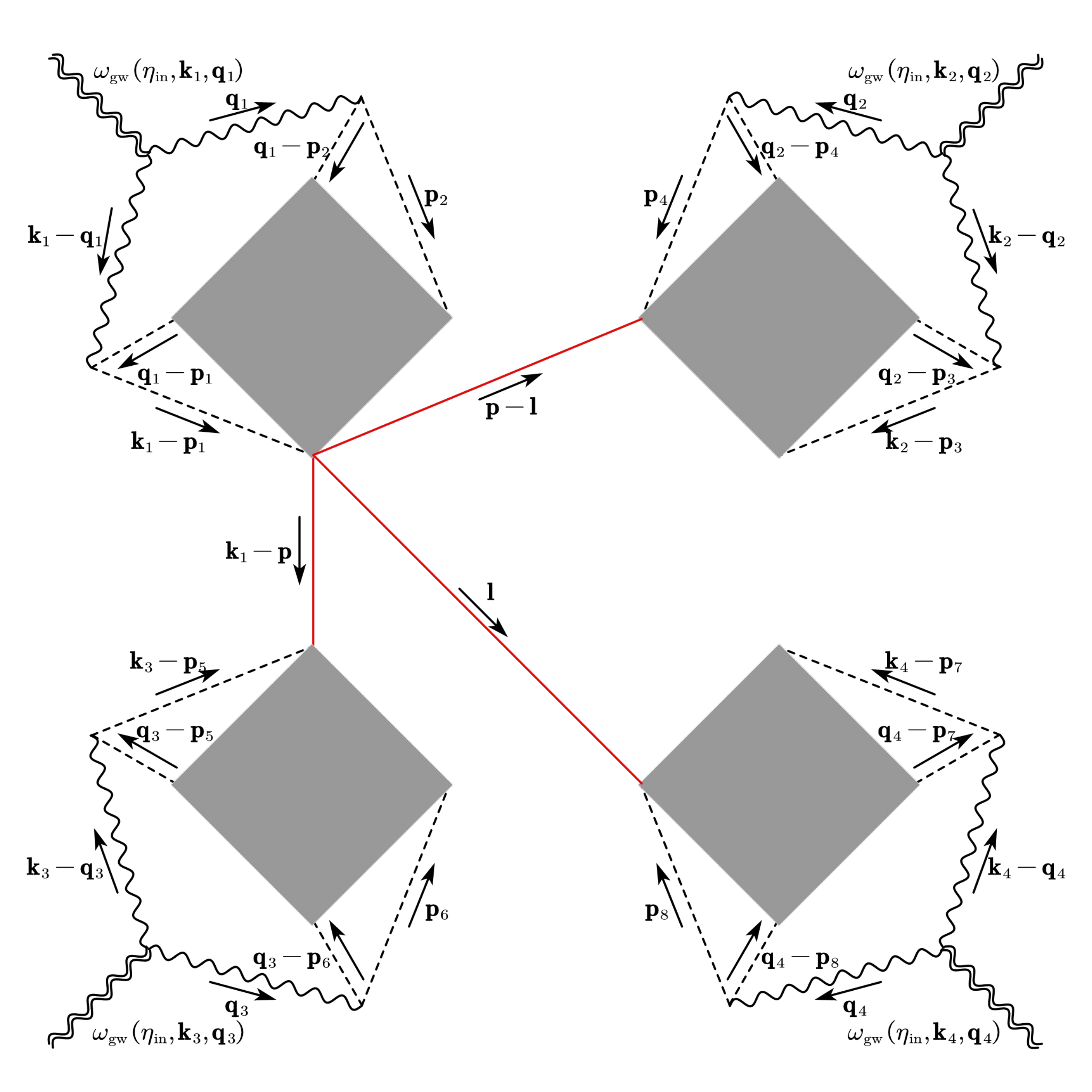}
    \caption{ Feynman-like diagrams for the four-point correlators of $\omega_\uGW (\eta_\uin,\bk,\bq)$ at leading order in $A_\uL$. The upper panels corresponds to $\langle\omega_{\uGW,\uin}^{(2)} \omega_{\uGW,\uin}^{(2)} \omega_{\uGW,\uin}^{(1)} \omega_{\uGW,\uin}^{(1)}\rangle$, and the lower panels corresponds to $\langle\omega_{\uGW,\uin}^{(3)} \omega_{\uGW,\uin}^{(1)} \omega_{\uGW,\uin}^{(1)} \omega_{\uGW,\uin}^{(1)}\rangle$. }\label{fig:4-correlator}
\end{figure*}

Based on our previous investigation \cite{Li:2024zwx} involving the angular trispectrum of \acp{SIGW} for \ac{PNG} up to $\Gnl$ order, we extend this work to incorporate \ac{PNG} up to all orders, providing a specific expression for \ac{PNG} up to $\Hnl$ order. 
The methodology for studying the \ac{SIGW} angular trispectrum is analogous to that for the \ac{SIGW} angular bispectrum. 
For the leading order in $A_\uL$, the reduced angular trispectrum $t_{\ell_1 \ell_2}^{\ell_3 \ell_4}(L,\nu)$ corresponds to the connected four-point correlators $\langle\delta_{\uGW,0}^{(3)} \delta_{\uGW,0}^{(1)} \delta_{\uGW,0}^{(1)} \delta_{\uGW,0}^{(1)}\rangle_\mathrm{c}$ and $\langle\delta_{\uGW,0}^{(2)} \delta_{\uGW,0}^{(2)} \delta_{\uGW,0}^{(1)} \delta_{\uGW,0}^{(1)}\rangle_\mathrm{c}$. 
This involves two types of four-point correlators: $\langle\omega_{\uGW,\uin}^{(2)} \omega_{\uGW,\uin}^{(2)} \omega_{\uGW,\uin}^{(1)} \omega_{\uGW,\uin}^{(1)}\rangle$ and $\langle\omega_{\uGW,\uin}^{(3)} \omega_{\uGW,\uin}^{(1)} \omega_{\uGW,\uin}^{(1)} \omega_{\uGW,\uin}^{(1)}\rangle$, which can be illustrated by the six diagrams in \cref{fig:4-correlator} according to \cref{fig:FD-Rules} and \cref{fig:FD_Frame}. 
These diagrams consist of four $\omega_{\uGW,\uin}$ components connected by three ``non-Gaussian bridges'', indicating that the leading contribution to $t_{\ell_1 \ell_2}^{\ell_3 \ell_4}(L,\nu)$ is of $\mathcal{O}(A_\uL^3)$ order. 
The three diagrams on the left, representing $\langle\omega_{\uGW,\uin}^{(2)} \omega_{\uGW,\uin}^{(2)} \omega_{\uGW,\uin}^{(1)} \omega_{\uGW,\uin}^{(1)}\rangle$, are equivalent to an additional ``non-Gaussian bridge'' connecting two diagrams of the two-point correlator in \cref{fig:2-correlator} and solely involves the large-scale modulations up to $\mathcal{O}(\zeta_{\ugL}^2)$ order. 
The expression for this correlator can be derived from Eq.~\eqref{eq:Ong1} and Eq.~\eqref{eq:Ong2}. 
In the three diagrams on the right, which represent $\langle\omega_{\uGW,\uin}^{(3)} \omega_{\uGW,\uin}^{(1)} \omega_{\uGW,\uin}^{(1)} \omega_{\uGW,\uin}^{(1)}\rangle$, the three ``non-Gaussian bridges'' connect one $\omega_{\uGW,\uin}$ diagram with three other $\omega_{\uGW,\uin}$ diagrams. 
These diagrams involve large-scale modulation at $\mathcal{O}(\zeta_{\ugL}^3)$ order, which is our goal to derive with an exact expression. 
We will subsequently evaluate these two types of four-point correlators while neglecting diagrams with more ``non-Gaussian bridges'', as they correspond to contributions of higher order in $A_\uL$.

Proceeding as outlined above, the definition of the reduced angular trispectrum $t^{\ell_1 \ell_2}_{\ell_3 \ell_4}(L,\nu)$ in Eqs.~(\ref{eq:Tl-def}, \ref{eq:Tl-Pl}, \ref{eq:tl-def}) indicates that it encompasses only a portion of the four-point correlator of $\delta_{\uGW,0}$. 
Specifically, ${T_\mathrm{c}}^{\ell_1 \ell_2}_{\ell_3 \ell_4}(L,\nu)$ contains three permutations of $P^{\ell_1 \ell_2}_{\ell_3 \ell_4}(L,\nu)$, while each permutation of $P^{\ell_1 \ell_2}_{\ell_3 \ell_4}(L,\nu)$ encompasses four permutations of $t^{\ell_1 \ell_2}_{\ell_3 \ell_4}(L,\nu)$. 
Thus, twelve permutations of $t^{\ell_1 \ell_2}_{\ell_3 \ell_4}(L,\nu)$ partition the two types of connected four-point correlators $\langle\delta_{\uGW,0}^{(3)} \delta_{\uGW,0}^{(1)} \delta_{\uGW,0}^{(1)} \delta_{\uGW,0}^{(1)}\rangle_\mathrm{c}$ and $\langle\delta_{\uGW,0}^{(2)} \delta_{\uGW,0}^{(2)} \delta_{\uGW,0}^{(1)} \delta_{\uGW,0}^{(1)}\rangle_\mathrm{c}$. 
According to the permutation symmetry, both $t^{\ell_1 \ell_2}_{\ell_3 \ell_4}(L,\nu)$ and $t^{\ell_1 \ell_4}_{\ell_3 \ell_2}(L,\nu)$ correspond to 
\begin{eqnarray}\label{eq:tl-delta-cor}
    && \frac{1}{2} \left\langle\delta_{\uGW,0}^{(2)} (\bq_1) \delta_{\uGW,0}^{(1)} (\bq_2) \delta_{\uGW,0}^{(2)} (\bq_3) \delta_{\uGW,0}^{(1)} (\bq_4)\right\rangle_\mathrm{c} \nonumber\\ 
    && + \frac{1}{6} \Bigl( \left\langle\delta_{\uGW,0}^{(3)} (\bq_1) \delta_{\uGW,0}^{(1)} (\bq_2) \delta_{\uGW,0}^{(1)} (\bq_3) \delta_{\uGW,0}^{(1)} (\bq_4)\right\rangle_\mathrm{c}
    + \left\langle\delta_{\uGW,0}^{(1)} (\bq_1) \delta_{\uGW,0}^{(1)} (\bq_2) \delta_{\uGW,0}^{(3)} (\bq_3) \delta_{\uGW,0}^{(1)} (\bq_4)\right\rangle_\mathrm{c} \Bigr) \nonumber\ .
\end{eqnarray} 
Similarly, one could enumerate the corresponding portions of the connected four-point correlator for the other ten permutations. 
Thus, to compute $t^{\ell_1 \ell_2}_{\ell_3 \ell_4}(L,\nu)$, we will analyze the two terms of Eq.~\eqref{eq:tl-delta-cor} individually. 
First, using the expressions of $\delta_{\uGW,0}^{(1)}$, $\delta_{\uGW,0}^{(2)}$, and $\delta_{\uGW,0}^{(3)}$ illustrated in Eqs.~(\ref{eq:delta0-1}, \ref{eq:delta0-2}, \ref{eq:delta0-3}), we will substitute their spherical harmonic coefficients obtained via Eq.~\eqref{eqs:spher-harm} into the four-point correlators, respectively. 
All the correlators reduce to six-point correlators of $\zeta_{\ugL}$, yielding factors proportional to $(\Delta_\uL^2)^3$ accompanied by three momentum-conservation factors. 
Two of these momentum-conservation factors vanish upon integration over the convolution momenta, leaving only one, $\delta^{(3)}(\bk_1 + \bk_2 + \bk_3 + \bk_4)$, to be integrated. 
Notably, this momentum-conservation factor is split into two Dirac delta functions, i.e., $\int \ud^3 \bK\, \delta^{(3)}(\bk_1 + \bk_2 + \bK) \delta^{(3)}(\bk_3 + \bk_4 - \bK)$ when calculating $\langle\delta_{\uGW,0}^{(2)} \delta_{\uGW,0}^{(1)} \delta_{\uGW,0}^{(2)} \delta_{\uGW,0}^{(1)}\rangle_\mathrm{c}$. 
In this way, we can handle the integration over $(\bk_1,\bk_2,\bn_1,\bn_2)$ and the integration over $(\bk_3,\bk_4,\bn_3,\bn_4)$ separately, followed by integrating over the other variables. 
The remaining derivation process is akin to that of the angular bispectrum discussed in Subsection~\ref{subsec:b}. 
We make use of the line-of-sight relation, the Fourier counterpart of Dirac delta functions in Eq.~\eqref{eq:Diracdelta-Fourier}, and the identity in Eq.~\eqref{eq:exp-jYY} to express the exponential functions with momentum-conservation factors in terms of spherical harmonics and spherical Bessel functions. 
Afterwards, we leverage the properties of spherical harmonics and Eq.~\eqref{eq:Glm-def} to manage the multiple spherical harmonics within the integral, along with Eqs.~(\ref{eq:jl-2-int}, \ref{eq:jl-closure}) to handle the spherical Bessel functions. 
Ultimately, we obtain the formula for the reduced angular trispectrum $t^{\ell_1 \ell_2}_{\ell_3 \ell_4}(L,\nu)$ as follows 
\begin{eqnarray}\label{eq:tl-res}
    t^{\ell_1 \ell_2}_{\ell_3 \ell_4} (L,\nu) 
    &=& \Biggl[
        \frac{t_1 (\nu)}{L (L + 1)} + \frac{t_2 (\nu)}{\ell_1 (\ell_1 + 1)} + \frac{t_2 (\nu)}{\ell_3 (\ell_3 + 1)} 
    \Biggr] 
    \frac{h_{\ell_1 \ell_2 L}}{\ell_2 (\ell_2 + 1)} \frac{h_{\ell_3 \ell_4 L}}{\ell_4 (\ell_4 + 1)}\ ,
\end{eqnarray} 
where $t_1(\nu)$ and $t_2(\nu)$ 
\footnote{There is a typo in our previous work \cite{Li:2024zwx}, where the denominator of the coefficient of $ t_2 (\nu) $ is $ 2\pi $ instead of $ 4 $; however, the deviation is minimal in the subsequent numerical results. } 
are introduced as follows 
\begin{eqnarray}
    t_1 (\nu) &=& 4 \cdot \bigl(2 \pi A_\uL\bigr)^3 
        \Biggl[
            \frac{\omega_{\ung,\uin}^{(1)} (q)}{\bar{\omega}_{\uGW,\uin} (q)}
            + \frac{3}{5} \bigl(6 - n_{\uGW} (\nu)\bigr)
        \Biggr]^2 \Biggl[
            \frac{\omega_{\ung,\uin}^{(2)} (q)}{\bar{\omega}_{\uGW,\uin} (q)}
            + \frac{3}{5} \Fnl \bigl(6 - n_{\uGW} (\nu)\bigr)
        \Biggr]^2 \ ,\label{eq:t1-def}\\
    t_2 (\nu) &=& \frac{\bigl(2 \pi A_\uL\bigr)^3}{2\pi}
        \Biggl[
            \frac{\omega_{\ung,\uin}^{(1)} (q)}{\bar{\omega}_{\uGW,\uin} (q)}
            + \frac{3}{5} \bigl(6 - n_{\uGW} (\nu)\bigr)
        \Biggr]^3 \Biggl[
            \frac{\omega_{\ung,\uin}^{(3)} (q)}{\bar{\omega}_{\uGW,\uin} (q)}
            + \frac{3}{5} \Gnl \bigl(6 - n_{\uGW} (\nu)\bigr)
        \Biggr] \ .\label{eq:t2-def}
\end{eqnarray}
The derivation process of Eq.~\eqref{eq:tl-res} closely resembles that of the \ac{CMB} angular trispectrum, with further specifics available in Refs.~\cite{Hu:2001fa,Komatsu:2010hc,Kogo:2006kh}. 
\footnote{Considering the scale variance of $ \Delta_\uL^2 (k) $, the explicit expressions for this case can be found in Ref.~\cite{Li:2024zwx}. 
}
Comparing Eq.~\eqref{eq:t1-def} with Eqs.~(\ref{eq:Ct}, \ref{eq:b-def}), we find that $t_1(\nu)$ can be expressed in terms of $\ell (\ell + 1) \tilde{C}_\ell(\nu)$ and $b(\nu)$ as follows 
\begin{equation}\label{eq:t1-C+b}
    t_1 (\nu) = \frac{b^2 (\nu)}{\ell (\ell+1) \tilde{C}_\ell (\nu)}\ .
\end{equation}
However, $t_2(\nu)$, as defined in Eq.~\eqref{eq:t2-def}, cannot be represented using these quantities, highlighting the unique frequency dependence of the reduced angular trispectrum. 
Among these equations, only the 3rd-order large-scale modulation $\omega_{\ung,\uin}^{(3)}$ remains unclear in this paper. 
Thus, it is essential to extend our diagrammatic approach to determine its specific expression.

As presented in the last three panels of \cref{fig:FD_Frame}, the diagrams of $\omega_{\uGW,\uin}^{(3)}$ can be viewed as attaching three extensional red solid lines to the diagram of $\bar{\omega}_{\uGW,\uin}$. 
These three solid lines can be connected to a single vertex, two vertices, or three vertices. 
From another perspective, the attachments can be classified into four groups. 
Among these, two groups involve attachments that exclusively transform one type of vertex into other types of vertices, while the other two groups pertain to attachments that involve two and three types of vertices, respectively. 
\begin{itemize}
    \item In the first group, $n$ $V_0^{[i]}$-vertices are attached with $l$ red solid lines, where the positive integers $n$, $l$, and $i$ satisfy $nl = 3$ and $i \leq o - l$. 
    The attachment transforms $n$ $V_0^{[i]}$-vertices into $n$ $V_{l}^{[i+l]}$ vertices, altering the symmetric factor by changing the $n$ permutations of $i!$ to $(i+1)!$, as well as selecting $n$ vertices from $N_i$ $V_0^{[i]}$-vertices. 
    \item In the second group, one $V_0^{[i]}$-vertex is attached with two red solid lines, and another $V_0^{[i]}$-vertex is attached with one red solid line, where the positive integer $i$ satisfies $i \leq o - 2$. 
    The attachment transforms two $V_0^{[i]}$-vertices into two $V_{l_1}^{[i+l_1]}$ vertices and a $V_0^{[i]}$-vertex into a $V_{l_2}^{[i+l_2]}$-vertex. 
    The symmetric factor is altered by changing two permutations of $i!$ to $(i+l_1)!$ and one permutation of $i!$ to $(i+l_2)!$, as well as selecting two vertices from $N_i$ $V_0^{[i]}$-vertices and one vertex from the remaining $(N_i - 2)$ $V_0^{[i]}$-vertices.
    \item In the third group, $n_1$ $V_0^{[i]}$-vertices are attached with $l_1$ red solid lines and $n_2$ $V_0^{[j]}$-vertices are attached with $l_2$ red solid lines, where the positive integers $n_1$, $l_1$, $n_2$, $l_2$, $i$, and $j$ satisfy $n_1 l_1 + n_2 l_2 = 3$, and $i \leq o-l_1$, $j \leq o-l_2$. 
    The attachment transforms $n_1$ $V_0^{[i]}$-vertices into $n_1$ $V_{l_1}^{[i+l_1]}$ vertices and $n_2$ $V_0^{[j]}$-vertices into $n_2$ $V_{l_2}^{[j+l_2]}$ vertices. 
    The symmetric factor is altered by changing $n_1$ permutations of $i!$ to $(i+l_1)!$ and $n_2$ permutations of $j!$ to $(j+l_2)!$, as well as selecting $n_1$ vertices from $N_i$ $V_0^{[i]}$-vertices and $n_2$ vertices from $N_j$ $V_0^{[j]}$-vertices.
    \item In the last group, the attachment transforms a $V_0^{[i]}$-vertex into a $V_{1}^{[i+1]}$-vertex, a $V_0^{[j]}$-vertex into a $V_{1}^{[j+1]}$-vertex, and a $V_0^{[k]}$-vertex into a $V_{1}^{[k+1]}$-vertex, where the positive integers $i$, $j$, and $k$ satisfy $i \neq j$, $i \neq k$, $j \neq k$, and $i, j, k \leq o-1$. 
    The symmetric factor is altered by changing the permutations of $i!$, $j!$, and $k!$ to $(i+1)!$, $(j+1)!$, and $(k+1)!$, respectively, while selecting one out of $N_i$ $V_0^{[i]}$-vertices, one out of $N_j$ $V_0^{[j]}$-vertices, and one out of $N_k$ $V_0^{[k]}$-vertices.
\end{itemize}
Consequently, for \ac{PNG} up to $\Hnl$ order (i.e., $o = 4$), the large-scale modulation at $\mathcal{O}(\zeta_{\ugL}^3)$, $\omega_{\ung,\uin}^{(3)}$, is given by 
\begin{subequations}\label{eqs:Ong3-dp-def}
\begin{eqnarray}
    &&\omega_{\ung,\uin}^{(3)} (q) 
    = \sum_{c=0}^4 \sum_{b=0}^{\lfloor 4-c \rfloor} \sum_{a=0}^{\lfloor 4-b-c \rfloor} \bar{\omega}_{\uGW,\uin}^{(a,b,c)} (q) \Bigl(\sT_1^{(3)} + \sT_2^{(3)} + \sT_3^{(3)} + \sT_4^{(3)}\Bigr)\ ,\\
    &&\sT_1^{(3)} = \sum_{l=1}^3 \sum_{n=1}^3 \sum_{i=1}^{o-l} \left(\frac{(i+l)!}{i!}\frac{V_l^{[i+l]}}{V_0^{[i]}}\right)^n \binom{N_i}{n} \delta_{ln,3}\ ,\\ 
    % &&\sT_2^{(3)} = \sum_{l_1=1}^2 \sum_{l_2=1}^{\min(l_1-1,2)} \sum_{n_1=1}^2 \sum_{n_2=1}^2 \sum_{i=1}^{\min(o-l_1,o-l_2)} \left(\frac{(i+l_1)!}{i!} \frac{V_{l_1}^{[i+l_1]}}{V_0^{[i]}}\right)^{n_1} \binom{N_i}{n_1} \left(\frac{(i+l_2)!}{i!} \frac{V_{l_2}^{[i+l_2]}}{V_0^{[i]}}\right)^{n_2} \binom{N_i-n_1}{n_2}\nonumber\\ 
    &&\sT_2^{(3)} = \sum_{i=1}^{o-2} \left(\frac{(i+2)!}{i!} \frac{V_{2}^{[i+2]}}{V_0^{[i]}}\right)^{1} \binom{N_i}{1} \left(\frac{(i+1)!}{i!} \frac{V_{1}^{[i+1]}}{V_0^{[i]}}\right)^{1} \binom{N_i-1}{1}\ ,\\ 
    &&\sT_3^{(3)} = \sum_{l_1=1}^2 \sum_{l_2=1}^2 \sum_{n_1=1}^2 \sum_{n_2=1}^2 \sum_{i=2}^{o-l_1} \sum_{j=1}^{\min(i-1,o-l_2)} \left(\frac{(i+l_1)!}{i!} \frac{V_{l_1}^{[i+l_1]}}{V_0^{[i]}}\right)^{n_1} \binom{N_i}{n_1} \\ 
    &&\hphantom{\sT_3^{(3)} = \sum_{l_1=1}^2 \sum_{l_2=1}^2 \sum_{n_1=1}^2 \sum_{n_2=1}^2 \sum_{i=2}^{o-l_1} \sum_{j=1}^{\min(i-1,o-l_2)}} \times \left(\frac{(j+l_2)!}{j!}\frac{V_{l_2}^{[j+l_2]}}{V_0^{[j]}}\right)^{n_2} \binom{N_j}{n_2} \delta_{l_1 n_1 + l_2 n_2,3}\ ,\nonumber\\
    &&\sT_4^{(3)} = \sum_{i=3}^{o-1} \sum_{j=2}^{\min(i-1,o-1)} \sum_{k=1}^{\min(j-1,o-1)} \left(\frac{(i+1)!}{i!}\frac{V_1^{[i+1]}}{V_0^{[i]}}\right)^1 \binom{N_i}{1} \left(\frac{(j+1)!}{j!}\frac{V_1^{[j+1]}}{V_0^{[j]}}\right)^1 \binom{N_j}{1} \nonumber\\ &&\hphantom{\sT_4^{(3)} = \sum_{i=3}^{o-1} \sum_{j=2}^{\min(i-1,o-1)} \sum_{k=1}^{\min(j-1,o-1)}} \times \left(\frac{(k+1)!}{k!}\frac{V_1^{[k+1]}}{V_0^{[k]}}\right)^1 \binom{N_k}{1}\ .
\end{eqnarray}
\end{subequations} 
Similar to Eq.~(\ref{eq:Ong1-dp-def}), this formulation is readily extensible to include higher-order \ac{PNG}. 
In particular, for scale-independent \ac{PNG} parameters, we have 
\begin{eqnarray}\label{eq:Ong3}
    \omega_{\ung,\uin}^{(3)}
    &=& 8\Fnl^3 \Bigl(4 \bar{\omega}_{\uGW,\uin}^{(0,0,0)} + \bar{\omega}_{\uGW,\uin}^{(0,1,0)}\Bigr) 
    + 12 \Fnl \Gnl \Bigl(12 \bar{\omega}_{\uGW,\uin}^{(0,0,0)} + 4 \bar{\omega}_{\uGW,\uin}^{(2,0,0)} + 6 \bar{\omega}_{\uGW,\uin}^{(0,1,0)} + 2 \bar{\omega}_{\uGW,\uin}^{(0,2,0)}  \nonumber\\
    &&\quad + 3 \bar{\omega}_{\uGW,\uin}^{(1,0,1)} + 2 \bar{\omega}_{\uGW,\uin}^{(0,0,2)}\Bigr)
    + 48 \Hnl \Bigl(2 \bar{\omega}_{\uGW,\uin}^{(0,0,0)} + 3 \bar{\omega}_{\uGW,\uin}^{(2,0,0)} + 3 \bar{\omega}_{\uGW,\uin}^{(0,1,0)} + 3 \bar{\omega}_{\uGW,\uin}^{(2,1,0)} \nonumber\\
    &&\quad + 3 \bar{\omega}_{\uGW,\uin}^{(0,2,0)} + 2 \bar{\omega}_{\uGW,\uin}^{(1,0,1)} + 2 \bar{\omega}_{\uGW,\uin}^{(0,3,0)} + 2 \bar{\omega}_{\uGW,\uin}^{(1,1,1)} + \bar{\omega}_{\uGW,\uin}^{(0,0,2)} + \bar{\omega}_{\uGW,\uin}^{(0,1,2)} \Bigr)\nonumber\\
    && + \frac{18\Gnl^2}{\Fnl} \Bigl(6 \bar{\omega}_{\uGW,\uin}^{(2,0,0)} + 3 \bar{\omega}_{\uGW,\uin}^{(2,1,0)} + 2 \bar{\omega}_{\uGW,\uin}^{(1,0,1)} + \bar{\omega}_{\uGW,\uin}^{(1,1,1)}\Bigr)\nonumber\\
    && + \frac{36\Gnl\Hnl}{\Fnl^2} \Bigl(2 \bar{\omega}_{\uGW,\uin}^{(2,0,0)} + 12 \bar{\omega}_{\uGW,\uin}^{(4,0,0)} + 3 \bar{\omega}_{\uGW,\uin}^{(2,1,0)} + 4 \bar{\omega}_{\uGW,\uin}^{(2,2,0)} + 6 \bar{\omega}_{\uGW,\uin}^{(3,0,1)} + 2 \bar{\omega}_{\uGW,\uin}^{(2,0,2)}\Bigr)\nonumber\\ 
    && + \frac{27\Gnl^3}{\Fnl^3} \Bigl(4 \bar{\omega}_{\uGW,\uin}^{(4,0,0)} + \bar{\omega}_{\uGW,\uin}^{(3,0,1)} \Bigr) 
    + \frac{48 \Hnl^2}{\Fnl \Gnl} \Bigl(2 \bar{\omega}_{\uGW,\uin}^{(2,1,0)} + 6 \bar{\omega}_{\uGW,\uin}^{(2,2,0)} + \bar{\omega}_{\uGW,\uin}^{(1,1,1)} + 3 \bar{\omega}_{\uGW,\uin}^{(1,2,1)} \Bigr)\nonumber\\
    && + \frac{16 \Fnl^2 \Hnl}{\Gnl} \Bigl(3 \bar{\omega}_{\uGW,\uin}^{(0,1,0)} + 2 \bar{\omega}_{\uGW,\uin}^{(0,2,0)}\Bigr) 
    + \frac{32 \Fnl \Hnl^2}{\Gnl^2} \Bigl(2 \bar{\omega}_{\uGW,\uin}^{(0,2,0)}+ 3 \bar{\omega}_{\uGW,\uin}^{(0,3,0)}\Bigr)\nonumber\\
    && + \frac{64 \Hnl^3}{\Gnl^3} \Bigl(\bar{\omega}_{\uGW,\uin}^{(0,3,0)} + 4 \bar{\omega}_{\uGW,\uin}^{(0,4,0)}\Bigr)\ .
\end{eqnarray}
Substituting the expression for $\omega_{\ung,\uin}^{(3)} $ along with Eqs.~(\ref{eq:Ong1}, \ref{eq:Ong2}) into Eqs.~(\ref{eq:t1-def}, \ref{eq:t2-def}), we derive the explicit formula for the reduced angular trispectrum $ t^{\ell_1 \ell_2}_{\ell_3 \ell_4} (L,\nu) $ presented in Eq.~\eqref{eq:tl-res}. 
We set aside the discussion of the multipole dependence of $t^{\ell_1 \ell_2}_{\ell_3 \ell_4} (L, \nu)$, which exhibits a behavior similar to that of the \ac{CMB} in the low-multipole regime, as previously reported in Ref.~\cite{Li:2024zwx}. 
Instead, the following analysis focuses on its frequency dependence, as described by Eqs.~(\ref{eq:t1-def}, \ref{eq:t2-def}).

\begin{figure*}[htbp]
    \centering
    \includegraphics[width = \textwidth]{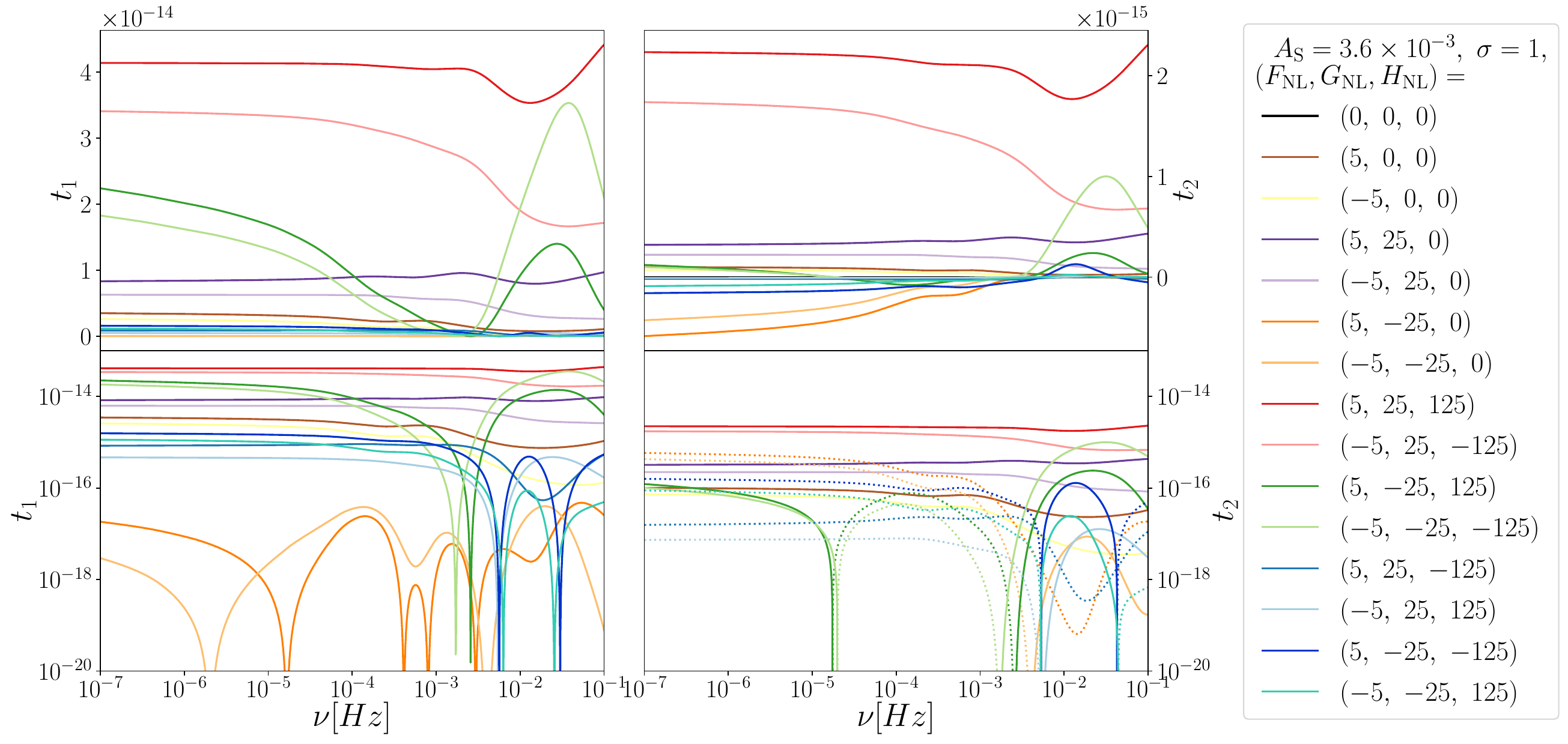}
    \caption{Frequency dependence of the reduced angular trispectrum of SIGWs, with the upper and lower panels displaying numerical results on linear and logarithmic vertical scales, respectively. 
    The dotted lines in the lower left panels indicate negative sign regions. }\label{fig:t_tilde}
\end{figure*}

As done in the analysis of the angular bispectrum, we analyze the frequency dependence of the reduced angular trispectrum across various scenarios characterized by different \ac{PNG} parameters in a similar manner. 
% While both $ t_1(\nu) $ and $ t_2(\nu)$, akin to $ \ell (\ell + 1) \tilde{C}_\ell(\nu) $ and $ b(\nu) $, exhibit nearly scale-invariant behavior in the infrared regime, we are particularly interested in their behavior at higher frequencies. 
First, for Gaussian $ \zeta $, it is evident that $ t_1(\nu) = t_2(\nu) = 0 $, as anticipated, indicating that a Gaussian $ \zeta $ yields a zero angular trispectrum of \acp{SIGW}. 
Second, in scenarios with minimal \ac{PNG}, where the Gaussian component $ \bar{\omega}_{\uGW,\uin}^{(0,0,0)} $ is dominant among the 19 categories of $ \bar{\omega}_{\uGW,\uin} $ in Eq.~\eqref{eq:omegabar-abc-total}, we consider only the leading-order terms of $ \omega_{\ung,\uin}^{(2)} $ and $ \omega_{\ung,\uin}^{(3)} $. 
The former has been discussed in the analysis of $ b(\nu) $ in the previous subsection, while the latter yields $ 16 (2\Fnl^3 + 9\Fnl\Gnl + 6\Hnl) \bar{\omega}_{\uGW,\uin}^{(0,0,0)} $. 
According to Eq.~\eqref{eq:t1-C+b}, $ t_1 (\nu) $ is determined by $ \ell (\ell + 1) \tilde{C}_\ell(\nu) $ and $ b(\nu) $, which are described by Eqs.~(\ref{eq:Ct-NG-min}, \ref{eq:b-NG-min}) for these scenarios. 
Therefore, we primarily focus on the behavior of $ t_2 (\nu) $, whose approximate expression reads 
\begin{eqnarray}
    % t_1 (\nu) &\approx& 4 \bigl(2 \pi A_\uL\bigr)^3 
    %     \biggl[
    %         8 \Fnl + \frac{3}{5} \bigl(6 - n_{\uGW} (\nu)\bigr)
    %     \biggr]^2  \biggl[
    %         24 \bigl(\Fnl^2 + \Gnl\bigr) + \frac{3}{5} \Fnl \bigl(6 - n_{\uGW}     (\nu)\bigr)
    %     \biggr]^2 \ ,\label{eq:t1-NG-min}\\
    t_2 (\nu) &\simeq& \bigl(2 \pi\bigr)^2 A_\uL^3 
        \biggl[
            8 \Fnl + \frac{3}{5} \bigl(6 - n_{\uGW} (\nu)\bigr)
        \biggr]^3 \nonumber\\
        && \times \biggl[
            16 (2\Fnl^3 + 9\Fnl\Gnl + 6\Hnl) + \frac{3}{5} \Gnl \bigl(6 - n_{\uGW} (\nu)\bigr)
        \biggr] \ .\label{eq:t2-NG-min}
\end{eqnarray} 
Under this approximation, $ t_2 (\nu) $ exhibits various characteristics for different magnitudes of the \ac{PNG} parameters. 
As discussed below Eq.~(\ref{eq:Ct-NG-min}), $ \bigl(6 - n_{\uGW} (\nu)\bigr) $ ranges from $3$ to $15$ as the frequency varies from $ 10^{-4} \nu_\ast $ to $ 10^2 \nu_\ast $. 
Similar to $ b (\nu) $ and $ \ell (\ell+1) \tilde{C}_\ell (\nu) $, $ t_2 (\nu) $ can be used to disrupt the sign degeneracy between $ \Fnl $ and $ \Hnl $ when $ |\Fnl| $ is not significantly larger than $ \bigl(6 - n_{\uGW} (\nu)\bigr) $. 
The expressions of Eqs.~(\ref{eq:Ct-NG-min}, \ref{eq:b-NG-min}, \ref{eq:t2-NG-min}) demonstrate that the \ac{PNG} parameter, $ \Hnl $, can only be discerned through the angular trispectrum in this approximation. 
In an extreme case, if the second terms in the square brackets dominate over the first term, $ t_2 (\nu) $ is found to be nearly proportional to $ \Gnl \bigl[\ell (\ell+1) \tilde{C}_\ell (\nu)\bigr]^{2} $. 
However, this condition is challenging to meet for an observable angular trispectrum since the contributions from the \ac{PNG} parameters in the first term of the second square brackets are significantly enhanced by the constant coefficients at $\cO (10^2)$ order, implying the second term dominates only when \ac{PNG} is very small or when the \ac{PNG} parameters cancel each other out. 
% Comparing Eq.~\eqref{eq:t2-NG-min} with Eqs.~(\ref{eq:Ct-NG-min}, \ref{eq:b-NG-min}), we can further examine the three \ac{PNG} parameters at the frequency around $ 4\times10^{-4} \nu_\ast $ where $ \bigl(6 - n_{\uGW} (\nu)\bigr) $ reaches zero. 
From Eq.~(\ref{eq:Ct-NG-min}) to Eq.~\eqref{eq:t2-NG-min}), the first terms in the square brackets, which represent the \ac{PNG}-induced inhomogeneity, become increasingly significant, suggesting that the angular trispectrum may improve the accuracy of the inferred model parameters. 
Last but not least, we illustrate several instances of substantial \ac{PNG} with numerical results depicted in \cref{fig:t_tilde}. 
The panels on the upper and lower sides of this figure demonstrate the same trispectra with model parameters identical to those in \cref{fig:C-Ct}, where the vertical scales are linear (upper) and logarithmic (lower), respectively. 
The convention of the dotted lines in the lower right panel is also the same as in \cref{fig:b_tilde}, indicating where the sign is negative. 
Although $ t_1 (\nu) $ is determined by $ \ell (\ell+1) \tilde{C}_\ell (\nu) $ and $ b (\nu) $, the distinct behaviors of $ t_2 (\nu) $ suggest that the analysis of the angular trispectrum would be useful for a deeper investigation of \acp{SIGW}. 
Comparing the left and right panels, it is apparent that the magnitude of $ t_1 (\nu) $ is usually larger than that of $ t_2 (\nu) $. 
However, given that the contributions from $ t_1 (\nu) $ and $ t_2 (\nu) $ to $ t^{\ell_1 \ell_2}_{\ell_3 \ell_4} (L,\nu) $ are modulated by different multipoles, the distinct behaviors between $ t_1 (\nu) $ and $ t_2 (\nu) $ indicate that one might distinguish them using various sets of multipoles. 
Thus, we can acquire a wealth of information from the reduced angular trispectrum, which will serve as an important complement to the information obtained from the energy-density fraction spectrum, reduced angular power spectrum, and angular bispectrum.

\begin{figure*}[htbp]
    \centering
    \includegraphics[width = 1 \textwidth]{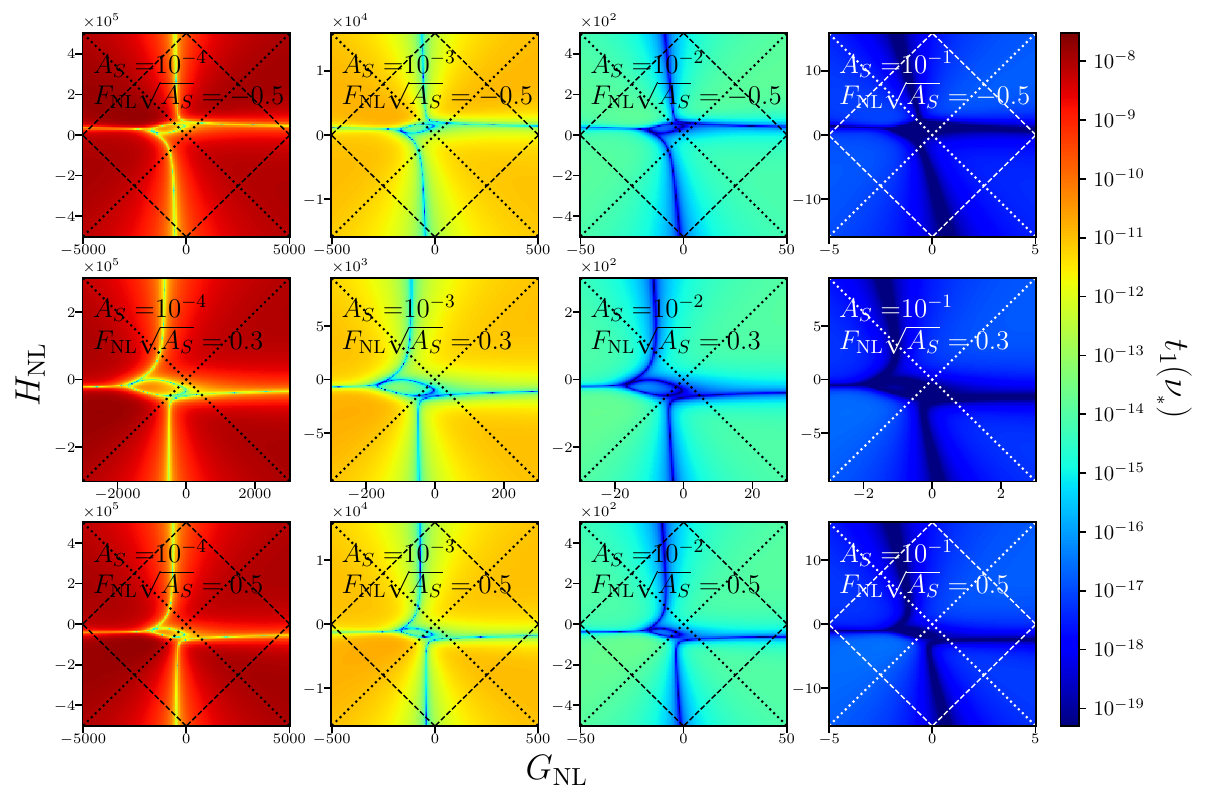}\\
    % \vspace{-8pt}
    \includegraphics[width = 1 \textwidth]{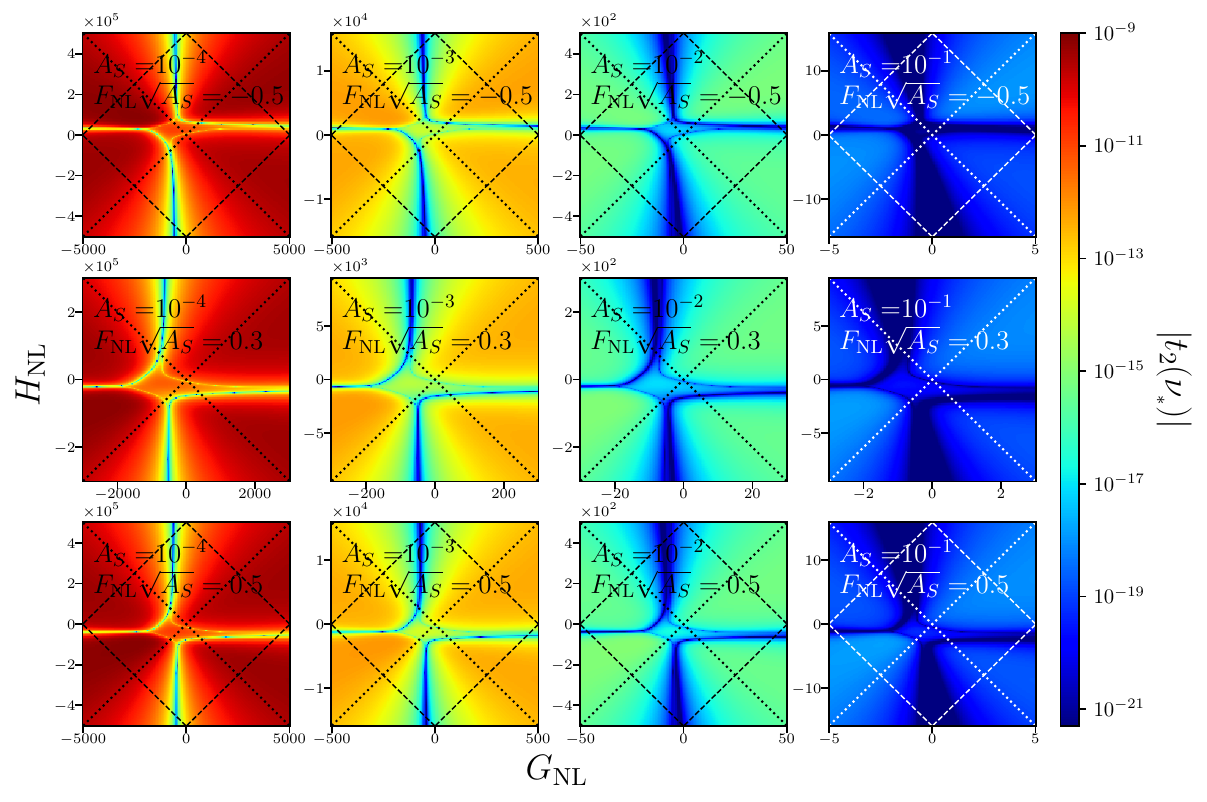}
    % \vspace{-8pt}
    \caption{Reduced angular trispectrum of SIGWs with respect to the $\Fnl$, $\Gnl$ and $\Hnl$. The conventions for the dotted and dashed lines are the same as those in \cref{fig:Ct_G-H}. }\label{fig:tt_G-H}
\end{figure*}

We also present two arrays of contour plots for $ t_1 (\nu) $ and $ |t_2 (\nu)| $ in \cref{fig:tt_G-H}, arranged identically to that of $ b (\nu) $. 
By combining these two arrays with Eqs.~(\ref{eq:tl-res}, \ref{eq:t1-def}, \ref{eq:t2-def}), we can easily discern the dependence of the \ac{SIGW} angular trispectrum on $ \Gnl $ and $ \Hnl $. 
The magnitudes of both $ t_1 (\nu) $ and $ |t_2 (\nu)| $ decrease as $ A_\uS $ increases. 
While the contour plots of $ t_1 (\nu) $ are similar to those of $ |b (\nu)| $ except for differences in magnitudes and signs, the contour plots of $ |t_2 (\nu)| $ exhibit distinct features. 
Notably, an irregular region located at the center, biased toward $ \Gnl < 0 $ and $ \Fnl\Hnl < 0 $, indicates where the sign of $ t_2 (\nu) $ is negative. 
This region extends a considerable distance into the $ \Gnl > 0 $ domain, suggesting that when $ \Fnl\Hnl $ is negative and small, $ t_2 (\nu) $ remains negative regardless of the magnitude of the positive $ \Gnl $. 
This implies that $ \omega_{\ung,\uin}^{(3)} / \bar{\omega}_{\uGW,\uin} $ typically dominates over $ 3 \Gnl \bigl(6 - n_\uGW (\nu)\bigr) / 5 $ at this parameter space. 
The comparison of the contour plots of $ t_2 (\nu) $ with those of $ t_1 (\nu) $ demonstrates that $ t_2 (\nu) $ is generally an order of magnitude smaller than $ t_1 (\nu) $, particularly when \ac{PNG} is significant. 
Since the multipoles are constrained by the triangular inequalities, the reduced angular trispectrum expressed by Eq.~\eqref{eq:tl-def} is unlikely to be negative, despite the possible negative values of $ t_2 (\nu) $ at certain frequencies. 
The differing dependencies on frequency, multipoles, and model parameters imply that the reduced angular trispectrum contains a wealth of information about the early Universe.

%% file: TeX/6Extended.tex
\section{Extended Study}\label{sec:discuss}

We have analyzed the statistics of the \ac{SIGW} background, considering the presence of \ac{PNG} up to $\Hnl$ order in the aforementioned study. 
This analysis encompasses the energy density fraction spectrum detailed in Section~\ref{sec:Omegabar}, the angular power spectrum outlined in Section~\ref{sec:Cl}, and the angular bispectrum and trispectrum discussed in Section~\ref{sec:bl&tl}. 
In our analyses, we employed a diagrammatic approach and neglected the scale dependence of \ac{PNG} for convenience. 
This section aims to extend and supplement the investigation by focusing on three key aspects. 
First, we will review the similar infrared scaling across all categories of the \ac{SIGW} energy-density spectrum, which results in the general behaviors of the energy density fraction spectrum and angular correlation functions. 
Next, we will elaborate on the impacts of the primordial trispectrum on the \ac{SIGW} background in the first subsection, which serves as an important supplement to the preceding investigation. 
Then, the following subsection will demonstrate how to extrapolate our study to address the scale dependence of \ac{PNG}. 
Finally, in the last subsection, a simplified example will illustrate the relationship between \ac{PBH} formation and the accompanying \ac{SIGW} background in the context of \ac{PNG}.

\subsection{Infrared scaling of the SIGW energy-density spectra}\label{subsec:IR}

\begin{figure}[htbp]
    \centering
    \includegraphics[width = 0.8 \columnwidth]{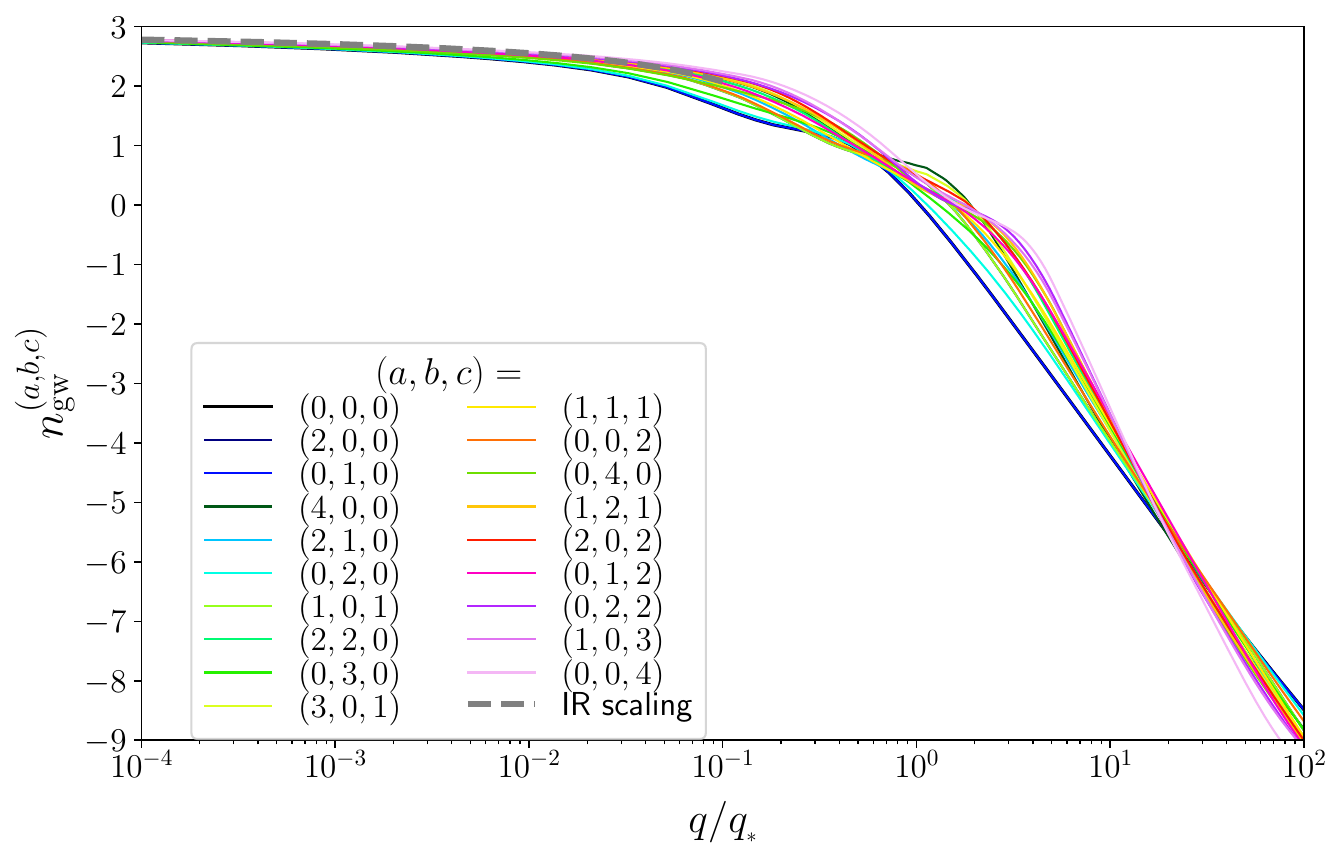}
    \caption{The spectral index for $\bar{\omega}_{\uGW,\uin}^{(a,b,c)}$. }\label{fig:ngw-abc}
\end{figure}

The generic infrared behavior of the \ac{SIGW} energy density spectrum has attracted considerable interest \cite{Yuan:2019wwo,Yuan:2023ofl,Cai:2018dig,Cai:2019cdl,Adshead:2021hnm,Luo:2025lgr,Han:2025wlo}. 
As illustrated in \cref{fig:Unscaled_Omegabar}, the energy density spectra across all categories exhibit a common logarithmic slope in the infrared region. 
Specifically, this shared feature can be characterized by 
\begin{equation}\label{eq:Omega-infrared}
    \bar{\omega}_{\uGW,\uin}^{(a,b,c)} (q) \Big|_\mathrm{IR} \propto \left(\frac{q}{q_\ast}\right)^3 \ln^2 \left(\frac{4q_\ast^2}{3 q^2}\right) \ ,
\end{equation}
where the subscript $\mathrm{IR}$ denotes the infrared region. 
Furthermore, akin to the spectral index for the energy density fraction spectrum $\bar{\Omega}_{\uGW,0}$ defined in Eqs.~\eqref{eq:ngw-def}, we can introduce the spectral index for each category $\bar{\omega}_{\uGW,\uin}^{(a,b,c)}$, as given by  
\begin{equation}\label{eq:ngw-infrared}
    n_\uGW^{(a,b,c)} (q)\Big|_\mathrm{IR} =  \frac{\partial\ln \bar{\omega}_{\uGW,\uin}^{(a,b,c)} (q)\Big|_\mathrm{IR}}{\partial\ln q} = 3 - \frac{4}{\ln^2 \left(\frac{4 q_\ast^2}{3 q^2}\right)} \ .
\end{equation}
The detailed verification of Eqs.~(\ref{eq:Omega-infrared}, \ref{eq:ngw-infrared}) is provided in the Appendix~\ref{sec:IR}. 
Naturally, the total energy density fraction spectrum, $\bar{\omega}_{\uGW,\uin}$, which is the sum of $\bar{\omega}_{\uGW,\uin}^{(a,b,c)}$ for all categories in Eq.~\eqref{eq:omegabar-abc-total}, exhibits the same logarithmic slope as that of Eq.~\eqref{eq:ngw-infrared}. 
In particular, in the extreme infrared region, we find that $n_\uGW (q \rightarrow 0) = 3$. 
Traditionally, this suggests that the source term of \acp{SIGW}, $S_\lambda (\eta,q)$ in Eq.~\eqref{eq:S}, exhibits a scale-invariant power spectrum for super-horizon scales, which corresponds to white noise due to the absence of causal correlations \cite{Maggiore:2018sht}. 
This feature is believed to generally exist in \acp{GW} sourced from cosmological mechanisms \cite{Maggiore:2018sht, Cai:2019cdl}.

This general infrared feature is crucial for investigating the infrared behaviors of the anisotropies in the \ac{SIGW} background. 
According to the analyses in \cref{sec:Cl} and \cref{sec:bl&tl}, the large-scale modulations, $\omega_{\ung,\uin}^{(1)}$, $\omega_{\ung,\uin}^{(2)}$, and $\omega_{\ung,\uin}^{(3)}$, can be expressed as linear combinations of the 19 categories $\bar{\omega}_{\uGW,\uin}^{(a,b,c)}$. 
Therefore, in the low-frequency limit, the common feature across all categories described by Eqs.~(\ref{eq:Omega-infrared}, \ref{eq:ngw-infrared}) indicates that the PNG-induced inhomogeneities $\omega_{\ung,\uin}^{(1)} / \bar{\omega}_{\uGW,\uin}$, $\omega_{\ung,\uin}^{(2)} / \bar{\omega}_{\uGW,\uin}$, and $\omega_{\ung,\uin}^{(3)} / \bar{\omega}_{\uGW,\uin}$ are independent of $q$. 
Since $6 - n_\uGW (q \rightarrow 0) = 3$ is also a constant, the reduced angular power spectrum expressed by Eq.~\eqref{eq:Ct}, the angular bispectrum in Eqs.~(\ref{eq:Bl-res}, \ref{eq:b-def}), and the reduced angular trispectrum in Eqs.~(\ref{eq:tl-res}, \ref{eq:t1-def}, \ref{eq:t2-def}) are nearly scale-invariant in the infrared region, regardless of the magnitude of \ac{PNG}. 
Different scale-independent \ac{PNG} parameters only alter the magnitude of the angular correlation functions but do not affect the features of the spectra. 
In other words, these characteristics may be beneficial for reducing noise in future \ac{GW} detection.

\subsection{Impacts of primordial trispectrum on the SIGW background}

In Section~\ref{sec:ED} and Section~\ref{sec:Omegabar}, we emphasized the significance of the four-point correlator of the primordial curvature perturbations $\zeta_\uS$ in calculating the energy-density fraction spectrum of \acp{SIGW} in the presence of \ac{PNG}.
Based on Eqs.~(\ref{eq:fnl-gnl-hnl-def}--\ref{eq:S-L-dec}), where $\zeta_\uS$ is expanded in a power series of its Gaussian component $\zeta_\ugS$, we employ the diagrammatic approach to compute this correlator using Wick's theorem.
The contraction terms are further decomposed into nine families in Subsection~\ref{subsec:Renorm-FD}. 
Alternatively, as emphasized in Refs.~\cite{Adshead:2021hnm,Garcia-Saenz:2022tzu,LISACosmologyWorkingGroup:2025vdz}, it is also conventional to separate this four-point correlator into the disconnected piece and the connected piece, which are expressed in terms of the primordial power spectrum and trispectrum of $\zeta_\uS$, respectively. 
In comparison to the nine families introduced in Subsection~\ref{subsec:Renorm-FD}, the disconnected piece corresponds to the $G$-like family, while the connected piece encompasses all other families. 
In other words, we utilize $\cP^{C-\mathrm{like}}$, $\cP^{Z-\mathrm{like}}$, etc., to represent the contribution of the primordial trispectrum to the energy-density fraction spectrum of \acp{SIGW} in Section~\ref{sec:Omegabar}. 
Traditionally, the primordial trispectrum of $\zeta_\uS$ is expressed in terms of $\gnl$ and $\tnl$, differing from the form we use. 
Thus, in this subsection, we first compare the primordial trispectrum represented by $\cP^{X-\mathrm{like}}$ ($X = C, Z, P, N, CZ, PZ, NC$) with that expressed in terms of $\gnl$ and $\tnl$, specifically focusing on their respective contributions to the energy-density full spectrum $\bar{\omega}_{\uGW,\uin}$. 
Furthermore, we discuss the pure impacts of the primordial trispectrum on the \ac{SIGW} background.

As is customarily done, the primordial trispectrum for local-type \ac{PNG} is expressed in the form outlined by \cite{Okamoto:2002ik,Boubekeur:2005fj,Suyama:2007bg}, i.e.,  
\begin{eqnarray}\label{eq:tnl-def}
    \cT_{\zeta_\uS}(\bk_1,\bk_2,\bk_3,\bk_4) 
    & = &\tnl \left(P_{\zeta_\uS}(|\bk_1 + \bk_3|) P_{\zeta_\uS}(k_3) P_{\zeta_\uS}(k_4) + 11 \mathrm{perms}\right)\nonumber\\
    && + \frac{54\gnl}{25} \left(P_{\zeta_\uS}(k_2) P_{\zeta_\uS}(k_3) P_{\zeta_\uS}(k_4) + 3 \mathrm{perms} \right)\ ,
\end{eqnarray} 
where $\tnl$ satisfies the condition $\tnl \geq \left(6 \fnl / 5\right)^2$ \cite{Suyama:2007bg,Huang:2009vk}. 
This condition, derived from the $\delta N$ formalism \cite{Sasaki:1995aw, Sasaki:1998ug, Lyth:2004gb, Lyth:2005fi}, holds as long as the \ac{PNG} arises from super-horizon evolution. 
The equality in this condition holds for single-field inflation models and certain special multi-field inflation models \cite{Byrnes:2006vq,Wands:2010af}. 
In fact, the power series expansion of $\zeta$ in terms of $\zeta_\ugS$, as presented in Eqs.~(\ref{eq:fnl-gnl-hnl-def}, \ref{eq:Fnl-Gnl-Hnl-def}), is a direct consequence of the $\delta N$ formalism for single-field and specific multi-field inflation models \cite{Byrnes:2006vq,Wands:2010af}. 
Therefore, we can regard $\tnl = \left(6 \fnl / 5\right)^2$ in the following discussion.
According to Eqs.~(\ref{eq:fnl-gnl-hnl-def} - \ref{eq:PgX-def}), the total power spectrum of $\zeta_\uS$, denoted as $P_{\zeta_\uS}$ in Eq.~\eqref{eq:tnl-def}, can be expressed as follows 
\begin{equation}
    P_{\zeta_\uS}(q) = (1 + 3 A_\uS \Gnl)^2 P^{[1]} (q) + 2 (\Fnl + 6 A_\uS \Hnl)^2 P^{[2]} (q) + 6 \Gnl^2 P^{[3]} (q) + 24 \Hnl^2 P^{[4]} (q) \ .
\end{equation} 
Similar to Eq.~\eqref{eq:PgX-def}, we can define its dimensionless form as follows 
\begin{equation}
    \Delta_{\zeta_\uS}^2 (q) = \frac{q^3}{2 \pi^2} P_{\zeta_\uS}(q) \ .
\end{equation}
Notably, the dimensionless power spectrum of $\zeta_\ugS$, denoted as $\Delta_\uS^2 (q)$ in Eqs.~(\ref{eq:PgX-def}, \ref{eq:Lognormal}), corresponds exactly to the tree-level of $\Delta_{\zeta_\uS}^2 (q)$, just as $P^{[1]}(q)$ is the tree-level of $P_{\zeta_\uS}$. 
In scenarios where $A_\uS$, representing the spectral amplitude of $\Delta_\uS^2 (q)$, is sufficiently small, $P_{\zeta_\uS}$ can be approximated as the tree-level, i.e., $P_{\zeta_\uS}(q) \simeq P^{[1]}(q)$. 
Ref.~\cite{Garcia-Saenz:2022tzu} argued that only the effects of the primordial trispectrum on the \ac{SIGW} background can be attributed to \ac{PNG}, since only the total primordial power spectrum is observable, while its tree-level contribution remains unobservable. 
Although the loop corrections in the power spectrum result from \ac{PNG}, we will discuss the impacts of the primordial trispectrum on the \ac{SIGW} background in the following paragraphs.

We are particularly interested in the contribution of the primordial trispectrum to the energy-density full spectrum $\omega_{\uGW,\uin}$ of \acp{SIGW}. 
Derived from Eqs.~(\ref{eq:Ph-def}, \ref{eq:h}, \ref{eq:omegabar-h}), this contribution can be expressed as $\cT_{\zeta_\uS}(\bq_1, \bq - \bq_1, -\bq_2, -\bq + \bq_2)$ in the integrand for $\omega_{\uGW,\uin}$, where the terms involving $\gnl$ in Eq.~\eqref{eq:tnl-def} vanish. 
The remaining terms that are proportional to $\tnl$ can be divided into two components, written in the form of  
\begin{equation}\label{eq:tT}
    \tilde{\cT}_{\zeta_\uS}(\bq_1,\bq-\bq_1,-\bq_2,-\bq+\bq_2) 
    = P^C + P^Z \ ,
\end{equation}
where $P^C$ and $P^Z$ are defined as 
\begin{eqnarray}\label{eqs:ttNL-PC-PZ}
    && P^C = 4 \tnl P_{\zeta_\uS}(q_2) P_{\zeta_\uS}(|\bq-\bq_2|) P_{\zeta_\uS}(|\bq_1-\bq_2|)\ , \\
    && P^Z = 4 \tnl P_{\zeta_\uS}(q_1) P_{\zeta_\uS}(|\bq-\bq_2|) P_{\zeta_\uS}(|\bq_1-\bq_2|) \ .
\end{eqnarray}
These two components remind us $P^{C-\mathrm{like}}$ defined in Eq.~\eqref{eq:C-like} and $P^{Z-\mathrm{like}}$ defined in Eq.~\eqref{eq:Z-like}. 
Furthermore, by substituting $\tnl = \left(6 \fnl / 5\right)^2$ into Eqs.~\eqref{eqs:ttNL-PC-PZ}, we compare Eqs.~(\ref{eq:C-like-HNL}, \ref{eq:Z-like-HNL}) with Eqs.~\eqref{eqs:ttNL-PC-PZ} and find that the leading-order terms are identical. 
Specifically, these leading-order contributions correspond to the $C$- and $Z$-diagrams shown in \cref{fig:C-like-FD} and \cref{fig:Z-like-FD}, respectively. 
However, the contributions of higher orders in $A_\uS$ differ. 
From this perspective, Eq.~\eqref{eq:tT} appears to be an approximation of the connected piece of the four-point correlator $\langle \zeta_\uS(\bk-\bq-\bq_1) \zeta_\uS(\bq_1) \zeta_\uS(\bq-\bq_2) \zeta_\uS(\bq_2) \rangle$, which is accurate at leading order but not at higher orders.

We thus utilize the sum of $\cP^{X-\mathrm{like}}$ ($X = C, Z, P, N, CZ, PZ, NC$) to represent the impacts of the primordial trispectrum on the \ac{SIGW} background. 
To explore these impacts without the influence of the disconnected component of the four-point correlator of $\zeta_\uS$, it is essential to reconstruct $\cP^{G-\mathrm{like}}$ to ensure equivalence across various sets of \ac{PNG} parameters. 
Notably, $\cP^{G-\mathrm{like}}$, as expressed by Eq.~\eqref{eq:G-like-HNL}, can be reformulated into 
\begin{equation}
    \cP^{G-\mathrm{like}} = 2 P_{\zeta_\uS}(q_1) P_{\zeta_\uS}(|\bq-\bq_1|) \ .
\end{equation} 
This expression indicates that the disconnected component characterizes the Gaussian statistics and loop corrections resulting from \ac{PNG}, while being irrelevant to the non-Gaussian statistics. 
This inspires us to construct the same $\cP^{G-\mathrm{like}}$ by eliminating the influence of loop corrections on $P_{\zeta_\uS}(q)$. 
In other words, we should construct identical $P_{\zeta_\uS}(q)$, or equivalently $\Delta_{\zeta_\uS}^2 (q)$, for different sets of \ac{PNG} parameters to investigate the pure impacts of the primordial trispectrum on the \ac{SIGW} background.

\begin{figure}[htbp]
    \centering
    \includegraphics[width = 0.8 \columnwidth]{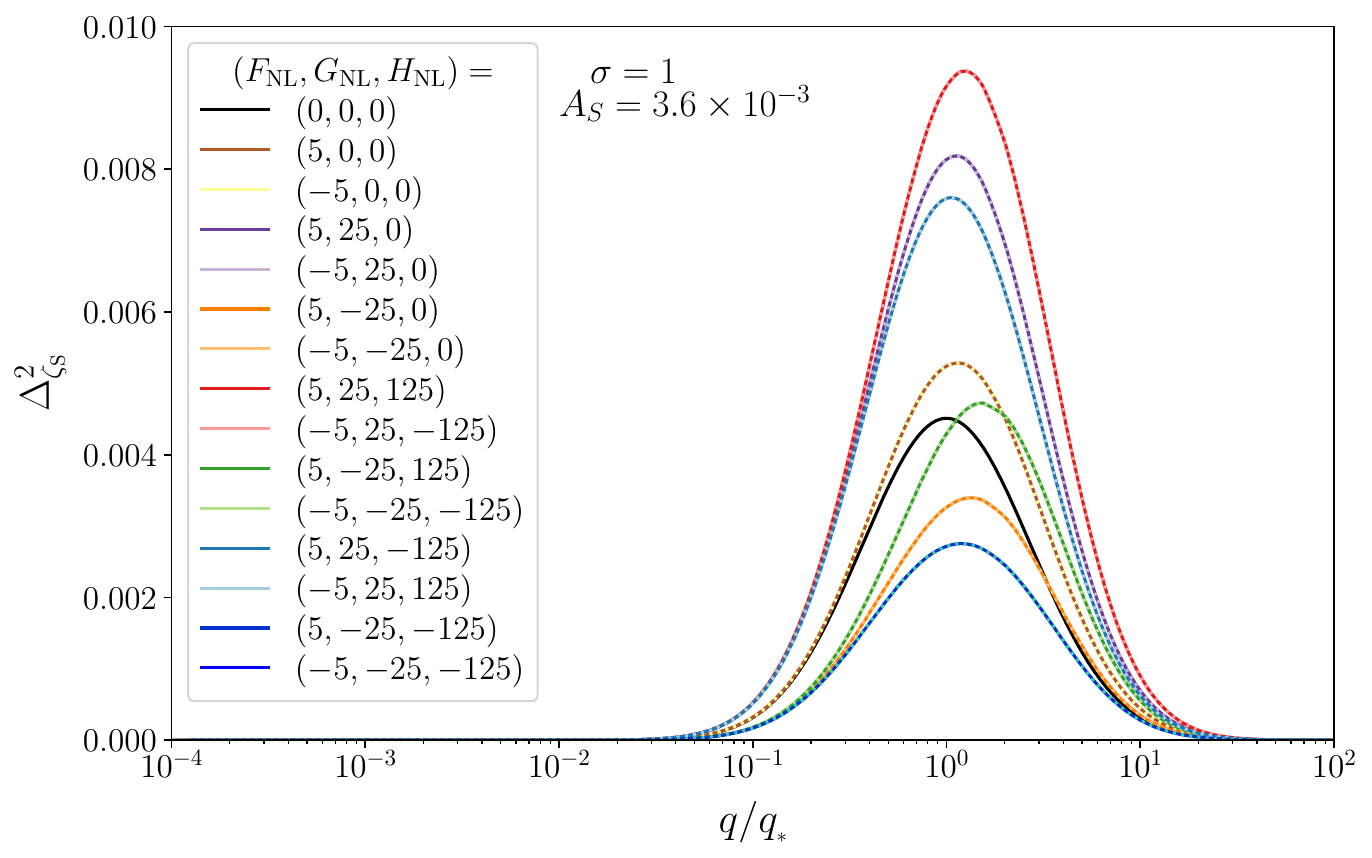}
    \caption{Dimensionless power spectra of $\zeta_\uS$ for various \ac{PNG} parameters. }\label{fig:Delta_zetaS}
\end{figure}

\begin{figure}[htbp]
    \centering
    \includegraphics[width = \textwidth]{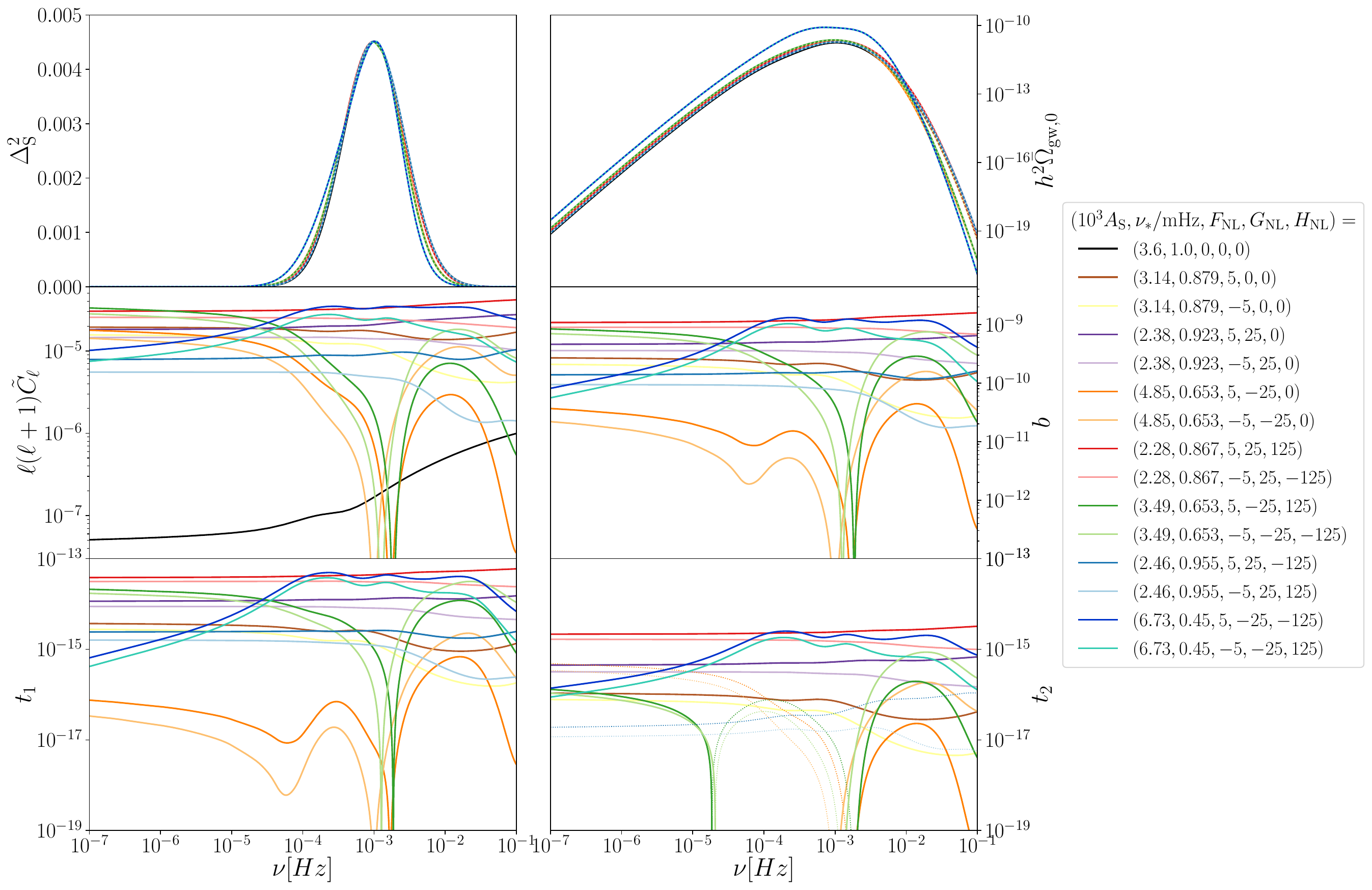}
    \caption{Imprints of the primordial trispectrum on the SIGW background, with $\Delta_{\zeta_\uS}^2 (q)$ adjusted to nearly overlap in the upper left panel.}\label{fig:renor_all}
\end{figure}

Here, we illustrate the impacts of the primordial trispectrum on the \ac{SIGW} background through figures derived from the same $\Delta_{\zeta_\uS}^2 (q)$. 
The previous numerical results, including \cref{fig:Omega-like}, \cref{fig:Total_Omegabar}, \cref{fig:C-Ct}, \cref{fig:b_tilde}, and \cref{fig:t_tilde}, are based on a consistent $\Delta_\uS^2 (q)$ across the same sets of \ac{PNG} parameters. 
However, the corresponding dimensionless power spectra of $\zeta_\uS$, denoted as $\Delta_{\zeta_\uS}^2 (q)$, differ across various non-Gaussian scenarios. 
As illustrated in \cref{fig:Delta_zetaS}, the primary differences among these $\Delta_{\zeta_\uS}^2 (q)$ are found in their magnitudes and peak frequencies. 
Therefore, we adjust $A_\uS$ and $\nu_\ast$ for each $\Delta_\uS^2 (q)$ to ensure that the spectral amplitudes and peak frequencies of $\Delta_{\zeta_\uS}^2 (q)$ are consistent across all sets of \ac{PNG} parameters. 
These modified power spectra are depicted in the upper left panel of \cref{fig:renor_all}, demonstrating that all the $\Delta_{\zeta_\uS}^2 (q)$ nearly overlap while exhibiting slight differences in their slopes. 
In comparison to the black curve, which signifies the Gaussian scenario $\Delta_\uS (q)$, the non-Gaussian scenarios depicted by colored curves are skewed due to the convolutions introduced in Eq.~\eqref{eq:Pxi-def}. 
Despite these discrepancies, it is evident that the power spectra $\Delta_{\zeta_\uS}^2 (q)$ can be regarded as nearly identical in the vicinity of the peak frequency ($10^4 \mathrm{Hz} - 5 \times 10^{-3} \mathrm{Hz}$). 
By utilizing these modified $\Delta_{\zeta_\uS}^2 (q)$, we recompute the corresponding $h^2 \bar{\Omega}_{\uGW,0} (\nu)$, $\ell (\ell + 1) \tilde{C}_\ell (\nu)$, $b (\nu)$, $t_1 (\nu)$, and $t_2 (\nu)$ for the respective non-Gaussian scenarios. 
All of these are additionally presented in the five other panels of \cref{fig:renor_all}. 
In each panel, the comparison between the Gaussian scenario, represented by the black curve, and the non-Gaussian scenarios illustrates the contributions of $T_{\zeta_\uS}$ to the corresponding spectra of \acp{SIGW} for various \ac{PNG}. 
For the energy-density fraction spectra $h^2 \bar{\Omega}_{\uGW,0} (\nu)$, the upper right panel reveals that they remain largely unchanged by the primordial trispectrum in most scenarios, except when both $\Gnl$ and $\Fnl \Hnl$ are negative. 
In these cases, $h^2 \bar{\Omega}_{\uGW,0} (\nu)$ is significantly enhanced due to an increased $A_\uS$, but is instead suppressed in \cref{fig:Total_Omegabar} when $A_\uS$ is held constant. 
Consequently, the contribution of $T_{\zeta_\uS}$ to the isotropic component of the \ac{SIGW} background is not always negligible, potentially augmenting the magnitude of $h^2 \bar{\Omega}_{\uGW,0} (\nu)$ for significant negative \ac{PNG}. 
Among the angular spectra, the varying behaviors across different \ac{PNG} suggest that $T_{\zeta_\uS}$ exerts notable effects on the anisotropies and non-Gaussianity in the \ac{SIGW} background. 
Thus, detecting the statistical properties of the \ac{SIGW} background holds promise as a probe of the primordial trispectrum in the future. 
Additionally, a common feature emerges: a dip appears around the peak frequency of $\Delta_{\zeta_\uS}^2 (q)$ in scenarios characterized by negative $\Gnl$ and non-negative $\Fnl\Hnl$. 
Based on the analyses of Eqs.~(\ref{eq:Ct}, \ref{eq:b-def}, \ref{eq:t1-def}, \ref{eq:t2-def}), this feature suggests that $\omega_{\uGW,\uin}^{(1)} (q) / \bar{\omega}_{\uGW,\uin} (q)$ cancels $3 (6 - n_\uGW (\nu)) / 5$ in these scenarios.

\subsection{Effects of scale-dependent PNG on the SIGW background}

In this study, we have proposed that all \ac{PNG} parameters be scale-independent for the sake of simplicity. 
Despite this assumption, the framework can be readily extended to accommodate a scale-dependent case. 
In this subsection, we aim to provide a brief discussion of this generalization concerning the analyses of the imprints of \ac{PNG} on the \ac{SIGW} background.

In the most general scenario, we reformulate Eq.~\eqref{eq:Fnl-Gnl-Hnl-def} in the Fourier modes as follows 
\begin{eqnarray}
    \zeta (\bq) &=& \zeta_\ug(\bq) + \int \frac{\ud^3 \bk}{(2\pi)^{3/2}} \Fnl (\bq,\bk,\bq-\bk) \zeta_\ug(\bk) \zeta_\ug(\bq-\bk) \\ 
    &&+ \int \frac{\ud^3 \bk\,\ud^3 \bp}{(2\pi)^{3}} \Gnl (\bq,\bp,\bk-\bp,\bq-\bk) \zeta_\ug(\bp) \zeta_\ug(\bk-\bp) \zeta_\ug(\bq-\bk) \nonumber\\
    &&+ \int \frac{\ud^3 \bk\,\ud^3 \bp\,\ud^3 \bl}{(2\pi)^{9/2}} \Hnl (\bq,\bl,\bp-\bl,\bk-\bp,\bq-\bk) \zeta_\ug(\bl) \zeta_\ug(\bp-\bl) \zeta_\ug(\bk-\bp) \zeta_\ug(\bq-\bk) \ ,\nonumber
\end{eqnarray}
where the \ac{PNG} parameters are no longer constants, but rather functions of the wave vectors. 
Therefore, when substituting this formula into Eq.~\eqref{eq:h} for the subsequent calculation of the energy-density full spectrum of \acp{SIGW}, these \ac{PNG} parameters cannot be extracted from the integrands.
The incorporation of these scale-dependent \ac{PNG} parameters into the integrands will complicate the integrals. 
Despite this increased complexity, the diagrammatic approach remains partially applicable in this scenario. 
We can still evaluate the symmetric factors for the diagrams of $\bar{\omega}_{\uGW,\uin} (q)$ using the Feynman-like rules in \cref{fig:G-like-FD} and the ``renormalized'' Feynman-like rules in \cref{fig:FD-Rules-new}, which is the primary objective of the diagrammatic approach. 
Specifically, the Feynman-like diagrams presented in \cref{fig:G-like-FD} - \cref{fig:PN-like-FD} and their corresponding symmetric factors remain unchanged in the scale-dependent scenario, allowing us to infer the symmetric factors based on Eqs.~(\ref{eq:G-like-HNL}, \ref{eq:C-like-HNL}, \ref{eq:Z-like-HNL}, \ref{eq:P-like-HNL}, \ref{eq:N-like-HNL}, \ref{eq:CZ-like-HNL}, \ref{eq:PZ-like-HNL}, \ref{eq:NC-like-HNL}, \ref{eq:PN-like-HNL}). 
However, the specific rules for the vertices and propagators must be modified. 
Minor adjustments should be made to the Feynman-like diagram rules in \cref{fig:G-like-FD}, as the vertices are modified to be scale-dependent while the propagators remain unchanged.
For the ``renormalized'' Feynman-like rules in \cref{fig:FD-Rules-new}, the loop integrals in Eqs.~(\ref{eq:Pxi-def}, \ref{eq:loop-int}) are no longer valid, since the scale-dependent \ac{PNG} parameters incorporated into the integrands involve other undetermined momenta in addition to the loop momentum. 
In this case, the convolved propagators defined in Eq.~\eqref{eq:P_redef} and the ``renormalized'' vertices described in Eqs.~\eqref{eqs:V_redef} can be regarded as operators, necessitating that all involved integrations be carried out ultimately. 
Thus, the ``renormalized'' Feynman-like rules can only be utilized to simplify the evaluation of the symmetric factors, but are not helpful for the integral calculations of each diagram of $\bar{\omega}_{\uGW,\uin} (q)$.
One could reformulate the integrals according to the modified Feynman-like rules in \cref{fig:FD-Rules}. 
Nevertheless, our method for calculating the large-scale modulations in Eqs.~(\ref{eq:Ong1-dp-def}, \ref{eqs:Ong2-dp-def}, \ref{eqs:Ong3-dp-def}) also proves ineffective, making the calculation of the multi-point correlations of $\delta_{\uGW,\uin}$ significantly more complicated. 
In this case, by constructing all the diagrams of the \ac{PNG}-induced inhomogeneities based on the modified Feynman-like rules and \cref{fig:FD_Frame}, it becomes necessary to compute the large-scale modulations for each diagram individually, rather than leveraging the entire categories of $\bar{\omega}_{\uGW,\uin}$ in this work. 
As a result, we have to perform thousands of integrals for \ac{PNG} up to quartic approximation. 
Even more challenging, since the loop integrals cannot be computed in advance, the expressions involve higher-dimensional integrals, presenting obstacles for accurate numerical calculations.

To address these difficulties, a compromise between scale-independent and scale-dependent \ac{PNG} may be viable. 
As discussed in Section~\ref{sec:ED}, we decompose the scalar perturbations into short- and long-wavelength modes in Eq.~\eqref{eq:S-L-dec}, where the short-wavelength mode is responsible for the production of \acp{SIGW}, while the long-wavelength mode modulates the distribution of their energy density on the \ac{CMB} scale. 
These two scales are distinctly different, while each scale region is relatively small in contrast. 
Thus, it is plausible to postulate that the \ac{PNG} parameters vary slowly within each scale region, allowing them to be regarded as graded across different scale regions \cite{Yu:2023jrs}. 
The Fourier mode of Eq.~\eqref{eq:Fnl-Gnl-Hnl-def} is revised to be 
\begin{eqnarray}\label{eq:fnl-def-k}
    \zeta (\bq) &=& \zeta_\ug(\bq) + \Fnl^{(0)} \int \frac{\ud^3 \bk}{(2\pi)^{3/2}} \zeta_{\ugS}(\bk) \zeta_{\ugS}(\bq-\bk) \\
    &&+ 2\Fnl^{(1)} \int \frac{\ud^3 \bk}{(2\pi)^{3/2}} \zeta_{\ugL}(\bk) \zeta_{\ugS}(\bq-\bk) 
    + \Fnl^{(2)} \int \frac{\ud^3 \bk}{(2\pi)^{3/2}} \zeta_{\ugL}(\bk) \zeta_{\ugL}(\bq-\bk) + \cdots\ .\nonumber
\end{eqnarray}
Similarly, we introduce $\Gnl^{(1)}$, $\Gnl^{(2)}$, $\Gnl^{(3)}$, $\Hnl^{(1)}$, $\Hnl^{(2)}$, $\Hnl^{(3)}$, $\Hnl^{(4)}$, and so forth for the couplings of the different wavelength modes, while the \ac{PNG} parameters representing the couplings among the short-wavelength modes are denoted as $\Gnl^{(0)}$ and $\Hnl^{(0)}$. 
The vertices in \cref{fig:FD-Rules} denote $V^{[2]}_j=\Fnl^{(j)}$, $V^{[3]}_j=\Gnl^{(j)}$, and $V^{[4]}_j=\Hnl^{(j)}$ in this context. 
In particular, the \textit{Planck} 2018 results provided the latest constraints on local-type \ac{PNG}, specifically $\fnl^\mathrm{CMB} = -0.9 \pm 5.1$ and $\gnl^\mathrm{CMB} = (-5.8 \pm 6.5) \times 10^4$ at a $68\%$ confidence level \cite{Planck:2019kim}, which correspond to $\Fnl^{(2)} = -0.54 \pm 3$ and $\Gnl^{(3)} = (-2.1 \pm 2.3) \times 10^4$. 
Moreover, the consistency relation for single-field inflation indicates that $\Fnl^{(1)}  = 3 (n_s - 1) / 5$, where $n_s$ is the power index of $\Delta^2_{\zeta_\uS} (k)$ \cite{Creminelli:2004yq}. 
In this compromise, our diagrammatic approach, including both the Feynman-like rules in \cref{fig:G-like-FD} and the ``renormalized'' Feynman-like rules in \cref{fig:FD-Rules-new}, remains effective. 
The energy-density fraction spectrum of \acp{SIGW} remains unchanged compared to the results of this work, while the modified large-scale modulations can be readily obtained via Eqs.~(\ref{eq:Ong1-dp-def}, \ref{eqs:Ong2-dp-def}, \ref{eqs:Ong3-dp-def}). 
These modified large-scale modulations may differ from Eqs.~(\ref{eq:Ong1}, \ref{eq:Ong2}, \ref{eq:Ong3}), because these expressions are applicable to the specific results for the scale-independent case. 
Thus, it is anticipated that the angular spectra could be enhanced or suppressed for different \ac{PNG} parameters.
Additionally, as mentioned in Subsection~\ref{subsec:SIGW&PNG}, the production of \acp{SIGW} is linked to $\zeta_\uS$, while the initial inhomogeneities arise from the coupling between short- and long-wavelength modes, as well as from large-scale scalar perturbations.
It appears that only the local-type \ac{PNG} could yield large anisotropies in \acp{SIGW}. 
However, \ac{PNG} of other shapes on small scales will alter the primordial trispectrum shown in Eq.~\eqref{eq:tnl-def}, which may leave significant imprints on the isotropic component of the \ac{SIGW} background.
Further analysis of this topic is reserved for future investigations.

\subsection{SIGW Background and PBH Formation}\label{sec:pbh}

In this study, we focus on the \acp{SIGW} associated with an enhanced power spectrum and the non-Gaussianity of primordial curvature perturbations during the radiation-dominated era, which is linked to the formation of \acp{PBH}. 
In particular, the \acp{PBH} within mass ranges where the accompanying \acp{SIGW} may be potentially detectable by space-borne detectors \cite{Saito:2009jt,Inomata:2018epa,Zhao:2022kvz} could constitute some or all of dark matter \cite{Carr:2020xqk,Carr:2023tpt}.
Therefore, it is crucial to investigate the relationship between \ac{PBH} formation and the \ac{SIGW} background while considering the presence of \ac{PNG}. 
Future detection of the \ac{SIGW} background will constrain \ac{PBH} abundance and \ac{PNG}, while the search for \acp{PBH} can also contribute to studying the statistics of primordial curvature perturbations \cite{Young:2014ana, Yoo:2018kvb, Yoo:2020dkz, Abe:2022xur, Pi:2024ert, Pi:2024jwt, Kalaja:2019uju, LISACosmologyWorkingGroup:2025vdz, Pritchard:2025yda, Zhou:2025djn}. 
A substantial amount of research \cite{Byrnes:2012yx, Young:2013oia, Franciolini:2018vbk, Yoo:2019pma, Inomata:2020xad, Nakama:2016gzw, Abe:2022xur, Perna:2024ehx, Ferrante:2022mui, Gow:2022jfb, vanLaak:2023ppj, Franciolini:2023pbf, Iovino:2024tyg, Kitajima:2021fpq, Inui:2024fgk, Young:2022phe, Franciolini:2023wun} has investigated the impact of \ac{PNG} on \ac{PBH} formation. 
Based on these studies, we will provide a brief discussion of this important topic.

To determine whether the cosmological perturbations collapse to form a \ac{PBH}, we utilize the compaction function \cite{Shibata:1999zs}, which can characterize the average density contrast of cosmological fluids over the Hubble horizon. 
During the radiation-dominated era, we consider a spherically symmetric region with an areal radius of $ R(r,t) = a(t) r e^{\zeta(t)} $, where $ r $ is the radial coordinate.
The compaction function can be expressed as 
\begin{equation}
    \cC (\bx,r) = \cC_1 (\bx,r) - \frac{3}{8} \cC_1^2 (\bx,r)\ ,
\end{equation}
where $ \cC_1 (\bx,r) $ is represented as 
\begin{equation}
    \cC_1 (\bx,r) = -\frac{4}{3} r \partial_r \zeta (r) 
    % = -\frac{4}{3} r \zeta_\ug ' (r) \frac{\ud \zeta}{\ud \zeta_\ug} 
    = \cC_\ug (\bx,r) \frac{\delta \zeta}{\delta \zeta_\ug} \ ,
    % &=& -\frac{4}{3} r \zeta_\ug ' (1 + 2\Fnl \zeta_\ug + 3\Gnl \zeta_\ug^2 + 4\Hnl \zeta_\ug^3 + \cdots )\ , \nonumber
\end{equation}
with the Gaussian component defined as $ \cC_\ug (\bx,r) = -4 r \partial_r \zeta_\ug (r) / 3 $. 
For local-type \ac{PNG} in the form of Eq.~\eqref{eq:Fnl-Gnl-Hnl-def}, the variational derivative of $\zeta$ with respect to $\zeta_\ug$, denoted as $\delta \zeta / \delta \zeta_\ug$, is given by 
\begin{equation}
    \frac{\delta \zeta}{\delta \zeta_\ug} = 1 + 2\Fnl \zeta_\ug + 3\Gnl \zeta_\ug^2 + 4\Hnl \zeta_\ug^3 + \cdots\ .
\end{equation}
We evaluate the compaction function at the time of horizon reentry, i.e., $ R(r_m,t) \cH = 1 $, where $r_m$ denotes the scale at which $\cC$ reaches its maximum value.
As shown in Ref.~\cite{Musco:2020jjb}, $ r_m $ depends on the shape of the power spectrum of $ \zeta_\ug $. 
For $ \Delta^2_\uS (q) $ as assumed in Eq.~\eqref{eq:Lognormal}, we have $ r_m k_\ast \simeq 1.75 $ according to Ref.~\cite{Musco:2020jjb}.

When the compaction function exceeds the threshold, a \ac{PBH} forms from primordial overdensity collapse, with its mass given by \cite{Niemeyer:1997mt, Niemeyer:1999ak} 
\begin{equation}
    M_\mathrm{PBH} = \cK M_H \left(\cC - \cC_\uth\right)^\gamma \ ,
\end{equation}
where $ M_H $ is the mass contained within a Hubble horizon in an unperturbed background, i.e., 
\begin{equation}
    M_H \simeq 17 M_\odot \left( \frac{r_m}{10^{-6} \Mpc }\right)^{2} \left(\frac{g_\ast }{10.75}\right)^{-1/6}\ ,
\end{equation}
with $M_\odot$ denoting the solar mass.
For simplicity, we take the prefactor to be $ \cK \simeq 4 $ and the critical exponent to be $ \gamma \simeq 0.36 $. 
Since the threshold $ \cC_\uth $ is robust against \ac{PNG} \cite{Kehagias:2019eil} and depends on the spectral width $ \sigma $ \cite{Musco:2018rwt, Musco:2020jjb, Escriva:2019phb, Musco:2023dak}, we utilize the result from Ref.~\cite{Musco:2020jjb}, which indicates that $ \cC_\uth \simeq 0.46 $ for $ \sigma = 1 $.

The mass fraction of \acp{PBH} at the formation epoch is characterized by $\beta$, expressed as \cite{Ferrante:2022mui,Gow:2022jfb,Young:2022phe, vanLaak:2023ppj, Franciolini:2023pbf,Iovino:2024tyg, Perna:2024ehx}
\begin{equation}\label{eq:PBH-abundance}
    \beta = \int_\cD \cK (\cC - \cC_\uth)^\gamma \uP_\ug (\cC_\ug,\zeta_\ug) \ud \cC_\ug \ud \zeta_\ug\ ,
\end{equation}
where the integration domain is given by $\cD = \{ \cC_\ug, \zeta_\ug \in \mathbb{R}; \cC > \cC_\uth \wedge \cC_1 < \frac{4}{3} \}$, and the probability distribution function  reads 
\begin{equation}
    \uP_\ug (\cC_\ug,\zeta_\ug) = \frac{e^{-\frac{1}{2(1-\gamma_{cr}^2)} \left(\frac{\cC_\ug}{\sigma_c}-\frac{\gamma_{cr}\zeta_\ug}{\sigma_r}\right)^2 - \frac{\zeta_\ug^2}{2\sigma_r^2}}}{2\pi \sigma_c \sigma_r \sqrt{1-\gamma_{cr}^2}}\ .
\end{equation}
In this probability distribution function, $\sigma_c$ and $\sigma_r$ represent the standard deviations of $\cC_\ug$ and $\zeta_\ug$, respectively, while $\gamma_{cr} = \sigma_{cr}^2 / (\sigma_c \sigma_r)$ denotes their coefficient of association. 
Using these three parameters, we can construct the covariance matrix between $\cC_\ug$ and $\zeta_\ug$.  
Considering the corrections from the non-linear radiation transfer function \cite{DeLuca:2023tun, Ferrante:2022mui, Iovino:2024tyg}, we can calculate the corresponding variances as follows 
\begin{eqnarray}
    \sigma_c^2 &=& \frac{16}{81} \int_0^\infty \frac{\ud k}{k}\, (k r_m)^4 W^2 (k,r_m) T^2 (k r_m) \Delta^2_\uS (k)\ ,\\
    \sigma_{cr}^2 &=& \frac{4}{9} \int_0^\infty \frac{\ud k}{k}\, (k r_m)^2 W (k,r_m) W_s (k,r_m) T^2 (k r_m) \Delta^2_\uS (k)\ ,\\
    \sigma_{r}^2 &=& \int_0^\infty \frac{\ud k}{k}\, W_s^2 (k,r_m) T^2 (k r_m) \Delta^2_\uS (k)\ ,
\end{eqnarray}
where the transfer function of the scalar perturbations, denoted as $ T(k r_m) $, is given by Eq.~\eqref{eq:T-RD}, and the top-hat window function $ W(k,r) $ and the spherical-shell window function $ W_s(k,r) $ are defined as 
\begin{eqnarray}
    W (k,r) &=& 3 \frac{\sin (kr) - kr \cos (kr)}{(kr)^3}\ ,\\
    W_s (k,r) &=& \frac{\sin (kr)}{kr}\ .
\end{eqnarray}
Specifically, with $\Delta^2_\uS(q)$ as assumed in Eq.~\eqref{eq:Lognormal} and the spectral width set to $\sigma=1$ in this study, numerical integration yields values of $\sigma_c^2 = 0.386 A_\uS$, $\sigma_{cr}^2 = 0.187 A_\uS$, and $\sigma_r^2 = 0.365 A_\uS$. 
By substituting these variances into the integral in Eq.~\eqref{eq:PBH-abundance}, we can perform the integral for $ \beta $ numerically for specific \ac{PNG} parameters. 
\iffalse
Furthermore, the present-day \ac{PBH} abundance is typically characterized by their total fraction $ f_\mathrm{PBH} $, which can be estimated through the mass fraction $ \beta $ as follows \cite{Nakama:2016gzw} 
\begin{equation}\label{eq:fPBH} 
    f_{\mathrm{PBH}} \simeq 2.5\times10^{8}\beta\left(\frac{g_{\ast,\rho}(T_\uin)}{10.75}\right)^{-\frac{1}{4}}\left(\frac{M_{\mathrm{PBH}}}{M_\odot}\right)^{-\frac{1}{2}}\ . 
\end{equation}
A more accurate expression for $ f_\mathrm{PBH} $ can be found in Refs.~\cite{Ferrante:2022mui,Gow:2022jfb,vanLaak:2023ppj,Franciolini:2023pbf,Iovino:2024tyg}. 
Nonetheless, since the impacts of \ac{PNG} on \ac{PBH} abundance are not the main focus of this study, the estimation in Eq.~\eqref{eq:fPBH} suffices to reflect the relationship between the \ac{SIGW} anisotropies and \ac{PBH} abundance.
\fi 

\begin{figure*}[htbp]
    \centering
    \includegraphics[width = 0.9 \textwidth]{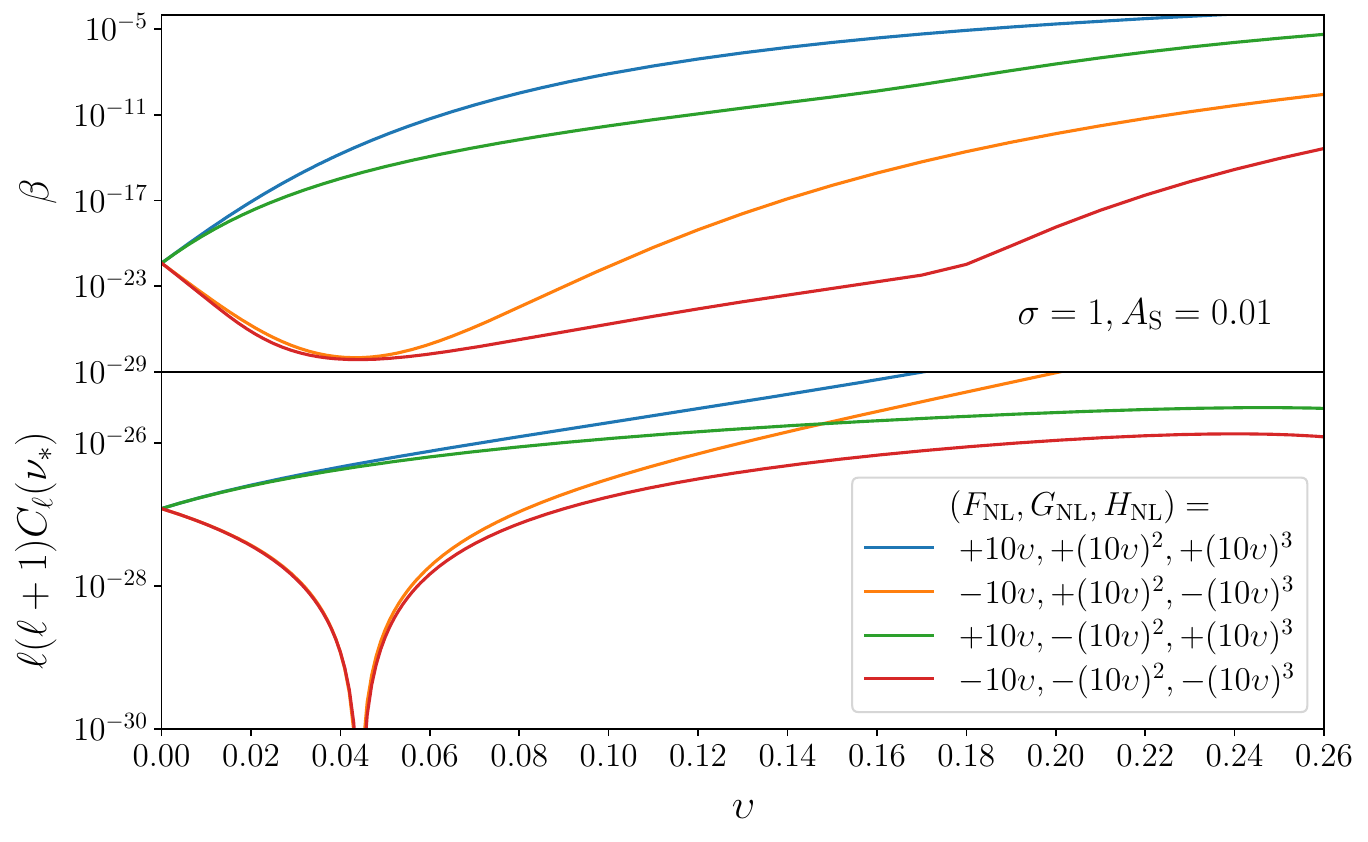}
    \caption{Dependence of the total fraction of \acp{PBH} and the angular power spectrum of SIGWs at the pivot frequency on PNG with the assumption $|\Hnl| A_\uS^{3/2} = \upsilon |\Gnl| A_\uS = \upsilon^2 |\Fnl| \sqrt{A_\uS} = \upsilon^3$.  
    }\label{fig:beta-Ct}
\end{figure*}

In Fig.~\ref{fig:beta-Ct}, we present a numerical example to illustrate the relationship between $ \beta$ and $ \ell (\ell + 1) C_\ell (\nu_\ast) $, assuming the spectral amplitude of $ \Delta^2_\uS (q) $ is $ A_\uS = 0.01 $.  
In this example, we define a quantity $\upsilon$ to characterize the ratios among the \ac{PNG} parameters. 
Specifically, we assume $|\Hnl| A_\uS^{3/2} = \upsilon |\Gnl| A_\uS = \upsilon^2 |\Fnl| \sqrt{A_\uS} = \upsilon^3$, such that $\upsilon(1 + \upsilon + \upsilon^2)$ quantifies the total amplitude of \ac{PNG}. 
Therefore, the perturbativity conditions require that $0 \leq \upsilon < 1$ and $\upsilon(1 + \upsilon + \upsilon^2) < 1$.
As demonstrated in this figure, $ \beta $ is significantly enhanced by substantial \ac{PNG}, regardless of the signs of the \ac{PNG} parameters. 
% While $ \beta $ is of order $ \cO (10^{-22}) $ in the Gaussian case, \acp{PBH} experience overproduction when $\upsilon(1 + \upsilon + \upsilon^2)$ increases to $ 0.05 \sim 0.26 $, with the precise value determined by the signs of the \ac{PNG} parameters. 
% Moreover, t
This figure indicates that the search of \acp{PBH} will aid in addressing the sign-degeneracy problem in \acp{SIGW} since $ \beta $ varies with the sign of any \ac{PNG} parameter. 
% although this behavior is not reflected in this figure because the sign degeneracy is broken by $ C_\ell (\nu) $ when \ac{PBH} is not overproduced.
Despite exhibiting a similar trend in the large $ \upsilon $ regime, $ \beta $ displays distinct behaviors for different signs of the \ac{PNG} parameters when \ac{PNG} is small. 
Notably, the formation of \acp{PBH} is suppressed when $ \Fnl $ is negative with a small absolute value. 
Observing the two panels of this figure, we can find that the anisotropies in \acp{SIGW} are altered in a remarkable way when \ac{PNG} increases in small $\upsilon$ regime, where \acp{PBH} only constitute a small fraction of the total energy density. 
As demonstrated in Section~\ref{sec:Omegabar}, the isotropic components of \acp{SIGW} remains nearly unchanged in this regime, suggesting that the detection of the anisotropies in the \ac{SIGW} background could provide crucial insights into \ac{PBH} formation in the early Universe. 
% search for \acp{PBH}.

%% file: TeX/7Conclusion.tex
\section{Conclusion and Discussion}\label{sec:conc}

The imprints of \ac{PNG} on the \ac{SIGW} background present a promising avenue for understanding the physics of the early Universe. 
In this study, we conduct a thorough analysis of the \ac{SIGW} background in the presence of \ac{PNG}, including its isotropic component, anisotropies, and non-Gaussianity. 
These properties are characterized by the energy-density fraction spectrum in Eq.~\eqref{eq:omegabar-zeta}, the reduced angular power spectrum in Eq.~\eqref{eq:Ct}, the angular bispectrum in Eqs.~(\ref{eq:Bl-res}, \ref{eq:b-def}), and the angular trispectrum in Eqs.~(\ref{eq:tl-res}, \ref{eq:t1-def}, \ref{eq:t2-def}), which constitute the main results of this work. 
In the derivation process, we develop a 
% diagrammatic approach, previously used in Refs.~\cite{Adshead:2021hnm,Ragavendra:2021qdu,Li:2023qua,Li:2023xtl,Ruiz:2024weh,Abe:2022xur,Perna:2024ehx}, into a more generic form. 
``renormalized'' diagrammatic approach, extending prior methods \cite{Adshead:2021hnm,Ragavendra:2021qdu,Li:2023qua,Li:2023xtl,Ruiz:2024weh,Abe:2022xur,Perna:2024ehx} to simplify calculations of the \ac{SIGW} energy density spectrum with high-order \ac{PNG}. 
Our approach is applicable for perturbative local-type \ac{PNG} up to all orders and can be extended to graded scale-dependent \ac{PNG}.
In the case of scale-independent \ac{PNG} up to $\hnl$ order, this approach significantly simplifies the calculation of the energy-density fraction spectrum, allowing us to derive the contractions of the multi-point correlators of $ \zeta_S $ directly using Eqs.~(\ref{eq:G-like}, \ref{eq:C-like}, \ref{eq:Z-like}, \ref{eq:P-like}, \ref{eq:N-like}, \ref{eq:CZ-like}, \ref{eq:PZ-like}, \ref{eq:NC-like}, \ref{eq:PN-like}) without the need for constructing specific diagrams. 
Moreover, we extend the computational technique of Ref.~\cite{Li:2024zwx} to calculate the \ac{PNG}-induced inhomogeneity up to $ \mathcal{O}(A_\uL^3) $ in Eqs.~(\ref{eq:Ong1-dp-def}, \ref{eqs:Ong2-dp-def}, \ref{eqs:Ong3-dp-def}), making it readily generalizable to \ac{PNG} of all orders. 
Consequently, for local-type \ac{PNG} expressed as a power-series expansion up to all orders, one can easily derive the aforementioned spectra of \acp{SIGW}, greatly simplifying the lengthy calculation process.

For \ac{PNG} up to $\hnl$ order, we have analyzed the behaviors of the energy-density fraction spectrum $ h^2 \bar{\Omega}_{\uGW,0} (\nu) $, the reduced angular power spectrum $ \tilde{C}_\ell (\nu) $, the angular bispectrum $ B_{\ell_1 \ell_2 \ell_3} (\nu) $, and the angular trispectrum $ t^{\ell_1 \ell_2}_{\ell_3 \ell_4} (L,\nu) $ across various scenarios characterized by different \ac{PNG} parameters. 
Our analyses indicate that \ac{PNG} may leave significant imprints on the \ac{SIGW} background, particularly regarding its anisotropies and non-Gaussianity.
For $ h^2 \bar{\Omega}_{\uGW,0} (\nu) $, the numerical results show that the $ G $-like family, representing the disconnected component of the four-point correlator of $ \zeta $, dominates the total spectrum.
Despite this finding, other families, which represent the contributions from the primordial trispectrum, should not be neglected, especially in the ultraviolet regime. 
Even if we eliminate the contributions from loop corrections to the power spectrum of $ \zeta $, the primordial trispectrum of $ \zeta $ can exert non-negligible impacts on $ h^2 \bar{\Omega}_{\uGW,0} (\nu) $ when both $ \fnl $ and $ \hnl $ are negative and their modes are large.
For the anisotropies and non-Gaussianity in the \ac{SIGW} background, we have derived explicit expressions for $ \tilde{C}_\ell (\nu) $, $ B_{\ell_1 \ell_2 \ell_3} (\nu) $, and $ t^{\ell_1 \ell_2}_{\ell_3 \ell_4} (L,\nu) $ at low multipoles, taking into account the \ac{SW} effect and the initial inhomogeneities induced by both the non-adiabaticity and \ac{PNG}. 
Although their multipole dependencies parallel those of the \ac{CMB} at low multipoles, they are useful for distinguishing \acp{SIGW} from \acp{GW} sourced from other cosmological or astrophysical phenomena. 
In contrast, their frequency dependencies exhibit different and non-trivial features, suggesting that the anisotropies and non-Gaussianity in the \ac{SIGW} background encompass valuable information about \acp{SIGW} beyond the isotropic component of the \ac{SIGW} background in their frequency dependencies. 
These angular spectra are more sensitive to \ac{PNG} than the energy-density fraction spectrum. 
Notably, Gaussian primordial curvature perturbations always yield a Gaussian \ac{SIGW} background, implying that the observation of a non-zero angular bispectrum of \acp{SIGW} can serve as compelling evidence for the presence of \ac{PNG}. 
The magnitude of $ \ell(\ell+1) C_\ell (\nu) $ is of order $ \mathcal{O}(10^{-7}) $ at the pivot frequency for the Gaussian case, while the presence of \ac{PNG} may significantly enhance this magnitude. 
Additionally, while a larger amplitude of the power spectrum of $ \zeta_\uS $ consistently yields an enhanced $ h^2 \bar{\Omega}_{\uGW,0} (\nu) $, it commonly results in a suppressed $ \tilde{C}_\ell (\nu) $, $ B_{\ell_1 \ell_2 \ell_3} (\nu) $, and $ t^{\ell_1 \ell_2}_{\ell_3 \ell_4} (L,\nu) $. 
Although the primordial trispectrum of $ \zeta_\uS $ leaves a minor imprint on $ h^2 \bar{\Omega}_{\uGW,0} (\nu) $ in most of cases, it has significant impacts on the angular spectra of \acp{SIGW}. 
The different behaviors of these angular spectra indicate that they contain a wealth of valuable information about the early Universe, which can be utilized to improve the accuracy of the inferred values of model parameters from $ h^2 \bar{\Omega}_{\uGW,0} (\nu) $ and to resolve the sign degeneracy between $ \fnl\hnl $ and $ \gnl $.

This study has significant implications for the search for \acp{PBH}, which are associated with the production of \acp{SIGW} and are considered potential candidates for cold dark matter. 
The model parameters inferred from the anisotropies in the future detected \ac{SIGW} background can impose constraints on the \ac{PBH} formation, particularly within the mass ranges corresponding to the frequency bands of the sensitivity regimes of space-borne interferometers \cite{Baker:2019nia,Smith:2019wny,Hu:2017mde,Wang:2021njt,Ren:2023yec,TianQin:2015yph,TianQin:2020hid,Zhou:2023rop,Seto:2001qf,Kawamura:2020pcg,Crowder:2005nr,Smith:2016jqs,Capurri:2022lze} and \ac{PTA} programs \cite{Hobbs:2009yy,Demorest:2012bv,Kramer:2013kea,Manchester:2012za,Sesana:2008mz,Thrane:2013oya,Janssen:2014dka,2009IEEEP..97.1482D,Weltman:2018zrl,Moore:2014lga}. 
Given that the presence of \ac{PNG} has a substantial impact on the \ac{PBH} formation, this work provides a brief discussion on the relationship between \ac{PBH} formation and the \ac{SIGW} anisotropies through a simplified example. 
Through this example, we demonstrate that the angular power spectrum of \acp{SIGW} is closely related to the \ac{PBH} mass fraction at the formation epoch. 
Notably, although the sign degeneracy between $\fnl\hnl$ and $\gnl$ cannot be resolved for large $\fnl$ in any of the aforementioned spectra of \acp{SIGW}, \ac{PNG} parameters with differing signs result in distinct \ac{PBH} mass fractions, suggesting that the exploration of \acp{PBH} can serve as a complementary tool in the search for \ac{PNG}. 
In summary, future observations of either the \ac{SIGW} background or the \ac{PBH} abundance can inform the detection of the other.

We anticipate that future advancements in \ac{GW} detectors will enable sensitivity levels at which \acp{SIGW} can be differentiated from other stochastic \ac{GW} backgrounds. 
Our theoretical predictions for the \ac{SIGW} background can be tested by networks of space-borne \ac{GW} detectors \cite{Baker:2019nia,Smith:2019wny,Hu:2017mde,Wang:2021njt,Ren:2023yec,TianQin:2015yph,TianQin:2020hid,Zhou:2023rop,Seto:2001qf,Kawamura:2020pcg,Crowder:2005nr,Smith:2016jqs,Capurri:2022lze} and \ac{PTA} programs \cite{Hobbs:2009yy,Demorest:2012bv,Kramer:2013kea,Manchester:2012za,Sesana:2008mz,Thrane:2013oya,Janssen:2014dka,2009IEEEP..97.1482D,Weltman:2018zrl,Moore:2014lga}. 
Notably, the \ac{GW} signals observed in recent \ac{PTA} experiments \cite{Xu:2023wog,EPTA:2023fyk,NANOGrav:2023gor,Reardon:2023gzh} can be interpreted as \acp{SIGW} \cite{Franciolini:2023pbf,Inomata:2023zup,Cai:2023dls,Wang:2023ost,Liu:2023ymk,Abe:2023yrw,Ebadi:2023xhq,Figueroa:2023zhu,Yi:2023mbm,Madge:2023cak,Firouzjahi:2023lzg,Wang:2023sij,You:2023rmn,Ye:2023xyr,HosseiniMansoori:2023mqh,Balaji:2023ehk,Das:2023nmm,Bian:2023dnv,Jin:2023wri,Zhu:2023gmx,Liu:2023pau,Yi:2023tdk,Frosina:2023nxu,Choudhury:2023hfm,Ellis:2023oxs,Kawasaki:2023rfx,Yi:2023npi,Harigaya:2023pmw,An:2023jxf,Gangopadhyay:2023qjr,Chang:2023ist,Inomata:2023drn,Choudhury:2023fwk,Choudhury:2023fjs,Domenech:2023jve,Chang:2023aba,Mu:2023wdt,Liu:2023hpw,Tagliazucchi:2023dai,Chen:2024fir,Choudhury:2024one,Chen:2024twp,Domenech:2024rks,Franciolini:2023pbf,Iovino:2024tyg,Choudhury:2024dzw,Choudhury:2024kjj,Zhou:2025djn,Wang:2025kbj,Ghaleb:2025xqn}. 
Although the source of this signal has not yet been confirmed, we expect that the improved angular resolution of \ac{SKA} \cite{2009IEEEP..97.1482D,Weltman:2018zrl,Moore:2014lga} holds promise for resolving this issue through analyses of the angular power spectrum, bispectrum, and trispectrum, besides the energy-density fraction spectrum. 
This work will be instrumental in overcoming this challenge and can be used to determine the model parameters if \acp{SIGW} contribute to this signal. 
Furthermore, this work can be easily extended to analyze the cross-correlation between the \ac{SIGW} background and the \ac{CMB} \cite{Dimastrogiovanni:2021mfs, Cusin:2018rsq, Ricciardone:2021kel, Malhotra:2020ket, Braglia:2021fxn, Capurri:2021prz, Dimastrogiovanni:2022eir, Galloni:2022rgg, Ding:2023xeg, Cyr:2023pgw, Zhao:2024gan}, which can facilitate the measurement of the \ac{SIGW} background due to the shared origin of anisotropies induced by propagation.

In conclusion, this comprehensive analysis of the imprints of \ac{PNG} on the \ac{SIGW} background is crucial not only for distinguishing \ac{SIGW} signals from the stochastic \ac{GW} background but also holds significant implications for exploring the early Universe, particularly regarding inflation models, the nature of cosmological fluids, and the \ac{PBH} mass fraction at the formation epoch. 
The incorporation of \ac{PNG} up to higher orders enables us to search for \ac{PNG} in a more generalized form, and the investigation of higher statistics of the \ac{SIGW} background aids in extracting additional information about the early Universe.

%% file: TeX/8Appendix.tex
\appendix
\section{Wigner symbols}\label{sec:Wigner}

Following Ref.~\cite{Komatsu:2001ysk}, we provide a brief summary of the basic properties of the Wigner 3-$j$ symbol and the Wigner 6-$j$ symbol in this appendix, which is useful in the derivation process of the angular bispectrum and trispectrum of \acp{SIGW} presented in Section~\ref{sec:bl&tl}.

The Wigner 3-$j$ symbol is used to describe the coupling of angular momenta. 
In this context, we use $\ell$ to denote the eigenvalue of the angular momentum operator, satisfying the equation $\mathbf{L}^2 Y_{\ell m} = \ell (\ell + 1) Y_{\ell m}$, and $m$ to represent the eigenvalue of the $z$-component of angular momentum, given by $L_z Y_{\ell m} = m Y_{\ell m}$. 
Thus, $\ell$ characterizes the magnitude of the angular momentum, while $m$ denotes its direction. 
The Wigner 3-$j$ symbol, 
\begin{equation}
    \begin{pmatrix}
        \ell_1 & \ell_2 & \ell_3 \\
        m_1    & m_2    & m_3 
    \end{pmatrix}\ , 
\end{equation}
specifically describes a triangle formed by three angular momenta, $\mathbf{L}_1 + \mathbf{L}_2 + \mathbf{L}_3 = 0$, indicating that the Wigner 3-$j$ symbol is meaningful only if the triples $(\ell_1, \ell_2, \ell_3)$ satisfy the triangle inequalities and $m_1 + m_2 + m_3 = 0$.

The Wigner 3-$j$ symbol changes sign under the interchange of any two columns when $\Sigma_\ell = \ell_1 + \ell_2 + \ell_3$ is odd, while it remains invariant under even permutations, such as 
\begin{eqnarray}
    \begin{pmatrix}
        \ell_1 & \ell_2 & \ell_3 \\
        m_1    & m_2    & m_3 
    \end{pmatrix}
    &=& \begin{pmatrix}
        \ell_2 & \ell_3 & \ell_1 \\
        m_2    & m_3    & m_1 
    \end{pmatrix}\ ,\\
    \begin{pmatrix}
        \ell_1 & \ell_2 & \ell_3 \\
        m_1    & m_2    & m_3 
    \end{pmatrix}
    &=& (-1)^{\Sigma_\ell} 
    \begin{pmatrix}
        \ell_2 & \ell_1 & \ell_3 \\
        m_2    & m_1    & m_3 
    \end{pmatrix}\ .
\end{eqnarray}
Moreover, the sign also changes under the transformation $(m_1, m_2, m_3) \rightarrow (-m_1, -m_2, -m_3)$ if $\Sigma_\ell$ is odd, i.e., 
\begin{equation}
    \begin{pmatrix}
        \ell_1 & \ell_2 & \ell_3 \\
        m_1    & m_2    & m_3 
    \end{pmatrix}
    = (-1)^{\Sigma_\ell} 
    \begin{pmatrix}
        \ell_1 & \ell_2 & \ell_3 \\
        -m_1   & -m_2   & -m_3 
    \end{pmatrix}\ .
\end{equation}
Therefore, if $m_1 = m_2 = m_3 = 0$ and $\Sigma_\ell$ is odd, the Wigner 3-$j$ symbol is zero, indicating that $h_{\ell_1 \ell_2 \ell_3} = 0$ for odd parity.
In addition, the orthogonality properties of the Wigner 3-$j$ symbol reads 
\begin{eqnarray}
    &&\sum_{\ell_3 m_3} (2\ell_3 + 1)
    \begin{pmatrix}
        \ell_1 & \ell_2 & \ell_3 \\
        m_1    & m_2    & m_3 
    \end{pmatrix}
    \begin{pmatrix}
        \ell_1 & \ell_2 & \ell_3 \\
        m_1'   & m_2'   & m_3' 
    \end{pmatrix}
    = \delta_{m_1 m_1'} \delta_{m_2 m_2'}\ ,\\
    && \sum_{m_1 m_2} 
    \begin{pmatrix}
        \ell_1 & \ell_2 & \ell_3 \\
        m_1    & m_2    & m_3 
    \end{pmatrix}
    \begin{pmatrix}
        \ell_1 & \ell_2 & \ell_3' \\
        m_1    & m_2    & m_3' 
    \end{pmatrix}
    = \frac{\delta_{\ell_3 \ell_3'} \delta_{m_3 m_3'}}{2\ell_3 + 1}\ ,\\
    && \sum_{m_1 m_2 m_3} 
    \begin{pmatrix}
        \ell_1 & \ell_2 & \ell_3 \\
        m_1    & m_2    & m_3 
    \end{pmatrix}^2
     = 1\ .
\end{eqnarray}

In our investigation, the Wigner 3-$j$ symbols appear in the definition of the \ac{SIGW} angular bispectrum and trispectrum, as discussed in Section~\ref{sec:bl&tl}, related to the Gaunt integral defined in Eq.~\eqref{eq:Glm-def}. 
The Gaunt integral, denoted as $\Glm$, represents the integration of three spherical harmonics over the same direction, resulting in a real number. 
Furthermore, by utilizing the completeness of spherical harmonics, we can express the integral of four spherical harmonics in terms of the Gaunt integrals as follows 
\begin{equation}
    \int \ud^2 \bn\, \prod_{i=1}^4 Y_{\ell_i m_i} (\bn)
    = \sum_{L M} (-1)^{M} \cG^{m_1 m_2 -M}_{\ell_1 \ell_2 L} \cG^{m_3 m_4 M}_{\ell_3 \ell_4 L}\ .
\end{equation}
This relation is beneficial for deriving the angular trispectrum associated with $\langle \delta_{\uGW,0}^{(3)} \delta_{\uGW,0}^{(1)} \delta_{\uGW,0}^{(1)} \delta_{\uGW,0}^{(1)} \rangle$.

The Wigner 6-$j$ symbol, related to the Wigner 3-$j$ symbol through  
\begin{eqnarray}
    \begin{Bmatrix}
        \ell_1  & \ell_2  & \ell_3 \\
        \ell_1' & \ell_2' & \ell_3'
    \end{Bmatrix}
    = (-1)^{\ell_1'+\ell_2'+\ell_3'} 
    \sum_{m_1 m_2 m_3} && (-1)^{m_1'+m_2'+m_3'}  \begin{pmatrix}
        \ell_1 & \ell_2 & \ell_3 \\
        m_1    & m_2    & m_3 
    \end{pmatrix}
    \begin{pmatrix}
        \ell_1 & \ell_2' & \ell_3' \\
        m_1    & m_2'    & -m_3' 
    \end{pmatrix}\nonumber\\ 
    && \times \begin{pmatrix}
        \ell_1' & \ell_2 & \ell_3' \\
        -m_1'   & m_2    & m_3' 
    \end{pmatrix}
    \begin{pmatrix}
        \ell_1 & \ell_2' & \ell_3' \\
        m_1    & -m_2'   & m_3' 
    \end{pmatrix} \ ,
\end{eqnarray}
expresses the relationship between two distinct couplings of three angular momenta. 
Geometrically, the Wigner 6-$j$ symbol represents a tetrahedron with the triangle ${\ell_1, \ell_2, \ell_3}$ as its base and edges $\ell_i'$ opposite to $\ell$. 
It is meaningful only if all four triplets satisfy the triangle inequalities. 
Moreover, the symbol is invariant under the interchange of any two columns and under the interchange of the upper and lower arguments in any two columns. 
Additionally, it obeys 
\begin{eqnarray}
    &&\sum_{L_1} (2L_1+1)
    \begin{Bmatrix}
        \ell_1 & \ell_2 & L_1 \\
        \ell_3 & \ell_4 & L_2
    \end{Bmatrix}
    \begin{Bmatrix}
        \ell_1 & \ell_2 & L_1 \\
        \ell_3 & \ell_4 & L_3
    \end{Bmatrix}
    = \frac{\delta_{L_2 L_3}}{2 L_2 + 1}\ ,\\
    &&\sum_{L_1} (-1)^{L_1+L_2+L_3} (2L_1+1)
    \begin{Bmatrix}
        \ell_1 & \ell_2 & L_1 \\
        \ell_3 & \ell_4 & L_2
    \end{Bmatrix}
    \begin{Bmatrix}
        \ell_1 & \ell_2 & L_1 \\
        \ell_4 & \ell_3 & L_3
    \end{Bmatrix}
    = \begin{Bmatrix}
        \ell_1 & \ell_3 & L_3 \\
        \ell_2 & \ell_4 & L_2
    \end{Bmatrix}\ .
\end{eqnarray}
One can employ these symmetric properties, as well as the properties of the Wigner 3-$j$ symbol, to determine that 
\begin{eqnarray}
    &&\sum_{LM} (2L+1) (-1)^{\ell_2+\ell_3+M}
        \begin{Bmatrix}
            \ell_1 & \ell_3 & L' \\
            \ell_4 & \ell_2 & L
        \end{Bmatrix}
        \begin{pmatrix}
            \ell_1 & \ell_2 & L \\
            m_1    & m_2   & -M 
        \end{pmatrix}
        \begin{pmatrix}
            \ell_3 & \ell_4 & L \\
            m_3    & m_4    & M
        \end{pmatrix}\nonumber\\
    &=& \sum_{M'} (-1)^{M'}
        \begin{pmatrix}
            \ell_1 & \ell_3 & L' \\
            m_1    & m_3    & -M'
        \end{pmatrix}
        \begin{pmatrix}
            \ell_2 & \ell_4 & L' \\
            m_2   & m_4    & M' 
        \end{pmatrix}\ .
\end{eqnarray}
This expression is useful for deriving Eq.~\eqref{eq:Tl-Pl}.

\section{Equation of motion for SIGWs}\label{sec:basic} 

In this appendix, we briefly review the derivation of the solution to the equation of motion for \acp{SIGW}, as presented in Eq.~\eqref{eq:h} \cite{Espinosa:2018eve,Kohri:2018awv,Adshead:2021hnm,Li:2023qua,Li:2023xtl}. 
For convenience, we omit the subscript $_\uin$, which denotes the time of \ac{SIGW} emission, where its omission does not cause confusion.
The equation of motion for \acp{SIGW} is derived from the metric of a perturbed spatially-flat \ac{FRW} spacetime in the conformal Newtonian gauge, which is given by  
\begin{equation}\label{metric} 
    \ud s^2 
    =  a^2 \left[ 
            - e^{2\Phi} \ud \eta^2 
            + \left( e^{-2\Phi}  \delta_{ij} + \frac{h_{ij}}{2} \right) \ud x^i \ud x^j 
    \right]\ .
\end{equation}
Here, $a(\eta)$ is the scale factor of the universe at conformal time $\eta$, $\Phi(\eta,\bx)$ denotes the linear scalar perturbations, and the transverse-traceless tensor perturbations $h_{ij}(\eta,\bx)$ represent \acp{SIGW}. 
We neglect possible stress anisotropy and other perturbations, as they are negligible in a standard radiation-dominated era filled with perfect and adiabatic cosmological fluids.

For the spatially flat \ac{FRW} spacetime, we introduce the conformal Hubble parameter $\cH(\eta)$ by defining $\cH(\eta) = \partial_{\eta} a / a$, which characterizes the expansion rate of the Universe. 
The time evolution of $\cH(\eta)$ is governed by the zeroth-order Einstein equation, also known as the Friedmann equation.
Assuming a perfect fluid and a fixed equation of state, $w = \bar{P} / \bar{\rho}$, we solve this equation and obtain
\begin{equation}\label{eq:cH}
    \cH (\eta) = \frac{2}{(1+3w) \eta} \ ,
\end{equation}
where $w$ is the equation-of-state parameter of the Universe, and $\bar{P}$ and $\bar{\rho}$ denote the background pressure and energy density, respectively.
Given the initial condition, the time evolution of $a(\eta)$ can be immediately obtained by solving Eq.~\eqref{eq:cH}. 

Under the perturbed metric in Eq.~\eqref{metric}, we use the Einstein equation to derive the evolution of the perturbations. 
At first order, the Einstein equation yields a master equation that governs the evolution of $\Phi(\eta, \bq)$. 
In particular, in adiabatic scenarios, this master equation can be written as  
\begin{equation}\label{eq:master}
    \partial_\eta^2 \Phi + 3\cH \left(1+c_s^2\right) \partial_\eta \Phi + 3\left(c_s^2 -w\right) \cH^2\Phi + c_s^2 q^2 \Phi  = 0\ ,
\end{equation}
where $c_s^2$ denotes the squared sound speed. 
Furthermore, $\Phi(\eta, \bq)$ is commonly expressed in terms of the primordial comoving curvature perturbation $\zeta(\bq)$ and the scalar transfer function $T(q\eta)$, i.e.,  
\begin{equation}\label{eq:T-zeta-def}
    \Phi(\eta, \bq)
        = \left(\frac{3+3w}{5+3w}\right) T(q \eta) \zeta(\bq)\ .
\end{equation}
By substituting Eq.~\eqref{eq:T-zeta-def} into Eq.~\eqref{eq:master}, we can solve the differential equation with respect to $q\eta$ to obtain the transfer function $T(q\eta)$. 
Specifically, in the case of a radiation-dominated era with $w = c_s^2 = 1/3$, the transfer function $T(q\eta)$ takes the form  
\begin{equation}\label{eq:T-RD}
    T (x) = \frac{9}{x^2} 
            \left(
                \frac{\sin (x/\sqrt{3})}{x/\sqrt{3}}
                -\cos (x/\sqrt{3})
            \right) \ ,
\end{equation}
where we use $x = q\eta$ for brevity.

For the second-order tensor perturbation $h_{ij}(\eta, \bq)$, the spatial components of the second-order Einstein equation yield its equation of motion \cite{Ananda:2006af,Baumann:2007zm}, namely,
\begin{equation}\label{eq:SIGW-motion} 
    \partial_\eta^2 h_{\lambda}(\eta, \bq)
    + 2\cH \partial_\eta h_{\lambda}(\eta,\bq) 
    +q^2 h_{\lambda}(\eta,\bq) 
    = 4 S_{\lambda}(\eta,\bq)\ , 
\end{equation} 
where the source term $S_{\lambda}(\eta, \bq)$ can be expressed in terms of quadratic combinations of the scalar perturbations by combining the lower-order Einstein equation, as given by  
\begin{align}\label{eq:source} 
    \cS_\lambda(\eta,\bq) 
    = \int\frac{\ud^3 \bq_a}{(2\pi)^{3/2}}
            & \epsilon_{ij}^{\lambda}(\bq) q_{a}^i q_{a}^j 
            \bigg[
                2 \Phi (\eta,\bq - \bq_a)\Phi(\eta,\bq_a)\\ 
                & +\frac{4 
                    \left(
                        \partial_\eta \Phi(\eta,\bq_a) 
                        + \cH\Phi(\eta,\bq_a)
                    \right)\left(
                        \partial_\eta \Phi(\eta,\bq -\bq_a)
                        + \cH\Phi(\eta,\bq -\bq_a)
                    \right)
                }{3(1+w)\cH^2}
            \bigg]\ ,\nonumber
\end{align} 
where the polarization tensor $\epsilon_{ij}^{\lambda}$ is introduced below Eq.~\eqref{eq:h-Fourier}.
We use the Green function method \cite{Espinosa:2018eve,Kohri:2018awv} to solve Eq.~\eqref{eq:SIGW-motion}, i.e.,
\begin{equation}\label{eq:h-G} 
    a(\eta) h_\lambda(\eta, \bq) 
    = 4 \int^{\eta}_{} \ud \eta' \,
        G_\bq(\eta,\eta') a(\eta') \cS_\lambda(\eta', \bq)\ , 
\end{equation}
where the Green function $G_\bq(\eta, \eta')$ satisfies the following differential equation 
\begin{equation}\label{eq:Green}
    \partial_{\eta}^2 G_\bq(\eta,\eta') 
        +  \left(q^2-\frac{\partial_\eta^2 a(\eta)}{a(\eta)}\right) G_\bq(\eta, \eta')
        = \delta(\eta - \eta')\ ,
\end{equation}
Since Eq.~\eqref{eq:cH} yields $\partial_\eta^2 a(\eta) / a(\eta) = 2 (1-3w) / [(1+3w) \eta]^2$, the solution during the radiation-dominated era is given by  
\begin{equation}\label{eq:G-RD}
    G_\bq (\eta,\eta')
        = \Theta (\eta -\eta') 
        \frac{\sin (q(\eta - \eta'))}{q}\ ,
\end{equation}
where $\Theta(x)$ is the Heaviside function.
Subsequently, we substitute this solution into Eq.~\eqref{eq:h-G} to obtain an integral for $h_\lambda(\eta, \bq)$. 
Furthermore, as the linear scalar perturbation $\Phi(\eta, \bq)$ is expressed in terms of $\zeta(\bq)$ and $T(q\eta)$ in Eq.~\eqref{eq:T-zeta-def}, we can reformulate $\mathcal{S}_{\lambda}(\eta, \bq)$ to simplify the integrand. 
By substituting Eq.~\eqref{eq:T-zeta-def} into Eq.~(\ref{eq:source}), we obtain
\begin{equation}\label{eq:S}
    \cS_\lambda(\eta, \bq)
        = \int \frac{\ud^3 \bq_a}{(2\pi)^{3/2}} q^2 Q_{\lambda}(\bq,\bq_a) F(\abs{\bq-\bq_a}, q_a, \eta) \zeta(\bq_a) \zeta(\bq-\bq_a)\ ,
\end{equation}
where $Q_{\lambda}(\bq, \bq_a)$ denotes the projection factor, defined as
where $Q_{\lambda}(\bq, \bq_a)$ denotes the projection factor, defined as 
\begin{equation}\label{eq:Q-def}
    Q_{\lambda}(\bq, \bq_a) = \epsilon_{ij}^{\lambda}(\bq) \frac{q_{a}^i q_{a}^j}{q^2}
    = \frac{\sin^2 \theta}{\sqrt{2}}
     \times
        \begin{cases}
            \cos(2\phi_a) &\lambda = + \\
            \sin(2\phi_a) &\lambda = \times 
        \end{cases}\ , 
\end{equation}
and the function $F(p_a, q_a, \eta)$ describes the time evolution of $\cS_\lambda(\eta, \bq)$, given by
\begin{eqnarray}\label{eq:F-def}
    F(p_a,q_a,\eta)
    & = & \frac{3 (1 + w)}{(5 + 3 w)^2}
        \biggl[
            2 (5 + 3 w) T(p_a \eta) T(q_a \eta)
            + \frac{4}{\cH^2} \partial_\eta T(p_a \eta) \partial_\eta T (q_a \eta)\nonumber\\ 
    &&\hphantom{ \quad \frac{3 (1 + w)}{(5 + 3 w)^2} \biggl[ }
            + \frac{4}{\cH} \left(
                T(p_a \eta) \partial_\eta T(q_a \eta) + \partial_\eta T(p_a \eta) T(q_a \eta)
            \right)
        \biggr]\ .
\end{eqnarray}
Here, $\theta$ denotes the separation angle between $\mathbf{q}$ and $\mathbf{q}_i$, $\phi_i$ is the azimuthal angle of $\mathbf{q}_i$ when $\mathbf{q}$ is aligned with the $\mathbf{z}$-axis, and we use $p_a = |\bq - \bq_a|$ for brevity.
By substituting Eq.~\eqref{eq:S} into Eq.~\eqref{eq:h-G}, we obtain an expression for $h_{\lambda}$ with integration variables $\eta'$ and $\bq_a$. 
We first integrate over $\eta'$, obtaining a kernel function of the form
\begin{eqnarray}\label{eq:I-def}
    \hat{I}(\abs{\bq - \bq_a},q_a,\eta)
    &=& \int^{\eta}_{} \ud \eta' \, q^2 G_\bq(\eta, \eta') \frac{a(\eta')}{a(\eta)}  F(\abs{\bq - \bq_a}, q_a, \eta')\ .
\end{eqnarray}
Finally, we can express $h_{\lambda}$ in terms of $Q_{\lambda}$, $\hat{I}$, and $\zeta$, namely,
\begin{eqnarray}\label{eq:h-app}
    h_\lambda(\eta, \bq) 
    &=& 4 \int \frac{\ud^3 \bq_a}{(2\pi)^{3/2}} 
        \zeta(\bq_a) \zeta(\bq-\bq_a) Q_{\lambda}(\bq,\bq_a) \hat{I} (\abs{\bq - \bq_a},q,\eta)\ .
\end{eqnarray}
This is exactly the expression given in Eq.~\eqref{eq:h}.

\section{Feynman-like diagrams}\label{sec:FD}

\begin{figure*}
    \centering
    \includegraphics[width=1\linewidth]{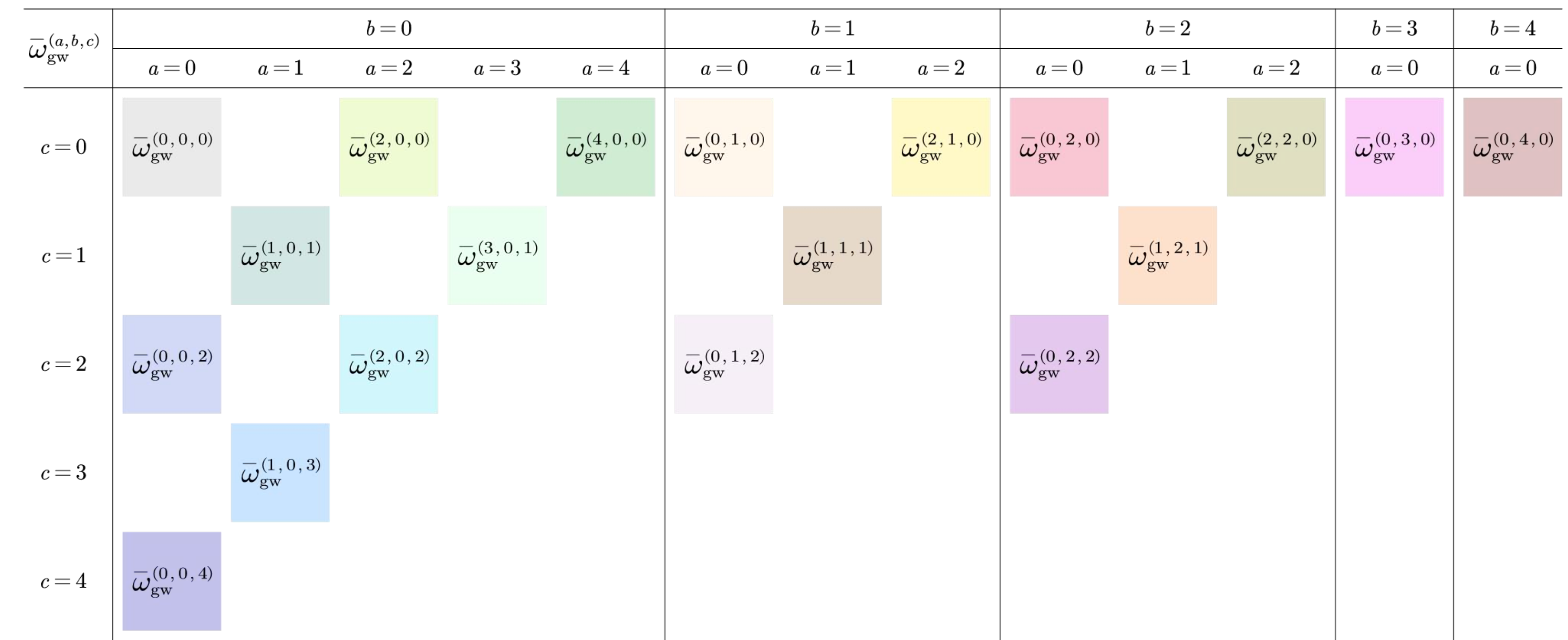}
    \caption{Colors of squares for different categories. }
    \label{fig:category}
\end{figure*}

\begin{figure*}
    \centering
    \includegraphics[width=1\linewidth]{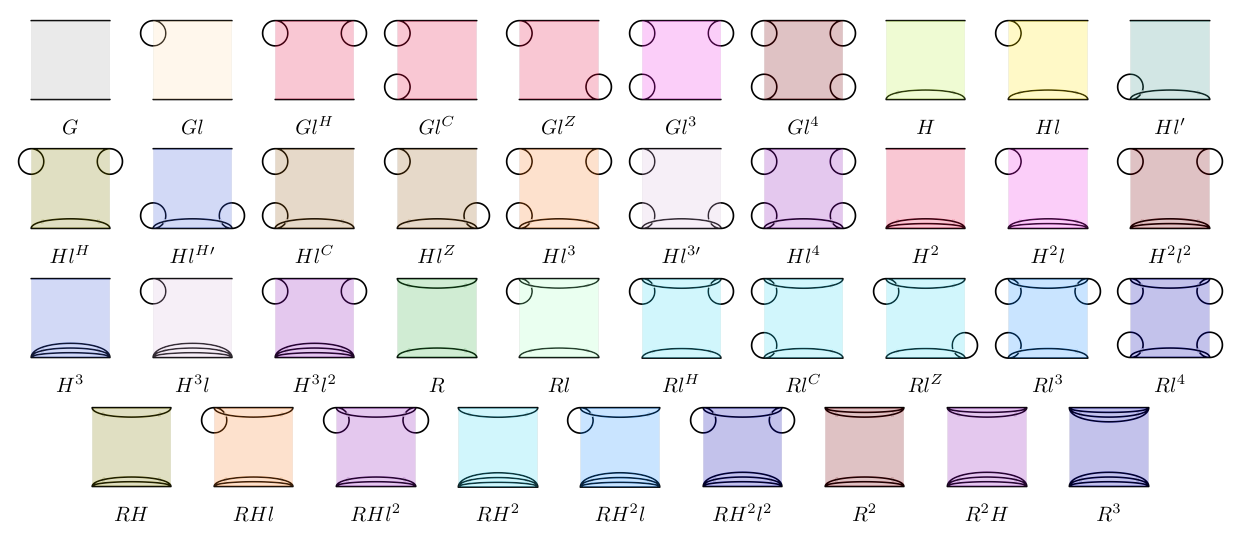}
    \caption{Specific diagrams for $G$-like family up to $\Hnl$ order. }
    \label{fig:G-like-FD}
\end{figure*}

\begin{figure*}
    \centering
    \includegraphics[width=1\linewidth]{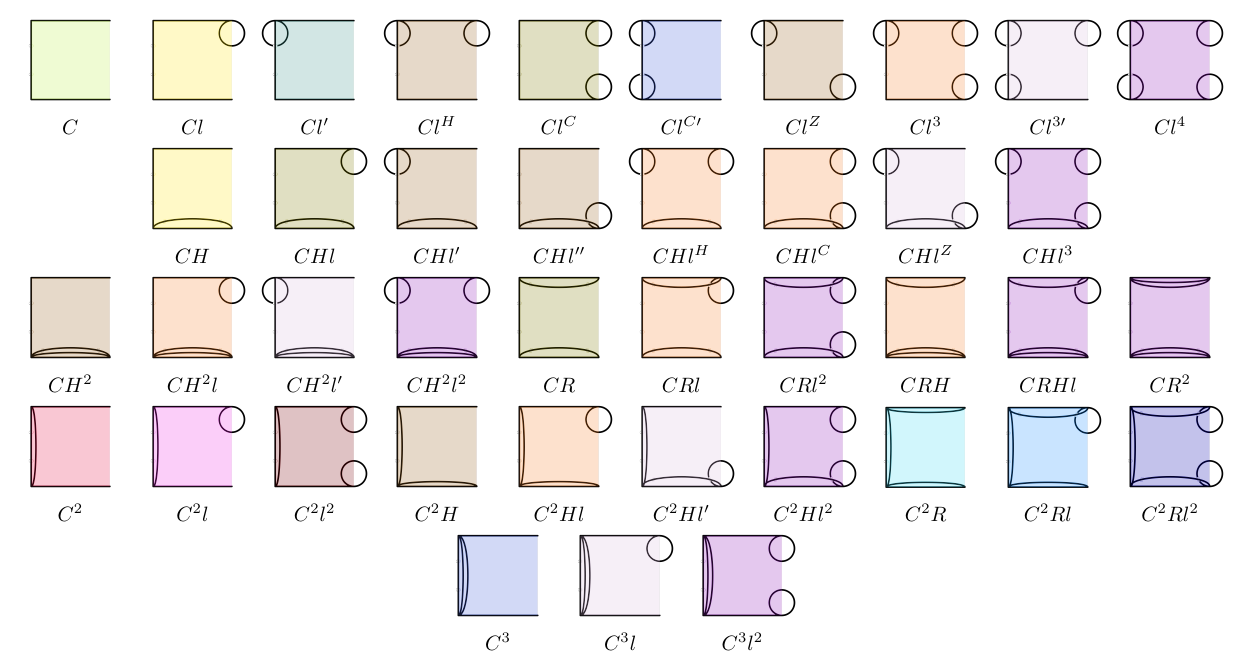}
    \caption{Specific diagrams for $C$-like family up to $\Hnl$ order. }
    \label{fig:C-like-FD}
\end{figure*}

\begin{figure*}
    \centering
    \includegraphics[width=1\linewidth]{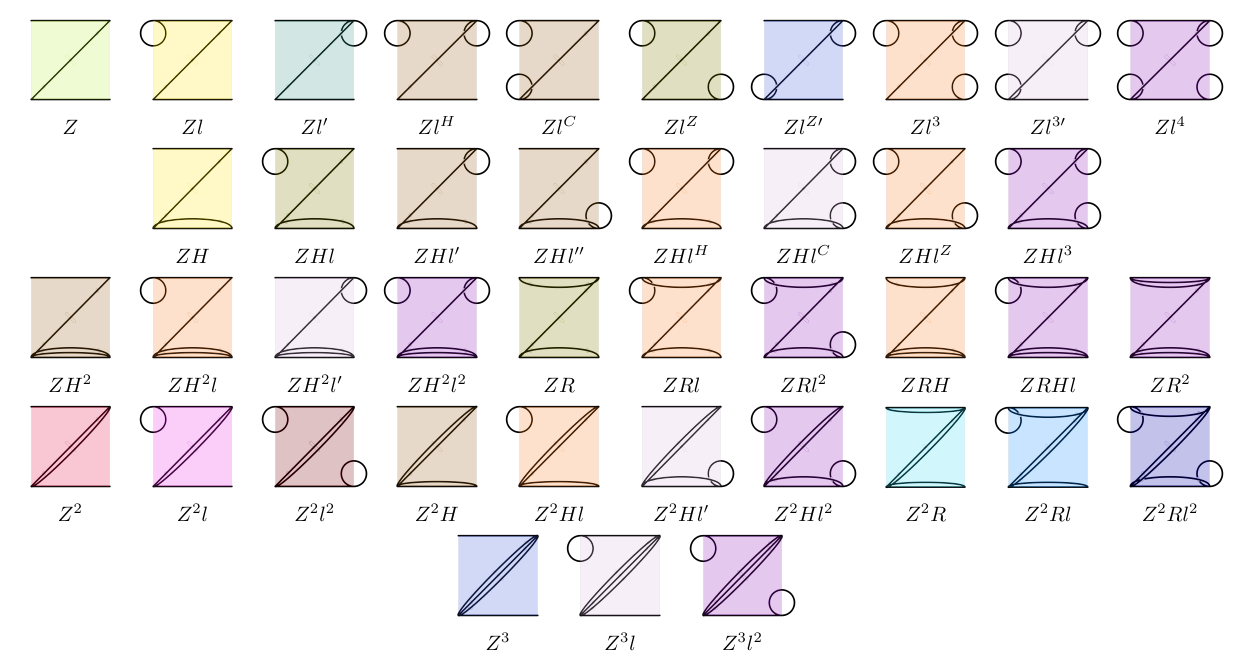}
    \caption{Specific diagrams for $Z$-like family up to $\Hnl$ order. }
    \label{fig:Z-like-FD}
\end{figure*}

\begin{figure*}
    \centering
    \includegraphics[width=1\linewidth]{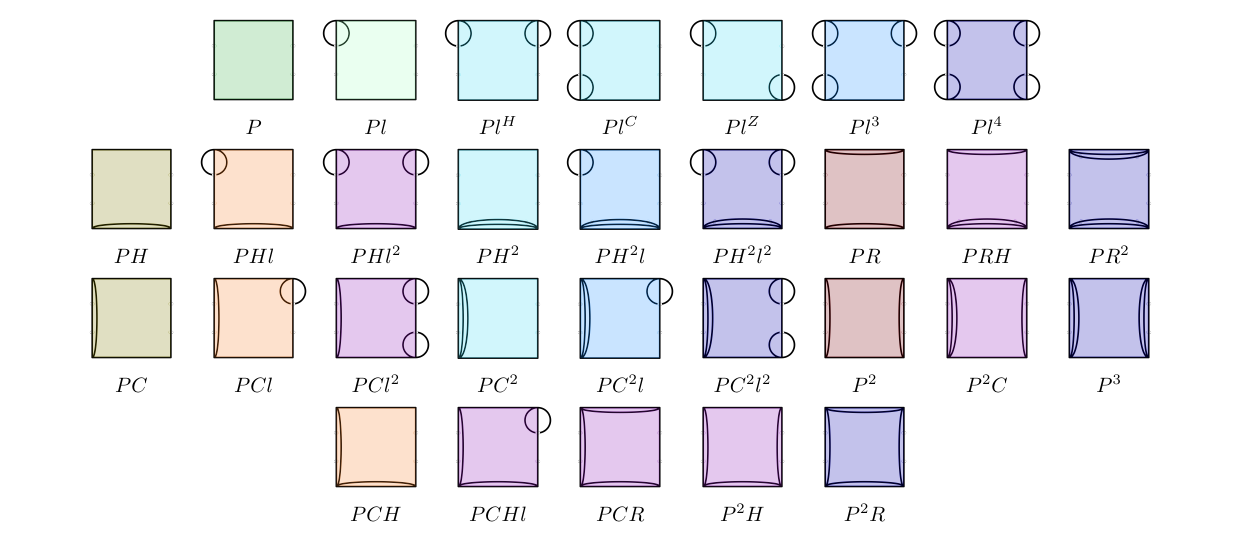}
    \caption{Specific diagrams for $P$-like family up to $\Hnl$ order. }
    \label{fig:P-like-FD}
\end{figure*}

\begin{figure*}
    \centering
    \includegraphics[width=1\linewidth]{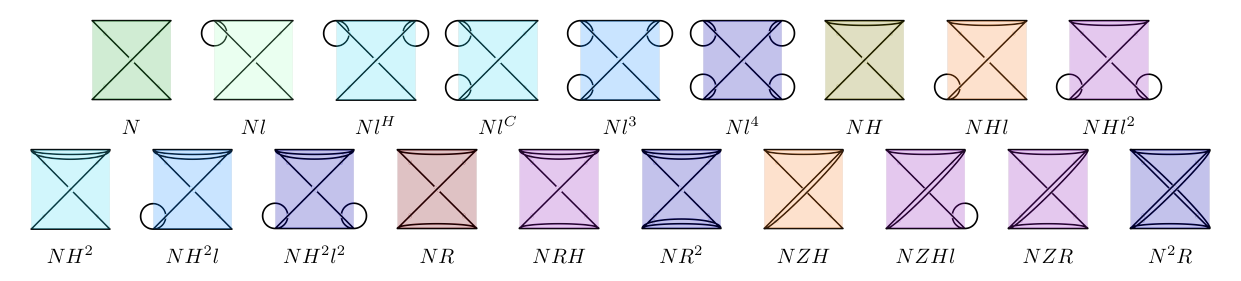}
    \caption{Specific diagrams for $N$-like family up to $\Hnl$ order. }
    \label{fig:N-like-FD}
\end{figure*}

\begin{figure*}
    \centering
    \includegraphics[width=1\linewidth]{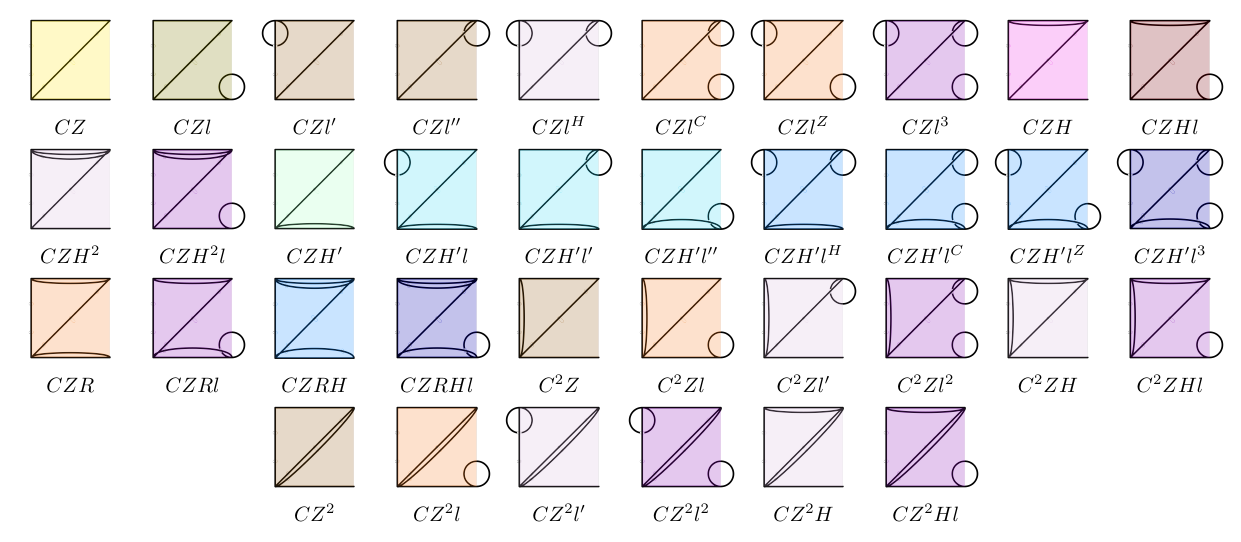}
    \caption{Specific diagrams for $CZ$-like family up to $\Hnl$ order. }
    \label{fig:CZ-like-FD}
\end{figure*}

\begin{figure*}
    \centering
    \includegraphics[width=1\linewidth]{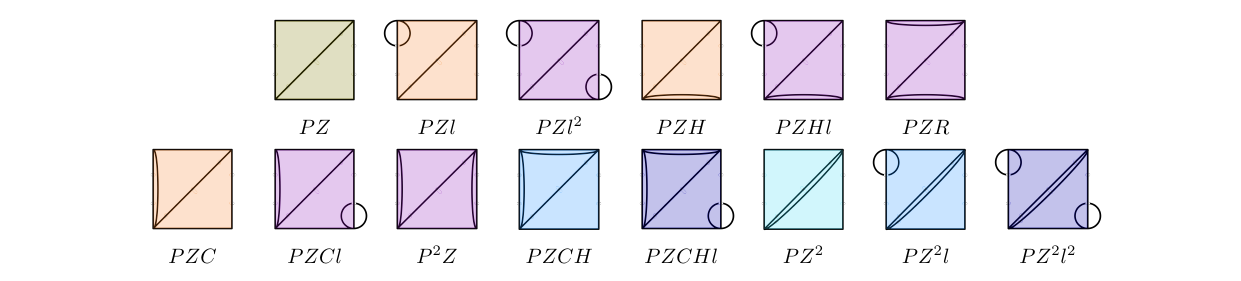}
    \caption{Specific diagrams for $PZ$-like family up to $\Hnl$ order. }
    \label{fig:PZ-like-FD}
\end{figure*}

\begin{figure*}
    \centering
    \includegraphics[width=1\linewidth]{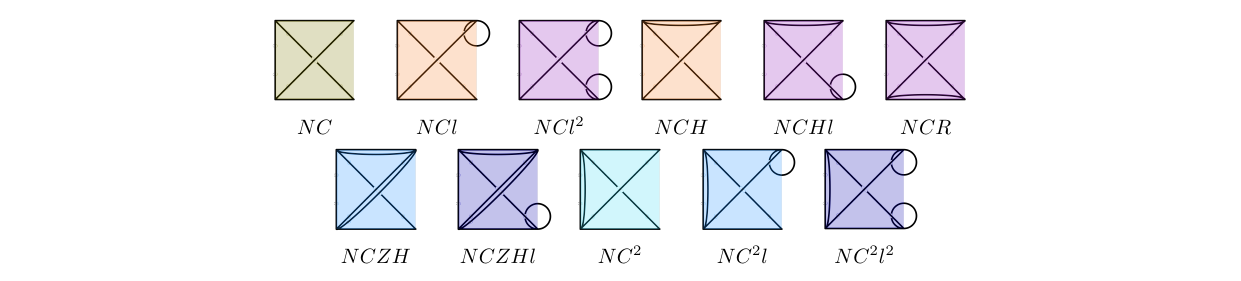}
    \caption{Specific diagrams for $NC$-like family up to $\Hnl$ order. }
    \label{fig:NC-like-FD}
\end{figure*}

\begin{figure*}
    \centering
    \includegraphics[width=1\linewidth]{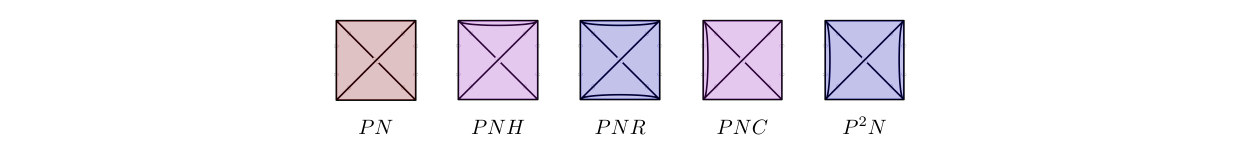}
    \caption{Specific diagrams for $PN$-like family up to $\Hnl$ order. }
    \label{fig:PN-like-FD}
\end{figure*}

In this appendix, we present specific Feynman-like diagrams of the average energy-density spectrum of \acp{SIGW} for local-type \ac{PNG} up to $\Hnl$ order. 
Based on the discussion in Section~\ref{sec:ED}, we construct all connections as colored squares within each family, with the figures displayed in \cref{fig:G-like-FD} - \cref{fig:PN-like-FD}. 
The shaded square in \cref{fig:omegabar-FD_Frame} can be substituted with any of the colored squares in \cref{fig:G-like-FD} - \cref{fig:PN-like-FD} to produce the corresponding complete Feynman-like diagrams, while the background colors of the squares indicate the categories to which the diagrams belong, as illustrated in \cref{fig:category}. 
Hence, each distinct square corresponds to an independent Feynman-like diagram for Wick's contractions of $\langle\zeta_\uS^4\rangle$, accounting for all equivalent contractions. 
As noted in Section~\ref{sec:ED}, only some of the categories contribute to the \ac{PNG}-induced inhomogeneities denoted by $\omega_{\uGW,\uin}^{(1)}$, $\omega_{\uGW,\uin}^{(2)}$ and $\omega_{\uGW,\uin}^{(3)}$, respectively. 
Diagrammatically, squares with specific background colors in \cref{fig:G-like-FD} to \cref{fig:PN-like-FD} can replace the shaded squares in \cref{fig:FD_Frame} to represent specific diagrams for \ac{PNG}-induced inhomogeneities. 
These diagrams can be further combined to generate all specific diagrams shown in \cref{fig:2-correlator}, \cref{fig:3-correlator}, and \cref{fig:4-correlator}. 
To provide clarity, we have created a table outlining the average energy-density spectrum for any specific category, $\bar{\omega}_{\uGW,\uin}^{(a,b,c)}$, in \cref{tab:order-FD}. 
This table includes information on the model parameters to which it is proportional, a list of all included diagrams, the background colors within these squares, and the total number of diagrams in this category. 
It also specifies whether each category contributes to the \ac{PNG}-induced inhomogeneities outlined in \cref{fig:FD_Frame}, using a checkmark (\checkmark) to denote contribution and a cross ($\times$) to indicate non-contribution. Importantly, although the red solid lines in the diagrams of $\omega_{\uGW,\uin,1}^{(2)}$ and $\omega_{\uGW,\uin,2}^{(3)}$ depicted in \cref{fig:FD_Frame} connect two vertices on one side, any two vertices are permissible as long as they collectively connect no more than four solid lines.

% \centering
\renewcommand\arraystretch{1.35}
\begin{landscape}
\begin{longtable}{c|c|c|c|c|cccccc@{ }}
    \caption{Table for illustration of the categories of SIGW energy-density fraction spectrum. The Parameter column shows the model parameters proportional to the category, while the Color column indicates the background color of the squares in the diagrams belonging to this category, as presented in \cref{fig:category}. The Diagram column lists all the diagrams included in each category, with the Num. column specifying the number of diagrams. Additionally, the last column indicates whether the category contributes to the \ac{PNG}-induced inhomogeneities in \cref{fig:FD_Frame}. }\label{tab:order-FD}\\
    \hline
    $\bar{\omega}_{\uGW,\uin}^{(a,b,c)}$ & Parameter & Color & Diagram & Num. & $\omega_\uGW^{(1)}$ & $\omega_{\uGW,1}^{(2)}$ & $\omega_{\uGW,2}^{(2)}$ & $\omega_{\uGW,1}^{(3)}$ & $\omega_{\uGW,2}^{(3)}$ & $\omega_{\uGW,3}^{(3)}$  \\
    \hline
    \endfirsthead
    \hline
    $\bar{\omega}_{\uGW,\uin}^{(a,b,c)}$ & Parameter & Color & Diagram-$D$ & Num. & $\omega_\uGW^{(1)}$ & $\omega_{\uGW,1}^{(2)}$ & $\omega_{\uGW,2}^{(2)}$ & $\omega_{\uGW,1}^{(3)}$ & $\omega_{\uGW,2}^{(3)}$ & $\omega_{\uGW,3}^{(3)}$  \\
    \hline
    \endhead
    \hline
    % \endfoot
    
    (0,0,0) & $A_\uS^2$                  & Grey        & $G$ & 1 & \checkmark & \checkmark & \checkmark & \checkmark & \checkmark & \checkmark \\
    \hline
    (2,0,0) & $\Fnl^2 A_\uS^3$           & Lime        & $H$, $C$, $Z$ & 3 & \checkmark & \checkmark & \checkmark & \checkmark & \checkmark & \checkmark \\
    \hline
    (0,1,0) & $\Gnl A_\uS^3$             & Apricot     & $Gl$ & 1 & \checkmark & \checkmark & \checkmark & \checkmark & \checkmark & \checkmark \\
    \hline
    (4,0,0) & $\Fnl^4 A_\uS^4$           & Green       & $R$, $P$, $N$ & 3 & \checkmark & \checkmark & \checkmark & \checkmark & \checkmark & $\times$ \\ 
    \hline
    (2,1,0) & $\Fnl^2 \Gnl A_\uS^4$      & Yellow      & $Hl$, $Cl$, $CH$, $Zl$, $ZH$, $CZ$ & 6 & \checkmark & \checkmark & \checkmark & \checkmark & \checkmark & \checkmark \\ 
    \hline
    (0,2,0) & $\Gnl^2 A_\uS^4$           & Red         & $Gl^H$, $Gl^C$, $Gl^Z$, $H^2$, $C^2$, $Z^2$ & 6 & \checkmark & \checkmark & \checkmark & \checkmark & \checkmark & \checkmark \\
    \hline
    (1,0,1) & $\Fnl \Hnl A_\uS^4$        & Teal        & $Hl'$, $Cl'$, $Zl'$ & 3 & \checkmark & \checkmark & \checkmark & \checkmark & \checkmark & \checkmark \\ 
    \hline
    \multirow{2}{*}{(2,2,0)} & \multirow{2}{*}{$\Fnl^2 \Gnl^2 A_\uS^5$} & \multirow{2}{*}{Olive} & $Hl^H$, $RH$, $Cl^C$, $CHl$, $CR$, $Zl^Z$, $ZHl$, & \multirow{2}{*}{14} & \multirow{2}{*}{\checkmark} & \multirow{2}{*}{\checkmark} & \multirow{2}{*}{\checkmark} & \multirow{2}{*}{\checkmark} & \multirow{2}{*}{\checkmark} & \multirow{2}{*}{$\times$} \\ & & & $ZR$, $PH$, $PC$, $NH$, $CZl$, $PZ$, $NC$ & & & & & & & \\ 
    \hline
    (0,3,0) & $\Gnl^3 A_\uS^5$           & Magenta     & $Gl^3$, $H^2 l$, $C^2 l$, $Z^2 l$, $CZH$ & 5 & \checkmark & \checkmark & \checkmark & \checkmark & \checkmark & \checkmark \\ 
          \hline
    (3,0,1) & $\Fnl^3 \Hnl A_\uS^5$      & Mint        & $Rl$, $Pl$, $Nl$, $CZH'$ & 4 & \checkmark & \checkmark & \checkmark & \checkmark & \checkmark & $\times$ \\ 
    \hline
    \multirow{3}{*}{(1,1,1)} & \multirow{3}{*}{$\Fnl \Gnl \Hnl A_\uS^5$} & \multirow{3}{*}{Brown} & $Hl^C$, $Hl^Z$, $Cl^H$, $Cl^Z$, $CHl'$, $CHl''$, & \multirow{3}{*}{18} & \multirow{3}{*}{\checkmark} & \multirow{3}{*}{\checkmark} & \multirow{3}{*}{\checkmark} & \multirow{3}{*}{\checkmark} & \multirow{3}{*}{\checkmark} & \multirow{3}{*}{\checkmark} \\ & & & $CH^2$, $C^2 H$, $Zl^H$, $Zl^C$, $ZHl'$, $ZHl''$, &  & & & & & & \\ & & & $ZH^2$, $Z^2 H$, $CZl'$, $CZl''$, $C^2 Z$, $CZ^2$ & & & & & & & \\ 
    \hline
    (0,0,2) & $\Hnl^2 A_\uS^5$           & Blue        & $Hl^{H\prime}$, $H^3$, $Cl^{C\prime}$, $C^3$, $Zl^{Z\prime}$, $Z^3$ & 6 & \checkmark & \checkmark & \checkmark & $\times$ & \checkmark & \checkmark \\ 
    \hline
    \multirow{2}{*}{(0,4,0)} & \multirow{2}{*}{$\Gnl^4 A_\uS^6$} & \multirow{2}{*}{Maroon} & $Gl^4$, $H^2 l^2$, $R^2$, $C^2 l^2$, $Z^2 l^2$, & \multirow{2}{*}{10} & \multirow{2}{*}{\checkmark} & \multirow{2}{*}{\checkmark} & \multirow{2}{*}{$\times$} & \multirow{2}{*}{\checkmark} & \multirow{2}{*}{$\times$} & \multirow{2}{*}{$\times$} \\ & & & $PR$, $P^2$, $NR$, $CZHl$, $PN$  & & & & & & & \\ 
    \hline
    \multirow{3}{*}{(1,2,1)} & \multirow{3}{*}{$\Fnl \Gnl^2 \Hnl A_\uS^6$} & \multirow{3}{*}{Orange} & $Hl^3$, $RHl$, $Cl^3$, $CHl^H$, $CHl^C$, $CH^2 l$, $CRl$, $CRH$, & \multirow{3}{*}{31} & \multirow{3}{*}{\checkmark} & \multirow{3}{*}{\checkmark} & \multirow{3}{*}{\checkmark} & \multirow{3}{*}{\checkmark} & \multirow{3}{*}{\checkmark} & \multirow{3}{*}{$\times$} \\ & & & $C^2 Hl$, $Zl^3$, $ZHl^H$, $ZHl^Z$, $ZH^2 l$, $ZRl$, $ZRH$, $Z^2 Hl$, & & & & & & & \\ & & & $PHl$, $PCl$, $PCH$, $NHl$, $NZH$, $CZl^C$, $CZl^Z$, & & & & & & & \\ & & & $CZR$, $C^2 Zl$, $CZ^2 l$, $PZl$, $PZH$, $PZC$, $NCl$, $NCH$ & & & & & & & \\ 
    \hline
    \multirow{3}{*}{(2,0,2)} & \multirow{3}{*}{$\Fnl^2 \Hnl^2 A_\uS^6$} & \multirow{3}{*}{Cyan} & $Rl^H$, $Rl^C$, $Rl^Z$, $RH^2$, $C^2 R$, $Z^2 R$, $Pl^H$, & \multirow{3}{*}{19} & \multirow{3}{*}{\checkmark} & \multirow{3}{*}{\checkmark} & \multirow{3}{*}{\checkmark} & \multirow{3}{*}{$\times$} & \multirow{3}{*}{\checkmark} & \multirow{3}{*}{$\times$} \\ & & &  $Pl^C$, $Pl^Z$, $PH^2$, $PC^2$, $Nl^H$, $Nl^C$, $NH^2$, &  & & & & & & \\ & & & $CZH'l$, $CZH'l'$, $CZH'l''$, $PZ^2$, $NC^2$ & & & & & & & \\ 
    \hline
    \multirow{3}{*}{(0,1,2)} & \multirow{3}{*}{$\Gnl \Hnl^2 A_\uS^6$} & \multirow{3}{*}{Lavender} & $Hl^{3\prime}$, $H^3 l$, $Cl^{3\prime}$, $CHl^Z$, $CH^2 l'$, $C^2 Hl'$, $C^3 l$, & \multirow{3}{*}{18} & \multirow{3}{*}{\checkmark} & \multirow{3}{*}{\checkmark} & \multirow{3}{*}{\checkmark} & \multirow{3}{*}{$\times$} & \multirow{3}{*}{\checkmark} & \multirow{3}{*}{\checkmark} \\ & & & $Zl^{3\prime}$, $ZHl^C$, $ZH^2 l'$, $Z^2 Hl'$, $Z^3 l$, $CZHl^H$, &  & & & & & & \\ & & & $CZH^2$, $C^2 Zl'$, $C^2 ZH$, $CZ^2 l'$, $CZ^2 H$ &  & & & & & & \\ 
    \hline
    \multirow{7}{*}{(0,2,2)} & \multirow{7}{*}{$\Gnl^2 \Hnl^2 A_\uS^7$} & \multirow{7}{*}{Purple} & $Hl^4$, $H^3 l^2$, $RHl^2$, $R^2 H$, $Cl^4$, $CHl^3$, $CH^2 l^2$, $CRl^2$, & \multirow{7}{*}{48} & \multirow{7}{*}{\checkmark} & \multirow{7}{*}{\checkmark} & \multirow{7}{*}{$\times$} & \multirow{7}{*}{$\times$} & \multirow{7}{*}{$\times$} & \multirow{7}{*}{$\times$} \\ & & & $CRHl$, $CR^2$, $C^2 Hl^2$, $C^3 l^2$, $Zl^4$, $ZHl^3$, $ZH^2 l^2$, &  & & & & & & \\ & & & $ZRl^2$, $ZRHl$, $ZR^2$, $Z^2 Hl^2$, $Z^3 l^2$, $PHl^2$, $PRH$, & & & & & & & \\ & & & $PCl^2$, $P^2 C$, $PCHl$, $PCR$, $P^2 H$, $NHl^2$, $NRH$, & & & & & & & \\ & & & $NZHl$, $NZR$, $CZl^3$, $CZH^2 l$, $CZRl$, $C^2 Zl^2$, & & & & & & & \\ & & & $C^2 ZHl$, $CZ^2 l^2$, $CZ^2 Hl$, $PZl^2$, $PZHl$, $PZR$, & & & & & & & \\ & & & $PZCl$, $P^2 Z$, $NCl^2$, $NCHl$, $NCR$, $PNH$, $PNC$ &  & & & & & & \\ 
    \hline
    \multirow{3}{*}{(1,0,3)} & \multirow{3}{*}{$\Fnl \Hnl^3 A_\uS^7$} & \multirowcell{3}{Dodger \\ Blue} & $Rl^3$, $RH^2 l$, $C^2 Rl$, $Z^2 Rl$, $Pl^3$, $PH^2 l$, $PC^2 l$, & \multirow{3}{*}{17} & \multirow{3}{*}{\checkmark} & \multirow{3}{*}{$\times$} & \multirow{3}{*}{\checkmark} & \multirow{3}{*}{$\times$} & \multirow{3}{*}{$\times$} & \multirow{3}{*}{$\times$} \\ & & & $Nl^3$, $NH^2 l$, $CZH'l^H$, $CZH'l^C$, $CZH'l^Z$, & & & & & & & \\ & & & $CZRH$, $PZCH$, $PZ^2 l$, $NCZH$, $NC^2 l$ &  & & & & & & \\ 
    \hline
    \multirow{4}{*}{(0,0,4)} & \multirow{4}{*}{$\Hnl^4 A_\uS^8$} & \multirow{4}{*}{Navy} & $Rl^4$, $RH^2 l^2$, $R^3$, $C^2 Rl^2$, $Z^2 Rl^2$, $Pl^4$, & \multirow{4}{*}{23} & \multirow{4}{*}{$\times$} & \multirow{4}{*}{$\times$} & \multirow{4}{*}{$\times$} & \multirow{4}{*}{$\times$} & \multirow{4}{*}{$\times$} & \multirow{4}{*}{$\times$} \\ & & & $PH^2 l^2$, $PR^2$, $PC^2 l^2$, $P^3$, $P^2 R$, $Nl^4$, $NH^2 l^2$, &  & & & & & & \\ & & & $NR^2$, $N^2 R$, $CZH'l^3$, $CZRHl$, $PZCHl$, &  & & & & & & \\ & & & $PZ^2 l^2$, $NCZHl$, $NC^2 l^2$, $PNR$, $P^2 N$ &  & & & & & & \\ 
    \hline 
\end{longtable}
\end{landscape}

\section{Numerical integrals}\label{sec:num-int}

In Subsection~\ref{subsec:ED-int}, we presented the integrals for all contributions to the \ac{SIGW} average energy density spectrum $\bar{\omega}_{\uGW, \uin}$. 
Although multiple integration over three-momenta is natural in theoretical analysis, it is inconvenient for numerical computation. 
Therefore, in this appendix, we provide methods to transform the integration variables, recasting the integrals into a form more suitable for numerical evaluation. 
The integrands, which consist of two projection factors, the oscillation average of two kernel functions, and several convolved propagators, are transformed accordingly.

Since the expressions for $\bar{\omega}_{\uGW,\uin}^{X-\mathrm{like}}$ in Eq.~\eqref{eq:omegabar-zeta} involve integration over at most five three-momenta, we first introduce five sets of variables $(u_i, v_i, \phi_i)$ with $i = 1, 2, 3, 4, 5$ to replace $\bq_i$, i.e.,  
\begin{equation}\label{eq:uv-def}
    v_i = \frac{q_i}{q}\ ,\qquad\qquad\quad
    u_i = \frac{\abs{\bq-\bq_i}}{q}\ ,
\end{equation}
and $\phi_i$ is the azimuthal angle of $\bq_i$, as defined below Eq.~\eqref{eq:Q-def}. 
We further transform $(u_i, v_i)$ to $(s_i, t_i)$ as follows,
\begin{equation}\label{eq:st-def}
    s_i = u_i - v_i\ ,\qquad\qquad\quad
    t_i = u_i + v_i -1\ ,
\end{equation}
ensuring that the integration limits for different variables are independent of each other. 
Hence, the integration measure transforms to  
\begin{equation}\label{eq:int-tf}
    \prod_i \int \frac{\ud^3 \bq_i}{(2\pi)^{3}} = \frac{q^3}{2(2\pi)^{3}} \prod_i \left(\int_0^\infty \ud t_i \int_{-1}^1 \ud s_i\, v_i u_i \int_0^{2\pi} \ud \phi_i\right)\ .
\end{equation}

Using the new variables, the $(\xi-1)$-fold convolved propagator $P^{[\xi]}(q)$, introduced in Eqs.~(\ref{eq:P1-def}, \ref{eq:Pxi-def}), becomes
\begin{eqnarray}
    P^{[1]} (q) &=& \frac{2\pi^2}{q^3} \frac{A_\uS}{\sqrt{2\pi\sigma^2}}\exp\left(-\frac{\ln^2 (q/q_\ast)}{2 \sigma^2}\right)\ ,\label{eq:P1} \\
    P^{[\xi]} (q) &=& \frac{q^3}{8 \pi^2} \int_0^\infty \ud \tilde{t} \int_{-1}^1 \ud \tilde{s}\, \tilde{v} \tilde{u} P^{[\xi-1]} (\tilde{v} q) P^{[1]} (\tilde{u} q)\ ,\ \text{where } \xi \geq 2, \xi \in \mathbb{N}_+\ ,\label{eq:Pxi}
\end{eqnarray}
where we have transformed $\tilde{\bq}$ into $(\tilde{v}, \tilde{u})$ in a manner similar to Eq.~\eqref{eq:uv-def}, and the transformations from $(\tilde{u}, \tilde{v})$ to $(\tilde{s}, \tilde{t})$ are analogous to Eq.~\eqref{eq:st-def}.
While $P^{[1]}(q)$ is analytic, the numerical results for $P^{[\xi]}(q)$ can be obtained directly by performing the numerical integration in Eq.~\eqref{eq:Pxi}. 
Next, we analyze the arguments involved in these convolved propagators. 
According to \cref{fig:renor_all}, the integral for $\bar{\omega}_{\uGW, \uin}^{G-\mathrm{like}}$ involves only $P^{[\xi]}(\bq_1)$ and $P^{[\xi]}(|\bq - \bq_1|)$, which can be expressed as $P^{[\xi]}(v_1 q)$ and $P^{[\xi]}(u_1 q)$ using Eqs.~\eqref{eq:uv-def}.
In other words, the integrand for $\bar{\omega}_{\uGW, \uin}^{G-\mathrm{like}}$ is independent of the azimuthal angles, resulting in a $2\pi$ factor after integrating over $\phi_1$. 
However, the integrals for other families in Eq.~\eqref{eq:omegabar-zeta} also involve $P^{[\xi]}(|\bq_i - \bq_j|)$ and $P^{[\xi]}(|\bq_i + \bq_j - \bq_k|)$. 
Thus, we introduce a new variable $y_{ij}$ to characterize the inner product of $\bq_i$ and $\bq_j$, i.e.,
\begin{eqnarray}\label{eqs:yij-def}
    y_{ij} &=& \frac{\bq_i \cdot \bq_j}{q^2} \nonumber\\
        &=& \frac{\cos\varphi_{ij}}{4}\sqrt{t_i (t_i + 2) (1 - s_i^2) t_j (t_j + 2) (1 - s_j^2)} 
        + \frac{1}{4}[1 - s_i (t_i + 1)][1 - s_j (t_j + 1)]\ ,
\end{eqnarray}
where we represent $\varphi_{ij} = \phi_i - \phi_j$ as the difference between the two azimuthal angles. 
Furthermore, we introduce quantities of the form  
\begin{eqnarray}\label{eqs:wij-def}
    w_{ij} &=& \frac{\abs{\bq_i-\bq_j}}{q} =  \sqrt{v_i^2 + v_j^2 - 2 y_{ij}}\ ,\\
    w_{ijk} &=& \frac{\abs{\bq_i+\bq_j-\bq_k}}{q}
        = \sqrt{v_i^2 + v_j^2 + v_k^2 + 2 y_{ij} - 2 y_{ik} - 2 y_{jk}}\ ,
\end{eqnarray}
so that $P^{[\xi]}(|\bq_i - \bq_j|)$ and $P^{[\xi]}(|\bq_i + \bq_j - \bq_k|)$ can be written as $P^{[\xi]}(w_{ij} q)$ and $P^{[\xi]}(w_{ijk} q)$, respectively. 
Consequently, $\cP^{X-\mathrm{like}}$ in Eq.~\eqref{eq:omegabar-zeta} can be expressed as a function of $v_i$, $u_i$, and $\phi_i$ for any family.

Ultimately, to rewrite the kernel function $\hat{I} \left(|\bq - \bq_i|, q_i, \eta\right)$ during the radiation-dominated era in terms of the variables $u_i$, $v_i$, and $x = q\eta$, we recast it in the following form
\begin{equation}\label{eq:Ihat-Iuv}
    \hat{I} \left(\abs{\bq - \bq_i}, q_i, \eta_\uin\right) 
        = I_{\uRD} \left(\frac{\abs{\bq - \bq_i}}{q}, \frac{q_i}{q}, q\eta_\uin\right)
        = I_{\uRD} (u_i, v_i, x)\ .
\end{equation}
The integral for $\hat{I} \left(|\bq - \bq_i|, q_i, \eta\right)$ in Eq.~\eqref{eq:I-def} can then be rewritten with respect to $u_i$ and $v_i$ accordingly. 
As elaborated in Refs.~\cite{Kohri:2018awv,Espinosa:2018eve}, this integral admits an analytical result on subhorizon scales ($x_\uin \gg 1$), given by
\begin{subequations}\label{eq:I-RD} 
\begin{eqnarray}
    I_{\uRD} (u_i, v_i, x \gg 1) &=& \frac{1}{x} I_A (u_i, v_i) 
            \left[I_B (u_i, v_i) \sin x - \pi I_C (u_i, v_i) \cos x\right]\ ,\\
    I_A (u_i, v_i) &=& \frac{3\left(u_i^2 + v_i^2 - 3\right)}{4 u_i^3 v_i^3}\ , \\
    I_B (u_i, v_i) &=& -4 u_i v_i  
        + \left(u_i^2 + v_i^2 - 3\right) \ln \abs{\frac{3 - (u_i + v_i)^2}{3 - (u_i - v_i)^2}}\ , \\
    I_C (u_i, v_i) &=& \left(u_i^2 + v_i^2 - 3\right)
        \Theta\left(u_i + v_i - \sqrt{3}\right)\ .
\end{eqnarray} 
\end{subequations} 
Furthermore, using this formula for $I_{\uRD}(u_i, v_i, x \gg 1)$, we can express the oscillation average of two kernel functions with the same $x$ as \cite{Adshead:2021hnm}
\begin{eqnarray}\label{eq:I-ave-12}
    && \overbar{I_\uRD (u_i, v_i, x \gg 1) 
        I_\uRD (u_j, v_j, x \gg 1)} \nonumber\\
        & = & \frac{I_A (u_i, v_i) I_A (u_j, v_j)}{2 x^2} 
            \left[
                I_B (u_i, v_i) I_B (u_j, v_j)
                + \pi^2 I_C (u_i, v_i) I_C (u_j, v_j)
            \right]\ .
\end{eqnarray}
As for the projection factor defined in Eq.~\eqref{eq:Q-def}, we can deduce $\cos \theta = (1 + v_i^2 - u_i^2) / (2 v_i)$ from Eq.~\eqref{eq:uv-def} and rewrite $Q_{\lambda}(\bq, \bq_i)$ in terms of $(u_i, v_i, \phi_i)$ as follows  
\begin{equation}\label{eq:Q-uv}
    Q_{\lambda}(\bq, \bq_i)
    = \frac{\bigl[(v_i + u_i)^2 - 1\bigr] \bigl[1 - (v_i - u_i)^2\bigr]}{\sqrt{2}}
     \times
        \begin{cases}
            \cos(2\phi_i) &\lambda = + \\
            \sin(2\phi_i) &\lambda = \times 
        \end{cases}\ . 
\end{equation}   
For brevity, we introduce a new function that combines the kernel function with the projection factor as follows \begin{equation}\label{eq:J-def}
    J (u_i, v_i, x) 
    = \frac{x}{8}\bigl[(v_i + u_i)^2 - 1\bigr] \bigl[1 - (v_i - u_i)^2\bigr] I_{\uRD} (u_i, v_i, x)\ . 
\end{equation}
Using this function, the two projection factors and two kernel functions in the integrand of Eq.~\eqref{eq:omegabar-zeta} can be expressed as  
\begin{eqnarray}
    \sum_\lambda Q_{\lambda}^2 (\bq, \bq_1) \overbar{\hat{I}^2 (\abs{\bq - \bq_1}, q_1, \eta)} &=& \frac{1}{x^2} \overbar{J^2 (u_1, v_1, x)}\ ,\label{eq:QI-J} \\
    \sum_\lambda Q_{\lambda}(\bq, \bq_1) Q_{\lambda}(\bq, \bq_2) \overbar{\hat{I} (\abs{\bq - \bq_1}, q_1, \eta) \hat{I} (\abs{\bq - \bq_2}, q_2, \eta)} 
    &=& \frac{\cos (2\varphi_{12})}{x^2} \overbar{J(u_1, v_1, x) J(u_2, v_2, x)} \ .\nonumber\\ \label{eq:QQII-JJ}
\end{eqnarray}
Consequently, the integrands in Eq.~\eqref{eq:omegabar-zeta} can be written in terms of $P^{[\xi]}$ and $J$, with the integration variables being $u_i$, $v_i$, and $\varphi_{ij}$ (or equivalently $s_i$, $t_i$, and $\varphi_{ij}$), where $1 \leq i, j \leq 5$.

\section{Infrared scaling of the SIGW energy-density spectra}\label{sec:IR}

In this appendix, we present a verification of the infrared scaling of the \ac{SIGW} energy density spectra discussed in Subsection~\ref{subsec:IR}. 
As shown in Appendix~\ref{sec:num-int}, the integrand of the energy density fraction spectrum in Eq.~\eqref{eq:omegabar-zeta} is primarily composed of two elements: $P^{[\xi]} (q)$, as defined in Eqs.~(\ref{eq:P1}, \ref{eq:Pxi}), and $J(u_i,v_i,x)$, given in Eq.~\eqref{eq:J-def}. 
We first analyze the behavior of $P^{[\xi]} (q)$. Subsequently, we examine the behavior of the integrals associated with the various diagrams contributing to the \ac{SIGW} energy density spectrum.

For the $(\xi-1)$-fold convolved propagator, while $P^{[1]}$ in Eq.~\eqref{eq:P1} is analytic, we focus on the infrared behavior of $P^{[\xi]}$ for $\xi \geq 2$. 
As indicated in Eq.~\eqref{eq:Pxi}, $P^{[\xi]}(q)$ is represented by convolution integrals involving $\tilde{u} P^{[1]}(\tilde{u} q)$. 
To examine the behavior of $P^{[\xi]}(q)$, we first introduce certain approximations in the integration involving $\tilde{u} P^{[1]}(\tilde{u} q)$. Since $\tilde{u} P^{[1]}(\tilde{u} q)$ is predominantly localized around the pivot wavenumber $\tilde{u} q = q_\ast e^{-2 \sigma^2}$, we introduce a parameter $\varepsilon$ to delineate the peak, infrared, and ultraviolet regions of $P^{[1]}(q)$. 
The region satisfying $q < q_- = q_\ast e^{-(2+\varepsilon)\sigma^2}$ is defined as the infrared region, while $q > q_+ = q_\ast e^{-(2-\varepsilon)\sigma^2}$ corresponds to the ultraviolet region. 
The intermediate region, $q_- \leq q \leq q_+$, is identified as the peak region. 
For $P^{[1]}(q)$ in the integrand of Eq.~\eqref{eq:Pxi}, we neglect the contributions from both the infrared and ultraviolet regions and retain only those from the peak region $q_- \leq q \leq q_+$, thereby adjusting the integration limits. 
Furthermore, we assume $\varepsilon \geq 5$ to ensure the accuracy of this approximation. 
For $\sigma=1$ and $\varepsilon=5$ in this work, the infrared region is given by $q/q_\ast \lesssim 9.1 \times 10^{-4}$. 
Nevertheless, to maintain generality, we leave $\varepsilon$ as a free parameter in the following analysis.

\begin{figure}
    \centering
    \includegraphics[width=1\linewidth]{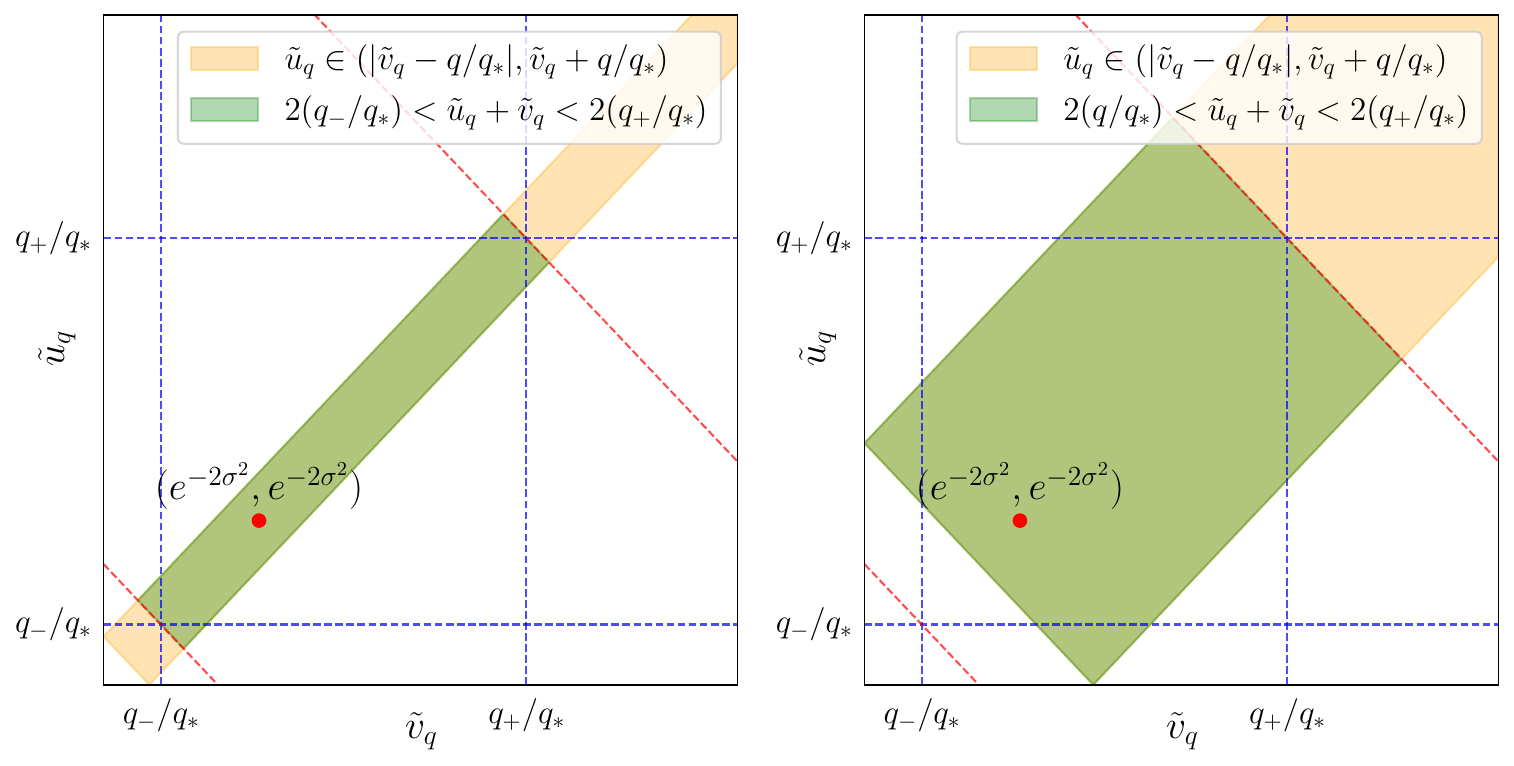}
    \caption{Approximation of the integration domain of Eq.~\eqref{eq:P2-approx}, where the left panel showcases the case where $q \leq 2q_-$, while the right panel illustrates the scenario where $2q_- < q < 2q_+$. The orange region represents the original integration domain and the green region indicates the approximated integration domain. The red point signifies the location where the integrand achieves its maximum value. The region between the two red dashed lines denotes the peak region $(2q_- / q_\ast < \tilde{v}_q + \tilde{u}_q < 2q_+ / q_\ast)$ for the integrand, with $\varepsilon$ taken as a very small constant for clarification. }
    \label{fig:Int_Domain}
\end{figure}

Using this approximation, the integral for $P^{[2]}(q)$ can be written as  
\begin{eqnarray}\label{eq:P2-approx}
    P^{[2]} (q) &\simeq& \frac{q^3}{8 \pi^2} \int_{\max(2 q_- / q - 1, 0)}^{\max(2 q_+ / q - 1, 0)} \ud \tilde{t} \int_{-1}^1 \ud \tilde{s}\, \tilde{v} \tilde{u} P^{[1]} (\tilde{v} q) P^{[1]} (\tilde{u} q) \nonumber\\ 
    &=& \frac{\pi}{4 \sigma^2 q_\ast^2 q} \int_{\max(2q_- / q_\ast - q / q_\ast, 0)}^{\max(2 q_+ / q_\ast - q / q_\ast,0)} \ud \tilde{t}_q \int_{-q / q_\ast}^{q / q_\ast} \ud \tilde{s}_q\, \frac{1}{(\tilde{v}_q \tilde{u}_q)^2} \exp \left(-\frac{\ln^2 \tilde{v}_q + \ln^2 \tilde{u}_q}{2\sigma^2}\right) \ ,
\end{eqnarray}
where $\tilde{u}_q = \tilde{u} q / q_\ast$, $\tilde{v}_q = \tilde{v} q / q_\ast$, $\tilde{s}_q = \tilde{s} q / q_\ast$, and $\tilde{t}_q = \tilde{t} q / q_\ast$, while the relationships between $(\tilde{u}, \tilde{v})$ and $(\tilde{s}, \tilde{t})$ are analogous to those between $(u_i, v_i)$ and $(s_i, t_i)$ in Eq.~\eqref{eq:st-def}.  
Since $\tilde{s}_q$ and $\tilde{t}_q$ are orthometric, the area of the approximated integration domain is  
\begin{equation}\label{eq:S-varepsilon}
    S(\varepsilon) 
    = 2 \frac{q}{q_\ast} \left(\max(2 q_+ - q,0) - \max(2q_- - q, 0)\right) 
    = \begin{cases}
        4 (q_+ - q_-) q \quad & q \leq 2 q_- \\ 
        2 (2q_+ - q) q & 2 q_- < q < 2 q_+ \\
        0  & q \geq 2 q_+
    \end{cases}\ .
\end{equation}
In particular, $S(\varepsilon) = 0$ indicates that the value of the integral is negligible and can be approximated as zero for $q \geq 2 q_+$.
We can then employ the first mean value theorem to estimate the value of the integral. 
Before proceeding, we adopt $\tilde{u}_q$ and $\tilde{v}_q$ in the subsequent analysis for simplification, transforming the integration domain to $|\tilde{v}_q - q/q_\ast| < \tilde{u}_q < \tilde{v}_q + q/q_\ast$ with $\tilde{v}_q > 0$. 
The approximated integration domain with respect to these variables is illustrated in \cref{fig:Int_Domain}, which visualizes the domain for different values of $q$ in the following analysis. 
Applying the first mean value theorem, there exists a specific point $(\tilde{v}_q^*, \tilde{u}_q^*)$ within the integration domain that allows the integral to be simplified as  
\begin{equation}\label{eq:P2-p}
    P^{[2]} (q) \simeq \frac{S(\varepsilon)}{2q} \frac{\pi}{2 \sigma^2 q_\ast^2} \frac{1}{(\tilde{v}_q^\ast \tilde{u}_q^\ast)^2} \exp \left(-\frac{\ln^2 \tilde{v}_q^\ast + \ln^2 \tilde{u}_q^\ast}{2\sigma^2}\right) \ .
\end{equation}
Here and hereinafter, we use the superscript $^*$ to denote a particular value of the integration variable within the integration interval. 
Moreover, as long as our approximation remains valid, the contributions from outside the peak region are negligible. Thus, we expect that this point $(\tilde{v}_q^*, \tilde{u}_q^*)$ lies within the square region $(q_-/q_\ast, q_+/q_\ast) \times (q_-/q_\ast, q_+/q_\ast)$. 
In this region, it is straightforward to see that the integrand attains its minimum at $\tilde{u}_q = \tilde{v}_q = q_+/q_\ast$ for $q < 2q_+$ (or equivalently, at $\tilde{u}_q = \tilde{v}_q = q_-/q_\ast$ for $q \leq 2q_-$), and achieves its maximum at $\tilde{u}_q = \tilde{v}_q = e^{-2\sigma^2}$.
Since the integrand is continuous in the domain, we further propose that $(\tilde{v}_q^*, \tilde{u}_q^*)$ can be located on the line $\tilde{u}_q = \tilde{v}_q$ within the approximated integration domain. 
Along this line, as $\tilde{u}_q = \tilde{v}_q$ decreases from $q_+/q_\ast$, the value of the integrand increases and reaches a maximum at $e^{-2\sigma^2}$. 
Therefore, for a given $q$, we further postulate that  
\begin{eqnarray}\label{eqs:vs-us-def}
    \tilde{v}_q^\ast = \tilde{u}_q^\ast = e^{-(2-c) \sigma^2}\ , \qquad
\end{eqnarray}
where $c \in (0, \varepsilon)$. 
Substituting Eqs.~(\ref{eq:S-varepsilon}, \ref{eqs:vs-us-def}) into Eq.~\eqref{eq:P2-p}, we can rewrite $P^{[2]}(q)$ as  
\begin{equation}\label{eq:P2}
    P^{[2]} (q) \simeq \frac{\pi e^{(2 - c^2) \sigma^2} }{2 \sigma^2 q_\ast^2} 
    \times 
    \begin{cases}
        2 q_\ast (e^{\varepsilon\sigma^2} - e^{-\varepsilon\sigma^2})  \quad & q \leq 2 q_- \\ 
        (2 q_\ast e^{\varepsilon\sigma^2} - q) & 2 q_- < q < 2 q_+ \\ 
        0  & q \geq 2 q_+
    \end{cases}\ .
\end{equation}
Strictly speaking, $c$ is a variable dependent on $q$, since the first mean value theorem is applied for a specific value of $q$. 
According to Eq.~\eqref{eq:P2-p}, we may use $2q P^{[2]}(q) / S(\varepsilon)$, which can be interpreted as the average line density of $P^{[2]}(q)$ with respect to $\tilde{t}_q$, to assess how $q$ influences the choice of the specific point $(\tilde{v}_q^*, \tilde{u}_q^*)$. 
In the region where $q < 2q_\ast e^{-2\sigma^2}$, the approximated integration domain spans the entire peak region of the integrand, as illustrated in the left panel of \cref{fig:Int_Domain}. 
Moreover, the integrand in Eq.~\eqref{eq:P2-approx} attains its maximum at $\tilde{s}_q = 0$ for any $\tilde{t}_q$ and varies slowly with respect to $\tilde{s}_q$. 
Hence, $2q P^{[2]}(q) / S(\varepsilon)$ decreases only slightly as $q$ increases. 
However, this effect is negligible in the infrared region where $q$ is very small. 
In other words, $P^{[2]}(q)$ is nearly scale-invariant for $q \leq 2q_-$.
For $2q_- < q < 2q_+$, as shown in the right panel of \cref{fig:Int_Domain}, the region between the two red dashed lines exceeds the size of the green region, implying that the approximated integration domain covers only part of the peak region. 
Furthermore, the occupied area decreases as $q$ increases, indicating that $P^{[2]}(q)$ decreases monotonically with increasing $q$. 
Specifically, $P^{[2]}(q)$ appears to decrease more rapidly with increasing $q$ when $q < 2q_\ast e^{-2\sigma^2}$.
Therefore, we may further approximate $P^{[2]}(q)$ as scale-invariant for $q \leq 2\sqrt{q_-q_+} = 2q_\ast e^{-2\sigma^2}$, and as zero for $q > 2\sqrt{q_-q_+} = 2q_\ast e^{-2\sigma^2}$, i.e.,
\begin{equation}\label{eq:P2-simplify}
    P^{[2]} (q) \simeq 
    \begin{cases}
        P^{[2]}_\mathrm{IR} \quad & q \leq 2 q_\ast e^{-2\sigma^2} \\
        0 & q > 2 q_\ast e^{-2\sigma^2}
    \end{cases} \ ,
\end{equation}
where $P^{[2]}_\mathrm{IR}$ is a constant representing the infrared limit of $P^{[2]}(q)$.

\begin{figure}
    \centering
    \includegraphics[width=1\linewidth]{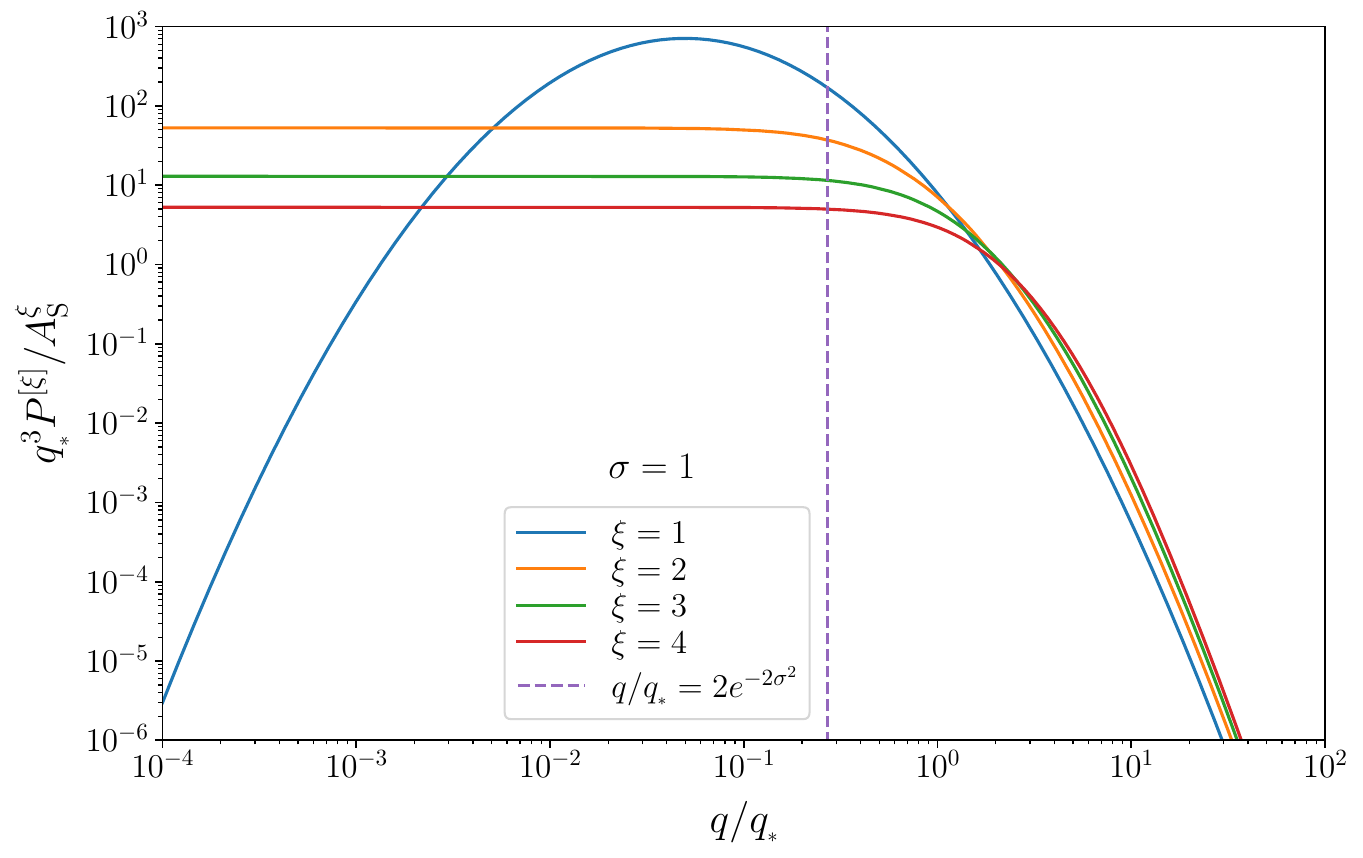}
    \caption{Numerical results for $P^{[\xi]} (q)$. }
    \label{fig:Pxi}
\end{figure}

According to the above approximation in Eq.~\eqref{eq:P2-simplify}, we can express $P^{[3]}(q)$ as  
\begin{eqnarray}
    P^{[3]} (q) &\simeq& \frac{q^3 P^{[2]}_\mathrm{IR}}{4 \pi^2} \int_{0}^{e^{-2\sigma^2} q_\ast / q} \tilde{v} \ud \tilde{v} \int_{|1-\tilde{v}|}^{1+\tilde{v}} \tilde{u} \ud \tilde{u}\, P^{[1]} (\tilde{u} q) \nonumber\\
    &=& \frac{e^{\sigma^2 / 2} P^{[2]}_\mathrm{IR}}{2 \sqrt{2 \pi} \sigma} \frac{q_\ast}{q} \int_{0}^{2 e^{-2\sigma^2}} \tilde{v}_q \ud \tilde{v}_q \int_{\ln |q / q_\ast -\tilde{v}_q|)}^{\ln (q / q_\ast + \tilde{v}_q)} \ud a \exp\left[-\left(\frac{a+\sigma^2}{\sqrt{2} \sigma}\right)^2\right]\ . 
\end{eqnarray}
The result of this integral can be written in terms of error functions. 
Despite its lengthy form, $P^{[3]}(q)$ remains nearly scale-invariant for $q/q_\ast < 2 e^{-2\sigma^2}$ and decreases rapidly for $q/q_\ast \geq 2 e^{-2\sigma^2}$. 
Thus, similar to the simplification of $P^{[2]}(q)$ in Eq.~\eqref{eq:P2-simplify}, $P^{[3]}(q)$ can be approximated in the same way.
In particular, the infrared limit of $P^{[3]}(q)$ can be written as  
\begin{equation}
    \frac{P^{[3]}_\mathrm{IR}}{P^{[2]}_\mathrm{IR}} 
    \simeq \frac{1}{2} + \frac{1}{\sqrt{\pi}} \int_{0}^{\ln(2 e^{-2\sigma^2})} e^{-\chi^2 / (2 \sigma^2)} \ud \chi\ . 
\end{equation}
For $\sigma = 1$ in this work, we have $P^{[3]}_\mathrm{IR} / P^{[2]}_\mathrm{IR} \simeq 0.1$. 
By employing induction, we conclude that for $\xi \geq 2$, the behavior of $P^{[\xi]}(q)$ given by Eq.~\eqref{eq:Pxi} exhibits similar features.
The above analyses can be validated by the numerical results, as shown in \cref{fig:Pxi}. 
In this figure, it is evident that $P^{[\xi]}(q)$ for $\xi \geq 2$, as anticipated, is nearly scale-invariant in the infrared region. 
Additionally, although our analysis predicts a step at $2e^{-2\sigma^2}$ for $\xi \geq 2$, the figure indicates that $P^{[\xi]}(q)$ begins to decrease rapidly at a higher wavenumber as $\xi$ increases. 
Hence, $P^{[\xi+1]}_\mathrm{IR} / P^{[\xi]}_\mathrm{IR}$ for $\xi \geq 2$ is expected to increase with increasing $\xi$. 
In fact, the numerical results show that $P^{[3]}_\mathrm{IR} / P^{[2]}_\mathrm{IR} \simeq 0.24$ and $P^{[4]}_\mathrm{IR} / P^{[3]}_\mathrm{IR} \simeq 0.41$. 
Despite these differences, this approximation remains sufficiently accurate for the subsequent analysis.

Next, we analyze the behavior of the full integrand in the infrared region. 
This analysis is analogous to those in previous works~\cite{Yuan:2019wwo,Yuan:2023ofl,Cai:2018dig,Cai:2019cdl}, which have discussed similar features in the infrared region for \acp{SIGW} without \ac{PNG} or with lower-order \ac{PNG}. 
As shown in Appendix~\ref{sec:num-int}, in the infrared region where $v_i q, u_i q \gg q_\ast$, $v_i$ and $u_i$ are of the same order due to the relation $|u_i - v_i| < 1$ for any $i$. 
Similarly, $w_{ij}$ and $w_{ijk}$, defined in Eqs.~(\ref{eqs:wij-def}), are also of $\cO(v_i)$. 
Moreover, in the integral for the \ac{SIGW} energy-density fraction spectrum, i.e., Eq.~\eqref{eq:omegabar-zeta}, each integration over a three-momentum contributes a factor of $\cO(v_i^2)$ according to Eq.~\eqref{eq:int-tf}.
Since $P^{[\xi]}(q)$ is nearly scale-invariant in the infrared region for $\xi \geq 2$, we define its dimensionless form as  
\begin{equation}
    \Delta^2_{[\xi]}(q) = \frac{q^3}{2 \pi^2} P^{[\xi]}(q) \ ,
\end{equation}
where $\Delta^2_{[\xi]}(q) = \Delta^2_\ugS(q)$ is given by Eq.~\eqref{eq:Lognormal}. 
For any $\xi$, $\Delta^2_{[\xi]}(q)$ exhibits a single peak. 
This peak is located at $q=q_\ast$ for $\xi=1$, and is situated near $2q_\ast e^{-2\sigma^2}$ for $\xi \geq 2$.
Therefore, for simplicity, we continue to use $q_- \leq v_i q \leq q_+$ as the region that predominantly contributes to the integral from $\Delta^2_{[\xi]}(v_i q)$. 
We do not further specify the values of $q_-$ and $q_+$, treating them as fixed parameters. 
With this approximation, the integration limit is set to $[q_-/q, q_+/q]$ with respect to the variable $v_i$. 
To eliminate $q$ from the integration limit, we transform the variable $v_i$ to $v_{q,i} = v_i q/q_\ast$, recasting the integration limit as $[q_-/q_\ast, q_+/q_\ast]$. 
Similarly, $u_i$ is transformed to $u_{q,i} = u_i q/q_\ast$. 
Both $u_{q,i}$ and $v_{q,i}$ are of order one when the integrand is sizable, regardless of $\xi$.

In the infrared region where $q \ll q_-$, we find that $v_i \geq q_-/q \gg 1$. 
Consequently, $u_i$ and all the relevant $w_{ij}$ and $w_{ijk}$ are much greater than one. 
Therefore, the kernel functions introduced in Eq.~\eqref{eq:I-RD} can be approximated as  
\begin{eqnarray}\label{eq:I-RD-IR}
    I_{\uRD}^\mathrm{IR} (u_i,v_i,x \gg 1) \simeq \frac{1}{x} \frac{3\left(u_i^2 + v_i^2\right)^2}{4 u_i^3 v_i^3} \left[\sin x \ln \frac{(u_i + v_i)^2}{3} - \pi \cos x\right]\ .
\end{eqnarray}
Then, $J(u_i, v_i, x)$ can be expressed as  
\begin{equation}\label{eq:J-IR}
    J_\mathrm{IR} (u_i,v_i,x) 
    = \frac{x}{8}(v_i+u_i)^2 I_{\uRD}^\mathrm{IR} (u_i,v_i,x)\ . 
\end{equation}
Notably, after the transformation of variables $(u_i, v_i) \rightarrow (u_{q,i}, v_{q,i})$, we have $J_\mathrm{IR}(u_i, v_i, x) = J_\mathrm{IR}(u_{q,i}, v_{q,i}, x)$. 
Therefore, by substituting Eqs.~(\ref{eq:I-RD-IR}, \ref{eq:J-IR}) into Eqs.~(\ref{eq:QI-J}, \ref{eq:QQII-JJ}), we find that for $u_{q,i} \sim v_{q,i} \sim 1$,  
\begin{eqnarray}\label{eq:J-IR-propto}
    J_\mathrm{IR} (u_{q,i},v_{q,i},x) \propto \ln^2 \frac{(u_{q,1} + v_{q,1})^2 q_\ast^2}{3 q^2} 
    \simeq \ln^2 \frac{4 q_\ast^2}{3 q^2} \ .
\end{eqnarray}

According to the above analysis, we can evaluate the infrared behavior of the \ac{SIGW} average energy-density spectrum $\bar{\omega}_{\uGW,\uin}$ in Eq.~\eqref{eq:omegabar-zeta} for different diagram families. 
For a diagram $X$ belonging to the $G$-like family, we approximate $\bar{\omega}_{\uGW,\uin}^{G-\mathrm{like}}$ and employ the first mean value theorem, yielding  
\begin{eqnarray}\label{eq:omegabar-G-IR}
    \bar{\omega}_{\uGW,\uin}^{X \in G-\mathrm{like}}\bigg|_\mathrm{IR} %(\eta_\uin,q) 
    &\simeq& \frac{\cF^X q^2 q_\ast^4}{24\pi} \int_{q_-/q_\ast}^{q_+/q_\ast} \ud v_q \int_{v_q - q/q_\ast}^{v_q + q/q_\ast} \ud u\, v_q u_q \frac{\Delta^2_{[\ca]} (v_q q_\ast)}{(v_q q_\ast)^3} \frac{\Delta^2_{[\cb]} (u_q q_\ast)}{(u_q q_\ast)^3} \overbar{J_\mathrm{IR}^2 (u_q,v_q,x)} \ ,\nonumber\\
    &\simeq& \cF^X \frac{q_+ - q_-}{12 \pi q_\ast} \left(\frac{q}{q_\ast}\right)^3 \frac{\Delta^2_{[\ca]} (v_q^\ast q_\ast)}{(v_q^\ast)^2} \frac{\Delta^2_{[\cb]} (u_q^\ast q_\ast)}{(u_q^\ast)^3} \overbar{J_\mathrm{IR}^2 (u_q^\ast,v_q^\ast,x)} \ ,
\end{eqnarray}
where $\cF^X$ is the product of the \ac{PNG} parameters multiplied by the symmetry factor of diagram $X$. 
It is reasonable to assume that both $v_q^*$ and $u_q^*$ lie within the interval $(q_-/q_\ast, q_+/q_\ast)$. 
Furthermore, by substituting Eq.~\eqref{eq:J-IR-propto} into Eq.~\eqref{eq:omegabar-G-IR}, we obtain  
\begin{equation}
    \bar{\omega}_{\uGW,\uin}^{X \in G-\mathrm{like}}\bigg|_\mathrm{IR}
    \propto \left(\frac{q}{q_\ast}\right)^3 \ln^2 \frac{4 q_\ast^2}{3 q^2} \ .
\end{equation}
Similarly, for a diagram $X$ in the $C$-like family,  
\begin{eqnarray}\label{eq:omegabar-C-IR}
    \bar{\omega}_{\uGW,\uin}^{X \in C-\mathrm{like}}\bigg|_\mathrm{IR} %(\eta_\uin,q) 
    &=& \frac{\cF^X q q_\ast^8}{96 \pi^2}
        \prod_{i=1}^2 \left[\int_{q_-/q_\ast}^{q_+/q_\ast} \ud v_{q,i} \int_{v_{q,i}-q/q_\ast}^{v_{q,i}+q/q_\ast} \ud u_{q,i}\, v_{q,i} u_{q,i} \right] \int_0^{2\pi} \ud \varphi_{12} \, \cos (2\varphi_{12})
        \frac{\Delta^2_{[\ca]} (v_{q,2} q_\ast)}{(v_{q,2} q_\ast)^3} \nonumber \\
        &&\hphantom{\frac{q^5}{3(2\pi)^3 \cH^2}} \times 
         \frac{\Delta^2_{[\cb]} (u_{q,2} q_\ast)}{(u_{q,2} q_\ast)^3} \frac{\Delta^2_{[\cb]} (w_{q,12} q_\ast)}{(w_{q,12} q_\ast)^3}
         \overbar{J_\mathrm{IR} (u_{q,1},v_{q,1},x) J_\mathrm{IR} (u_{q,2},v_{q,2},x)}\ ,\nonumber\\
    &\simeq& \frac{\cF^X}{12 \pi} \left(\frac{q_+ - q_-}{q_\ast}\right)^2       
        \left(\frac{q}{q_\ast}\right)^3 \cos (2\varphi_{12}^\ast)
        \frac{\Delta^2_{[\ca]} (v_{q,2}^\ast q_\ast)}{(v_{q,2}^\ast)^3} \frac{\Delta^2_{[\cb]} (u_{q,2}^\ast q_\ast)}{(u_{q,2}^\ast)^3} \frac{\Delta^2_{[\cb]} (w_{q,12}^\ast q_\ast)}{(w_{q,12}^\ast)^3} \nonumber \\
        &&\times \overbar{J_\mathrm{IR} (u_{q,1}^\ast,v_{q,1}^\ast,x) J_\mathrm{IR} (u_{q,2}^\ast,v_{q,2}^\ast,x)} \ .
\end{eqnarray}
By substituting Eq.~\eqref{eq:J-IR-propto} into this expression, we find  
\begin{equation}\label{eq:omegabar-C-IR-propto}
    \abs{\bar{\omega}_{\uGW,\uin}^{X \in C-\mathrm{like}}\bigg|_\mathrm{IR}}
    \propto \left(\frac{q}{q_\ast}\right)^3 \ln^2 \frac{4 q_\ast^2}{3 q^2} \ ,
\end{equation}
where the absolute value sign arises from the sign ambiguity of $\cos (2\varphi_{12}^*)$.
By the same method, we prove that $\bar{\omega}_{\uGW,\uin}^X\big|_\mathrm{IR}$ for diagrams in other families all share a common logarithmic slope, i.e.,
\begin{equation}\label{eq:ngw-infrared-X}
    n_\uGW^{X} (q) = \frac{\partial\ln \abs{\bar{\omega}_{\uGW,\uin}^{X} (q)}}{\partial\ln q} = 3 - \frac{4}{\ln^2 \left(\frac{4\nu_\ast^2}{3 \nu^2}\right)} \ .
\end{equation}
According to \cref{tab:order-FD}, we can sum $\bar{\omega}_{\uGW,\uin}^X$ over a certain category to obtain $\bar{\omega}_{\uGW,\uin}^{(a,b,c)}$. The above relation indicates that $\bar{\omega}_{\uGW,\uin}^{(a,b,c)}\big|_\mathrm{IR}$ in all categories exhibits the same behavior in the infrared region, i.e.,
\begin{equation}
    \bar{\omega}_{\uGW,\uin}^{(a,b,c)}\bigg|_\mathrm{IR}
    \propto \left(\frac{q}{q_\ast}\right)^3 \ln^2 \frac{4 q_\ast^2}{3 q^2} \ ,
\end{equation}
which is precisely Eq.~\eqref{eq:Omega-infrared} in Subsection~\ref{subsec:IR}.

\section{Present-day density contrast of SIGWs}\label{sec:Boltz}

In this appendix, we present the derivation of the present-day \ac{SIGW} energy density contrast $\delta_{\uGW,0}$, as described by Eq.~\eqref{eq:delta-0}. 
This contrast consists of two components: the initial inhomogeneities and the propagation effects. Specifically, the initial inhomogeneities comprise contributions from both \ac{PNG}-induced fluctuations and those induced by large-scale effects. 
Since the main text focuses on the \ac{PNG}-induced inhomogeneity, we first provide additional details regarding the initial inhomogeneity arising from large-scale effects. 
The graviton propagation, derived using the Boltzmann equation for gravitons, will be discussed later.

We begin by providing an overview of the initial inhomogeneity arising from non-adiabaticity. 
In this study, we consider \acp{SIGW} produced on subhorizon scales, and the \ac{GW} energy density derived from Eq.~\eqref{eq:rho-def} is indeed only valid on such scales. 
However, when examining the initial density contrasts along two observational directions, we must account for the \ac{GW} energy density modulated by large-scale perturbations.
Following the approach in Refs.~\cite{Mierna:2024pkh, ValbusaDallArmi:2023nqn,ValbusaDallArmi:2024hwm}, the modulated \ac{GW} energy density $\tilde{\rho}_\uGW$ is derived as follows. 
As shown in Eq.~\eqref{eq:S-L-dec}, we have separated the primordial curvature perturbations $\zeta$ into short- and long-wavelength modes. 
The short-wavelength mode is responsible for the production of \acp{SIGW}, while the long-wavelength mode modulates their energy-density distribution on large scales. 
Consequently, the linear scalar perturbations $\Phi$, which are proportional to $\zeta$ in Eq.~\eqref{eq:T-zeta-def}, can also be decomposed into these two modes as follows
\begin{equation}
    \Phi (\eta_\uin, \bx_\uin) = \Phi_\uS (\eta_\uin, \bx_\uin) + \Phi_\uL (\eta_\uin, \bx_\uin)\ .
\end{equation}
Notably, $\Phi_\uS (\eta_\uin, \bx_\uin)$ (or $\Phi_\uL (\eta_\uin, \bx_\uin)$) is related to $\zeta_\uS$ (or $\zeta_\uL$) in a manner analogous to the relationship between $\Phi(\eta_\uin, \bx_\uin)$ and $\zeta$ in Eq.~\eqref{eq:T-zeta-def}. 
Moreover, because the wavelength of $\Phi_\uL$ is much larger than the horizon scale, we neglect its spatial derivatives. 
In other words, $\Phi_\uL$ can be treated as independent of conformal time $\eta$ during \ac{SIGW} emission. 
Therefore, we can absorb $\Phi_\uL$ into the scale factor and coordinates in the metric given by Eq.~\eqref{metric} through the redefinition 
\begin{eqnarray}
    \tilde{a} &=& e^{-\Phi_\uL} a \ ,\nonumber\\
    \ud \tilde{\eta} &=& e^{2\Phi_\uL} \ud \eta \ ,\\
    \tilde{h}_{ij} &=& e^{2\Phi_\uL} h_{ij}\nonumber\ .
\end{eqnarray}
The metric from Eq.~\eqref{metric} can then be written as
\begin{equation}\label{metric-tilde} 
    \ud s^2 
    =  \tilde{a}^2 \left[ 
            - e^{2\Phi_\uS} \ud \tilde{\eta}^2 
            + \left( e^{-2\Phi_\uS}  \delta_{ij} + \frac{\tilde{h}_{ij}}{2} \right) \ud x^i \ud x^j 
    \right]\ .
\end{equation}
Thus, the large-scale effects are naturally incorporated if we evaluate \ac{SIGW} production under this metric. 
By substituting $\tilde{h}_{ij}$ into Eq.~\eqref{eq:rho-def}, the modulated energy density of \acp{SIGW} at the emission moment is given by 
\begin{equation}\label{eq:rho-tilde-def}
    \tilde{\rho}_\uGW (\eta_\uin,\bx_\uin) 
    = \frac{\mpl^2}{16 \tilde{a}^2} \overbar{\partial_{\tilde{\eta}_{\uin}} \tilde{h}_{ij}(\tilde{\eta}_{\uin},\bx_\uin) \partial_{\tilde{\eta}_{\uin}} \tilde{h}_{ij}(\tilde{\eta}_{\uin},\bx_\uin)}
    = \frac{\mpl^2 e^{2\Phi_\uL}}{16 a^2(\eta)} \overbar{\partial_{\eta_\uin} h_{ij}(\eta_\uin,\bx_\uin) \partial_{\eta_\uin} h_{ij}(\eta_\uin,\bx_\uin)}\ .
\end{equation}
We use a tilde above these quantities to indicate the inclusion of the non-adiabaticity. 
Using Eq.~\eqref{eq:rho-tilde-def}, the total density contrast can be obtained via Eqs.~(\ref{eq:omega-def}, \ref{eq:omega-delta}), respectively. 
By excluding the \ac{PNG}-induced inhomogeneity $\delta_{\uGW,\uin}$, we define the initial inhomogeneity arising from non-adiabaticity (NAD) as follows
\begin{equation}\label{eq:delta-NL}
    \delta_{\uGW,\uin}^\mathrm{NAD} 
    = \tilde{\delta}_{\uGW,\uin} - \delta_{\uGW,\uin}
    = e^{2\Phi_\uL} - 1 \simeq 2\Phi_\uL (\bx_\uin) \ .
\end{equation}
Additionally, large-scale effects exert no influence on the \ac{SIGW} average energy-density spectrum $\bar{\omega}_{\uGW,\uin}$. 
Since $\langle\Phi_\uL\rangle = 0$, $\bar{\omega}_{\uGW,\uin}$ remains unchanged when Eq.~\eqref{eq:rho-tilde-def} is used to derive it using Eqs.~(\ref{eq:omegabar-h}).

Next, we analyze the propagation effects using the Boltzmann equation for gravitons, following Refs.~\cite{Contaldi:2016koz,Bartolo:2019oiq,Bartolo:2019yeu}. 
The distribution function $f(\eta, \bx, \bq)$ for gravitons is introduced in connection with the expression for the energy density $\rho_\uGW(\eta, \bx)$, i.e.,
\begin{equation}\label{eqn:rho-f}
    \rho_\uGW (\eta,\bx) 
        = \frac{1}{a^4} \int \ud^3 \bq\, q f(\eta,\bx,\bq)\ .
\end{equation} 
Comparing with Eq.~\eqref{eq:omega-def}, the relation between $f(\eta, \bx, \bq)$ and $\omega_\uGW(\eta, \bx, \bq)$ is given directly by
\begin{equation}\label{eq:f-omega}
    f(\eta,\bx,\bq) 
        = \rho_\uc \left(\frac{a}{q}\right)^4 \omega_\uGW (\eta,\bx,\bq) \ .
\end{equation}
As we have decomposed $\omega_\uGW$ into the average component $\bar{\omega}_\uGW$ and the fluctuation $\delta_\uGW$ in Eq.~\eqref{eq:omega-delta}, we may similarly decompose $f(\eta, \bx, \bq)$ into a background component $\bar{f}(\eta, q)$ and perturbations $\Gamma(\eta, \bx, \bq)$, namely,
\begin{equation}\label{eq:Gamma-def}
    f(\eta,\bx,\bq)
        =\bar{f}(\eta,q)
        - q\frac{\partial \bar{f}}{\partial q}\Gamma(\eta,\bx,\bq)\ .
\end{equation}
By comparing with Eq.~\eqref{eq:omega-delta}, we obtain the relation between $\bar{f}$ and $\bar{\omega}_\uGW$, and between $\Gamma$ and $\delta_\uGW$. 
Further employing Eq.~\eqref{eq:O-o}, the former can be written as
\begin{equation}\label{eq:fbar-Omegabar}
    \bar{f}(\eta,q)
        = \frac{\rho_\uc}{4\pi} \left(\frac{a}{q}\right)^4 \bar{\Omega}_\uGW (\eta,q) \ . 
\end{equation}
Using this formula and Eq.~\eqref{eq:ngw-def}, one can express $\partial \bar{f} / \partial q$ in relation to $n_\uGW$. 
This allows $\Gamma(\eta, \bx, \bq)$ to be represented in terms of $\delta_\uGW(\eta, \bx, \bq)$, i.e.,
\begin{eqnarray}\label{eq:Gamma-delta}
    \Gamma (\eta,\bx,\bq)
        &=& \left(
            4-\frac{\partial \ln \bar{\Omega}_\uGW (\eta,q)}{\partial\ln q}
        \right)^{-1} \delta_\uGW (\eta,\bx,\bq)\nonumber\\
        &=& \left(4-n_{\uGW} (\nu)\right)^{-1} \delta_\uGW (\eta,\bx,\bq)\ .
\end{eqnarray}

The distribution function $f(\eta, \bx, \bq)$ for any particle species evolves according to the Boltzmann equation, i.e.,
\begin{equation}\label{eq:Boltzmann-def} 
    \frac{\ud f}{\ud \eta}=\mathcal{I}(f) + \mathcal{C}(f)\ ,
\end{equation}
where $\mathcal{I}$ denotes the emissivity term and $\mathcal{C}$ stands for the collision term. 
Collisions between gravitons are generally negligible, i.e., $\mathcal{C} = 0$ \cite{Contaldi:2016koz,Bartolo:2018igk,Flauger:2019cam}. 
Moreover, for \acp{GW} sourced by cosmological mechanisms, the emissivity term can be regarded as the initial condition, implying $\mathcal{I} = 0$ \cite{Bartolo:2019oiq,Bartolo:2019yeu}. 
Therefore, gravitons propagate freely once produced, and the Boltzmann equation simplifies to
\begin{equation}\label{eq:f-total-derivation}
    \frac{\ud f}{\ud \eta}
        = \frac{\partial f}{\partial \eta}
        +\frac{\partial f}{\partial x^i}\frac{\ud x^i}{\ud \eta}
        +\frac{\partial f}{\partial q}\frac{\ud q}{\ud \eta} 
        +\frac{\partial f}{\partial n^i}\frac{\ud n^i}{\ud \eta}=0\ .
\end{equation}
To linear order, the perturbed geodesics of massless gravitons yield the following relations
% \begin{subequations}
\begin{eqnarray}\label{geodesics}
    \frac{\ud x^i}{\ud \eta} = n^i \ ,\quad
    \frac{\ud q }{ \ud \eta} = \left(\partial_{\eta}\Phi - n^i \partial_i \Phi\right) q \ ,\quad
    \frac{\ud n^i }{ \ud \eta} = 0 \ .
\end{eqnarray}
% \end{subequations}
% These relations indicate that the gravitons propagate along a straight line and the direction remains unchange. 
% Moreover, the comoving momentum $q$ is also a constant at the background level, implying that the present-day frequency is $\nu = q / (2 \pi)$. 
By substituting Eq.~\eqref{eq:Gamma-def} and Eqs.~\eqref{geodesics} into Eq.~\eqref{eq:f-total-derivation}, we obtain the evolution equations for the background and perturbation terms individually, i.e.,
\begin{eqnarray}
    \partial_{\eta} \bar{f} & = & 0\ ,\label{eq:Boltzmannfbar}\\
    \partial_{\eta}\Gamma + n^i \partial_i \Gamma 
    & = & \partial_{\eta}\Phi - n^i\partial_i \Phi\ . \label{eq:Boltzmann-1st}
\end{eqnarray}
Eq.~(\ref{eq:Boltzmannfbar}) implies that the background part $\bar{f}$ remains invariant throughout cosmological evolution.
Regarding the perturbations, we can express Eq.~(\ref{eq:Boltzmann-1st}) in Fourier space, denoted as
\begin{equation}\label{eq:Boltzmann-k}
    \partial_{\eta}\Gamma + i k \mu \Gamma 
    = \partial_{\eta}\Phi - i k \mu \Phi\ , 
\end{equation}
where we introduce $\mu = \bk \cdot \bn / k$ as the cosine of the angle between $\bk$ and $\bn$.
The solution to this equation is given by \cite{Contaldi:2016koz,Bartolo:2019oiq,Bartolo:2019yeu}
\begin{eqnarray}\label{eq:Gamma-solution}
    \Gamma (\eta_0,\bk,\bq) 
    & = & e^{i k \mu (\eta_\uin - \eta_0)} 
        \left(
            \Gamma (\eta_\uin, \bk, \bq) + \Phi (\eta_\uin, \bk)
        \right)
        + 2 \int_{\eta_\uin}^{\eta_0} \ud \eta\,
            e^{i k \mu (\eta - \eta_0)}
                \partial_{\eta} \Phi (\eta,\bk)\ ,
\end{eqnarray}
where the monopole term $\Phi(\eta_0, \bk)$ has been neglected. 
This formula describes the perturbations in the graviton distribution in the present-day Universe.
By employing the line-of-sight condition $\bx_0 - \bx_\uin = (\eta_0 - \eta_\uin)\bn$, this result can be recast as the density contrast in real space as follows \cite{Li:2023qua}
\begin{eqnarray}\label{eq:deltaGW-CGW}
    \delta_{\uGW,0} (\bq) 
    &=& \tilde{\delta}_{\uGW,\uin} (\bq) + 
        \left(4 - n_{\uGW}(\nu)\right)\Phi (\eta_\uin, \bx_\uin)\nonumber\\ 
            && + 2 \left(4 - n_{\uGW}(\nu)\right) 
            \int \frac{\ud^3 \bk}{(2\pi)^{3/2}} e^{i\bk\cdot\bx_0}
            \int_{\eta_\uin}^{\eta_0} \ud \eta \,  
            e^{i k \mu (\eta - \eta_0)}
                \partial_{\eta} \Phi (\eta,\bk)\ .
\end{eqnarray}
The three terms on the right-hand side represent, from left to right, the initial density contrast, the \ac{SW} effect \cite{Sachs:1967er}, and the \ac{ISW} effect \cite{Sachs:1967er}. 
In addition, since the \ac{SW} and \ac{ISW} effects are generated by long-wavelength scalar modes that re-entered the Hubble horizon during matter domination, the \ac{ISW} effect is typically subdominant and can be neglected \cite{Bartolo:2019zvb,ValbusaDallArmi:2020ifo}.
Notably, we use $\tilde{\delta}_{\uGW,\uin}(\bq)$ rather than $\delta_{\uGW,\uin}(\bq)$ to represent the initial density contrast, as it more comprehensively incorporates the inhomogeneity arising from non-adiabaticity in Eq.~\eqref{eq:delta-NL}. 
By separating $\delta_\uGW(\eta_\uin, \bx_\uin, \bq)$ into $\delta_{\uGW,\uin}(\bq)$ and $\delta_{\uGW,\uin}^\mathrm{NAD}$ and substituting Eq.~\eqref{eq:delta-NL} into Eq.~\eqref{eq:deltaGW-CGW}, we have
\begin{eqnarray}
    \delta_{\uGW,0} (\bq) 
    &=& \delta_{\uGW,\uin} (\bq) + 
        \left(6 - n_{\uGW}(\nu)\right)\Phi_\uL (\bx_\uin)\ .
\end{eqnarray}
This result is precisely Eq.~\eqref{eq:delta-0}. 
Additionally, since we are interested in low multipoles, we ignore the gravitational lensing effects, which are expected to mainly affect $\delta_{\uGW,0}$ at higher multipoles \cite{Dodelson:2003ft}.

\section{Comparison with previous works}\label{sec:compare}

The energy-density fraction spectrum of \acp{SIGW} in the presence of local-type non-Gaussianity has attracted extensive interest in recent years, with many works \cite{Adshead:2021hnm,Ragavendra:2021qdu,Abe:2022xur,Yuan:2023ofl,Perna:2024ehx,Li:2023qua,Li:2023xtl,Ruiz:2024weh,Wang:2023ost,Yu:2023jrs,Iovino:2024sgs,Zeng:2024ovg,Cai:2018dig,Unal:2018yaa,Atal:2021jyo,Yuan:2020iwf,Chang:2023aba,Nakama:2016gzw,Garcia-Bellido:2017aan,Ragavendra:2020sop,Zhang:2021rqs,Lin:2021vwc,Chen:2022dqr,Cai:2019elf,Papanikolaou:2024kjb,He:2024luf,Rey:2024giu,Pi:2024jwt,Zhou:2025djn} dedicated to this topic. 
In this appendix, we compare our work with previous studies \cite{Unal:2018yaa,Atal:2021jyo,Adshead:2021hnm,Ragavendra:2021qdu,Abe:2022xur,Li:2023qua,Li:2023xtl,Perna:2024ehx,Ruiz:2024weh} that also employ a diagrammatic approach to compute $\Omega_{\uGW,0}$, highlighting the specific advances presented here. 

\begin{table}[htbp]
\centering
\begin{tabular}{c|c|c}
    \hline
    \multirow{2}{*}{References} & Basic & ``Renormalized'' \\
    & diagrammatic approach & diagrammatic approach \\
    \hline 
    Refs.~\cite{Unal:2018yaa,Atal:2021jyo,Adshead:2021hnm,Ragavendra:2021qdu,Li:2023qua,Li:2023xtl,Perna:2024ehx, Ruiz:2024weh} & \checkmark & $\times$ \\
    \hline
    \multirow{2}{*}{Ref.~\cite{Abe:2022xur}} & \multirow{2}{*}{\checkmark} &  Conceptual introduction without \\ & & concrete Feynman-like rules \\
    \hline
    \multirow{3}{*}{This work} & \multirow{3}{*}{\checkmark}
    & First systematic proposal \\ & & of concrete rules and \\ & & application to calculation \\
    \hline 
\end{tabular}
\caption{Comparison of the diagrammatic approaches utilized in Refs.~\cite{Unal:2018yaa,Atal:2021jyo,Adshead:2021hnm,Ragavendra:2021qdu,Abe:2022xur,Li:2023qua,Li:2023xtl,Perna:2024ehx,Ruiz:2024weh} and the present work. 
The ``basic diagrammatic approach'' denotes the framework established in Ref.~\cite{Adshead:2021hnm}.}\label{tab:comparison-approach}
\end{table}

To clarify the distinction between the ``basic'' and ``renormalized'' diagrammatic approaches, we first compare the methodologies employed in Refs.~\cite{Unal:2018yaa,Atal:2021jyo,Adshead:2021hnm,Ragavendra:2021qdu,Abe:2022xur,Li:2023qua,Li:2023xtl,Perna:2024ehx,Ruiz:2024weh} with the framework developed in this work, as summarized in Table~\ref{tab:comparison-approach}. 
The basic diagrammatic approach, first fully established in Ref.~\cite{Adshead:2021hnm}, can be extended to incorporate higher-order \ac{PNG} vertices. 
All studies listed in Table~\ref{tab:comparison-approach} utilize this approach to calculate the \ac{SIGW} energy-density fraction spectrum. 
In contrast, the ``renormalized'' diagrammatic approach—conceptually proposed in Ref.~\cite{Abe:2022xur} and fully developed herein—directly incorporates loop structures into the propagators and vertices, as detailed in Subsection~\ref{subsec:Renorm-FD}. 
Notably, while the approach in Ref.~\cite{Ruiz:2024weh} accounts for intrinsic non-Gaussianity and field renormalization \cite{Ballesteros:2024zdp} via inflaton self-interaction vertices, it remains an implementation of the basic approach and is distinct from the ``renormalized'' framework presented here. 
The latter provides a systematic alternative for treating multi-point correlators of $\zeta$ for power-series-expanded \ac{PNG} to arbitrary order, with a formulation that is independent of specific inflationary models.

While these works all employ the basic diagrammatic approach, we next compare the number of diagrams constructed via this methodology with the corresponding number of independent integrals required for numerical computation. 
The latter is often smaller than the former, because for scale-independent \ac{PNG}, contributions from diagrams containing ``self-closed loops'' are proportional to those without such loops and thus do not necessitate the formulation of new integrals. 
Since the Feynman-like rules utilized in Ref.~\cite{Ruiz:2024weh} differ substantially from the other works discussed here, and the rules in early works~\cite{Unal:2018yaa,Atal:2021jyo} are incomplete (specifically leading to a conflation of diagrams $C$ and $Z$ \cite{Adshead:2021hnm}), these studies are excluded from the following comparison, as illustrated in Table~\ref{tab:comparison-diagram}.
\begin{table}[htbp]
\centering
\begin{tabular}{c|c|c|c|c}
    \hline
    References & Refs.~\cite{Adshead:2021hnm,Ragavendra:2021qdu,Li:2023qua} & Refs.~\cite{Abe:2022xur, Perna:2024ehx} & Ref.~\cite{Li:2023xtl} & This work \\ \hline
    Diagram Numbers  & 7 & 24 & 49 & 236 \\
    \hline
    Involved Integrals & 7 & 13 &  27 & 81 \\
    \hline
\end{tabular}
\caption{
Number of diagrams and independent integrals in Refs.~\cite{Adshead:2021hnm,Ragavendra:2021qdu,Abe:2022xur,Li:2023qua,Li:2023xtl,Perna:2024ehx} and the present work. 
For Ref.~\cite{Perna:2024ehx}, diagrams with vanishing contributions to $\bar{\Omega}_\uGW$ (e.g., those nulled by azimuthal integration \cite{Adshead:2021hnm}) are omitted. 
While diagrams $Gl^H$, $Gl^C$, and $Gl^Z$ are treated as identical in Ref.~\cite{Perna:2024ehx} due to their equivalence in the scale-independent \ac{PNG} regime, they are counted here as three distinct diagrams to maintain consistent comparison.
}\label{tab:comparison-diagram}
\end{table}

In this comparison, it is important to note that Refs.~\cite{Perna:2024ehx, Abe:2022xur} and other studies adopt different computational approximations. 
The former retain contributions up to $\cO(A_\uS^4)$. 
In contrast, the latter retain all terms up to a specified \ac{PNG} order: specifically, up to order $\fnl$ in Refs.~\cite{Adshead:2021hnm,Ragavendra:2021qdu,Li:2023qua}, up to order $\gnl$ in Ref.~\cite{Li:2023xtl}, and up to order $\hnl$ in this work. 
Consequently, the diagrams appearing in Refs.~\cite{Perna:2024ehx, Abe:2022xur} are not strictly a subset of those in Ref.~\cite{Li:2023xtl} or the present work, as they include a diagram involving an $\i_\mathrm{NL}$ vertex. 
Nevertheless, the 13 independent integrals required in Refs.~\cite{Perna:2024ehx, Abe:2022xur} constitute a subset of the 27 integrals in Ref.~\cite{Li:2023xtl} and the 81 integrals required here.

The ``renormalized'' diagrammatic approach introduced in this work is essential for calculating high-order \ac{PNG} contributions to the \ac{SIGW} energy density. 
While the basic diagrammatic approach is theoretically general and extensible, its practical application becomes increasingly prohibitive at higher orders. 
For instance, computing all contributions up to order $\hnl$ requires the construction and evaluation of 236 distinct diagrams and their associated symmetry factors, which is a formidable computational task that grows exponentially for even higher orders. 
The ``renormalized'' framework presented here circumvents this difficulty. As illustrated in Fig.~\ref{fig:renor_all}, only 9 ``renormalized'' diagrams are required to fully determine the \ac{SIGW} energy-density spectrum. 
From these, all relevant integrals, including symmetry factors and powers of non-Gaussian parameters, can be derived directly. 
Consequently, the 236 basic diagrams listed in Appendix~\ref{sec:FD} are not utilized in the primary derivation, but are provided solely to verify the correspondence established by the ``renormalized'' approach. 
All integrals for contributions up to order $\hnl$ are expressed analytically and compactly by substituting Eqs.~(\ref{eq:G-like}, \ref{eq:C-like}, \ref{eq:Z-like}, \ref{eq:P-like}, \ref{eq:N-like}, \ref{eq:CZ-like}, \ref{eq:PZ-like}, \ref{eq:NC-like}, \ref{eq:PN-like}) with $o=4$. 
This formulation allows for a straightforward extension to higher \ac{PNG} orders by simply increasing $o$, an operation that requires minimal coding in \texttt{Mathematica} or \texttt{Python}. 
Similarly, for analyses restricted to a specific order in $A_\uS$ (e.g., Refs.~\cite{Perna:2024ehx, Abe:2022xur}), results up to $\mathcal{O}(A_\uS^N)$ can be efficiently generated by setting $o = N+1$ and truncating the series accordingly. 
In this framework, no further diagrammatic construction is required for \ac{PNG} up to arbitrary order.